# Chapter 134 :

# First-Principles Modeling of
# Ferroelectric Oxides Nanostructures

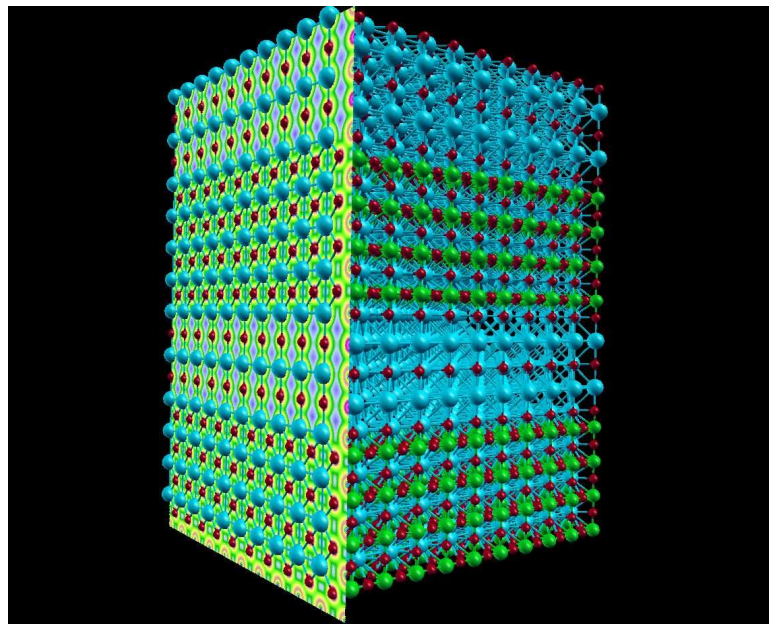

by

**Philippe GHOSEZ (Université de Liège, Belgium),**
and **Javier JUNQUERA (Universidad de Cantabria, Spain)**

Edited by M. Rieth and W. Schommers

*To Sophie and Julia,*
*to Arthur, Marie, and Javi.*

# Contents

# Introduction.

The aim of this Chapter is to provide an account of recent advances in the first-principles modeling of ferroelectric oxide nanostructures. Starting from a microscopic description of ferroelectricity in bulk materials and considering then, successively, different kinds of nanostructures (films, multilayers, wires, and particles), we try to identify the main trends and to provide a coherent picture of the role of finite size effects in ferroelectric oxides.

The last decades have seen a rapid evolution in the atomistic modeling of materials, driven both by the fast and recurrent increase of the computational power (hardware), and by important progresses in the development of more efficient algorithms (software). Nowadays, it is possible to describe very accurately the properties of materials using methods directly based on the fundamental laws of quantum mechanics and electrostatics. Even if the study of complex systems requires some practical approximations, these methods are free of empirically adjustable parameters. For this reason, they are referred to as "first-principles" or "*ab-initio*" techniques. Since 1990, ferroelectric oxides have been intensively studied from first-principles, and significant advances in the microscopic understanding of their properties have been achieved.

To be considered as ferroelectric, a material must satisfy two conditions [1]. First, it has to be a piezoelectric with two or more stable polarization states in the absence of an electric field. Second, it must be possible to switch from one to another of these states by applying a sufficiently large electric field (nevertheless smaller than the breakdown field). Ferroelectricity, first discovered in the Rochelle salt by Valasek [2], has been reported in different families of compounds, including hydrogen bonded systems such as KDP (potassium dihydrogen phosphate; $KH_2PO_4$), polymeric systems such as PVDF (poly-vinylidene fluoride; [-$CH_2$-$CF_2$-]$_n$), or the wide family of $ABO_3$ compounds. If selected first-principles calculations have been reported rencently on KDP [3, 4] and PVDF [5], the great majority of the theoretical works concern the $ABO_3$ family (mainly cubic perovskites but also trigonal [6] and hexagonal phases [7, 8]).

In this Chapter, we will only explicitly consider the class of $ABO_3$ compounds of perovskite structure. On the one hand, the simplicity of the perovskite structure makes of them ideal candidates for fundamental studies. On the other hand, they present unusually high functional properties (dielectric, piezoelectric, and electro-optic among others) and are important for numerous technological applications (sensors, transducers, memories, and optical devices) [9, 10].

A prototypical ferroelectric perovskite is barium titanate, $BaTiO_3$[1]. At high temperature, $BaTiO_3$ is paraelectric and has the high-symmetry cubic perovskite structure illustrated in Fig. 1.1(a): Ba atoms are located at the corners of the cubic unit-cell while the Ti atoms are at the center and are surrounded by an octahedra of O atoms, themselves located at the center of each face of the cube. Below 130 °C and still at room temperature, $BaTiO_3$ has a tetragonal structure in which the Ti sublattice and oxygen atoms have been shifted in opposite direction with respect to Ba atoms, taken as reference. These atomic shifts are accompanied with a small relaxation of the unit cell that becomes tetragonal and produce a stable polarization of 26 $\mu$C/cm$^2$. In the tetragonal phase, the cubic symmetry is broken, resulting in six symmetry equivalent variants with polarization along the [100], [010] and [001] directions. The two variants along [001] ("up" and "down" states) are represented in Fig. 1.1(a).

---

[1]$BaTiO_3$ crystallizes in two distinct polymorphic forms. One is the well known and most extensively investigated perovskite structure. The other is a hexagonal phase first observed by Megaw [11] in 1947, and characterized by Burbank and Evans [12]. Each polymorph undergoes its own sequence of phase transition. Both of them are ferroelectric [13], but exhibit different ferroelectric properties. All along this work, we will be only concerned with the perovskite structure.

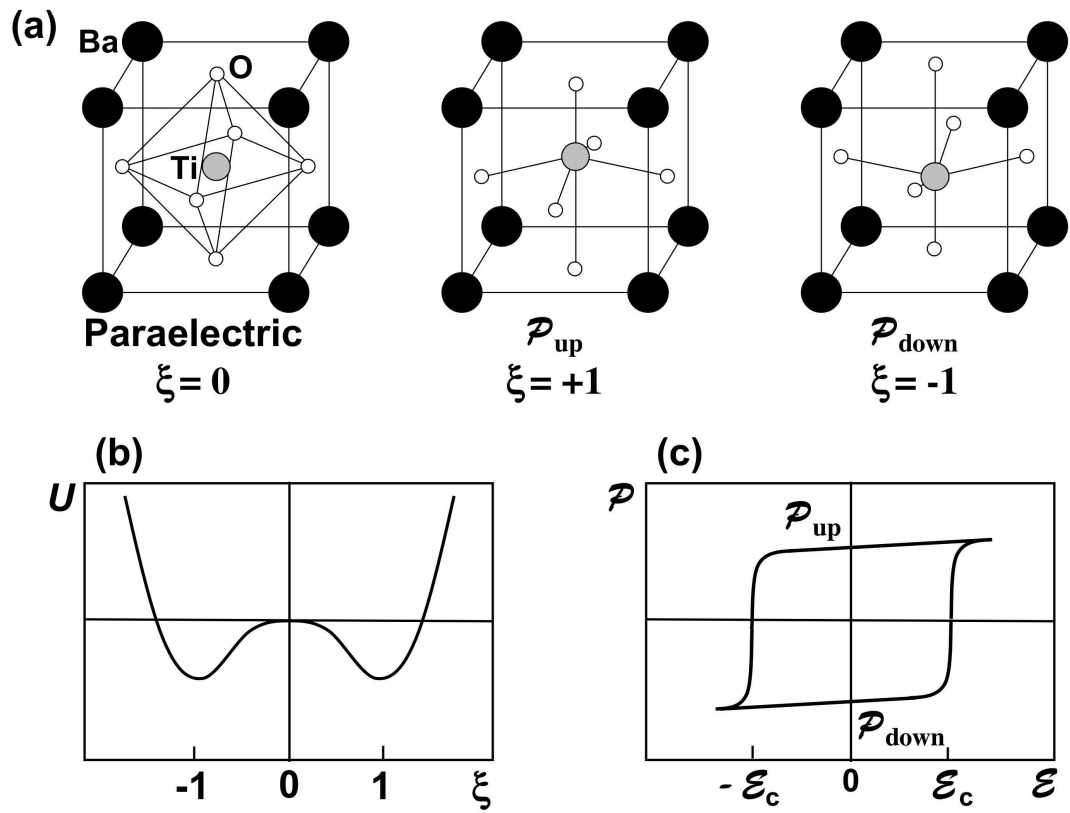

Figure 1.1: (a) Crystal structure of BaTiO$_3$ in its high temperature paraelectric cubic perovskite structure and in its room temperature tetragonal structure (for "up" and "down" polarization states). (b) Typical double-well shape for the internal energy of BaTiO$_3$ in terms of $\xi$ (see text). (c) Hysteretic behavior of the polarization-electric field curve.

All along this Chapter, the pattern of cooperative atomic displacements associated to the phase transition [2] will be referred to as $\xi$ so that paraelectric, up and down states respectively can be labeled as $\xi$ equal to 0, 1 and $-1$ respectively. For continuous evolution of $\xi$, the internal energy has a typical double-well shape as illustrated in Fig. 1.1(b). In BaTiO$_3$, the well depth is of the order of 30 meV/cell. What makes BaTiO$_3$ ferroelectric is that it is possible to switch from one polarization state (i.e. one double-well minimum) to the other by applying an electric field larger than the coercive field $\mathcal{E}_c$. Because of this switchability, the relationship between polarization and electric field in ferroelectrics is hysteretic, as illustrated in Fig. 1.1(c). For bulk ferroelectrics, the experimental coercive field is usually relatively modest, on the order of 10-100 kV/cm, and much smaller than what would be needed for uniform switching through the high symmetry structure ($\xi$=0) as it can be estimated from the double-well energy barrier. This arises from the fact that bulk ferroelectrics usually break into domains (to decrease the electrostatic energy) and that switching occurs through nucleation and growth of domains, which are processes with lower energy barriers [14].

Some ferroelectric perovskites have only one ferroelectric phase transition from the cubic perovskite structure to a ground-state structure of tetragonal symmetry, as presented above. This is the case of PbTiO$_3$ but it is not a general rule. For BaTiO$_3$, below room temperature, the structure evolves successively from tetragonal to orthorhombic (around 5°C) and finally rhombohedral (around -90°C) symmetry, with the polarization rotating from the [001] to the [011] and finally [111] directions. Also, not all of the ABO$_3$ perovskites are ferroelectric. Polar and non-polar instabilities compete in these systems producing various kinds of phase transitions (ferroelectric, antiferroelectric, antiferrodistortive) and a wide variety of ground-states. Some systems such as SrTiO$_3$ or KTaO$_3$ are nearly ferroelectric: precursor signs of a ferroelectric phase transition are observed at very low temperature (close to zero Kelvin) but the transition is suppressed by quantum fluctuations and never appears. These materials are referred to as incipient ferroelectrics.

Looking at classic textbooks [1, 15], ferroelectricity appears as a collective phenomenon, arising from the competition between short-range covalent repulsions and long-range Coulomb interactions. Going from bulk to the nanoscale, these interactions will be modified and it is therefore expected that the ferroelectric properties will be strongly affected. For many years, it was thought that ferroelectricity was suppressed at small sizes in thin films and fine particles. However, as experimental techniques improved to grow nanostructures with a control at the atomic scale and to probe locally their properties, it appeared that ferroelectricity might be preserved at very small sizes, even well below the initially expected limit. A timely and crucial question concerns therefore the possible existence (or the eventual absence) of a critical size for ferroelectricity. As it will become clear in this Chapter, the answer to this question is difficult and not unique. In fact, ferroelectricity is strongly affected by mechanical, electrical and chemical boundary conditions. For this reason, a ferroelectric object cannot be considered in isolation and the previous question must be answered individually for each specific system, considering the ferroelectric and its environment (interfaces, electrodes, substrate) as a whole.

In this Chapter, we try to cover various aspects of the ferroelectricity in perovskite oxide nanostructures. Nevertheless, we restrict ourselves to the fundamental questions that are accessible and have been addressed from first-principles. So, important questions for technological applications such as switching, finite conductivity or fatigue will not discussed here. They have however been recently addressed in independent reviews (see, for instance, Ref. [16] and references therein). Also, in spite of the fact that it is closely related and was the subject of recent first-principles studies, the emerging field of magneto-electric and multiferroic nanostructures (see for instance Ref. [17, 18, 19] and references therein) has not been included in this Chapter.

---

[2]In the next sections, depending of the context, $\xi$ will refer either to the pattern of atomic displacements between paraelectric and ferroelectric phases, as explained here, or to the eigenvector of the unstable mode in the cubic phase. In BaTiO$_3$, both definitions are almost equivalent, with an overlap for $\xi$ between these two choices close to 100%.

In Sec. 2, we provide a summary of the main first-principles techniques that have been recently applied to ferroelectrics. In Sec. 3, we provide a microsocsopic picture of the ferroelectric instability at the bulk level, introducing key quantities for the understanding of nanostructures. In Sec. 4, we discuss the case of ultrathin films. In Sec. 5, we consider ferroelectric multilayers. In Sec. 6 we discuss some issues related to the inclusion of ferroelectric oxides within the Si technology. In Sec. 7, we report a few studies of ferroelectric nanoparticles and nanowires. In all these Sections, the electrical and mechanical boundary conditions will be seen as the main factors affecting the ferroelectric properties.

# Theoretical background.

## 2.1 Overview.

Numerous properties of materials can now be determined directly from first-principles, providing new insights into critical problems in physics, chemistry and materials sciences. Amongst the first-principles approaches, density functional theory (DFT) [20, 21] has become a reference and is one of the most widely used method (for reviews see [22, 23, 24, 25, 26]). During the last decade, it has been successfully applied to various ferroelectric oxides, yielding several breakthroughs in the understanding of the behavior of bulk crystals [27, 28, 29] and more recently nanostructures [30, 31, 32, 33, 34, 35, 36]. The successes achieved by DFT methods in the field of ferroelectrics since the early 1990's rely on different theoretical advances.

The major advance that promoted the use of DFT for ferroelectrics is the *modern theory of the polarization* [37, 38, 39]. The electric polarization is a central quantity in the study of ferroelectric materials. Incredible as it may seem, the formulation of a proper quantum mechanical approach for the calculation of the electronic polarization in periodic solids has remained a tricky and challenging problem until the early 1990s. The modern theory of the polarization provided an elegant solution to this problem and associated the polarization of continuous periodic electronic charge densities to a Berry phase of the Bloch functions [37]. This technique is now routinely applied to access various quantities such as the spontaneous polarization, the Born effective charges or the piezoelectric tensors. It also allowed subsequent theoretical advances: it was at the origin of the theory of electron localization localization [40, 41, 42, 43, 44] and opened the door to the study of crystals under finite electric field [45, 46]. The Berry phase method is not restricted to DFT and was also implemented in the Hartree-Fock approach [47, 48, 49].

A second important advance concerns the development of a technique to compute *maximally localized Wannier functions* [50, 51, 52, 53]. On the one hand, Wannier functions provide an insightful picture of the nature of the chemical bonds in solid that is missing in the Bloch picture. On the other hand, they offer a physically appealing interpretation of the modern theory of the polarization and of the theory of electron localization. Wannier functions can also be used as basis functions in order-N methods [26] or for the construction of effective Hamiltonians, allowing to study the transport properties of nanostructures [54].

A third significant advance consists in the recurrent progresses (in particular concerning the response to electric fields) of *density functional perturbation theory* (DFPT) [55, 56, 57, 58, 59, 60, 61, 62, 63, 64], a merge of DFT and perturbation theory. Nowadays, this formalism provides systematic [65] and easy access to an increasing number of important functional properties such as phonon frequencies, infra-red and Raman spectra, elastic constants, optical and static dielectric constants, proper piezoelectric tensor, non-linear optical susceptibilities and electro-optic coefficients among others. It constitutes therefore a tool of choice for the characterization of the various functional properties of ferroelectrics. Even if most quantities can be alternatively accessed by finite difference techniques, it provides a coherent and systematic approach, particularly suited for the study of complex properties or for the efficient fitting of effective Hamiltonians as it is discussed below.

These recent theoretical developments made possible the microscopic and first-principles modeling of ferroelectrics. DFT was successfully applied to various $ABO_3$ compounds [66, 27, 29]. However, it has also its limitations. The most important one is the limited number of atoms that can be handled in a first-principles calculation, considering explicitly the electronic degrees of freedom. Currently, this limit is of the order of hundred atoms (eventually few hundreds atoms for basic ground-state calculations). This imposes serious restrictions to the applicability of pure DFT calculations for the study of ferroelectrics. First, many technologically important ferroelectrics are complex solid solutions such as $Ba_{1-x}Sr_xTiO_3$ (BST) which emerges as a leading candidate dielectric material for the memory cell capacitor in dynamical random access memories (DRAMs), $PbZr_{1-x}Ti_xO_3$ (PZT) which is widely used in transducers and actuators, or the new class of relaxor ferroelectrics (i.e. $PbMg_{1/3}Nb_{2/3}O_3$-$PbTiO_3$) with remarkably large piezoelectric constants. In spite of the rapid increase of computational power, these complex materials are not directly accessible to DFT or their study is limited to few artificial ordered supercells. Second, ferroelectrics exhibit structural phase transitions and their properties strongly evolves with temperature. Meaningful predictions and direct comparison with the experiment requires therefore to estimate the properties at finite temperature which is nowadays unaffordable within DFT, mainly because random thermal vibrations cannot be describe properly with too small simulation box.

In this context, a significant breakthrough consisted in the development of a microscopic effective Hamiltonian approach for ferroelectrics. Initially designed by Rabe and Joannopoulos for GeTe [67, 68, 69], the technique was generalized and first applied to ferroelectric perovskites by Zhong, Vanderbilt and Rabe [70, 71]. In this approach the soft mode is considered as the driving force for the phase transitions. The Hamiltonian is constructed from a Taylor expansion of the energy around the paraelectric phase in terms of the soft mode ionic degree of freedom and the strains. All parameters that appear in the expansion are determined from DFT total energy and linear-response calculations. The effective Hamiltonian makes it possible to study the structural phase transitions of ferroelectrics [71, 72] and the temperature dependence of their properties (dielectric [73, 74], piezoelectric [75], and optical [76]). It provides also access to complex disordered solid solutions and to nanostructures [77]. We note that alternative methods have been reported for the study of ferroelectrics: shell-model calculations fitted on first-principles results similarly allow to access to their finite temperature properties with a good accuracy [78] and a phenomenological model based on the chemical rules obtained from DFT allowed to reproduce the behavior of PZT structures [79]. However, nowadays, the effective Hamiltonian remains the most popular and most widely used approach.

The remaining part of this Section is devoted to a rapid survey of the recent theoretical advances highlighted above. A comprehensive explanation of density functional theory and related methods might be found in the excellent book by Richard M. Martin [26].

## 2.2   Density functional theory.

The description of macroscopic solids from first-principles is based on the determination of the quantum mechanical ground-state associated to their constituting electrons and nuclei. It consists in calculating the quantum mechanical total energy of the system and in the subsequent minimization of that energy with respect to the electronic and nuclear coordinates (variational principle). This defines a complex many-body problem of interacting particles. It is unaffordable in practice and requires some approximations.

A first natural simplification arises from the large difference in mass between electrons and nuclei, allowing separate treatment of their dynamics using the so-called Born-Oppenheimer approximation. It reduces the many-body problem to the study of interacting electrons in some frozen-in configuration of the nuclei whose positions $\mathbf{R}_\kappa$ are considered as parameters [1].

In spite of this first simplification, the problem remains complex because of the electron-electron interaction. Density Functional Theory (DFT), proposed in the 1960's by Hohenberg and Kohn [20] and

---

[1] All along this chapter we will use atomic units ($\left| e^- \right| = \hbar = m_{e^-} = 1$). Greek subindices refer to the nuclei whereas latin subindices refer to the electrons.

Kohn and Sham [21], provided a simple method for describing the effects of electron-electron interactions. Hohenberg and Kohn [20] first proved that the total energy of an electron gas is a unique functional of the electronic density. This means that instead of seeking directly for the complex many-body wave function of the system, we can adopt an intrinsically different point of view and consider the electronic density $n(\mathbf{r})$, a simple scalar function of position, as the fundamental quantity of the problem. The minimum value of the total energy density functional is the ground-state energy of the system and the density yielding this minimum value is the exact ground-state density.

The theorem of Hohenberg and Kohn demonstrated the *existence* of such a functional but did not provide any clue on its explicit form. Their finding would remain a minor curiosity today if it were not for the ansatz made by Kohn and Sham [21] which has provided a way to make useful, approximate functionals for real systems. Kohn and Sham showed that it is possible to map the interacting electronic system onto another system of non-interacting particles moving in an external potential, with the *same* exact ground-state electronic density. In this context, the electronic density can be obtained from one-body wave functions $\psi_i$, self-consistent solution of a set of one-particle equations describing the behavior of an electron in an effective potential. Again, the form of this potential is *a priori* unknown but, as it will be discussed below, can be efficiently approximated.

## 2.2.1 Kohn-Sham energy functional.

In practical DFT studies, the Born-Oppenheimer total energy of the system (electrons + ions) is decomposed as

$$E_{e+i}[\mathbf{R}_\kappa, \psi_i] = E_{ion}[\mathbf{R}_\kappa] + E_{el}[\mathbf{R}_\kappa, \psi_i]. \qquad (2.1)$$

The potential energy of the ions, $E_{ion}$, can be trivially deduced from the ionic positions. Within the one-particle Kohn-Sham framework, the electronic energy $E_{el}$ for a set of doubly occupied states [2] $\psi_i$ can be written as

$$\begin{aligned} E_{el}[\mathbf{R}_\kappa, \psi_i] &= \sum_i^{occ} \left\langle \psi_i \left| -\frac{1}{2}\nabla^2 \right| \psi_i \right\rangle + \int v_{ext}(\mathbf{r}) \, n(\mathbf{r}) \, d\mathbf{r} \\ &+ \frac{1}{2} \int \frac{n(\mathbf{r}_1) \, n(\mathbf{r}_2)}{|\mathbf{r}_1 - \mathbf{r}_2|} \, d\mathbf{r}_1 d\mathbf{r}_2 + E_{xc}[n], \end{aligned} \qquad (2.2)$$

where the successive terms represents, respectively, the electronic kinetic energy, the interaction between the electrons and the static electron-ion potential $v_{ext}$, the self-interaction Coulomb energy of the electron density treated as a classical charge density (Hartree energy $E_H[n]$), and the exchange-correlation energy $E_{xc}[n]$, that contains all the electron-electron interactions going beyond the classical Coulomb term. The electronic density, $n(\mathbf{r})$, is given by

$$n(\mathbf{r}) = \sum_i^{occ} \psi_i^*(\mathbf{r}) \cdot \psi_i(\mathbf{r}). \qquad (2.3)$$

For a given set of atomic positions $\mathbf{R}_\kappa$, the ground-state is obtained by minimizing Eq. (2.2) under the following orthonormalization constraints

$$\langle \psi_i | \psi_j \rangle = \delta_{ij}.$$

This gives access to the total energy of the system and the associated ground-state electronic density.

In practice, the minimization of Eq. (2.2) with the previous constraint can be achieved using the Lagrange multiplier method. The problem turns into the minimization of

$$F_{e+i}[\mathbf{R}_\kappa, \psi_i] = E_{e+i}[\mathbf{R}_\kappa, \psi_i] - \sum_{i,j}^{occ} \Lambda_{ji} \left( \langle \psi_i | \psi_j \rangle - \delta_{ij} \right), \qquad (2.4)$$

---

[2]Along the rest of the chapter we will assume spin-degeneracy

where $\Lambda_{ji}$ are the Lagrange multipliers. The corresponding Euler-Lagrange equation is

$$H_{el}\left|\psi_i\right\rangle = \sum_j \Lambda_{ji}\left|\psi_j\right\rangle, \qquad (2.5)$$

where the Hamiltonian is

$$H_{el} = -\frac{1}{2}\nabla^2 + v_{\text{ext}}(\mathbf{r}) + v_{\text{H}}(\mathbf{r}) + v_{\text{xc}}(\mathbf{r}) \qquad (2.6)$$

with the Hartree and exchange-correlation potentials, respectively, defined as the functional derivative of the Hartree and exchange-correlation energy with respect the density,

$$v_H(\mathbf{r}) = \frac{\delta E_H[n]}{\delta n(\mathbf{r})}, \qquad (2.7)$$

and

$$v_{xc}(\mathbf{r}) = \frac{\delta E_{xc}[n]}{\delta n(\mathbf{r})}. \qquad (2.8)$$

The solution of Eq. (2.5) is not unique. In fact, we can always apply a unitary transformation $U$ to the wave functions of the occupied states

$$\left|\psi_i\right\rangle \rightarrow \sum_j^{occ} U_{ji}\left|\psi_j\right\rangle \qquad (2.9)$$

without affecting neither the energy nor the density. Such a transformation is called a *gauge transformation*. Since the Hamiltonian is a hermitian operator, it is always possible to work within the so-called *diagonal gauge* where the Lagrange multiplier matrix is diagonal

$$\Lambda_{ij} = \left\langle\psi_i\left|H_{el}\right|\psi_j\right\rangle = \epsilon_j\delta_{ji}. \qquad (2.10)$$

Within this diagonal gauge, the minimization of Eq. (2.2) is equivalent to solve self-consistently the following set of Kohn-Sham equations

$$\begin{cases} \left[-\frac{1}{2}\nabla^2 + v_s\right]\left|\psi_i\right\rangle = \epsilon_i\left|\psi_i\right\rangle, \\[2mm] v_s(\mathbf{r}) = v_{\text{ext}}(\mathbf{r}) + \int \frac{n(\mathbf{r_1})}{|\mathbf{r_1}-\mathbf{r}|}\,d\mathbf{r_1} + \frac{\delta E_{xc}[n]}{\delta n(\mathbf{r})}, \\[2mm] n(\mathbf{r}) = \sum_i^{occ}\psi_i^*(\mathbf{r})\cdot\psi_i(\mathbf{r}). \end{cases} \qquad (2.11)$$

The above-described Kohn-Sham approach consists in a mapping of the interacting many-electron system onto a system of non-interacting fictitious particles moving in an effective potential $v_s$ produced by the ions and all the other electrons. If the exchange-correlation energy functional were known exactly, taking the functional derivative with respect to the density would provide an exchange-correlation potential that included the effects of exchange and correlation exactly. In practice, however, the form of $E_{xc}[n]$ is unknown and must be approximated.

### 2.2.2 Usual approximate functionals.

The exchange-correlation energy, $E_{xc}[n]$, is expected to be a universal *functional* of the density *everywhere*. However, Hohenberg and Kohn theorem [20] provides some motivation for using approximate methods to describe the exchange-correlation energy as a simple *function* of the electron density.

The first, and most widely used approach in this sense is the Local Density Approximation (LDA) [21]. It assumes (i) that the exchange-correlation energy per particle at point $\mathbf{r}$, $\epsilon_{xc}(\mathbf{r})$, only depends on the density at this point, and (ii) that it is equal to the exchange-correlation energy per particle of a homogeneous electron gas of density $n(\mathbf{r})$ in a neutralizing background

$$E_{xc}[n] = \int n(\mathbf{r}) \, \cdot \, \epsilon_{xc}^{LDA}(\mathbf{r}) \ d\mathbf{r}, \tag{2.12}$$

with

$$\epsilon_{xc}^{LDA}(\mathbf{r}) = \epsilon_{xc}^{hom}[n(\mathbf{r})]. \tag{2.13}$$

The form of $\epsilon_{xc}^{hom}[n]$ used in the calculation may be borrowed from various sources. The *exchange* part can be obtained analytically from the Hartree-Fock technique. It can be shown that it scales like

$$\epsilon_{x}^{hom}[n] = -\frac{3}{4\pi}(3\pi^2)^{1/3} \ n^{1/3}. \tag{2.14}$$

For the *correlation* part, one may rely on accurate values obtained by Ceperley-Alder [80] from Monte-Carlo simulations of the energy of the homogeneous electron gas. Various formulations are available (Wigner, X-alpha, Gunnarson-Lundqvist, Perdew-Zunger, Perdew-Wang, Teter ...) that are referred to as local density approximations. They rely on the same exchange part but consider slightly different treatments of the correlation term.

The LDA is probably one of the crudest approximation that can be done. It has however the advantage of the simplicity. Moreover, in many cases, it allows to describe structural and dynamical properties with surprising accuracy [23, 22] [3] : atomic positions and lattice constants reproduces the experiment within $\approx 1$ %; phonon frequencies are usually obtained within $\approx 5$ %. Well known exceptions are the bandgap, the cohesive energy and the optical susceptibilities.

An accuracy of 1% on the lattice constant might be considered as a success for a method without any adjustable parameter. However, it appears in some cases as a serious limitation. In perovskite oxides the ferroelectric instability is known to be very sensitive to pressure, and thus to the lattice constant. In this context, an error of 1% can have dramatic consequences (see for instance Fig. 3.1 in Sec. 3.2). As it will be discussed more extensively later, to avoid this problem, it has been accepted in some circumstances to fix the lattice parameters to their experimental values, when they are known [4].

Different techniques are going beyond the LDA. A first alternative, but connected approach, is to build a "semi-local" functional that does not only depend on the density at $\mathbf{r}$ but also on its gradient, or on higher order gradient expansion. Different forms have been proposed that are summarized under the label of Generalized Gradient Approximations (GGA). They are based on a functional of the type [81, 82]

$$E_{xc}^{GGA}[n] = \int n(\mathbf{r}) \, . \, \epsilon_{xc}^{GGA} \left[ n(\mathbf{r}); |\nabla n(\mathbf{r})|; \nabla^2 n(\mathbf{r}) \right] \ d\mathbf{r}. \tag{2.15}$$

This kind of approximation improves the computed value of the cohesive energy. It can also improve the description of bond lengths and lattice parameters even if the gradient correction usually overcorrects the LDA [83, 84] yielding longer values than the experimental ones and is therefore of little help for ferroelectrics. Also, the correction has a rather limited effect on the dielectric constant [83]. This is due to the fact that GGA remains a quasi-local approximation that cannot include any long-range density dependency of $E_{xc}[n]$ [85, 86].

Different other functionals also exist like the average density approximation (ADA) [87] or the weighted density approximation (WDA) [87]. It was sometimes argued that WDA should be intrinsically unable to improve LDA results [88]. However, for $ABO_3$ compounds, it was identified as the most promising alternative to the LDA [89] that might allow to avoid the typical volume underestimate, problematic for a quantitative description of the ferroelectric instability. Up to now, it was however only marginally applied.

---

[3]The LDA exchange-correlation hole integrates to $-1$. This simple feature should be a first intuitive argument to explain its success.

[4]This is equivalent to apply a negative pressure in the calculation, that compensates for the LDA error.

As an alternative to approximate DFT, the Hartree-Fock method can also be consider. It has been applied to ferroelctrics for bulk and thin films geomerty [47, 48, 49]. Without being exhaustive, let us finally mention that another interesting scheme consists in a mixing of Hartree-Fock and local density functionals as justified from the adiabatic connection formula [90]. The use of so-called *hybrid functionals* is quite popular in quantum chemistry but was only marginally applied to ferroelectrics [91].

A comparison of the results obtained with different functionals for ferroelectric perovskite oxides can be found in Ref. [92].

### 2.2.3 Periodic solids.

In the study of periodic solids, such as the ferroelectric crystals considered here, it is usual to consider Born-von Karman periodic boundary conditions. For such infinite periodic systems, the electronic wave functions $\psi_i(\mathbf{r})$ are Bloch functions characterized by their wave vector $\mathbf{k}$ and a band index $n$

$$\psi_{n,\mathbf{k}}(\mathbf{r}) = e^{i\mathbf{k}\cdot\mathbf{r}} u_{n,\mathbf{k}}(\mathbf{r}), \qquad (2.16)$$

where $u_{n,\mathbf{k}}(\mathbf{r})$ is a periodic function that has the same periodicity as the crystal lattice. Different methods are available for accurate representation of the electronic wave-function and calculations on ferroelectric oxides have been reported using virtually all of them: all-electrons (LAPW and FLAPW) [93, 94], norm-conserving [95] and ultra-soft pseudopotentials [66] with plane-waves [66, 95], a local basis set [96], or projector-augmented wavefunction method [97] among others.

Periodic boundary conditions are well suited for the study of bulk materials and superlattices (Fig. 2.1). The computational cost is nevertheless directly related to the number of atoms in the primitive unit cell, nowadays limited to about 80-100 atoms. For perovskite oxide alloys or superlattices, this limits in practice the study to unit cells smaller than 15-20 $ABO_3$ basic units.

Periodic boundary conditions are less appropriate for the study of individual nano-objects (isolated thin films, nanowires, nano-particles) for which they require the use of a *supercell* technique: the nano-object is embedded into a supercell in which it is surrounded by a vacuum region sufficiently large to avoid artificial interactions with its periodic images (see Fig. 2.1). For such isolated systems, the computational cost becomes rapidly prohibitive. Again, it scales with the number of atoms in the supercell that, for instance, grows as the cube of the radius in the case of nanoparticles. Also, the treatment of the vacuum region, free when expanding the wave-function in a local basis set, is restrictive when using plane-waves.

Finally, the use of periodic boundary conditions can generate some additional difficulties for isolated finite systems with a polarization perpendicular to the surface, whose discussion will be postponed to Sec. 4.4.

### 2.2.4 Density functional perturbation theory.

Many properties of materials are related to the variations, induced by various perturbations, of the total energy $E_{e+i}$ (or a related thermodynamical potential, see Sec. 2.2.8) around the equilibrium configuration. DFT provides access to the total energy of the system and offers therefore the possibility of calculating its derivatives with respect to perturbations (mainly atomic displacements, electric fields and strains). This can be done by carrying out full calculations when freezing into the structure finite values of the perturbation potential and extracting derivatives from a finite difference formula. However, it is also possible to extract derivatives directly using perturbation theory in a convenient and systematic way. Density functional pertubation theory (DFPT) was first reported by Baroni, Giannozzi and Testa [55] for the linear response to atomic displacements and then generalized to various other perturbations [56, 57, 58]. A different algorithm, based on a variational principle, was then proposed by Gonze, Allan and Teter [98], providing more accurate expressions for the energy derivatives, and giving also access to non-linear responses thanks to the $(2n+1)$ theorem [98].

Let us consider a small external perturbation, characterized by the parameter $\lambda$. We can expand perturbed quantities $X$ (where $X$ stands for the energy $E_{e+i}$, the wave function $\psi_\alpha(\mathbf{r})$, the charge density $n(\mathbf{r})$, the hamiltonian $H$, or anything else) in terms of $\lambda$ around their unperturbed values $X^{(0)}$

$$X(\lambda) = X^{(0)} + X^{(1)}\lambda + X^{(2)}\lambda^2 + X^{(3)}\lambda^3 ... \tag{2.17}$$

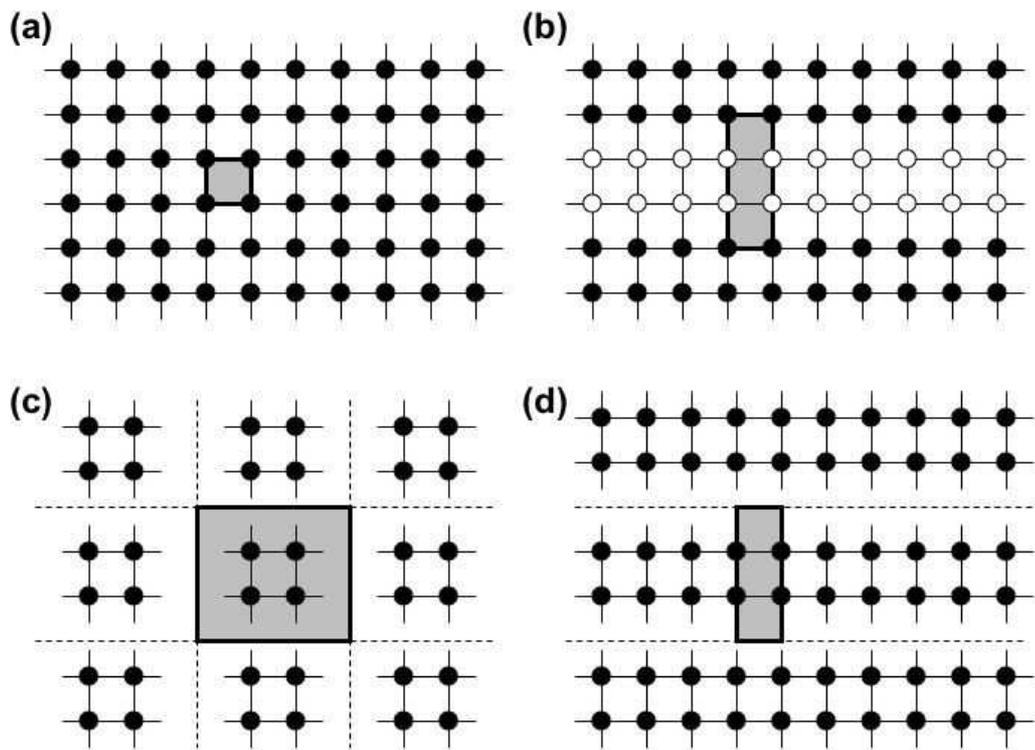

Figure 2.1: Schematic view of the the unit cell for a bulk crystal (a), a superlattice (b), an isolated nanoparticle (c) and an isolated slab (d).

Because $E_{e+i}$ satisfies a variational principles, it is possible to derive two major theorems [60, 61].

First, a variational principle can be established for even order of perturbations. It states that the $n^{th}$-order derivatives of the wave functions, $\psi_\alpha^{(n)}$, can be obtained by minimizing a variational expression of $E_{e+i}^{(2n)}$. At its minimum, this expression provides both the value of $E_{e+i}^{(2n)}$ and related $\psi_\alpha^{(n)}$. From the computational point of view, this formulation is valuable since the minimization of $E_{e+i}^{(2n)}$ in terms of $\psi_\alpha^{(n)}$ can be achieved using the same minimization algorithm as for the ground-state. Alternatively, applying pertubation theory to the Kohn-Sham equations [Eq. (2.11)] as done in the seminal work of Baroni *et al.*, provides a set of so-called Sternheimer equations that also give access to $\psi_\alpha^{(n)}$ when solved self-consistently. Both approaches are commonly used.

Second, a "$2n+1$" theorem can be demonstrated for the odd order of perturbation. It states that the derivatives of the energy up to order $2n+1$ can be computed from the derivatives of the wave functions to all order 0 through $n$. This means that first-order energy derivatives such as forces and stresses can be directly deduced from the ground-state wavefunctions [5]. It means also that the additional computation of the first-order wavefunctions only is sufficient to provide access to the whole set of second and third derivatives of the energy and therefore to numerous linear and non-linear properties of the crystal, as described in Sec. 2.2.8.

A review of DFPT by Baroni *et al.* is provided in Ref. [59]. Explicit expressions of even- and odd-order energy derivatives can be found in Ref. [62, 63, 64].

## 2.2.5 The modern theory of polarization.

The spontaneous polarization of a ferroelectric crystal is related to a first-derivative of its energy with respect to an electric field. From the previous discussion, it should be therefore directly accessible from the ground-state wave-functions, in a way similar to atomic forces and stresses. The formulation is, however, less trivial. Part of the problem is that the perturbation created by a homogeneous electric field corresponds to a linear potential, incompatible with periodic boundary conditions.

Let us first consider a finite piece of matter of volume $\Omega$. The polarization of this system $\mathcal{P}$ [6] is well defined and corresponds to the dipole moment per unit volume

$$\mathcal{P} = \frac{e}{\Omega}\left[\sum_\kappa Z_\kappa \mathbf{R}_\kappa - \int_\Omega \mathbf{r} n(\mathbf{r}) d\mathbf{r}\right]. \tag{2.18}$$

where $Z_\kappa$ and $\mathbf{R}_\kappa$ are atomic numbers and positions of atom $\kappa$, $e$ is the absolute value of the electronic charge and $n(\mathbf{r})$ the electronic charge density defined in Eq. (2.3).

In infinite periodic systems, the previous formula is no more valid and suffers from the difficulty that the factor $\mathbf{r}$ is unbounded. Many textbooks relate the polarization of a periodic crystal to the dipole moment of its unit-cell of volume $\Omega_0$. However, it is easy to see that such a quantity is ill-defined and strongly depends on an arbitrary choice of the unit-cell. The problem is illustrated in Fig. 2.2 for a simple model of localized charges. It is even more stringent for delocalized electronic densities where the polarization evolves continuously when translating smoothly the unit cell. The correct choice of the unit cell only becomes clear when specifying the surface of the sample. For infinite solids with periodic boundary conditions, there is no surfaces any more and it was not clear during many years that the polarization was a well-defined bulk quantity.

The modern theory of polarization solved this problem [37, 38, 39]. It shows that it is impossible to find the polarization or changes of polarization of infinite periodic systems simply from the periodic charge density. Instead, *changes of polarization* between two crystal states must be related to the polarization current going through the bulk during the transformation from one state to the other and can be formulated as a Berry phase of the Bloch functions.

---

[5] For the specific case $n = 0$, the $2n + 1$ theorem is know as the Hellmann-Feynman theorem.

[6] The polarization vector will be denoted by $\mathcal{P}$ in this chapter. A subindex will be added when referring to a cartesian component of the vector.

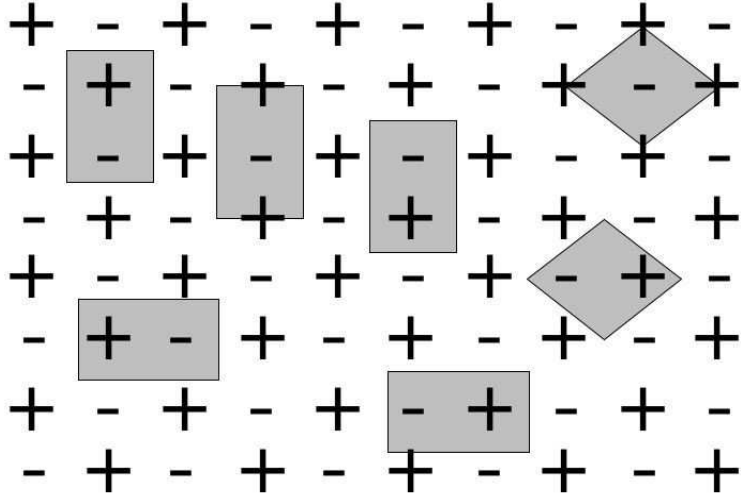

Figure 2.2: Two dimensional system of classical point charges denoted by + and - signs. Depending on the choice of the unit cell, the dipole moment of the unit cell is different.

Let us consider an insulating crystal with $N$ doubly occupied bands. We consider two states $\lambda_1$ and $\lambda_2$ of the crystal connected by a continuous adiabatic transformation of the ionic potential, described by the evolution of the variable $\lambda$[7]. The change in polarization induced by this transformation can be expressed as

$$\Delta \mathcal{P} = \int_{\lambda_1}^{\lambda_2} d\lambda \frac{\partial \mathcal{P}}{\partial \lambda} = \mathcal{P}(\lambda_2) - \mathcal{P}(\lambda_1). \tag{2.19}$$

The polarization $\mathcal{P}(\lambda)$ can be decomposed into ionic and electronic contributions,

$$\mathcal{P}(\lambda) = \mathcal{P}_{ion}(\lambda) + \mathcal{P}_{el}(\lambda). \tag{2.20}$$

The ionic term can be computed in a way similar to the first term in Eq. (2.18),

$$\mathcal{P}_{ion}(\lambda) = \frac{e}{\Omega_0} \sum_{\kappa}^{cell} Z_\kappa \mathbf{R}_\kappa, \tag{2.21}$$

while the electronic polarization can be formulated as a Berry phase of the occupied bands [37],

$$\mathcal{P}_{el}(\lambda) = -\frac{2ie}{8\pi^3} \sum_{n=1}^{N} \int_{BZ} d\mathbf{k} \left\langle u_{n,\mathbf{k}}^\lambda | \nabla_k | u_{n,\mathbf{k}}^\lambda \right\rangle, \tag{2.22}$$

when the wave-functions are chosen to satisfy the periodic gauge condition

$$u_{n,\mathbf{k}}^\lambda(\mathbf{r}) = e^{i\mathbf{G} \cdot \mathbf{R}} u_{n,\mathbf{k}+\mathbf{G}}^\lambda(\mathbf{r}), \tag{2.23}$$

---

[7]It is assumed that the system remains an insulator all along the transformation.

where $\mathbf{G}$ is any vector of the reciprocal lattice. Taken independently at independent $k$-points, the matrix elements in Eq. (2.22) are ill-defined but the integral over the Brillouin zone provides a well-defined quantity which has the form of a Berry phase of band $n$ as discussed by Zak [99]. In practical calculations, integration over the Brillouin zone and differentiation with respect to $k$-points have to be performed on a finite grid of $k$-points. A discretization of the formula suitable for practical calculations was provided in Ref. [37].

We note that, associated to the fact that a phase is only defined modulo $2\pi$, Eq. (2.19) only provides the change of polarization modulo a quantum $(2e/\Omega_0)\mathbf{R}$. In practice, $\lambda_1 - \lambda_2$ must be chosen sufficiently small for the change of polarization to be uniquely defined or the path of integration must be explicitly specified.

A review of the modern theory of polarization is provided in Ref. [39]. Using the previous formalism, the spontaneous polarization of a ferroelectric can be computed as the change of polarization when the crystal is transformed from its highly symmetric paraelectric structure to a ferroelectric phase. First applied to $KNbO_3$ [100] and then to various perovskites [101] to compute the spontaneous polarization and the Born effective charges, it provides also easy access to the piezoelectric tensor [102, 103, 104]. The method is now accessible in most conventional total energy codes and routinely applied.

### 2.2.6   Wannier functions.

Wannier functions, $W_n(\mathbf{r} - \mathbf{R})$ are orthogonal *localized* functions that span the same space as the Bloch states of a band or a group of bands [105, 106]. They are characterized by two quantum numbers: a band index $n$ and a lattice vector $\mathbf{R}$. They are defined as Fourier transforms of the Bloch eigenstates

$$W_n(\mathbf{r} - \mathbf{R}) = \frac{\Omega_0}{8\pi^3} \int_{BZ} d\mathbf{k} e^{i\mathbf{k}\cdot(\mathbf{r}-\mathbf{R})} u_{n,\mathbf{k}}(\mathbf{r}). \tag{2.24}$$

Since many years, Wannier functions are considered as a convenient support in formal proofs. However, the major drawback of the Wannier representation is that the functions are not uniquely defined but can vary strongly both in shape and range. This is a consequence of the intrinsic phase indeterminacy [gauge freedom, Eq. (2.9)] of the Bloch functions entering in Eq. (2.24).

Marzari and Vanderbilt proposed a possible way to solve this problem and developed a method to construct Wannier functions that are maximally localized around their centers [50, 53, 52]. There are various possible ways to define "maximally localized". What they proposed is to choose the phase of the Bloch functions in order to minimize the mean square spread of the Wannier functions [50]

Wannier functions are the solid state equivalent of "localized molecular orbitals" [107, 42, 40]. Being localized in real space, they provide an insightful picture of the nature of the chemical bond, otherwise hidden within the fully delocalized Bloch representation. For instance, they were computed for $BaTiO_3$ and allowed to visualize transfer of charges related to changes of hybridization at the ferroelectric phase transition [51].

The Wannier functions also provide an interesting interpretation of the Berry phase formalism presented in the previous Sections. Using Eq. (2.24), the change of electronic polarization in Eq. (2.19) can be recast as

$$\Delta\mathcal{P}_{el} = \frac{-2e}{\Omega_0} \sum_{n=1}^{N} \left[ \int d\mathbf{r} \ \mathbf{r} |W_n^{\lambda 2}(\mathbf{r})|^2 - \int d\mathbf{r} \ \mathbf{r} |W_n^{\lambda 1}(\mathbf{r})|^2 \right]. \tag{2.25}$$

Physically, this means that the change of polarization of a solid can be deduced from the displacement of the center of charge of the Wannier functions of the occupied bands that is induced by the adiabatic change in the Hamiltonian. In other words, for the purpose of determining the polarization, the true quantum mechanical electronic system can be considered as an effective system of quantized point charges, located at the Wannier centers associated with the occupied bands in each unit cell. In general the center of each function is not unique. However, the sum of the moments of all the functions is unique as demonstrated by Blount [108]. This is a consequence of the gauge invariance of the Berry phase.

Finally, let us mention that the modern theory of polarization also allowed to define an electronic localization tensor, providing a rigorous measure of the degree of localization of the electrons in a solid and to discriminate between metal and insulator [41]. At first sight, such a quantity is difficultly accessible from fully delocalized Bloch functions. It has been shown by Souza *et al.* [43] that the spread of the maximally localized Wannier functions introduced by Marzari *et al.* is related to an invariant quantity providing a physical measure of the localization. From the relationship between Wannier and Bloch functions [Eq. (2.24)], the so-called localization tensor can also be deduced from Bloch functions [42]. It has been computed for different ferroelectric oxides, giving interesting complementary insight into their electronic structure [44].

### 2.2.7  The electric field perturbation.

The interactions between the electrons and a homogeneous electric field $\mathcal{E}$ is described by a potential $V(r) = e\mathcal{E} \cdot \mathbf{r}$ (in the scalar-potential gauge). Although such a potential is convenient to study the response of finite systems (such as molecules) to electric fields, it is problematic for extended systems. On the one hand, this linear potential is non-periodic and therefore not compatible with the Born-von Karman boundary conditions. On the other hand, it is unbounded from below. The presence of a finite field will bend the energy bands so that the conduction band in a given region will always be lower in energy than the valence band in another sufficiently distant region (at a distance $L$ such that $e\mathcal{E} \cdot L > E_{gap}$). This will give rise to inter-band tunneling (Zener effect, see Fig. 2.3) so that an infinite crystal in an electric field has no true ground-state.

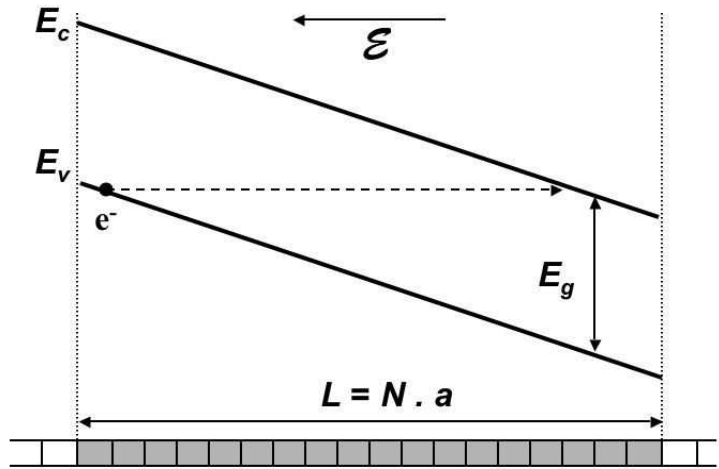

Figure 2.3: Valence ($E_v$) and conduction ($E_c$) energy band edges along the Born-von Karman supercell for a ferroelectric insulator of bandgap $E_g$ in presence of a finite macroscopic electric field $\mathcal{E}$. When, the drop of potential along the supercell is larger than the gap ( $e\mathcal{E}Na > E_g$), transfer of electron from the valence band on one side to the conduction band on the other side will decrease the energy of the crystal. This phenomenon is known as *Zener breakdown*.

However, for sufficiently small fields [8], the tunneling current through the band-gap can be neglected and the system can be described by a set of electric field dependent Wannier functions. As shown by Nunes and Vanderbilt [109], these Wannier functions minimize the energy functional

$$\mathcal{F}_{e+i}[\mathbf{R}_\kappa, W_n, \mathcal{E}] = E_{e+i}[\mathbf{R}_\kappa, W_n] - \Omega_0 \mathcal{E} \cdot \mathcal{P}, \tag{2.26}$$

where $E_{e+i}$ is the energy under zero field and $\mathcal{P}$ is the macroscopic polarization. The Wannier functions, $W_n$, do not correspond to the true ground-state of the system but rather to a long lived metastable state (a resonance).

---

[8]In practice, calculations are restricted to field smaller than $\mathcal{E}_c = E_{gap}/eNa$ where N is the number of unit cell in the Born-von Karman supercell associated to the k-point mesh used in the calculation. In most cases, this still allows the interesting range of electric fields to be covered by the calculations.

The previous energy functional $\mathcal{F}$ can be equivalently written in terms of Bloch functions in which case the first term correspond to Eq. (2.1) and the polarization $\mathcal{P}$ is obtained as a Berry phase of the Bloch functions

$$\mathcal{F}_{e+i}[\mathbf{R}_\kappa, u_{n,\mathbf{k}}, \mathcal{E}] = \mathbf{E_{e+i}}[\mathbf{R}_\kappa, \mathbf{u_{n,k}}] - \mathbf{\Omega_0} \mathcal{E} \cdot \mathcal{P}. \tag{2.27}$$

Different schemes have been proposed in the literature to perform calculations under finite electric field [110, 45, 46]. Pratical calculations are cumbersome and such methods were not yet applied to complex oxides such as ferroelectric. Instead, simplified approaches have been reported. Partially inspired from a pioneer work of Fu and Cohen [111], an approximate method based on an expansion of $\mathcal{F}_{e+i}$ limited to first-order in the electric field was proposed by Sai *et al.* [112]. This method allows structural optimization under the constraint of fixed polarization and is particularly interesting for the study of ferroelectric superlattices with broken inversion symmetry [112]. It has been recently improved by Diéguez and Vanderbilt [113] to account with the exact polarization, properly incorporated using the theory of insulators under finite electric fields [45]. An alternative easy-to-implement approach for determining the structural geometry of insulators under finite electric fields was also reported by Fu and Bellaiche [114].

### 2.2.8   Energy derivatives and multifunctional properties.

As previously stated, various properties of materials are related to derivatives, with respect to selected perturbations, of their total energy $E_{e+i}$ or, more generally in presence of finite electric fields, of the related thermodynamic potential $\mathcal{F}_{e+i}$ [115]. $\mathcal{F}_{e+i}$ depends on distinct parameters : atomic positions $\mathbf{R}_\kappa$, homogeneous strains $\eta$ and homogeneous electric fields $\mathcal{E}$. For the sake of simplicity, we can summarize these three kinds of perturbations within a single parameter $\lambda$, defined as $\lambda = (\mathbf{R}_\kappa, \eta, \mathcal{E})$.

The energy functional $\mathcal{F}_{e+i}$ can be expanded in a power series in $\lambda$ around the equilibrium structure $\lambda = 0$

$$\mathcal{F}_{e+i}(\lambda) = \mathcal{F}_{e+i}(0) + \sum_i \frac{\partial \mathcal{F}_{e+i}}{\partial \lambda_i}\bigg|_0 \lambda_i + \frac{1}{2}\sum_{i,j} \frac{\partial^2 \mathcal{F}_{e+i}}{\partial \lambda_i \partial \lambda_j}\bigg|_0 \lambda_i \lambda_j + \frac{1}{6}\sum_{i,j,k} \frac{\partial^3 \mathcal{F}_{e+i}}{\partial \lambda_i \partial \lambda_j \partial \lambda_k}\bigg|_0 \lambda_i \lambda_j \lambda_k + ... \tag{2.28}$$

The successive derivatives in the previous equations are directly linked to important physical quantities. As can be seen in Table 2.1 , first-order derivatives are related to the forces on the atoms ($F$), the stress tensor ($\sigma$) and the spontaneous polarization ($\mathcal{P}^s$). Forces and stresses are readily accessible in all DFT codes as they are needed for structural optimizations. The polarization is also currently computed within the Berry phase approach.

Second-order derivatives (in Table 2.1) characterize the *linear response* of the solid. They are connected to the interatomic force constants ($C$) providing access to the full phonon dispersion curves, the optical dielectric tensor ($\epsilon^\infty$), the clamped-ion elastic constants ($c^0$), the Born effective charges ($Z^*$), the clamped-ion piezoelectric tensor ($e^0$) and the internal strain coupling parameters ($\gamma$). Combining the previous quantities, additional properties can still be accessed such as the static dielectric tensor ($\epsilon^0$) or the relaxed-ion elastic and piezoelectric tensors.

Third-order derivatives characterize the *non-linear response* of the solid and have been more marginally computed. In the case of the ferroelectrics, let us point out the calculation of selected non-linear optical responses such as Raman tensors, non-linear optical susceptibilities and electro-optic coefficients [116, 64].

Table 2.1: *Physical quantities related (within a factor of normalization) to first- and second-order derivatives of $\mathcal{F}_{e+i}$.*

| $\mathcal{F}_{e+i}$ | $1^{\text{st}}$-order derivatives | $2^{\text{nd}}$-order derivatives | | |
|---|---|---|---|---|
| | | $\partial/\partial R$ | $\partial/\partial\eta$ | $\partial/\partial\mathcal{E}$ |
| $\partial/\partial R$ | $F$ | $C$ | $\gamma$ | $Z^*$ |
| $\partial/\partial\eta$ | $\sigma$ | $\gamma$ | $c^0$ | $e^0$ |
| $\partial/\partial\mathcal{E}$ | $\mathcal{P}^s$ | $Z^*$ | $e^0$ | $\epsilon^\infty$ |

Nowadays, first- and second-order derivatives are routinely computed using either finite-difference techniques or density functional perturbation theory. The finite-difference approach presents the advantage that it is readily accessible in any total energy DFT package, without any additional computational effort required beyond the implementation of the Berry-phase formalism for the computation of $\mathcal{P}$. At the opposite, density functional perturbation theory require a major effort of implementation for each new quantity that is considered. However, as soon as it is implemented, it allows to build an energy derivative database in a systematic way, giving direct access to a large set of properties (full tensors directly provided in appropriate units and in cartesian coordinates independently of the crystal structure and symmetry) without any additional human effort. For this reason, it is usually preferred when available.

DFPT allowed to study the dielectric and dynamical properties of various ferroelectric perovskites [117, 118, 119, 120, 121, 122, 95, 123, 124, 116]. It also allows to envisage the computation of complex properties (such as the tuning of the dielectric constant [125]) or complex optimization procedures (such as optimization under constraint polarization [112, 113]) by providing "in the flight" all the required quantities.

## 2.3 First-principles effective Hamiltonian.

Part of the fascination for the family of perovskite ferroelectric oxides is related to their ability to undergo with temperature different sequences of structural phase transitions [1, 15]. A major challenge for theoretitians is therefore to see to which extent they are able to predict from first-principles the temperature behavior of these compounds. The interest is not only academic. For these oxides, the various functional properties (dielectric, piezoelectric, and optical properties) of direct interest for applications are particularly large but also strongly dependent of temperature, specially around the phase transition temperatures where they exhibit a divergent character. To be helpful and reliable, theoretical predictions must therefore take explicitly into account the temperature and be able to reproduce the correct sequence of phase transitions.

At first sight, we might consider doing first-principles DFT molecular dynamics simulations [26]. However, such an approach is computationally intensive so that, nowadays, it is still restricted to relatively small systems and time-scale. As it will be emphasized later, because of the long-range character of the Coulomb forces that are at the origin of the ferroelectric instability, the accurate modeling of ferroelectric phase transitions requires simulation boxes containing thousands of atoms. This is clearly beyond the scope of present DFT capabilities so that more approximate approaches must be considered. We will see that is not necessarily a drawback: because those simplified approaches include the physics in a transparent way, and are also very helpful to clarify the microscopic mechanisms responsible for the properties of ferroelectric oxides. They appear therefore has useful tools to discuss with the experimentalists.

### 2.3.1 Phenomenological Landau-Devonshire theory.

Until the advent of the recent atomistic simulations, the most popular approach to describe the temperature behavior of ferroelectrics at the macroscopic level was the phenomenological Devonshire-Ginzburg-Landau (DGL) theory [126, 127].

We do not intend to review the vast literature on the phenomenological Devonshire-Ginzburg-Landau theory and simply emphasize here the common points with the model hamiltonian approach that will be presented below.

The first step in the development of a thermodynamic theory is the identification of the relevant degrees of freedom of the problem under consideration. Three independent variables have to be chosen among the conjugate pairs temperature-entropy $(T, S)$, stress-strain $(\sigma, \eta)$, and electric field-electric displacement $(\mathcal{E}, \mathcal{D}$ or eventually $\mathcal{P})$. Since there are eight different ways of combining these magnitudes, there are eight different thermodynamic potentials (see [1], chapter 3). All of them contain the same information, i. e. there is no "best" thermodynamic potential but only "most suitable" functional for the particular boundary conditions. They are related by the proper Legendre transformations between the independent variables.

Once the most suitable thermodynamic potential has been selected, the following step is the choice of a prototype state. For perovskite oxides this state usually corresponds to a non-polar phase. Then, a polynomial expansion of the thermodynamic potential as a function of the independent variables is done and is assumed to remain valid before and after the phase transition. Coupling terms between the independent variables are also considered. The coefficients of the expansion are fitted from experiment, usually performed in a regime close to the phase transitions. Finally, the thermodynamic potential is minimized to find the equilibrium thermodynamic states. The effect of the temperature arises from a temperature dependence of some parameters assumed *a priori*.

Let us exemplify the approach considering polarization and strains as independent degrees of freedom. The expression for the thermodynamic potential around the unpolarized cubic crystal, as reported by Devonshire in the pioneering works [126, 127] is given by

$$
\begin{aligned}
F(T, \eta, \mathcal{P}) &= \tfrac{1}{2} C_{11} \left( \eta_{11}^2 + \eta_{22}^2 + \eta_{33}^2 \right) + C_{12} \left( \eta_{22}\eta_{33} + \eta_{33}\eta_{11} + \eta_{11}\eta_{22} \right) + \tfrac{1}{2} C_{44} \left( \eta_{12}^2 + \eta_{23}^2 + \eta_{31}^2 \right) + \\
&\quad a_1 \left( P_1^2 + P_2^2 + P_3^2 \right) + a_{11} \left( P_1^4 + P_2^4 + P_3^4 \right) + a_{12} \left( P_1^2 P_2^2 + P_1^2 P_3^2 + P_2^2 P_3^2 \right) + \\
&\quad a_{111} \left( P_1^6 + P_2^6 + P_3^6 \right) + g_{11} \left( \eta_{11} P_1^2 + \eta_{22} P_2^2 + \eta_{33} P_3^2 \right) + \\
&\quad g_{12} \left[ \eta_{11} \left( P_2^2 + P_3^2 \right) + \eta_{22} \left( P_1^2 + P_3^2 \right) + \eta_{33} \left( P_1^2 + P_2^2 \right) \right] + \\
&\quad g_{44} \left( \eta_{23} P_2 P_3 + \eta_{31} P_3 P_1 + \eta_{12} P_1 P_2 \right),
\end{aligned}
\tag{2.29}
$$

where $C_{11}$, $C_{12}$, and $C_{44}$ stand for the elastic stiffness of the material, and the subindices refer to the cartesian directions. The expansion has been cut at 6th order in $\mathcal{P}$ for simplicity, and it does not contain odd terms in polarization since the prototype state is centrosymmetric. The last three terms in the expansion clearly couple the strain and the polarization. As it will appear clearly later, this expression shows remarkable similitude with the effective Hamiltonian that corresponds also to an expansion in terms of polarization and strain with appropriate coupling terms.

In order to simplify a little bit more the discussion, let us suppose that all the strains are zero and the polarization is directed along one of the cartesian directions. Then

$$
F(T, \eta, \mathcal{P}) = a_1 \mathcal{P}^2 + a_{11} \mathcal{P}^4 + a_{111} \mathcal{P}^6.
\tag{2.30}
$$

As previously mentioned, the coefficients of the expansion should be temperature dependent. However, for practical purposes, most of them are usually considered as constants. For instance, to deal with second order phase transitions, it is enough to consider the quadratic coefficient as $T$-dependent, while $a_{11}$ and $a_{111}$ are fixed to positive constant values. It is assumed that $a_1$ has a linear dependency of the form $a_1 = \beta (T - T_0)$ with $\beta$ a positive constant and $T_0$ the temperature of the ferroelectric-to-paraelectric phase transition. For $T > T_0$ the quadratic term is positive and the thermodynamic functional is a single well with its minimum corresponding to the non-polar $\mathcal{P} = 0$ phase. For $T < T_0$ the quadratic coefficient is negative, and the thermodynamic potential displays the typical double well shape. There are two minima for $\mathcal{P} \neq 0$ (see Fig. 2.4).

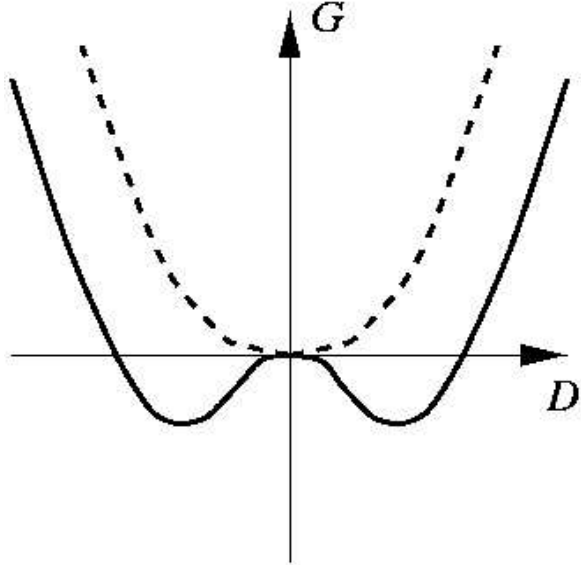

Figure 2.4: Elastic Gibbs energy, $G_1$ as a function of the displacement field $\mathcal{D}$ for a paraelectric ($T > T_0$, dotted line) and a ferroelectric ($T < T_0$, full line) material. From Ref. [32].

In a similar way, a first-order phase transition can be reproduced assuming a constant negative 4-th order term [1, 128].

From the previous examples, it clearly appear that DGL theory offers a simple and coherent framework to classify and describe structural phase transitions at a phenomenological level. For this reason it was widely used in the past to study ferroelectric systems and still remain a valuable approach.

### 2.3.2 Effective Hamiltonian for ferroelectrics.

At the microscopic level, a first-principles effective Hamiltonian approach had been proposed in the 1980's by Rabe and Joannopoulos to study the phase transition of GeTe [67, 68, 69]. Their approach was (i) to identify the important degrees of freedom for describing the transition though the local mode approximation of Lines [129], and (ii) to perform a low-order expansion of the energy in terms of these degrees of freedom, with coefficients directly determined from total energy DFT calculations. In the 1990's this model was successfully generalized to ferroelectric perovskite oxides [70, 71]. In what follows, we briefly summarize the effective Hamiltonian approach as it was developed for bulk perovskites. We will describe later how it can be generalized to nanostructures.

#### General framework.

To build an effective Hamiltonian with a simple form and a reduced number of parameters, it is worth noticing that ferroelectric phase transitions are driven by a restricted number of degrees of freedom. In simple ferroelectrics such as prototype $BaTiO_3$, the structural distortion at the phase transition nearly exactly corresponds to the freezing into the structure of one of the transverse optic modes at the $\Gamma$ point, usually referred to as the soft-mode (see Sec. 3.2), and to a subsequent strain relaxation. A reasonable approximation for the study of the phase transition is therefore to only consider explicitly (i) the ionic degree of freedom of the soft-mode thanks to a local mode (i.e. a local cooperative pattern of atomic displacement) $\xi_i$ that will be associated to each unit cell $i$ and (ii) the strains (homogeneous and eventually inhomogeneous). This corresponds to the assumption that only the lower part of the phonon spectrum (soft-modes and acoustic modes) is relevant for the phase transition and does not significantly couples with higher energy modes. As the amplitude of the ferroelectric distortion is relatively small, the Hamiltonian can then be written as a Taylor expansion in terms of $\xi_i$ and strains limited to low orders.

**Local modes and lattice Wannier functions.**

The first-step in the construction of the effective Hamiltonian is to identify the relevant ionic degree of freedom, $\xi_i$. In their pioneer work [70, 71], Zhong, Vanderbilt and Rabe considered a *local mode* (LM) confined to one unit cell and directly deduced from the unstable ferroelectric mode eigenvector at the $\Gamma$ point [9]. There is however no unique way to decompose the unstable mode at $\Gamma$ into a sum of local atomic displacements and there is no obligation to confine the local mode to one unit cell. A more systematic approach was then proposed by Waghmare and Rabe [72]. Starting from the full phonon dispersion curves, they proposed to construct the *lattice Wannier function* (LWF) [130] associated to the phonon branch containing the $\Gamma$ unstable mode. Similarly to the electronic Wannier functions, these LWF correspond to a pattern of displacement localized in real space [10]. In comparison to the initial local mode it presents the advantage that it is not fitted at $\Gamma$ only but include informations from the whole Brillouin zone. Even if the LWF approach is conceptually more rigorous, both LM and LWF approaches are currently used and allow to achieve similar results.

When different kinds of instabilities are present (ferroelectric, ferrodistortive, ...) and compete with each other, the model can be readily generalized in order to include properly all the relevant ionic degrees of freedom [132, 133].

**The effective Hamiltonian.**

The effective Hamiltonian is then written as a low-order Taylor expansion around the paraelectric cubic phase in terms of the selected degrees of freedom. Depending of the authors and of the compounds, slightly different expressions have been reported. Following the approach of Waghmare and Rabe [72], the effective Hamiltonian is written in terms of the ferroelectric degree of freedom $\xi_i$ and the homogeneous strains $\eta$ (inhomogeneous strains can also be explicitely included through an independent degree of freedom $v_i$ [71] but are neglected here) as

$$\mathcal{H}^{eff}(\{\xi_i\}, \eta) = \mathcal{H}^{ions}(\{\xi_i\}) + \mathcal{H}^{elastic}(\eta) + \mathcal{H}^{int}(\{\xi_i\}, \eta). \tag{2.31}$$

The first term, $\mathcal{H}^{ions}$, describes the evolution of the energy in terms of $\{\xi_i\}$. In the limit of a uniform configuration $\xi_i = \xi$, it simply reproduces the double-well energy of Fig. 1.1. But, it must also properly describe the evolution of the energy when the $\xi_i$ fluctuate independently in the different unit cells. To that end it is decomposed as

$$\mathcal{H}^{ions}(\{\xi_i\}) = \underbrace{[\mathcal{H}^{SR}(\{\xi_i\}) + \mathcal{H}^{DD}(\{\xi_i\})]}_{\mathcal{H}^{harm}} + \mathcal{H}^{anh}(\{\xi_i\}). \tag{2.32}$$

The first part in brackets correspond to the *harmonic* term. It is adjusted to reproduce the DFT harmonic energy surface of the crystal within the subspace spanned by the $\xi_i$. It is decomposed into short-range (SR) and long-range dipole-dipole (DD) interactions, the parameters of which are adjusted to reproduce the unstable phonon branch (in practice, at few selected point of the Brillouin zone).

---

[9] The Bloch eigenvector of the ferroelectric mode at $\Gamma$ corresponds to a specific pattern of atomic displacements, identical from one unit cell to the other in order to provide a uniform distortion. The idea is to decompose this homogeneous atomic distortion of the crystal into individual local contributions $\xi_i$ affected to each individual unit cell $i$. The local mode $\xi_i$ is therefore a local pattern of atomic displacements such that it reproduces the homogeneous ferroelectric distortion when frozen with the same amplitude in all the unit cells.

[10] Wannier functions are not unique and a systematic way to build LWF was also proposed by Íñiguez *et al.* [131].

For the short-range interaction between $\xi_i$, quadratic interactions up to third nearest neighbors with the most general form allowed by the space group symmetry are considered,

$$
\begin{aligned}
\mathcal{H}_{SR}(\{\xi_i\}) &= A|\xi_i|^2 + \sum_i \sum_{\hat{d}=nn1} \left\{ a_L(\xi_i \cdot \hat{d})(\xi_i(\hat{d}) \cdot \hat{d}) + a_T[\xi_i \cdot \xi_i(\hat{d}) - (\xi_i \cdot \hat{d})(\xi_i(\hat{d}) \cdot \hat{d})] \right\} \\
&+ \sum_i \sum_{\hat{d}=nn2} \left\{ b_L(\xi_i \cdot \hat{d})\xi_i(\hat{d}) \cdot \hat{d} + b_{T1}(\xi_i \cdot \hat{d}_1)(\xi_i(\hat{d}) \cdot \hat{d}_1) \right. \\
&+ \left. b_{T2}(\xi_i \cdot \hat{d}_2)(\xi_i(\hat{d}) \cdot \hat{d}_2) \right\} \\
&+ \sum_i \sum_{\hat{d}=nn3} \left\{ c_L(\xi_i \cdot \hat{d})(\xi_i(\hat{d}) \cdot \hat{d}) \right. \\
&+ \left. c_T[\xi_i \cdot \xi_i(\hat{d}) - (\xi_i \cdot \hat{d})(\xi_i(\hat{d}) \cdot \hat{d})] \right\}.
\end{aligned}
\tag{2.33}
$$

The sums over $\hat{d}$ in Eq. (2.33) are taken over the first (nn1), second (nn2) and third (nn3) nearest neighbors of site $i$ that are located respectively along the $\langle 100 \rangle$, $\langle 110 \rangle$ and $\langle 111 \rangle$ directions. $\xi_i(\hat{d})$ denotes the LWF at a neighbor of site $i$ in $\hat{d}$ direction. The second neighbor sites are located along the diagonal of a square of side $a$, where $a$ is the lattice constant of the cubic unit cell. The unit vector $\hat{d}_1$ is in the plane of the square perpendicular to this diagonal, while $\hat{d}_2$ is perpendicular to the plane of the square.

The dipole-dipole interaction is parametrized thanks to the Born effective charge $\overline{Z}^*$ associated to $\xi_i$ and the electronic dielectric constant $\varepsilon_\infty$,

$$
\mathcal{H}_{DD}(\{\xi_i\}) = \sum_i \sum_{\mathbf{d}} \frac{(\overline{Z}^*)^2}{\varepsilon_\infty} \frac{(\xi_i \cdot \xi_i(\mathbf{d}) - 3(\xi_i \cdot \hat{d})(\xi_i(\hat{d}) \cdot \hat{d}))}{|\mathbf{d}|^3}.
\tag{2.34}
$$

The sum over $\mathbf{d}$ in Eq. (2.34) is taken over all neighbors of site $i$ and $\mathcal{H}_{DD}$ is evaluated using Ewald summation technique. The value of $\overline{Z}^*$ is obtained from the individual Born effective charges ($Z_\kappa^*$) and the eigendisplacements of the unstable mode at $\Gamma$ ($U_\kappa^{SM}$) using $\overline{Z}^* = \sum_\kappa Z_\kappa^* U_\kappa^{SM}$.

The last part includes the anharmonicities mandatory to stabilize the crystal in a given polar configuration. It is restricted to a simple form without intersite coupling and directly adjusted on the double-well curve corresponding to $\Gamma$-type displacements,

$$
\mathcal{H}_{anh}(\{\xi_i\}) = \sum_i \left[ B|\xi_i|^4 + C(\xi_{ix}^4 + \xi_{iy}^4 + \xi_{iz}^4) + D|\xi_i|^6 + E|\xi_i|^8 \right].
\tag{2.35}
$$

The second term in Eq. (2.31), $\mathcal{H}^{elastic}$, is the elastic energy that takes into account the evolution of the crystal energy in terms of the macroscopic strains (isotropic, uniaxial, and shear). It is written in terms of the elastic constants $C_{ij}$ as a second-order expansion of the energy with respect to the homogeneous strain variables,

$$
\mathcal{H}_{elas}(\eta) = Nf \sum_{\mu=1}^3 \eta_{\mu\mu} + \frac{N}{2}C_{11} \sum_{\mu=1}^3 \eta_{\mu\mu}^2 + \frac{N}{2}C_{12} \sum_{\substack{\mu,\nu=1 \\ \mu \neq \nu}}^3 \eta_{\mu\mu}\eta_{\nu\nu} + \frac{N}{4}C_{44} \sum_{\substack{\mu,\nu=1 \\ \mu \neq \nu}}^3 \eta_{\mu\nu}^2.
\tag{2.36}
$$

The last term in Eq. (2.31), $\mathcal{H}^{int}$, includes the coupling between $\{\xi_i\}$ and $\eta$. It is particularly important in ferroelectrics and, for instance, at the origin of the piezoelectric response. It is determined from the strain dependence of the unstable phonon frequency at $\Gamma$,

$$
\mathcal{H}_{int}(\{\xi_i\}, \eta) = g_0 \left( \sum_{\mu=1}^3 \eta_{\mu\mu} \right) \sum_i |\xi_i|^2 + g_1 \sum_{\mu=1}^3 \left( \eta_{\mu\mu} \sum_i \xi_{i\mu}^2 \right)
$$

$$+g_2 \sum_{\substack{\mu,\nu=1 \\ \mu<\nu}}^{3} \eta_{\mu\nu} \sum_i \xi_{i\mu}\xi_{i\nu}. \tag{2.37}$$

The parameters $g_0$, $g_1$, and $g_2$ give the strength of the coupling of strain with the local atomic distortion. Such an effective Hamiltonian has been fitted on DFT calculations for numerous perovskite oxides. In principle, this should be done at the theoretical lattice constant. In practice, however, it has been observed that this provides relatively bad results and that the results can be significantly improved when applying a negative pressure in order to correct the typical LDA underestimate of the lattice constant [71]. When it is known, it is therefore not unusual to fit the effective Hamiltonian at the experimental lattice constant in order to avoid further a posteriori correction of the volume. In Table 2.2, we report typical parameters for BaTiO$_3$. The model has been fitted at the experimental volume by Ghosez *et al.* [134] using the LWF approach and the Hamiltonian form reported above.

Table 2.2: Parameters in the effective Hamiltonian of BaTiO$_3$ (units eV per unit cell). The model has been fitted at the experimental volume ($a_{cell} = 4.00$ Å) using the LWF approach by Ghosez *et al.* [134].

| $A$ | 2.9080 | $a_L$ | 0.3718 | $C_{11}$ | 123.0243 |
|---|---|---|---|---|---|
| $B$ | 11.5242 | $a_T$ | -0.4832 | $C_{12}$ | 47.1910 |
| $C$ | 23.2260 | $b_L$ | 0.2302 | $C_{44}$ | 192.6313 |
| $D$ | -53.1421 | $b_{T1}$ | 0.0354 | $g_0$ | -7.2916 |
| $E$ | 169.9803 | $b_{T2}$ | -0.1047 | $g_1$ | -51.8323 |
| $\overline{Z}^*$ | 1.9220 | $c_L$ | 0.2094 | $g_2$ | -2.2036 |
| $\varepsilon_\infty$ | 6.7467 | $c_T$ | -0.0389 | $f$ | 3.0611 |

**Accessible properties.**

The effective Hamiltonian or a related thermodynamical potential [11] can then be used to investigate the temperature behavior of ferroelectrics thanks to classical Monte-Carlo [72] or molecular dynamics simulations [135] [12]. Quantum Monte-Carlo simulations have also been reported in order to investigate the role of quantum fluctuations on the behavior of incipient ferroelectrics at low temperatures [136, 137].

Such calculations are performed with periodic boundary conditions and typically require a $12\times12\times12$ supercell (equivalent to 8640 atoms). As an output, they provide the mean values $<\xi>$ and $<\eta>$ in terms of temperature, external pressure and electric field. The macroscopic polarization of the crystal $<\mathcal{P}>$ is also readily accessible since the local polarization $\mathcal{P}_i$ is directly proportional to the amplitude of the local mode $\xi_i$,

$$\mathcal{P}_i = \frac{1}{\Omega_0} \bar{Z}^* \xi_i, \tag{2.38}$$

where $\bar{Z}^*$ is the Born effective charge of the local mode and $\Omega_0$ the unit cell volume.

On the one hand, the effective Hamiltonian simulations allow to predict ferroelectric phase transitions from the temperature dependence of $<\xi>$ and $<\eta>$ [70, 71] and to construct temperature–pressure phase diagram [138]. An illustration is provided in Fig. 2.5 for the case of BaTiO$_3$ with the parameters of Table 2.2. In Table 2.3, the results are compared to those reported in the pioneer work of Zhong *et al.* using the local mode (LM) approximation and to shell-model results (see Sec. 2.4). When the temperature is lowered, the model predicts the correct sequence of transitions from cubic paraelectric to successively tetragonal, orthorhombic and rhombohedral ferroelectric phases. The main features are nicely reproduced even if the absolute values of the transition temperatures are not quantitatively predicted. This is a common drawback of the effective Hamiltonian that was related primarily to the improper treatment of thermal expansion and also to errors inherent in the underlying DFT approach itself [139].

---

[11]In presence of applied electric field and stress, the quantity to be considered is no more $\mathcal{H}^{eff}(\xi_i,\eta)$ but $\mathcal{F}^{eff}(\xi_i,\eta,\mathcal{E},\sigma) = \mathcal{H}^{eff}(\xi_i,\eta) - \Omega_0\mathcal{E}\cdot\mathcal{P} - \Omega_0\ \sigma\cdot\eta$ [75].

[12]Molecular dynamics simulations are restricted to Hamiltonians fitted on the dynamical matrix of the system instead of the force constant matrix usually considered in most studies.

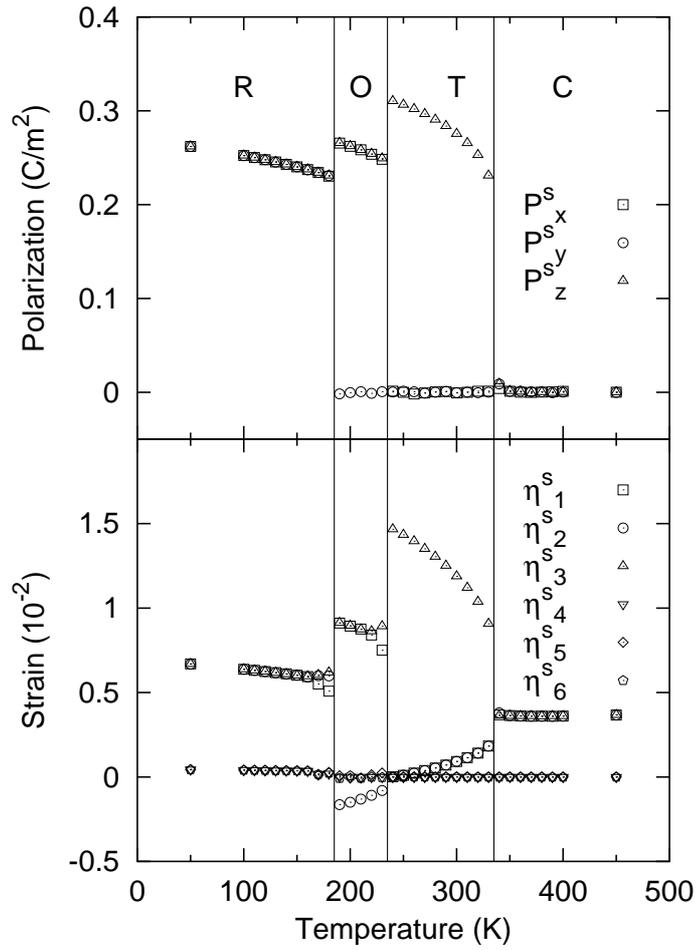

Figure 2.5: Temperature dependence of the spontaneous polarization and the spontaneous strain in the cubic (C), tetragonal (T), orthorhombic (O) and rhombohedral (R) phases of BaTiO₃ as obtained with the effective Hamiltonian parameters of Table 2.2 (LWF approach).

Table 2.3: Magnitude of the spontaneous polarizations ($\mathcal{P}_s$, in $\mu C\ cm^{-2}$) and phase transition temperatures ($T_c$, in $K$) of $BaTiO_3$ as predicted using different models.

| Approach | | Experiment | $\mathcal{H}_{eff}$ | | Shell-model |
|---|---|---|---|---|---|
| | | | LWF | LM | |
| | | Ref. [70] | Table 2.2 | Ref. [70] | Ref. [78] |
| Cubic | $\mathcal{P}_s$ | 0 | 0 | 0 | 0 |
| ↓ | $T_c$ | *403* | *335* | *297* | *190* |
| Tetragonal | $\mathcal{P}_s$ | 27 | 30 | 28 | 17 |
| ↓ | $T_c$ | *278* | *240* | *230* | *120* |
| Orthorhombic | $\mathcal{P}_s$ | 36 | 37 | 35 | 14 |
| ↓ | $T_c$ | *183* | *190* | *200* | *90* |
| Rhombohedral | $\mathcal{P}_s$ | 33 | 45 | 43 | 12.5 |

On the other hand, simulations give direct access to the temperature dependence of various functional properties. This is illustrated in Figs. 2.6–2.7. The dielectric [73, 74] and piezoelectric [75] properties are directly accessible from finite differences or a correlation function approach [74]. Recently, the Hamiltonian was generalized to also describe the temperature evolution of the electro-optic coefficients [76]. Direct comparison with the experiment is *a priori* difficult since the model do not reproduce the absolute value of the phase transition temperature while the property strongly evolve with T. To correct for this problem, it has been proposed to rescale linearly the theoretical temperature in order to adjust the theoretical and experimental (when available) phase transition temperatures [75] . As can be seen in Fig. 2.6 (from parameters of Table 2.2), the agreement with experimental data is good when the temperatures are properly rescaled.

### 2.3.3 Mixed compounds.

The effective Hamiltonian approach is not restricted to pure $ABO_3$ compounds but was generalized for solid solution of direct technological interest. It was so applied to $PbZr_{1-x}Ti_xO_3$ (PZT) [77, 142] or to the relaxor $PbSc_{1-x}Nb_xO_3$ (PSN) [143, 144, 33]. For such $A(B'B")O_3$ alloys, the method consists in starting from an averaged Hamiltonian within the virtual crystal approximation [145, 146] (usually written in terms of local mode $\{\xi_i\}$, inhomogeneous strain $\{v_i\}$ and homogeneous strains $\eta$) and including first-order correction terms that takes into account the real nature of the atoms at site $i$ (B' or B") through a set of variables $\sigma_i$ ($\sigma_i = 1$ for B' and $-1$ for B"),

$$\mathcal{H}^{eff}(\{\xi_i\}, \{v_i\}, \eta, \{\sigma_i\}) = \mathcal{H}^{eff}_{ave}(\{\xi_i\}, \{v_i\}, \eta) + \mathcal{H}^{eff}_{loc}(\{\xi_i\}, \{v_i\}, \{\sigma_i\}). \quad (2.39)$$

The average Hamiltonian, $\mathcal{H}^{eff}_{ave}$, is fitted as previously described for pure compounds but from DFT calculations on a virtual $ABO_3$ crystal where $B$ is a virtual atom averaging between B' and B" [145, 146]. The local corrections, $\mathcal{H}^{eff}_{loc}$ include (i) on-site effects of alloying on the local-mode energy at site $i$ and (ii) intersite contributions linear in $\xi_i$ and $v_i$,

$$\mathcal{H}^{eff}_{loc}(\{\xi_i\}, \{v_i\}, \{\sigma_i\}) = \sum_i [\Delta\alpha(\sigma_i)\ \xi_i^4 + \Delta\gamma(\sigma_i)\ (\xi_{ix}^2\xi_{iy}^2 + \xi_{iy}^2\xi_{iz}^2 + \xi_{iz}^2\xi_{ix}^2)]$$

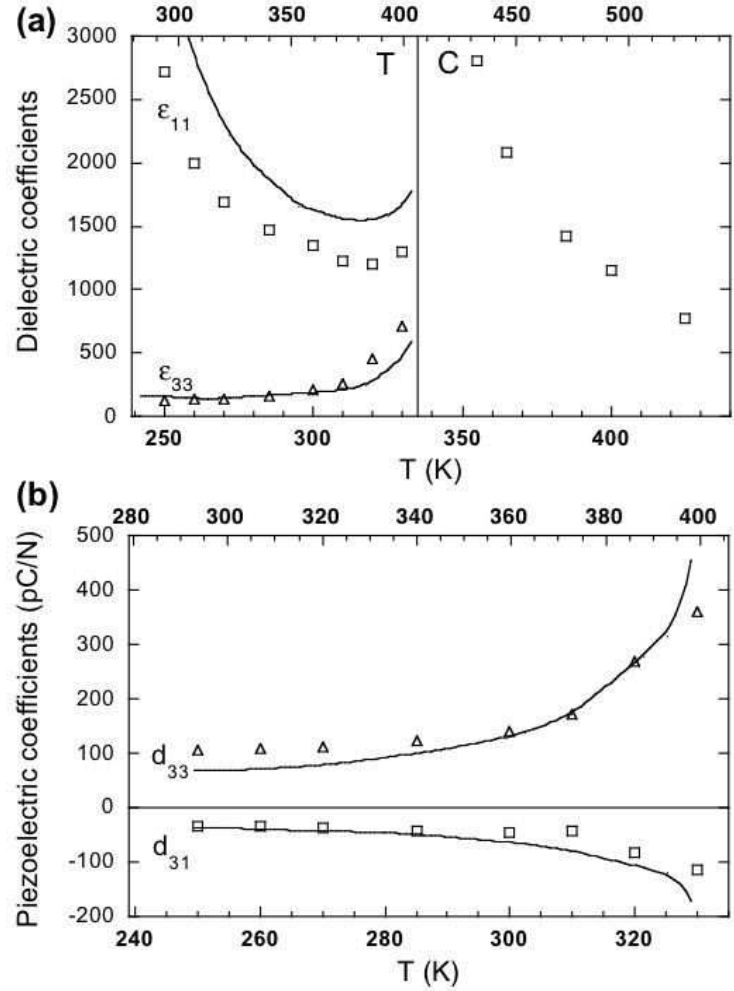

Figure 2.6: Temperature dependence of the static dielectric constants (a) and piezoelectric constants (b) in the cubic (C) and tetragonal (T) phases of BaTiO$_3$. These results have been obtained with the effective Hamiltonian parameters of Table 2.2 (LWF approach). The bottom and top $x$-axes correspond respectively to the theoretical and experimental temperatures (see text).

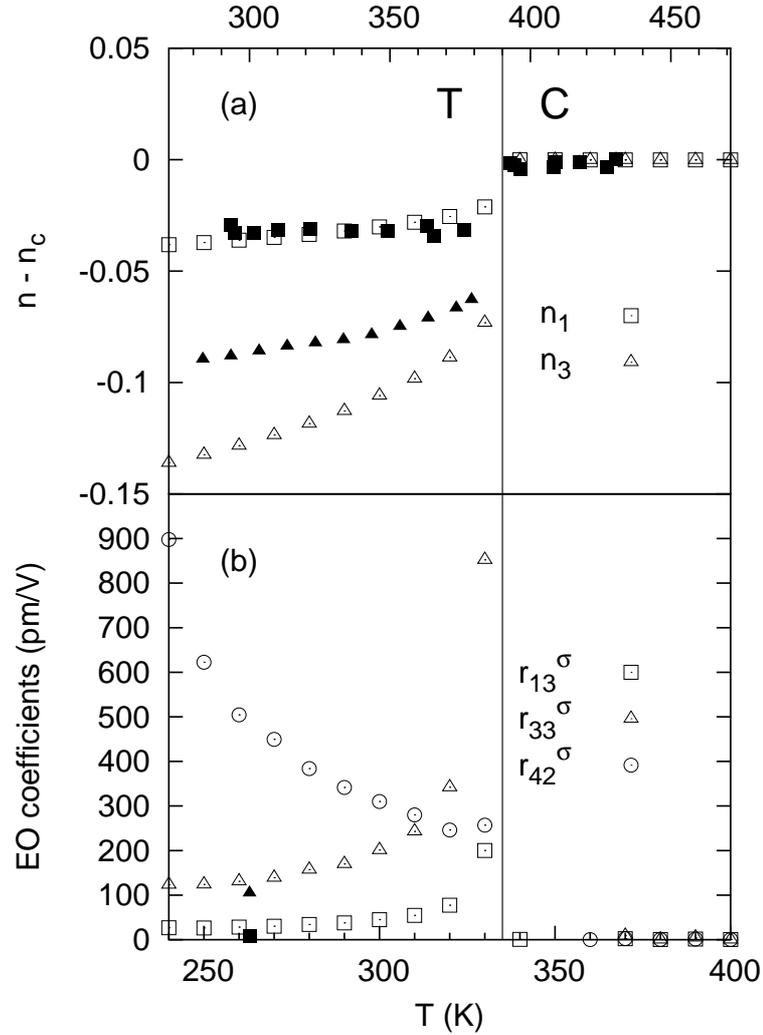

Figure 2.7: Temperature dependence of the refractive indices (a) and EO coefficients (b) in the cubic (C) and tetragonal (T) phases of BaTiO$_3$. The open (solid) symbols correspond to the theoretical (experimental [140, 141]) values. These results have been obtained with the effective Hamiltonian parameters of Table 2.2 (LWF approach) as described in Ref. [76]. The bottom and top $x$-axes correspond respectively to the theoretical and experimental temperatures (see text). (From Ref. [76] by Veithen *et al.*)

$$+ \sum_{ij}[Q_{ji}(\sigma_j)\,\mathbf{e}_{ji}\cdot\xi_i + R_{ji}(\sigma_j)\,\mathbf{f}_{ji}\cdot\mathbf{v}_i], \qquad (2.40)$$

where $\mathbf{e}_{ji}$ and $\mathbf{f}_{ji}$ are unit vectors joining site $j$ to the center of $\xi_i$ and $\mathbf{v}_i$. The parameters $\Delta\alpha(\sigma_i)$, $\Delta\gamma(\sigma_i)$, $Q_{ji}(\sigma_j)$ and $R_{ji}(\sigma_j)$ are fitted from DFT calculations on small supercells in which a real atom is surrounded by virtual atoms.

Such an approach is restricted to relatively small variations of composition around 50/50. This was not a problem for PZT where many interesting properties are close to the morphotropic phase boundary around 50/50 [77].

### 2.3.4 Comparison with the Landau-Devonshire approach.

Since the local polarization is directly related to the local mode $\xi_i$, the effective Hamiltonian can be viewed has an expansion of the energy in terms of local polarization and strains (also including their coupling). From this viewpoint, its form strongly resembles that of the Landau-Devonshire theory with polarization and strain as primary and secondary order parameters (see Sec. 2.3.1). This provides a valuable connection between the first-principles approach and popular phenomenological models.

However, let us emphasize here that the effective Hamiltonian remains a microscopic theory in which the polarization can fluctuate locally. It is also a truly "first-principles" approach with parameters directly fitted on DFT simulations and not adjusted to reproduce the experiment. Finally, there is no necessary assumption on the temperature-dependence of the parameters as in Landau-Devonshire theory, leading to dependences different from those usually assumed [147].

## 2.4 Shell-model calculations.

In parallel to effective Hamiltonians, interatomic potentials in the framework of a shell-model and directly fitted on first-principles have also been developed for perovskite oxide systems. The use of shell-models for ferroelectrics dates back to the 1960's. Distinct approaches have been considered recently, including the well-known non-linear polarizability model of Migoni, Bilz and Baüerle [148] and Buckingham potentials [149]. The main novelty of some recent works, respect to former calculations, is to fit the model directly on DFT results instead than on experimental measurements [78]. In the recent years, shell-models (directly fitted on DFT calculations, or at least in agreement with them) combined with molecular dynamics simulations revealed powerful to predict the qualitative behavior of pure compounds [78, 150] and alloys [151], including nanostructures [152, 153, 154, 155].

Contrary to the effective Hamiltonian limited to the subspace spanned by $\xi_i$, the shell-model is an atomic description including all the ionic degrees of freedom and adjusted to reproduce the full phonon dispersion curves. Also, it properly accounts for the thermal expansion [78] improperly described by the effective Hamiltonian [139]. However, as it is illustrated in Table 2.3, shell-models directly fitted on DFT calculations does not improve the description compared to the effective Hamiltonian: the quantitative agreement of the predicted transition temperatures and spontaneous polarizations with experiment is even significantly worst. The origin of the shell-model errors were not explicitly discussed but might stay in the limited number of adjustable parameters. Let us emphasize that contrary to the effective Hamiltonian in which most of the parameters (such as Born effective charges, elastic constants, ...) have a physical meaning and are determined *independently* from DFT calculations, the shell-model parameters are determined indirectly through a *global* adjustment that realizes the best compromise between different properties determined *ab initio*.

## 2.5 Perspectives.

Nowadays, the combination of first-principles DFT calculations with effective Hamiltonian, or shell-model techniques offers a multiscale approach to investigate the various functional properties of ferroelectric oxides in terms of pressure, temperature, composition, and electric field. It offers also the possibility to direclty link the properties to the atomic arrangement and boundary conditions. It constiues therefore a powerful tool for the design of new structures with tailor-made properties. Usually, the search for optimized structures proceeds by first specifying an atomic arrangement and then computing its properties. It can also follow an inverse approach in looking for the specific atomic arrangement leading to prefixed properties as first reported by Íñiguez *et al.* [144].

In spite of many success of present available methods, the door is still open to alternative approaches for large scale modeling. In this context, Grinberg *et al.* [79] successfully developed for PZT a phenomenological model based on chemical rules and with parameters directly adjusted on first-principles calculations. Also it is worth noticing that in spite of numerous successes, many things still need to be done. For instance, virtually nothing has been done concerning defects (such as oxygen vacancies, dislocations, or grain boundaries) and this constitutes certainly a challenge for the future.

# Ferroelectricity in bulk materials.

## 3.1 Overview.

A transparent discussion of ferroelectric finite-size effects must rely on a preliminary clear understanding of the microscopic origin and main features of the ferroelectric instability at the bulk level. Different relevant questions can be formulated that, historically, have been addressed through distinct models and, recently, were reinvestigated from first-principles.

A first question concerns the global *mechanism* of the structural phase transition. To that respect, the first important step dates back to 1950 and was performed by Slater [156] who suggested that the ferroelectric behavior of $BaTiO_3$ should be caused by long-range dipolar forces which (via the Lorentz local effective field) tend to destabilize the high symmetry configuration favoured by more localized forces. The concept of "rattling" Ti ion was introduced in models that consider motion of the Ti atom in an otherwise rigid lattice. It was a first neat picture, however questionable as all the atoms were actually displaced after the transition. A second breakthrough was reported in 1959 by Cochran [157] who realized that the theory describing the instability should be cast within the framework of lattice dynamics, considering one of the lattice mode as the basic variable. His theory was exhibited in the context of a shell-model approach. The notion of soft-mode was introduced. The competition between short-range and Coulomb forces previously highlighted by Slater reappeared coherently in this framework to explain the origin of the softening of a particular transverse optic mode. Consequently to Cochran's contribution, numerous experimental data were accumulated concerning frequency and temperature dependent properties of the soft-mode [1]. A new step in the microscopic understanding of the ferroelectricity in $ABO_3$ compounds arose from the fit of these experimental data within a shell-model approach. In 1976, Migoni, Bilz and Bäuerle [158] suggested that the ferroelectric instability should originate in a non-linear and anisotropic polarizability of the oxygen atoms. This gave rise to the "polarizability-model" [148, 159] that was widely used to describe the dynamics of $ABO_3$ compounds. The unusual polarizability of the oxygen atom was discussed [158, 148, 160] and related to dynamical hybridization between oxygen p-states and transition metal d-states [158, 148]. It was recently argued that what was considered as an unusual polarizability in the shell-model corresponds in fact to dynamical transfer of charges [122].

A second question concerns the kind of *correlation* between the atomic displacements that is needed to produce the instability. At the end of the sixties, Comes, Lambert and Guinier [161, 162] performed X-ray diffraction measurements on $BaTiO_3$ and $KNbO_3$ crystals. They reported diffuse scattering in three sets of planes normal to the cubic axis. When a scattering is observed outside the directions of diffraction, it must originate in a defect in the crystal periodicity. Clearly, the pattern observed by Comes *et al.* was the fingerprint of a *linear* disorder in real space. To explain their results they proposed what is now usually refered to as the 8-sites model [162]. In this model, it is suggested that the equilibrium position of the Ti atom is not at the center of the cubic unit cell but is slightly displaced along one of the <111> directions. The Ti atom may therefore occupy 8 equivalent positions. In this context, the diffuse scattering was explained by a strong correlation of the Ti ferroelectric displacements along <100>

chains. It was suggested that the correlation should propagate through the subsequent displacement of the O atoms. It was not clear however whether this correlation was static or dynamic [163].

In what follows, we do not report an exhaustive review of DFT studies on bulk ferroelectrics that can be find elsewhere [27, 28, 29]. Instead, we focus on the two specific questions above that are particularly important for the understanding of nanostructures. We highlight how first-principles simulations allowed to achieve a better microscopic understanding of the ferroelectricity in $ABO_3$ compounds. We also briefly discuss the origin of the unusually large functional properties of this class of compounds.

## 3.2   Origin of the ferroelectric instability.

Following the initial idea of Cochran, the ferroelectric instability of $ABO_3$ compounds can be explained by a close competition between short-range and long-range dipolar forces. Going beyond the qualitative picture, first-principles simulations allowed to reinvestigate this model and to provide a coherent and quantitative description, making the bridge between their electronic and dynamical properties.

A first intrinsic property of the ferroelectric $ABO_3$ compounds is the mixed ionic-covalent nature of their bonding. Indeed, in spite of a dominant ionic character, they exhibit well-known covalent interactions of the A and B metal atoms with oxygen. On the one hand, the hybridization between O 2p and B-metal d states is a common feature of this class of compounds. It was already inherent to pioneer LCAO calculations [164, 165, 166] and was further confirmed from first-principles [93]. On the other hand, in most cases, the A atom has a much more ionic character, except Pb that has a significant covalent interaction with oxygen. This makes the Pb-based perovskites such as $PbTiO_3$ different from other typical ferroelectrics such as $BaTiO_3$ or $KNbO_3$.

Table 3.1: Born effective charges of various $ABO_3$ compounds in their cubic structure. The Born effective charges of the A and B atoms are compared to the *nominal* ionic charges $Z_A$ and $Z_B$. The theoretical lattice parameter $a_o$ is also mentioned. (Adapted from Ref. [123])

| $ABO_3$ | $a_o$ | $Z_A^*$ | $Z_B^*$ | $Z_{O\parallel}^*$ | $Z_{O\perp}^*$ | $Z_A^*/Z_A$ | $Z_B^*/Z_B$ | Reference |
|---|---|---|---|---|---|---|---|---|
| nominal | | 2 | 4 | -2 | -2 | | | |
| $CaTiO_3$ | 7.19 | 2.58 | 7.08 | -5.65 | -2.00 | 1.29 | 1.77 | Ref. [101] |
| $SrTiO_3$ | 7.30 | 2.56 | 7.26 | -5.73 | -2.15 | 1.28 | 1.82 | Ref. [123] |
| | | 2.54 | 7.12 | -5.66 | -2.00 | 1.27 | 1.78 | Ref. [101] |
| | | 2.55 | 7.56 | -5.92 | -2.12 | 1.28 | 1.89 | Ref. [121] |
| | | 2.4 | 7.0 | -5.8 | -1.8 | 1.2 | 1.8 | Ref. [167] |
| $BaTiO_3$ | 7.45 | 2.77 | 7.25 | -5.71 | -2.15 | 1.39 | 1.81 | Ref. [123] |
| | | 2.75 | 7.16 | -5.69 | -2.11 | 1.38 | 1.79 | Ref. [101] |
| $BaZrO_3$ | 7.85 | 2.73 | 6.03 | -4.74 | -2.01 | 1.37 | 1.51 | Ref. [101] |
| $PbTiO_3$ | 7.35 | 3.90 | 7.06 | -5.83 | -2.56 | 1.95 | 1.77 | Ref. [101] |
| $PbZrO_3$ | 7.77 | 3.92 | 5.85 | -4.81 | -2.48 | 1.96 | 1.46 | Ref. [101] |
| nominal | | 1 | 5 | -2 | -2 | | | |
| $NaNbO_3$ | 7.40 | 1.13 | 9.11 | -7.01 | -1.61 | 1.13 | 1.82 | Ref. [101] |
| $KNbO_3$ | 7.47 | 0.82 | 9.13 | -6.58 | -1.68 | 0.82 | 1.83 | Ref. [100] |
| | | 1.14 | 9.23 | -7.01 | -1.68 | 1.14 | 1.85 | Ref. [101] |
| | | 1.14 | 9.37 | -6.86 | -1.65 | 1.14 | 1.87 | Ref. [168] |
| nominal | | - | 6 | -2 | -2 | | | |
| $WO_3$ | 7.05 | - | 12.51 | -9.13 | -1.69 | - | 2.09 | Ref. [169] |

A second important and related feature of the $ABO_3$ compounds is the anomalously large value of their Born effective charges. The Born effective charge tensor $Z^*_{\kappa,\beta\alpha}$ of atom $\kappa$ is a dynamical concept. It is defined as the coefficient of proportionality relating, at linear order and under the condition of vanishing macroscopic electric field, the macroscopic polarization created along the direction $\beta$ and the collective displacement along direction $\alpha$ of the atoms belonging to the sublattice $\kappa$ :

$$Z^*_{\kappa,\beta\alpha} = \Omega_0 \frac{\partial \mathcal{P}_\beta}{\partial \tau^{\mathbf{q}=0}_{\kappa,\alpha}} \tag{3.1}$$

Introduced by Born, $Z^*$ is fundamental quantity in the study of the dynamics of polar crystals and monitors for instance the so-called "LO-TO" splitting (for a review concerning $Z^*$ see Ref. [123]). Based on the mixed ionic-covalent character of the bonding in ABO$_3$ compounds (essentially B-O covalency), Harrison [166] suggested that the sensitivity of the hybridizations to the bond length should produce dynamical transfers of electrons when the atoms are displaced, resulting in an anomalous contribution to the Born effective charges $Z^*$. When developing his model, he was however unaware of earlier experimental estimate of $Z^*$ by Axe [167] corroborating his results. The discussion on the unusual amplitude of $Z^*$ in ABO$_3$ crystals came back only recently when accurate predictions were obtained from first principles [100, 170, 101, 169] (see Table 3.1). The computations confirmed that the Born effective charges of these compounds are anomalously large in the sense that their amplitude can reach close to twice the value of the nominal ionic charges $Z_\kappa$. For instance, as first reported by Resta $et$ $al.$ [100], the effective charge on the Nb atom in cubic KNbO$_3$ is equal to $+9.13e$, while that on the oxygen is highly anisotropic and equal to $-6.58e$ or $-1.68e$, depending if the O atom is displaced along the Nb-O direction or perpendicularly to it. Similar anomalous values of $Z^*_B$ and $Z^*_O$ were reported for various ABO$_3$ compounds [101, 170] (see Table 3.1). They were explicitly related to the hybridizations between O 2p and B-metal d states [171, 95, 123]. In most cases, $Z^*_A$ is close to its nominal value, except for Pb that has more covalent interactions with oxygen (see Table 3.1). The sensitivity of $Z^*$ to structural features was also investigated. It was shown that the amplitude of the anomalous contribution usually decreases in the ferroelectric phases [95, 119].

Table 3.2: Frequencies (cm$^{-1}$) and mode effective charges ($\|e\|$) of IR-active transverse optical (TO) phonon modes of various perovskite oxides as computed by Zhong $et$ $al.$ in Ref. [101]. The mode effective charge has been defined there as $\bar{Z}^*_j = \sum_m M_m^{1/2} Z^*_m \xi^{TO}_{jm}$ where $\xi$ is the mode eigenvector. (Adapted from Ref. [101])

|  | TO1 | | TO2 | | TO3 | |
|---|---|---|---|---|---|---|
|  | $\omega$ | $\bar{Z}^*$ | $\omega$ | $\bar{Z}^*$ | $\omega$ | $\bar{Z}^*$ |
| BaTiO$_3$ | $178i$ | 8.95 | 177 | 1.69 | 468 | 1.37 |
| SrTiO$_3$ | $41i$ | 7.37 | 165 | 3.22 | 546 | 3.43 |
| CaTiO$_3$ | $153i$ | 6.25 | 188 | 4.94 | 610 | 4.50 |
| KNbO$_3$ | $143i$ | 8.58 | 188 | 1.70 | 506 | 4.15 |
| NaNbO$_3$ | $152i$ | 6.95 | 115 | 2.32 | 556 | 5.21 |
| PbTiO$_3$ | $144i$ | 7.58 | 121 | 4.23 | 497 | 3.21 |
| PbZrO$_3$ | $131i$ | 4.83 | 63 | 4.86 | 568 | 4.30 |
| BaZrO$_3$ | 95 | 5.57 | 193 | 5.57 | 514 | 3.84 |

A third common feature of ferroelectric perovskites is the existence, in their cubic phase, of an unstable phonon transverse optic (TO) mode at the $\Gamma$ point, coherently with their typical double-well energy (Fig. 3.1). First-principles provide direct access to phonon frequencies and properly reproduces this features. Calculations of phonon frequencies at the $\Gamma$ point were reported for various perovskites, providing results in close agreement with the experimental data [101, 172]. They highlight the existence of a $unstable$ TO mode, hereafter also referred to as the $soft$ mode, that is highly polar and exhibits a giant LO-TO splitting [101, 173] (see Table 3.2). Within the harmonic approximation, the unstable mode appears with an imaginary frequency $\omega$, related to the negative curvature at the origin ($\omega^2 < 0$) of the total energy surface along the line of atomic displacements associated with the ferroelectric mode [1]. The amplitude of the unstable frequency does not directly provide the double-well depth (that additionally depends on anharmonic effects) but measures to some extent the strength of the instability. On top of the frequency of the soft mode, the calculations also provide the related eigen-displacement vector or in other words the pattern of atomic displacements that is able to decrease the energy when it is frozen in the structure. This pattern of displacement is directly related to the ionic degree of freedom $\xi$ included in the model Hamiltonian.

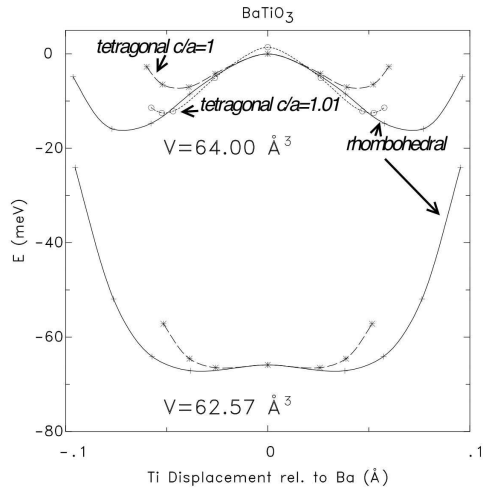

Figure 3.1: Energy as a function of soft-mode distortion $\xi$ in cubic phase of BaTiO$_3$, from first-principles DFT calculations [93]. Tetragonal (dashed line) and rhombohedral (black line) curves respectively correspond to $\xi$ frozen along the [001] and [111] directions respectively. Calculations have been performed at the experimental volume ($V = 64\text{\AA}$) as well as with an isotropic compressive strain of 0.75% ($V = 62.57\text{\AA}$) and with a [001] uniaxial tensile strain of 1% (c/a=1.01). In all cases, the energy has a typical double-well shape. The negative curvature of the energy at the origin ($\xi = 0$) is directly related to the imaginary frequency of the soft mode. The well-depth is larger for the rhombohedral distortion, in agreement with the rhombohedral ground-state of BaTiO$_3$, but the curvature at the origin (frequency of the soft mode) is the same for the tetragonal and rhombohedral distortions in agreement with the cubic symmetry. The calculations emphasize the strong sensitivity of the ferroelectric instability to strain (strong polarization-strain coupling) : (i) isotropic pressure decreases the well-depth and the ferroelectric instability is nearly suppressed for an underestimate of the experimental volume of the order of the usual LDA error; (ii) uniaxial tensile strain enhances the well-depth and the ferroelectric instability. (From Ref. [93] by Cohen)

Combining the previous results, it was also possible to investigate from first-principles the balance of forces that, as suggested by Slater and reformulated by Cochran, is at the origin of the ferroelectric instability [172, 173]. Within Cochran's model, the partial contributions from short-range (SR) forces and long-range dipolar (DD) forces to the frequencies of the transverse modes were separated from each others ($\omega^2 = \omega_{SR}^2 + \omega_{DD}^2$) and the structural instability ($\omega^2 < 0$) was attributed to the possible *cancellation* of the two terms. The different quantities involved in Cochran's model are directly accessible from first-principles calculations, offering the possibility for validating this explanation. In BaTiO$_3$, the decomposition confirms that the instability of the ferroelectric mode ($\omega = 113i$ cm$^{-1}$) [2] is induced by the compensation of stabilizing short-range forces ($\omega_{SR} = 783$ cm$^{-1}$) by a larger destabilizing dipolar interaction ($\omega_{DD} = 791i$ cm$^{-1}$) [172]. The unusually large value of the latter is directly related to the anomalous value of $Z^*$. The close competition between forces is a specific feature of the unstable TO mode (that is highly polar, see Table 3.2) and is not observed for the other modes. It is also worth noticing for further discussions that the instability results from a very *delicate* balance between SR and DD forces : it is observed to be very sensitive to small changes such as slight modifications of the volume or of the Born effective charges [172]. This explains why the strength of the instability can vary from one perovskite to another and why some of them are not even ferroelectric.

---

[1]In finite temperature simulations, such unstable modes appear as anharmonically stabilized soft modes [135]

[2]This frequency reported by Ghosez *et al.* [172] at the theoretical volume, slightly differs from the value of Zhong *et*

The main features of ferroelectric perovskites highlighted above provide a coherent picture of the interplay between electronic and dynamical properties in generating the ferroelectric instability. The origin of the instability can be summarized as follows [174]. Because of the mixed ionic-covalent character of their bonding, these materials exhibit unusual transfers of charge when the atoms are displaced, that result in anomalous effective charges. In turns, these charges are responsible for a giant Coulomb interaction, able to compensate the short-range forces and to produce a structural ferroelectric instability. This picture is in line with the initial thoughts of Cochran. It is currently accepted for the ferroelectric perovskites and also applies to trigonal ferroelectric such as LiNbO$_3$ [6]. It is however not universal and in manganites that may combine ferromagnetic and ferroelectric order, it was proposed that ferroelectricity has a different origin [175, 176].

The existence of an unstable phonon mode at $\Gamma$ and the related negative curvature at the origin of the total energy surface for a given polar pattern of atomic displacements are the *fingerprint* of the ferroelectric instability. The search for these characteristics offers therefore a practical way to check for the existence and survival of ferroelectricity in nanostructures.

## 3.3 A collective phenomenon.

It is commonly accepted that ferroelectricity is a collective phenomenon [1]. For instance, it can be checked that an isolated atomic displacement in bulk cubic perovskite will never decrease the energy of the system. When an atom is individually displaced from its cubic perovskite position, it feels a force that brings it back in its initial position (Table 3.3). The cubic structure is only unstable against specific collective displacements. An important question concerns therefore the kind of atomic correlation needed to produce the ferroelectric phase transition.

Table 3.3: Self interatomic force constant $C_{\alpha,\alpha}(l\kappa, l\kappa)$ (Ha/bohr$^2$) for the different atoms in the cubic unit cell of various perovskite compounds [124]. When atom $\kappa$ is individually displaced by $\Delta\tau_\alpha(l\kappa)$ along direction $\alpha$ in unit cell $l$, it experiences a force $F_\alpha(l\kappa) = -C_{\alpha,\alpha}(l\kappa, l\kappa).\Delta\tau_\alpha(l\kappa)$. The fact that all the values are *positive* means that the force is always opposed to the displacement and attests for the absence of instability for *isolated* atomic displacements. (Adapted from Ref. [124] by Ghosez *et al.*)

| Atom | Position | Direction | BaTiO$_3$ | PbTiO$_3$ | PbZrO$_3$ |
|------|----------|-----------|-----------|-----------|-----------|
| A | (0, 0, 0) | x=y=z | +0.0806 | +0.0247 | +0.0129 |
| B | (0.5, 0.5, 0.5) | x=y=z | +0.1522 | +0.1393 | +0.2302 |
| O$_1$ | (0.5, 0.5, 0) | x=y | +0.0681 | +0.0451 | +0.0166 |
| | | z | +0.1274 | +0.1518 | +0.2758 |

The first-principles approach offers the opportunity to address this question. Calculations of the $\Gamma$ phonons correctly reproduces the presence of an instability. The full phonon dispersion curves have been computed for various ABO$_3$ compounds (see Fig. 3.2) and their inspection allow to get new insight into the type of atomic correlation associated to the ferroelectric instability.

*al.* [101] reported in Table 3.2. Nevertheless, both frequencies agree with the existence of an instability and are of the same order of magnitude. The origin of the small discrepancy might be found in the different pseudopotentials and LDA

Figure 3.2: Calculated phonon dispersion relations of BaTiO$_3$, PbTiO$_3$ and PbZrO$_3$ along various high symmetry lines in the simple cubic Brillouin zone. High symmetry points of the cubic Brillouin zone are labeled as follows (in reduced coordinates): $\Gamma \equiv (0,0,0)$, $X \equiv (0.5,0,0)$, $M \equiv (0.5,0.5,0)$ and $R \equiv (0.5,0.5,0.5)$. (From Ref. [124] by Ghosez *et al.*)

In BaTiO$_3$ (Fig. 3.2), the ferroelectric unstable phonon mode at $\Gamma$ remains unstable at the $X$ and $M$ points. The mode shows essentially no dispersion within the $\Gamma - X - M$ plane but rapidly stabilizes when going to the $R$-point. This behavior is the fingerprint of a linear instability as it was first reported for KNbO$_3$ [118]. This more clearly appears in Fig. 3.3 where we have plotted the zero-frequency isosurface limiting the region of the Brillouin zone in which there is an unstable mode. It is shown that the unstable phonon modes of BaTiO$_3$ are localized in three perpendicular interpenetrating slab-like regions of the Brillouin zone containing the $\Gamma$ point [122]. The planar character of the unstable region in reciprocal space must be related to a *linear* correlation of the atomic displacements in real space. Moreover, the finite thickness of the unstable region allows to estimate the length of the smallest unstable chain to $\approx$ 4-5 unit cells. In other words, an instability only appears when Ti and O displacements along Ti–O chains are correlated over at least 4-5 unit cells along the direction of the mode polarization.

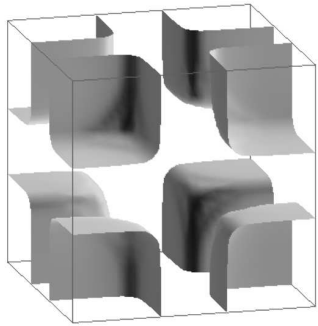

Figure 3.3: Zero-frequency isosurface of the lowest unstable phonon branch over the Brillouin zone. $\Gamma$ is located at the center of the cube, $X$ at the center of each face, $M$ at the middle of each line and $R$ at the corners of the cube. The mode is unstable in the three perpendicular planar regions (containing $\Gamma$, $X$ and $M$) defined by the nearly flat surfaces. (From Ref. [122] by Ghosez *et al.*)

In agreement with this picture, it was observed [122] that both short-range and dipolar parts of the interatomic force constants (IFC) in real space are highly anisotropic in BaTiO$_3$: they are associated to a much stronger correlation of the atomic displacements along the Ti–O chains, coherent with the partial covalent character of the bonding along these directions. From the analysis of the IFC in real space, it was pointed out that the cooperative displacements of Ti and O along a single isolated Ti–O chain of 10-12 atoms ($\approx$ 5 unit cells) is enough to induce the ferroelectric instability, coherently with the estimate made from the inspection of the unstable region in reciprocal space. Also, it is worth noticing that both the analysis of the IFC and the absence of dispersion in the $\Gamma - X - M$ plane point out that the different Ti–O chains behave essentially independently.

The previous findings can be related to the experimental measurements of Comes *et al.* [161]. From molecular dynamics simulations based on a first-principles model Hamiltonian, Krakauer *et al.* [135] demonstrated that local polar distortions with short-range chain-like correlation in KNbO$_3$ and BaTiO$_3$ are compatible with the diffuse x-ray scattering patterns observed by Comes *et al.* [161]. Clarifying an old debate, they conclusively showed that the linear disorder is not static as described in the 8-sites model, but has a dynamic origin as previously proposed by Hüller [163]. The local distortions suggest an order-disorder character for the transition. However, the molecular dynamics simulations also points out the softening of optical phonon branches in the cubic phase, not only at $\Gamma$, suggesting a displacive character for the transition, but also over large region of the Brillouin zone. This provided a unified framework

---

functional used in these calculations.

for understanding the combined displacive and order-disorder characteristics of the ferroelectric phase transition [135] .

Let us emphasize that $BaTiO_3$ and $KNbO_3$ are prototype examples but that other compounds can exhibit a more complex behavior (Fig. 3.3). In $PbTiO_3$, due to the fact that Pb is much strongly involved than other A atoms, the instability is more isotropic [124]. Also, other instabilities can be present such as the antiferrodistortive instability (rotation of the oxygen octahedra) that has a two-dimensional character in real space. In $SrTiO_3$ it competes with the ferroelectric instability that is consecutively confined in a much more reduced region of the Brillouin zone [121]. In $PbZrO_3$, there are unstable modes all over the Brillouin zone and the analysis is much more complex (Fig. 3.2).

## 3.4   Functional properties.

Ferroelectric oxides exhibit various important functional properties. Beyond their switchable spontaneous polarization, they present unusually large dielectric, piezoelectric and non-linear optical coefficients, the amplitude of which are directly related to their ferroelectric character.

Taking into account electronic and ionic contributions, the static dielectric constant takes the form [63]

$$\epsilon^0_{\alpha\beta} = \epsilon^\infty_{\alpha\beta} + \frac{4\pi}{\Omega_0} \sum_m \frac{p_{m\alpha} \cdot p_{m\beta}}{\omega_m^2}, \tag{3.2}$$

where $\epsilon^\infty$ is the optical dielectric constant. The sum in the second term is performed on the different TO phonon modes $m$ of frequency $\omega_m$ and $p_{m\alpha}$ is the mode polarity defined as ($U^m$ are the phonon eigendisplacements)

$$p_{m\alpha} = \sum_{\kappa\beta} Z^*_{\kappa,\alpha\beta} U^m_{\kappa\beta}. \tag{3.3}$$

From the previous formula, it clearly appears that materials with TO modes combining a high polarity and a low frequency will exhibit a large dielectric response. As previously emphasized, in perovskite oxides, the soft mode of the cubic phase is highly polar. In the ferroelectric phases, it gives rise to a highly polar mode of low frequency that can strongly couple with an electric field and is mainly responsible for the huge dielectric coefficients.

In the same way, the clamped electro-optic tensor can be decomposed into electronic and ionic contributions [64]

$$r_{ij\gamma} = \frac{-8\pi}{n_i^2 n_j^2} \chi^{\infty(2)}_{ijk}\bigg|_{k=\gamma} - \frac{4\pi}{\sqrt{\Omega_0} n_i^2 n_j^2} \sum_m \frac{\alpha^m_{ij} p_{m\gamma}}{\omega_m^2}, \tag{3.4}$$

where $\chi^{\infty(2)}$ is the non-linear optical susceptibility, $n_i$ are the principal refractive indices, and $\alpha^m$ is the Raman susceptibility of mode $m$. Here again, the successor of the soft mode in the ferroelectric phases can give rise to a giant contribution if, on top of a low frequency and a large polarity it also exhibits a significant Raman activity [116]. This is for instance the case in the tetragonal phase of $BaTiO_3$.

The piezoelectric coefficient can also be decomposed into a clamped-ion part and a contribution arising from ionic relaxations

$$e_{\alpha,\mu\nu} = e^0_{\alpha,\mu\nu} - \frac{1}{\Omega_0} \sum_m \frac{p_{m\alpha} \cdot g_{m\mu\nu}}{\omega_m^2}, \tag{3.5}$$

where $g_{m\mu\nu} = \sum_{\kappa\beta} \gamma^{\mu\nu}_{\kappa\beta} U^m_{\kappa\beta}$ and depends on the internal strain parameter $\gamma$ (Sec 2.2.8). In the same spirit as for dielectric and electro-optic coefficients, the formula teaches us that the ionic contribution to the piezoelectric coefficient can be larger if, on top of a low frequency and a large polarity, the successor of the soft mode in the ferroelectric phase exhibit strong coupling with strain. The problem of a proper definition of the piezoelectric tensor in crystal with a spontaneous polarization has been discussed by Saghi-Szabo [102, 103] and also by Vanderbilt [177]. The piezoelectric properties of ferroelectric oxides are probably those that have been the most widely studied because of their current interest for technological applications. A recent review by L. Bellaiche can be found in Ref. [28]. Let us simply emphasize here that, as a main result obtained recently by Fu and Cohen [111], high piezoelectric coupling in ferroelectric perovskites can be related to the ease with which the polarization can rotate in presence of an applied electric field. In many cases, the existence of intermediate phases in the phase diagram in which the polarization can rotate easily (such as the monoclinic phase making the bridge between the tetragonal and rhombohedral phases around the morphotropic phase boundary in PZT) appears as an important feature to exhibit unusually high piezoelectric responses [77].

The previous functional properties are directly accessible from DFT calculations (Sec. 2.2.8). These calculations provide some estimate of the intrinsic properties of the compound but are nevertheless restricted to zero Kelvin. In practice, the functional properties strongly evolves with temperature and can diverge around the phase transition temperature (from the previous equations, they diverge when the frequency of the soft-mode is going to zero). Meaningful comparison with the experiment requires therefore to access them at finite temperatures. As previously discussed (Sec. 2.3.2), this can be achieved with the effective Hamiltonian. In that case, the response is limited to the soft-mode contribution (i.e. the only polar ionic degree of freedom within $\mathcal{H}^{eff}$) that is expected to dominate.

## 3.5 Stress and strain effects.

The ferroelectric instability of perovskite compounds is dependent on the volume and more generally strongly affected by strains and stresses. This is known at the experimental level. It was also first highlighted by Cohen [93] in the first-principles context (see Fig. 3.1).

When the volume is reduced under hydrostatic pressure, the ferroelectric instability is progressively suppressed [93]. In reference [172], this feature was related for $BaTiO_3$ to a modification of the balance between SR and DD interactions: in compressed cubic $BaTiO_3$, the Born effective charges remain the same as at the optimized volume, the destabilizing role of the DD interaction slightly increases and the suppression of the ferroelectric instability must be assigned to a strong increase of the stabilizing SR forces. At the opposite, a negative hydrostatic pressure can in some cases induce a large tetragonal strain and an increase of the spontaneous polarization as it was pointed out for $PbTiO_3$ by Tinte *et al.* [178].

The range of stability of the ferroelectric phases and the amplitude of the spontaneous polarization of $ABO_3$ compounds can be strongly affected by strains. Defects can produce inhomogeneous strain and, in that way, affect the ferroelectric behavior. Also, epitaxial strains play a major role. They will induce new phase and affect the phase transition temperature. In-plane strain phase diagrams have been reported for selected perovskites. This is discussed in Sec. 4.3.

Because ferroelectric and functional properties are linked to each others, strains offer a way to tune dielectric, piezoelectric and optical properties. Overall, the dielectric and piezoelectric responses should be reduced by clamping to a substrate. However, as functional properties strongly evolves around the phase transition temperature, an appropriate strain-induced shift of $T_c$ can dramatically enhance the properties at a given temperature of interest.

## 3.6 Expected behavior of nanostructures.

The origin of the ferroelectric instability at the bulk level already allows to anticipate the existence of size effects in ferroelectric nanostructures. It was highlighted that the ferroelectric instability results from a delicate balance between SR and DD interactions. Both will be modified in nanostructures. The SR forces will be modified at surface and interfaces. The same is true for the Born effective charges that cannot keep their bulk value as highlighted by Ruini, Resta and Baroni [179] and illustrated on $BaTiO_3$ thin films by Fu [49]. Beyond that, the DD interaction that has a long-range character will be affected by the finite size of the sample and will be strongly dependent on the electrical boundary conditions. As stated above, the ferroelectric instability is also strongly sensitive to strain and will be influenced by mechanical boundary conditions such as epitaxial strains. These different factors can act independently to either enhance or suppress ferroelectricity. They will compete with each other in such a way that it is difficult to predict the ferroelectric properties of nanostructures without making explicit calculations.

# Ferroelectric ultrathin films.

## 4.1 Overview.

There is no doubt that the understanding of finite-size effects in ferroelectric thin films is nowadays a very hot and challenging topic of research [180]. As previously discussed (Sec. 3.3), ferroelectricity is a *collective* effect. As such, it was thought for a long time that it is suppressed in small particles and thin films below a critical size of about 100 Å, so preventing the use of ferroelectric materials in devices smaller than this threshold, and reducing a priori their impact in the technology of the future. However, recent improvements in the synthesis and characterization of ferroelectric oxide thin films with a control at the atomic scale, have allowed the observation of ferroelectricity well below the long-accepted critical thickness of 10 nm. This brougth the field to a high level of excitement and a huge research activity was recently devoted to the understanding of ferroelectricity in ultrathin films [180, 16].

The investigations in the field are also pushed by the possible use of ferroelectric materials, particularly thin films, into various microelectronic devices that take advantage of their multifunctional properties as illustrated in Fig. 4.1 [181]. The existence of a switchable spontaneous polarization is at the basis of the design of non-volatile ferroelectric random access memories (FERAMs), where one bit of information can be stored by assigning one value of the Boolean algebra ("1" or "0") to each of the possible polarization states. These memories have the advantage of being non-volatile (i.e. of retaining the stored values in the absence of any refreshing current) [182]. The high dielectric permittivity of ferroelectrics (and incipient ferroelectrics) makes them possible candidates to replace silica as the gate dielectric in metal-oxide-semiconductor field-effect-transistors (MOSFETs). Their piezoelectric behaviour enables them to convert mechanical energy in electrical energy and vice versa, as required in transducers such as sonar detectors or piezoelectric actuators (including those required in inkjet printers and video-cassette-recording head positioning). Their pyroelectric properties are the basis for highly sensitive infrared room temperature detectors. Finally, their non-linear optical properties make them useful materials for optical devices such as electro-optic modulators.

The ongoing miniaturization of microelectronic devices, imposed by the semiconductor industry, raises the question of possible size-effects on the properties of the active components [183]. Except for the case of the gate dielectric problem in MOSFETs, the thickness of the films used in contemporary applications (of the order of 1200 Å) is still far away from the thickness-range where size effects become highly noticeable, so that the question at this moment is merely academic. Nevertheless, it is possible that the fundamental limits of materials might be reached in the future.

A lot of efforts have been devoted to ferroelectric finite-size effects during the last few years (Fig. 4.2), both experimentally and theoretically (for recent reviews on the literature, see [180] and [16]), and many issues have already been sorted out. Nevertheless, many questions still remain open and are the subject of interesting debates in the literature. The main reason for the poor understanding of some of the size-effects on ferroelectricity is the vast amount of different effects that compete and might modify the delicate balance between the long-range dipole-dipole (DD) electrostatic interactions and the short-range (SR) forces, whose subtle equilibrium is known to be at the origin of the ferroelectric instability (see Sec. 3.2).

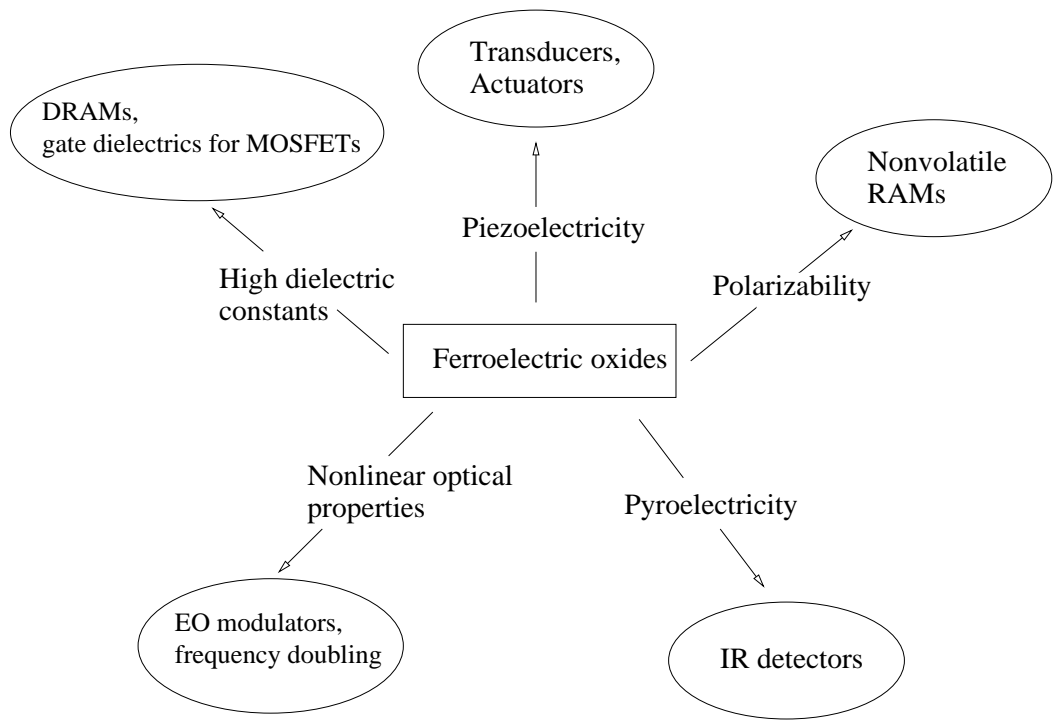

Figure 4.1: Most important functional properties of ferroelectric oxide and related technological devices (Courtesy of M. Veithen. Adapted from Ref. [181].)

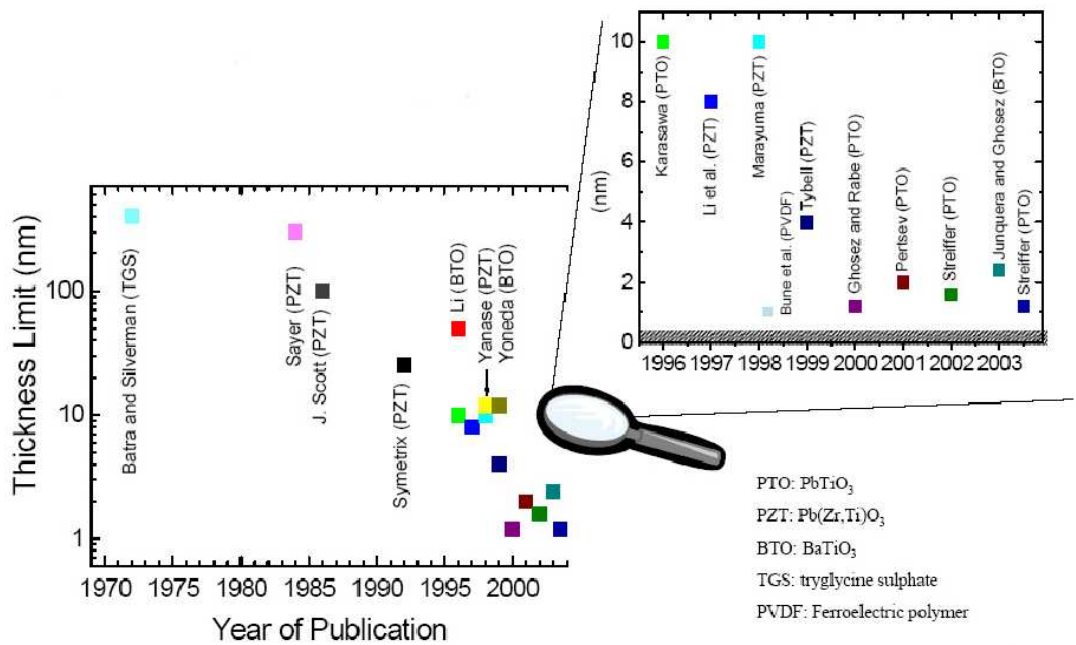

Figure 4.2: Minimum critical thickness (either theoretically or experimentally) versus publication year. The value of the the critical thickness decreases with time, as a consequence of the recent breakthroughs on the synthesis of complex oxides. Note the change of scale from logarithmic in the left to linear in the right. Updated from Fig. 1 of Ref. [184]. Courtesy of H. Kohlstedt.

Some of the most important effects to take into account while dealing with the thickness dependence of ferroelectric properties in thin films, together with some questions to be clarified from first-principles, can be classified as follows.

1. *Surface effects.* In many ferroelectric nanostructures one or more faces of the ferroelectric material are free surfaces (not covered by any kind of electrode or other capping material). At the surface, where some of the interactions are missing due to the lack of neighboring atoms, the hybridizations between Ti $3d$ and O $2p$ orbitals, directly related to the emergence of ferroelectricity in bulk, are affected so that the SR and DD forces can be significantly modified. The smaller the nanostructure, the more important the surface effects due to the increasing surface/volume ratio. For an understanding of the behaviour of ferroelectric thin films it is therefore important to explore how the ferroelectric order parameter couples with the surface in order to answer the following questions: Is the degradation of the ferroelectric properties sometimes observed in ferroelectric particles and thin films an *intrinsic effect* caused only by the presence of the surface, or is it monitored by other *extrinsic effects*? How does the polarization couple with the relaxations and the eventual reconstruction of the surface? What is the influence of the surface termination on the evolution of the ferroelectric properties?

2. *Mechanical effects.* Ferroelectric thin films are usually epitaxially grown on top of a thick substrate, that imposes its in-plane lattice constant and symmetry. The epitaxial strain resulting from the lattice mismatch between the film and the substrate is known to strongly couple with the ferroelectric polarization. We can wonder whether (i) new phases, not present in the bulk phase diagram can be epitaxially stabilized, (ii) strain-engineering is possible in order to tune specific properties to desired values, or (iii) misfit dislocations that appear in thin films to relief the strain energy will affect the ferroelectric properties [185, 186, 187]

3. *Electrostatic effects.* A finite polarization normal to the surface will give rise to a depolarizing field in ferroelectric films, and to a consequent huge electrostatic energy, able to suppress the ferroelectric instability. To preserve ferroelectricity the depolarizing field must be screened, either by free charges coming from metallic electrodes or by breaking into a domain structure. On the one hand, it is supposed that the level of screening that can be achieved changes substantially using different electrodes [188]. Does this difference come only from the density of free carriers in the electrodes, or is there any influence of the chemical environment at the interface? At the microscopic level, the polarization charge might be screened in part within the ferroelectric material due to the interpenetration of electrode and ferroelectric wavefunctions (metallic eigenstates decay exponentially within the ferroelectric and vice versa). To what extent does this interpenetration depend on the bonding and hybridization of states at the ferroelectric/electrode interface? Other quantities, as the electrostatic barriers at the interface (Schottky barriers) depend on the amount of charge transferred from one side to the other after the heterostructure has been grown, and this transfer is directly related to the chemical bonding at the interface. On the other hand, monodomain polarizations have been observed in ferroelectric capacitors uncovered by a top electrode. Which mechanism is providing the screening of in such situations? Also, what is the ultimate lateral size of the domains when they are observed, and do they form independently of the temperature?

4. *Finite conductivity.* Ferroelectric materials are a priori insulators but finite conductivity is often reported in ferroelectric thin films. To which extent does this finite conductivity interfere with the ferroelectric properties? Also, the leakage current mechanism [189] changes depending on the kind of interface, from the interface-limited Schottky emission in *blocking* contacts (typically produced in noble-metal/ferroelectric-perovskite interfaces [190, 191]), to the bulk limited Poole-Frenkel emission in *non-blocking* contacts (observed mainly with conducting perovskite electrodes [190, 191, 192, 193, 194]). The former shows long-term leakage current and poor fatigue and imprint properties while the latter has higher leakage currents, but solves the fatigue and imprint problem. An atomistic explanation for such behaviour is still missing.

5. *Defects.* Extrinsic effects, such as impurities, oxygen vacancies and other defects in the chemical composition of the surface, play also a critical role [195, 196]. Vacancies act as donors of charge carries after ionization. That explains the final conductivity of bulk perovskite oxides, considered sometimes as semiconducting materials despite the large band gap (above 3 eV) [197]. But, can these internal charges act as an internal source to produce further screening of the depolarizing field? Also, what is the influence of local strains in the vicinity of defects on the ferroelectric behavior?

Experimentally, determination of the influence of all these individual factors on the ferroelectricy is difficult since their effects are intertwined: the measured properties of real thin films are *global* results, combining all the different effects. In this context, first-principles calculations appear as a valuable tool, allowing to discriminate between the different effects and to isolate their respective influence on the ferroelectric properties.

Good progresses have so been reported in the last few years that mainly concern (i) the understanding of surfaces in free-standing slabs; (ii) the role of the mechanical boundary conditions (polarization strain coupling); and (iii) the influence of the electrical boundary conditions and screening. We will devote the three next Sections to review some important works concerning these three topics. It is worth noticing that other topics, such as the discussion of the role of defects and free carriers, are also of primary importance. Nevertheless, they are beyond the scope of current computational capabilities and will not be further discussed here.

## 4.2 Free surfaces.

Historically, pioneer first-principles works on $ABO_3$ perovskite oxides thin films considered isolated slabs in vacuum with free surfaces. From a fundamental point of view, these calculations were a perfect scenario to ascertain whether the mere presence of a surface is enough to suppress the spontaneous polarization or, in other words, to assign a completely intrinsic origin to the degradation of the ferroelectric properties observed experimentally in small particles and thin films [198, 199, 200, 201]. From an applied point of view, the interest came mainly from the catalytic properties of $SrTiO_3$ and $BaTiO_3$ surfaces, and the common use of $SrTiO_3$ as a substrate for the epitaxial growth of other oxides and high-$T_c$ superconductors [202].

At any surface, the translational symmetry is broken: the surface can remain periodic *in the plane*, but the crystal is no more periodic in the direction *perpendicular* to the surface. The atoms in the outermost layers (those which are closer to the surface) will feel the absence of neighboring atoms to interact with. These missing bonds affect the structural and electronic properties of the materials. In the particular case of $ABO_3$ perovskite oxides, the hybridization between Ti $3d$ and O $2p$ orbitals, responsible for the giant effective charges and, ultimately, for the ferroelectric instability is modified (see Sec. 3) The absence of some of the atoms leaves highly energetic and electronically active dangling bonds at the surface. In most cases this leads to structural relaxations, and even reconstructions of the surface geometry to produce a more satisfactory bonding configuration, that might affect significantly the physical properties of the ferroelectric slab. The energy scale of the surface relaxations and reconstructions is orders of magnitude higher than the depth of the bulk ferroelectric double well, so it is expected to affect the ferroelectric ground state of the system.

The determination of electronic and structural properties is a delicate problem, that reveals strongly dependent on the nature of the material. Only changing one of the cations in a $ABO_3$ perovskite surface can lead to very different qualitative and quantitative behaviour. Even more, for a given material the results change also with the particular orientation and atomic termination of the surface. Therefore, individual atomistic simulations using parametrized shell-models [203, 204, 152], model hamiltonians [31] or fully first-principles calculations (using a large variety of methods from LAPW [205, 206], plane-waves with ultrasoft-pseudopotentials [207, 208, 209, 210], plane-waves with norm-conserving pseudopotentials [211, 212], or localized atomic orbitals [49, 213, 91]) are required for an accurate prediction of the properties of each specific geometry and compound.

The basic geometry to study surface properties of materials is that of a monocrystalline planar slab, made of some layers of the material under study surrounded by vacuum. Each slab has two surfaces, and it might be symmetrically (both termination equivalent by symmetry), or asymmetrically-terminated. The thickness of the slab, controlled by the number of material layers, should be large enough to avoid surface-surface interactions.

Such an *isolated slab* seems the ideal geometry to isolate the surface-related physical properties. However, most of the *ab-initio* schemes used to study surface properties are typically based on plane-wave computational methods and other techniques that rely on the fast fourier transform (FFT) algorithm [214]. These methodologies assume periodic boundary conditions in all three directions, so the supercell containing both the slab and the vacuum previously described is periodically repeated in space (see Fig. 2.1) [25]. This approach generate two major problems, not present for isolated slabs. First, there are interactions between periodic replicas of the slab, even for symmetric non-polar slabs. Second, when considering systems with different value of the electrostatic potentials on the two sides of the slab (such as asymmetric slabs in which the two surfaces have different work functions, or slabs with a non-vanishing polarization perpendicular to the surface, that gives rise to a surface charge density of different sign at each surface), the appearance of a fictitious electric field is required to enforce the periodic boundary conditions on the potential at the level of the whole supercell. The self-consistency on the charge density is achieved under the presence of this unphysical field (the amplitude of which moreover depends on the thickness of the vacuum region) that influences the atomic relaxations, energetics and electronic band structures. A special treatment, developed by Neugebauer and Scheffler [215] and lately refined by Bengtsson [216] is required to eliminate its effects (see Sec. 4.4). Both problems might be reduced by enlarging the amount of vacuum in the supercell, although the price to pay is the increase of computational burden (except when a localized basis set is used [49, 91]).

In the first-principles study of thin films or heterostructures, the first step is to choose an initial symmetry for the simulations and to define a *reference ionic configuration* that is obtained by cutting the bulk material along a particular direction. The reference configuration to which we will refer when discussing surface atomic relaxation is therefore the atomic arrangement of an unrelaxed truncated bulk structure. Usually, the bulk structure that is truncated is the theoretically relaxed structure. In some cases, however, in order to include the effect of the epitaxial strain induced by a substrate, the reference configuration is defined by cutting a strained bulk phase, with the same strain contraints than those expected in the thin film.

The three most important orientations for $ABO_3$ surfaces (and also those which have been most investigated theoretically) are (001), (110) [203, 217], and (111) [205, 206]. In the (001) cut, the $ABO_3$ perovskite structure can be considered as an alternating stack of $AO$ and $BO_2$ layers. For II-IV perovskites, where atoms A and B are divalent and tetravalent respectively, in the ionic limit, the structure is a sequence of *neutral* sheets $\left(A^{2+}O^{2-}\right)^0$ and $\left[Ti^{4+}\left(O^{2-}\right)_2\right]^0$. The surface might have two possible different terminations, with the outermost layer being a AO layer or a $TiO_2$ layer. By contrast, both the (110) and the (111) orientations are *polar* irrespectively of the cation valence states. In the (110) cut, the perovskite structure is composed of $(ABO)^{4+}$ and $(O_2)^{4-}$ stacks, whereas the atomic planes in the (111) cut are $AO_3$ and B, and they are both charged. The highly polar nature of the (110) and (111) cuts makes them unstable and highly reactive in comparison to the (001) surface [218]. In this review we will focus on the (001) orientation of the II-IV perovskite structures. A review on the polar surfaces of complex oxides has been recently reported by Noguera *et al.* [219].

If the original bulk from which the reference structure is cleaved had a perfect cubic symmetry, then the resulting "cubic" surface would display mirror symmetry planes $M_x$, $M_y$ and, if the slab is symmetrically terminated, $M_z$ (this last one relative to the central layer of the slab) [1]. A tetragonal reference structure, more suitable to study the ground state at room temperature, can also be constructed for BaTiO$_3$ and PbTiO$_3$ by cleaving the bulk tetragonal material. As it will be discussed in Sec. 4.4, a polarization normal to the surface is strongly suppressed by the depolarization field in free slabs in vacuum. Assuming that the ferroelectric axis lies parallel to the surface along the $x$-axis, then the $M_x$ symmetry is broken in the tetragonal reference configuration, whereas the system preserves the $M_y$ symmetry and, additionally for symmetric slabs, the $M_z$ symmetry with respect to the center of the slab.

---

[1]To set up the notation, the plane parallel to the surface will be referred to as the $(x, y)$ plane while the perpendicular direction will be denoted as the $z$-axis.

Bulk properties are recovered a few atomic layers away from the outermost surface layers. This allows the use of thin slabs of the order of ten layers thick.

Typically $(1 \times 1)$ surface geometries are simulated. This constraint allows atomic relaxations but prevents the study of more complicated reconstructions involving for instance the rotation of oxygen octahedra, essential to describe the antiferrodistortive instability that appears in bulk SrTiO$_3$ under 105 K, and observed in PbTiO$_3$ surfaces [220]. Calculations relaxing this constraint will be presented in Sec. 4.2.4.

### 4.2.1 Structural relaxations.

As it has been already mentioned before, the existence of highly energetics dangling bonds at the surface yields non-vanishing forces on the atoms in the layers close to the surface. Therefore atomic rearrangements (including relaxations and reconstructions) occur at the surface in order to strengthen the remaining bonds, and to minimize the effects of the lack of hybridization with the missing atoms.

Starting from the truncated bulk reference ionic configuration previously defined, a relaxation of the atomic coordinates, while preserving the full set of symmetries, is carried out until the maximum component of the force on any atom is smaller than a given threshold, of the order of a few meV/Å.

In order to characterize the atomic relaxation, we define $\delta_z(M_i)$ [respectively $\delta_z(O_i)$] as the displacement of the cation (respectively oxygen) along $z$ in layer $i$, with respect to the initial reference configuration. We introduce the displacement of the mean position of each atomic plane as

$$\beta_i = \left[ \delta_z(M_i) + \delta_z(O_i) \right]/2, \qquad (4.1)$$

and the change in the interplanar distance between consecutive planes $i$ and $j$ as

$$\Delta d_{ij} = \beta_i - \beta_j. \qquad (4.2)$$

The rumpling parameter of layer $i$ describes the movement of the ions with respect to the mean position of each atomic plane and corresponds to

$$\eta_i = \left[ \delta_z(M_i) - \delta_z(O_i) \right]/2. \qquad (4.3)$$

It is positive when the cation $M_i$ is above the oxygen, and negative otherwise.

Figures 4.3 and 4.4 show a schematic view of the atomic relaxations for the "cubic", symmetric, $(1 \times 1)$ TiO$_2$-terminated and AO-terminated free-standing slabs of SrTiO$_3$ [208], BaTiO$_3$ [207], and PbTiO$_3$ [209], all of them obtained within the LDA approximation to DFT, using ultrasoft pseudopotentials and a plane-wave basis set. In each case, the in-plane lattice constant was constrained to the theoretical one of each material. This choice ensures that the relaxations are intrinsic to the surface and not induced by mechanical strains, whose effects will be reviewed in Sec. 4.3.

Some common important features for all the surfaces are as follows. (i) As expected, the larger relaxations are on the surface atoms. (ii) The surface layer contracts substantially inwards (i.e. toward the bulk) with both metal and oxygen atoms displacing in the same direction with respect to reference positions (with the only exception for the top O ion of SrO-terminated SrTiO$_3$ surface). (iii) The relaxations of the metal atoms are much larger, leading to a rumpling of the layers, and to the appearance of an ionic surface dipole. (iv) All the surfaces display a similar oscillating relaxation pattern with a reduction of the interlayer distance $\Delta d_{12}$, an expansion of $\Delta d_{23}$, and a reduction again of $\Delta d_{34}$. Compared to BaTiO$_3$ and SrTiO$_3$, the amplitudes of the relaxation of PbTiO$_3$ are significantly larger; (v) A similar oscillating behaviour is observed for $\eta_i$ (successive $+$ and $-$ signs). The amplitude of $\eta$ rapidly decreases showing that the major relaxations are localized at the surface. (vi) The outward relaxation of the A-cation of the second layer on the TiO$_2$ surface is noteworthy, at odds with an usual experimental approximation, made to simplify the refinement procedure of the data, that assumes the absence of buckling in the second and third layers.

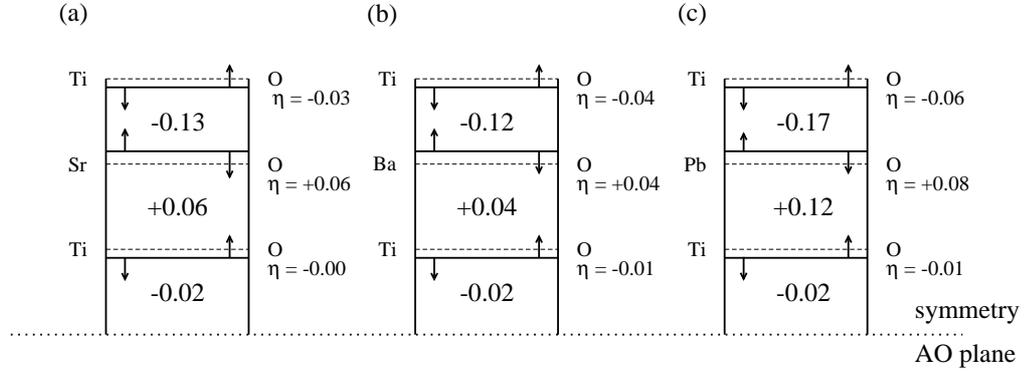

Figure 4.3: Schematic view of the atomic relaxation for the top half of TiO$_2$-terminated "cubic" and symmetric free-standing slab of SrTiO$_3$ [208] (a), BaTiO$_3$ [207] (b), and PbTiO$_3$ [209] (c). Dashed lines correspond to the reference positions of the atomic planes, and the full lines are the mean position in the relaxed structure. Changes in the interplanar distance are written in Å. The atoms (A or Ti, depending on the layer, at the left and O at the right) move in the direction indicated by the arrow. The rumpling parameter, $\eta$, is expressed in Å. The in-plane lattice constant correspond to those computed theoretically for the corresponding bulk material (3.86 Å for SrTiO$_3$, 3.95 Å for BaTiO$_3$ and 3.89 Å for PbTiO$_3$). Symmetric seven layer slabs were used in the simulations. Bottom half is equivalent by M$_z$ mirror symmetry.

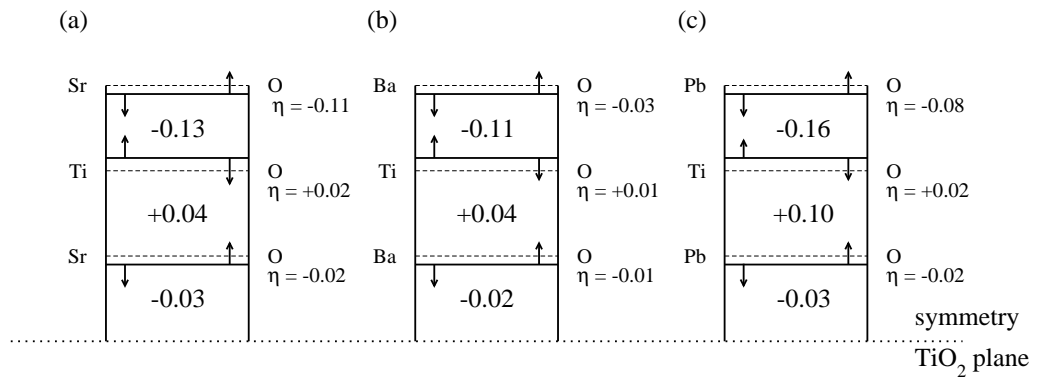

Figure 4.4: Schematic view of the atomic relaxation for the top half of AO-terminated "cubic" and symmetric free-standing slab of SrTiO$_3$ [208] (a), BaTiO$_3$ [207] (b), and PbTiO$_3$ [209] (c). Meanings of the lines as in Fig. 4.3.

The same kind of relaxation pattern for free-standing BaTiO$_3$ slabs was obtained using a first-principles parametrized shell-model [204, 152], and a "hybrid-functional" method [213, 91], with the only difference of the direction of the displacement of the top O atom in the TiO$_2$-terminated surface of PbTiO$_3$ slab.

Starting from the tetragonal phase with the polarization lying in a plane parallel to the surface, the most important change is the lack of symmetry of the O atoms of the TiO$_2$ layer with respect to their position perpendicular to the surface. Almost no change in the relaxations along $z$ for BaTiO$_3$ is observed with respect to those of the cubic phase. For PbTiO$_3$, the changes in the interplanar distances and rumpling parameters are strongly reduced in the tetragonal phase [209]. Influence of the surface on in-plane polarization will be discussed in Sec. 4.2.4.

## 4.2.2 Surface energetics.

One of the goals of the first-principles calculations in surfaces is the prediction of the most stable phases under realistic experimental conditions. How much energy is required to cleave the surface? Which termination of the (001) ABO$_3$ surface is more stable, AO or BO$_2$? Although the questions might seem trivial to answer, there are some subtleties to consider before.

A direct comparison with the bulk energy is only possible for stoichiometric asymmetric slabs, because these simulation boxes contain an integer number of ABO$_3$ unit cells. However, asymmetric slabs in combination with periodic boundary conditions give rise to fictitious electric fields (due to the different chemical potentials at each interface) that might affect the results.

A comparison of the total energy for two symmetric slabs with different termination to elucidate which one is more energetically favorable is not possible, since the number of atoms of each type in the AO and BO$_2$-terminated slabs is not the same. To predict the stability of the different phases, the surface energy must be determined with respect to the chemical potential $\mu_I$ for each type of atom $I$ [26]. Formally, the chemical potential $\mu_I$ is defined as the derivative of the Gibbs free energy with respect to the number of particles of type $I$ [221].

In order to simplify the following analysis, the discussion will not be performed in terms of A, B and O atoms, but AO and BO$_2$ units will be taken as the independent constituents of the slabs [208]. Therefore, from now on we will refer to the chemical potential for BO$_2$, $\mu_{BO_2}$, and the chemical potential for AO, $\mu_{AO}$.

First of all, let us define the formation energy $E_f$ of bulk ABO$_3$ as the energy per formula unit required to form a ABO$_3$ unit cell from the AO and BO$_2$ constituents,

$$-E_f = E_{ABO_3} - E_{BO_2} - E_{AO}, \tag{4.4}$$

where $E_{ABO_3}$, $E_{BO_2}$, and $E_{AO}$ are the cohesive energies of the bulk crystals. By convention $E_f$ is positive.

Let us consider now that the bulk ABO$_3$ solid is a reservoir which can exchange constituents with the surface. If the surface is in equilibrium with the bulk, the number of constituents that are leaving the surface is equal to the number of constituents coming back from the bulk reservoir. For this to be true, the sum of the chemical potential of the consituents must be equal to the bulk formation energy, i. e. we gain the same energy by adding a BO$_2$ and AO unit to the surface as forming a bulk ABO$_3$ unit cell (if this is condition does not holds, one of the processes will be energetically more favourable and the equilibrium will be broken),

$$\mu_{BO_2} + \mu_{AO} = -E_f. \tag{4.5}$$

Eq. (4.5) imposes a constraint on the chemical potentials of the constituents; only one of the two is an independent variable. The zero of the chemical potential can be set up in such a way that $\mu_{BO_2}$

= 0 represents a surface in equilibrium with a reservoir of bulk crystalline $BO_2$, and the corresponding relationship for $\mu_{AO}$. Since the chemical potential can never exceed the energy of the condensed pure element, there are some limits on the allowable range of values

$$-E_f \leq \mu_{BO_2} \leq 0. \tag{4.6}$$

The value of the chemical potential allow us to simulate the experimental chemical conditions. If $\mu_{BO_2}$ = 0 the system is in equilibrium with bulk crystalline $BO_2$, i. e. we are under very rich $BO_2$ conditions. In the other extreme, $\mu_{BO_2} = -E_f$, the system is in equilibrium with bulk crystalline AO (very rich AO conditions).

The equilibrium state of the surface as a function of the composition is determined by minimizing the grand potential [221]

$$F = \frac{1}{2} \left[ E_{slab} - N_{BO_2} \left( \mu_{BO_2} + E_{BO_2} \right) - N_{AO} \left( \mu_{AO} + E_{AO} \right) \right]. \tag{4.7}$$

where $E_{slab}$ is the energy of the slab computed from first principles and $N_{BO_2}$ and $N_{AO}$ are, respectively, the number of $BO_2$ and AO planes included in the slab. The factor 1/2 appears because in one slab we have two surfaces. The grand thermodynamic potential as a function of the chemical potential of $TiO_2$ is plotted in Fig. 4.5 for the two surface terminations of a $BaTiO_3$ free standing slab. From the Figure we can see that both kinds of terminations have a similar range of stability, indicating that both of them could be formed depending on the chemical conditions during growth (rich Ti or rich Ba conditions).

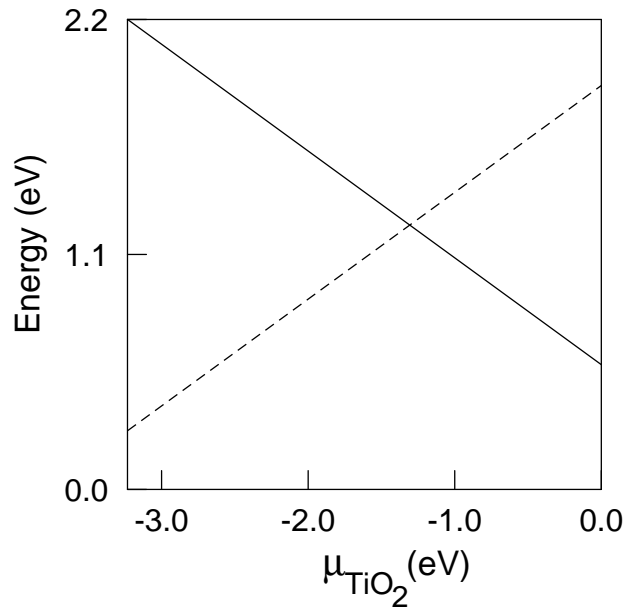

Figure 4.5: Grand potential as a function of the chemical potential $\mu_{TiO_2}$ for the two types of surfaces of $BaTiO_3$. Dashed line represents the AO-terminated surface. Solid line corresponds to the $TiO_2$ terminated surface. From Ref. [207].

A similar phase diagram was obtained for $SrTiO_3$ [208]. For $PbTiO_3$ only the PbO-terminated surface can be obtained in thermodynamic equilibrium [209].

An extension of this theory on the chemical potential was done by Zhang and Demkov to study the relative stability of surface steps on (001) $SrTiO_3$ surfaces [211]. The stepped $SrTiO_3$ prefers the O step edge under extremely oxygen rich conditions, and Sr or TiO edge under extremely oxygen deficient conditions. However, under the majority of the chemical environments, including typical growth conditions, the termination is mixed and it is the stoichiometry that drives terrace termination.

Sometimes, it is more appropiate for a direct comparison between different surfaces to define a surface energy independently of the chemical potential. After cleaving the bulk, surfaces with the two types of terminations arise simultaneously. Assuming that the cleavage energy is the same for both terminations, we can define [222]

$$E_{surf} = \frac{1}{4}\left[E_{slab}(AO) + E_{slab}(BO_2) - nE_{bulk}\right] \qquad (4.8)$$

where $E_{slab}(AO)$ and $E_{slab}(BO_2)$ respectively stands for the energy of symmetrically terminated AO and $BO_2$ slabs respectively, $E_{bulk}$ is the energy of the bulk unit cell and $n$ is the number of $ABO_3$ unit cells considering together the atomic layer of both AO and $BO_2$ slabs. The factor $1/4$ comes from the fact that, after the cleaving procedure, we create four surfaces. Equation (4.8) might be considered as an average of the grand potential for the two kind of terminations.

If the energy of the slab in the previous formula is taken for the unrelaxed configuration we define an average unrelaxed surface energy $E_{surf}^{unrelax}$, whereas if we take the slab energies for the relaxed geometry, we are introducing an average relaxed surface energy, $E_{surf}^{relax}$. The difference between both of them is giving us the average relaxation energy

$$E_{relax} = E_{surf}^{unrelax} - E_{surf}^{relax}. \qquad (4.9)$$

A comparison of the surface energies for $SrTiO_3$, $BaTiO_3$, and $PbTiO_3$ using the same plane wave ultrasoft-pseudopotential method might be found in Table 4.1. Hybrid-functional calculations tends to understimate slightly these energies [91], while they are overestimated by the shell-model [203].

Table 4.1: Surface cleavage energy $E_{surf}$ and surface relaxation energies $E_{relax}$ of $SrTiO_3$, $BaTiO_3$, and $PbTiO_3$ free-standing slabs. Units in eV/unit cell. From Ref. [209].

|             | $SrTiO_3$ | $BaTiO_3$ |            | $PbTiO_3$ |            |
|-------------|-----------|-----------|------------|-----------|------------|
|             | cubic     | cubic     | tetragonal | cubic     | tetragonal |
| $E_{surf}$  | 1.26      | 1.24      | 1.24       | 0.97      | 0.97       |
| $E_{relax}$ | 0.18      | 0.13      |            | 0.21      | 0.22       |

The unrelaxed cleavage energy is very similar for $BaTiO_3$ and $SrTiO_3$, while it is significantly lower for $PbTiO_3$. The relaxation energy is much larger than the bulk double well depth, estimated to be around 0.03 eV for $BaTiO_3$ and 0.05 eV for $PbTiO_3$, indicating that the surface relaxation might influence strongly the surface polarization. This eventual coupling will be discussed in Sec. 4.2.4.

### 4.2.3  Surface electronic structure.

Surface electronic band structure of $SrTiO_3$ [223, 208], $BaTiO_3$ [205, 206, 207, 224] and $PbTiO_3$ [209] have been computed within the LDA approximation to DFT. Although it is well known that DFT is a ground-state theory and therefore unreliable for excitation properties, such as band gaps, in practice it is observed that the valence energy bands are usually well reproduced within DFT, and that the LDA and quasi-particle valence wave functions are nearly identical. In addition, many studies focus only on the comparison of the band structures of different materials, or between different surface terminations for the same material. In such studies a cancellation of errors is expected and the conclusions drawn from the LDA results should be, at least, qualitatively correct.

The (001) surface band structure of $SrTiO_3$, $BaTiO_3$ and $PbTiO_3$ have been recently reported using "hybrid functionals" (see Sec. 2.2.2). They show the best agreement with experimental data for both bulk geometry and optical properties of materials. Figures 4.7, and 4.8 display the (001) surface electronic structure of $BaTiO_3$ and $PbTiO_3$ (the band structure of $SrTiO_3$ is quite similar to that of $BaTiO_3$ and is not shown here).

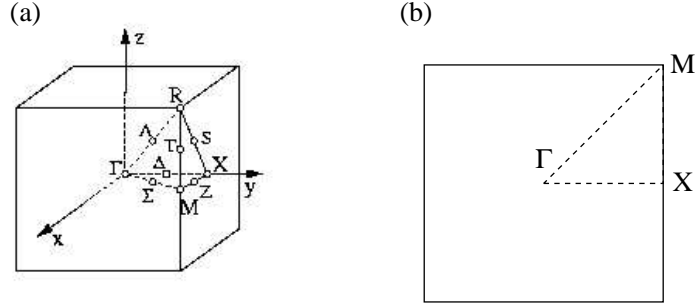

Figure 4.6: High symmetry points of the three dimensional Brillouin zone for a bulk unit cell of cubic symmetry (a), and the correspoding surface Brillouin zone along the (001) direction.

Before starting the study on the surface-induced effects on the electronic band structure, it is important to look at the projected bulk structure on the surface Brillouin zone (SBZ). Due to the lack of symmetry of a surface in the perpendicular direction, the Bloch theorem only holds for lattice vectors parallel to the surface, where the slab should maintain some periodicity. In the isolated slab limit, we can consider that the lattice vector perpendicular to the surface has an infinite length. The corresponding lattice vector in reciprocal space vanishes or, in other words, the SBZ has only two dimensions (2D) (see Fig. 4.6). Each row in the bulk three dimensional (3D) Brillouin zone projects on a point in the SBZ. The process of plotting for a given point of the SBZ all the eigenvalues for all the bands and **k**-points of the associated row in the 3D BZ is referred to as the projection of the bulk band structure onto the surface. This projection is shown for BaTiO$_3$ and PbTiO$_3$ in Fig. 4.7(a) and 4.8(a) respectively.

For BaTiO$_3$ and SrTiO$_3$ the highest valence bands of the projected bulk structure on the surface are quite flat, with the top located at the M point of the SBZ (the topmost valence state in bulk is at the R point of the 3D BZ, that projects on M in the SBZ). The projected density of states (PDOS) on the different atoms reveals a O $2p_x$, $2p_y$ character. The bottom of the conduction band, a Ti $3d$ threefold degenerated level, lies at the $\Gamma$ point, and disperse very little (very flat band) between $\Gamma$ and X.

Due to the hybridization between the Pb $6s$ and the O $2p$ orbitals in PbTiO$_3$, the projected bulk band structure and PDOS have a different behaviour compared to BaTiO$_3$ and SrTiO$_3$. For PbTiO$_3$, the top of the valence band has a substantial contribution of the Pb $6s$ orbitals. This hybridization leads to a lift of the upper valence band states near the X point of the SBZ. The bottom of the valence band is a flat band between $\Gamma$ and X of Ti $3d$ character.

Once the bulk projected band structure is fully characterized, the changes introduced by the presence of a surface can be isolated and analyzed.

The AO terminated surface of BaTiO$_3$ and SrTiO$_3$ shows a clean gap with a complete absence of in-gap states. Hybrid-functionals calculations report a reduction of the band gap of 0.3 eV, whereas in pure DFT simulations the gap remains almost unchanged with respect the bulk value [207, 209]. The main contribution for the top of the valence band (at M) comes from O $2p$ orbitals from the central bulk-like plane. The band is very flat across the XM line. The contribution to the bottom of the conduction band is mainly peaked on the Ti $3d$ states of the subsurface layer.

The TiO$_2$ terminated surface of SrTiO$_3$ and BaTiO$_3$ free-standing slabs shows a rather similar behaviour. The most important feature is the tendency of the upper valence bands to intrude into the lower part of the band gap near the M point. Therefore, the top of valence band is not so flat as for the AO-terminated slab, and the indirect gap is reduced more significantly. Nevertheless, no gap state is clearly seen in the center of the gap. The surface Ti bond collapses, and instead of dangling, the charge moves back on to the Ti and the Ti-O surface bonds [205, 224]. This self-healing enhances the surface reactivity and highlights the application of BaTiO$_3$ surfaces as a substrate for epitaxial growth and surface catalysis.

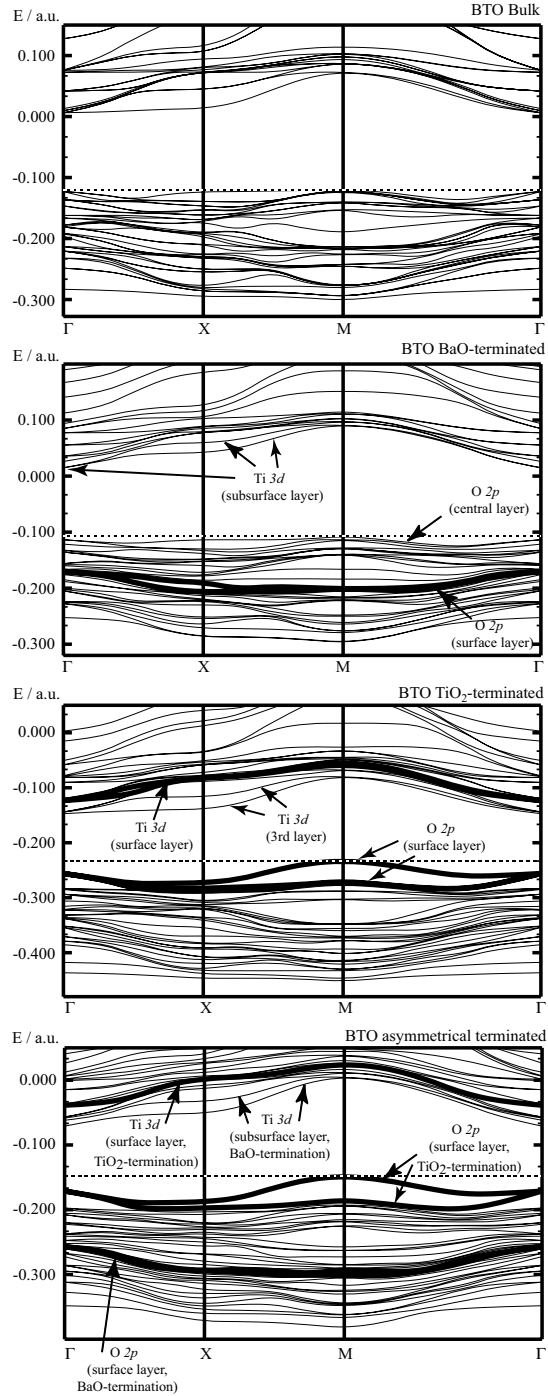

Figure 4.7: Surface band structure of a (001) free-standing isolated BaTiO$_3$ slab obtained using a "hybrid functional approach". Results for (a) projection of the bulk band structure onto the surface Brillouin zone, (b) a seven-layer BaO symmetrically-terminated slab, (c) a seven-layer TiO$_2$ symmetrically-terminated slab, and (d) and a eight-layer asymmetrical termination of the surface are shown. From Ref. [91]

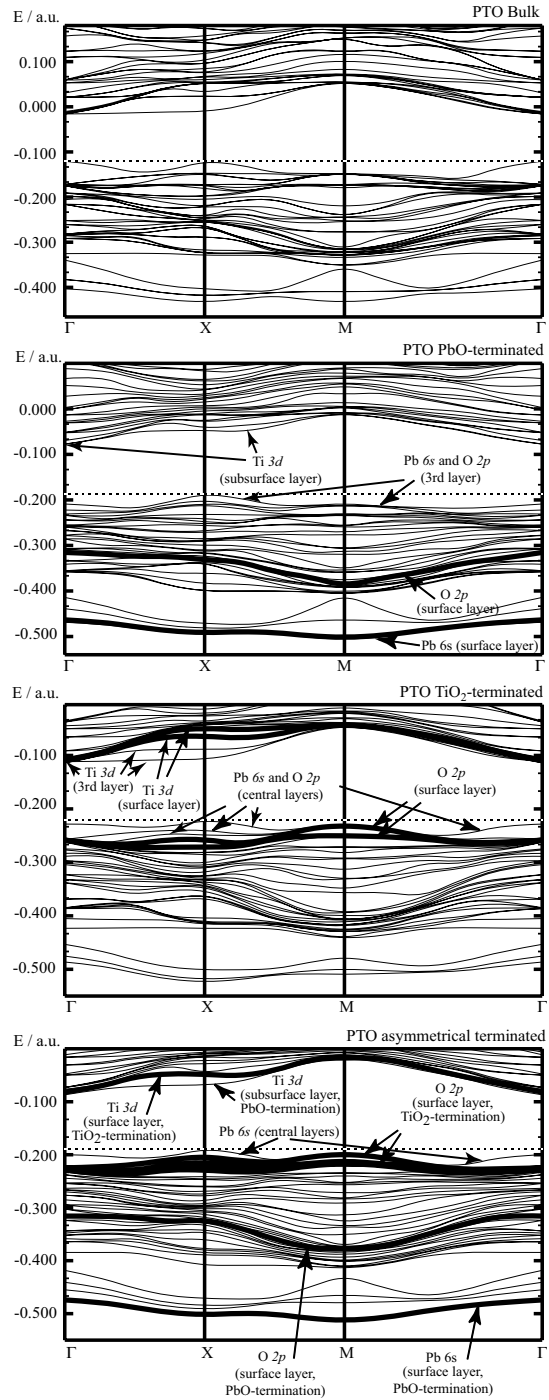

Figure 4.8: Surface band structure of a (001) free-standing isolated PbTiO$_3$ slab obtained using a "hybrid functional approach". Results for (a) projection of the bulk band structure onto the surface Brillouin zone, (b) a seven-layer PbO symmetrically-terminated slab, (c) a seven-layer TiO$_2$ symmetrically-terminated slab, and (d) and a eight-layer asymmetrical termination of the surface are shown. (From Ref. [91])

The explanation for these changes is provided by the different behaviour of the A-cation orbitals and the Ti-orbitals at the surface. As mentioned above, the Ba or Sr atoms can give up two electrons to O and are highly ionized. They can be considered, to a good approximation as spherical ions charged with +2 electrons. On the other hand, the O atom in the AO-terminated surface are always directly above a Ti atom and strongly hybridizes with it. Therefore, they are quite insensitive to the absence of neighbours at the surface, and no tendency for the formation of surface states is expected. However, in the TiO$_2$ termination, the hybridization of Ti $3d$ and O $2p$ orbitals, responsible for the lowering of the O $2p$ states in bulk, is no longer possible. At the surface, the energy of the $2p$ states of O are higher than in bulk, and lie slightly above the top of the valence band. The top most occupied state is a pure O-$p$ state at the TiO$_2$ surface.

The same behaviour is observed in the TiO$_2$-terminated PbTiO$_3$ surface. However, the movements of the eigenvalues are always below the top of the valence band, that is now located at X and not at M as it happened for BaTiO$_3$ and SrTiO$_3$. Therefore, the band gap remains almost unchanged, being this the most important difference between those surfaces.

Bands structures for asymmetrically terminated slabs have been computed [205, 91] showing a mixture of the band structure of the two symmetrical slabs.

All the previous band structures have been obtained for paralectric states. Krčmar and Fu [225] have extended the study to the case of an isolated free-standing BaTiO$_3$ single slab in vacuum with a polarization perpendicular to the surface. Starting from the paraelectric coordinates of a nine-layer TiO$_2$ terminated slab, the symmetry is atificially broken, and the atoms are allowed to relax starting from the induced ferroelectric configuration. The relaxation is stopped at an intermediate point, that still exhibits a polarization perpendicular to the surface (a full relaxation of the coordinates will end up with the atoms back in the paraelectric configuration). The O $2p$ states located at the top of the valence band split. Some of the states move down into the valence band, whereas the others move up into the bulk midgap, crossing the Fermi level at $M$. The resonant Ti $3d$ states that constitute the bottom of the conduction band for the paraelectric slab move down into the bulk midgap and crosses the Fermi level at $\Gamma$. This crossing of the Fermi energy produces a transfer of charge from the valence bands to the conduction bands, that become partially occupied and conducting. For a polarization along the [001] direction, there is an accumulation of electrons at the top surface and an accumulation of holes at the bottom surface. In such a configuration, the depolarizing field contributes to the movement of electrons from the bottom to the top surfaces. The PDOS on the surface Ti atom and subsurface O atom do not vanishes at the Fermi level, supporting the metallization of the slab. The carrier layers in the vicinity of the surface layers might help to explain why uncoated (no top electrode) ferroelectric thin films may exhibit a non zero polarization perpendicular to the surface [226]. The density of surface carriers increases as the magnitude of the surface layer buckling decreases.

### 4.2.4 Coupling of structural relaxations with ferroelectricity.

ABO$_3$ perovskite structures are subject to several competing structural instabilities from ferroelectric (FE) to non-polar antiferrodistortive (AFD) or antiferroelectric (AFE) distortions. Polar and non-polar instabilities compete in a delicate way and tend to suppress one another, with the result of the rich phase diagrams showing a large variety of different structures. The presence of a surface might affect the subtle balance between them and new phases, not found in bulk, might be observed in the phase diagrams. Two examples of particular importance have been studied theoretically from first-principles: the influence of surface relaxations on in-plane ferroelectricity and the appearance of AFD reconstructions on PbTiO$_3$ surfaces.

As it has been shown in the previous section, the surface relaxation energies are of the order of hundreth of meV, one order of magnitude larger than the bulk ferroelectric double well depth, estimated to be around 30 meV for BaTiO$_3$ and 50 meV for PbTiO$_3$ when the coupling with the tetragonal distortion is included [93]. Thus, it might be thought that the relaxation-induced atomic displacements and the ferroelectric soft-mode distortion compete at the surface, with the former prevailing to the later due to the larger gain in energy and, therefore strongly affect the ferroelectric order at the surface. However, first-principles calculations [207, 209] show a very small influence of the surface relaxation on ferroelectricity, when the polarization lies in a plane parallel to the surface (the case of an out-of-plane polarization deserves an special treatment of the electrical boundary conditions, and will be discussed on Sec. 4.4). The in-plane polarization increases slightly for the PbO-terminated $PbTiO_3$ surface and $TiO_2$-terminated $BaTiO_3$ surface, while the two other surfaces ($TiO_2$-terminated $PbTiO_3$ and BaO-terminated $BaTiO_3$) the ferroelectric distortions are slightly decreased when going from the bulk-like layers in the middle of the slab to the surface. This small interaction between in-plane polarization and surfaces contrasts with the large polarization-strain coupling that will be reviewed in Sec. 4.3.

$SrTiO_3$ is an incipient ferroelectric; it nearly becomes ferroelectric at very low T, but quantum zero-point fluctuations prevents to undergo the phase transition. For some time, it was thought that the presence of a surface might be a driving force that helps stabilizing a ferroelectric state. Results based on empirical interatomic potentials [227] suggested that the SrO-terminated surface of $SrTiO_3$ reconstructs to a ferroelectric monolayer phase. However, more accurate first-principles calculations [208] predict very modest ferroelectric distortions, not greater in surface layer than in deeper layers, that is likely to be destroyed by thermal or quantum fluctuations, preventing its appearance at room temperature.

Almost invariably, when a surface is cleaved, atoms with dangling bonds are left at the surface and a rearrangement of the position takes place to produce an energetically more favourable configuration. Beyond simple atomic relaxations previously described, atomic reconstruction of the surface also appear in many cases. (001)-$ABO_3$ perovskites surfaces are not an exception. Using a combination of high-resolution electron microscopy and first-principles DFT calculations, a $(2 \times 1)$ reconstruction on $TiO_2$-terminated (001)-$SrTiO_3$ has been identified by Erdman and coworkers [228, 229]. In the reconstructed structure, the surface is terminated with a double $TiO_2$ layer. The polyhedra at the surface share edges, and not only corners as in the bulk, with the result of the formation of a row of 5-coordinated $TiO_5$ units at the surface. More recent first-principles calculations by Johnston *et al.* Johnston *et al.* [230] reported surface energies and atomic structures as a function of the chemical potential of $TiO_2$, oxygen partial pressure and temperature. They concluded that this $(2 \times 1)$ reconstruction is only the most stable in a $TiO_2$ rich condition with a partial pressure of oxygen greater than $10^{-8}$ atm. At standard pressure conditions and temperatures about 1000 K, the $(1 \times 1)$ configuration is stable, while all the $(2 \times 1)$ reconstructions are energetically unstable. Despite this fact, the $(1 \times 1)$ configuration is very difficult to observe using scanning tunneling microscopy (STM) due to the small corrugation of the surface charge.

Two surface phases were observed by x-ray diffraction of (001)-$PbTiO_3$ surfaces, having $c(2 \times 2)$ and $(1 \times 6)$ reconstructions [220]. The $c(2 \times 2)$ phase, appearing under most conditions of temperature and PbO pressure, displays sharp diffraction peaks, signature of a very well ordered phase, and consists of a single layer AFD structure, with alternating clockwise and anticlockwise rotation of the Ti-centered O squares in the $TiO_2$ layers. First-principles calculations on double unit cells, confirm this finding [231] with a substantial enhancement of the AFD distortion for the PbO-terminated surface. The driving force for the $c(2 \times 2)$ reconstruction is the formation of shorter PbO bonds, because of the tendency of Pb to form stronger covalent bonds with O, underlining the important role played by Pb in the bonding of $PbTiO_3$. This explains also why the reconstruction does not happen in $BaTiO_3$, with a more ionic BaO bond. The large AFD surface reconstruction observed at the PbO-termination does not occur at the $TiO_2$ termination.

### 4.2.5   Effective charges at the surface.

Born effective charges, $Z^*$, are central quantities in the study of ferroelectrics. As discussed in Sec. 3.2, their anomalous values in $ABO_3$ perovskite compounds is at the origin of a giant destabilizing dipole-dipole (DD) interaction, itself responsible for the ferroelectric phase transition. Moreover, the ferroelectric instability was shown to be very sensitive to the amplitude of $Z^*$ [172].

It was shown by Ruini *et al.* in a very general context that $Z^*$ cannot keep their bulk values at the vicinity of a surface [179] or of an interface [232]. It is therefore important to quantify this effect in perovskites and to see to which extent it could influence the ferroelectric instability.

First-principles calculations of dynamical charges have been reported for $BaTiO_3$ isolated slabs by Fu *et al.* [49], using the Hartree-Fock technique. They studied 7-layers slabs with a (001) truncated bulk geometry, without relaxation. The slabs were symmetric with either BaO or $TiO_2$ terminations. In the direction perpendicular to the surface, the dynamical charge to be considered is not more the usual Born effective charge $Z^*$ but, instead, a related quantity called the Callen or longitudinal charge $Z_L^*$. In bulk perovskites, both are related through $Z_L^* = Z^*/\epsilon^\infty$ [123]. The results are summarized in Table 4.2 where individual charges are reported as well as the total surface charge of each layer.

Table 4.2: Longitudinal charge $Z_L^*$ in the surface region of $BaTiO_3$ slabs as computed by Fu *et al.* within the Hartree-Fock approach. Individual charges and total charges of individual BaO and $TiO_2$ layers are reported for BaO terminated (bold) and $TiO_2$ terminated (italic) slabs as well as for bulk. (Adapted from Ref. [49] by Fu *et al.*)

| | Ba | Ti | $O_I$ | $O_{II}$ | Ba+$O_I$ | Ti+2$O_{II}$ |
|---|---|---|---|---|---|---|
| First layer | **1.104** | *2.093* | **-1.354** | *-0.919* | **-0.250** | *0.255* |
| Second layer | *0.911* | **2.119** | *-1.473* | **-0.767** | *-0.562* | **0.585** |
| Bulk | 0.883 | 2.187 | -1.543 | -0.765 | -0.660 | 0.657 |

We observe that the total surface charge in layers close to the surface are strongly different than in the bulk. In the surface region, they sum up to close to one half the total charge (0.66) in each layer in bulk ($-0.250 + 0.585 = 0.335$ and $0.255 + -0.562 = 0.307$ for BaO and $TiO_2$ surfaces respectively), in agreement with the sum rule of Ruini *et al.* [179]. In contrast the modifications of individual charges are less spectacular and usually of the order of 10%. An important feature is that the effective charges rapidly converge to the bulk value when going to the interior of the film. This convergence could be even increased when taking into account the surface atomic relaxation.

If the evolution of the surface dynamical charge reported above will modify slightly the dipolar interaction in the surface region, no clear indication has been reported yet attesting that it play a significant role in modifying the ferroelectric properties of thin films.

## 4.3 Strain effects

### 4.3.1 General features.

A second important factor monitoring the structural, ferroelectric and multifunctional properties of nanometric ferroelectric heterostructures is the macroscopic epitaxial strain imposed by the substrates.

Recent breakthroughs in the methods to synthesize complex oxides have allowed the fabrication of oxide heterostructures with a control at the atomic scale [180]. Nowadays, single-crystalline perovskite oxides can be grown with atomically abrupt interfaces on carefully prepared, termination controlled and atomically flat on terraces hundreths of nanometer wide substrates [233]. In these high-crystalline quality heterostructures, thin films are coherently matched to the substrate, i. e., the epilayers are forced to grow with the same in-plane symmetry as the underlying substrate, although this might be not their most stable phase in bulk. New phases, not observed in bulk, might even be stabilized. Since differences with respect to the unstrained bulk material appear already at the structural level, it is expected that other physical properties, such as the dielectric and piezoelectric responses, or the paraelectric-to-ferroelectric transition temperature of two-dimensional (2D) clampled thin films, could be significantly different from those of the mechanically free material. This fact opens the door to the design of artificial structures with improved taylor-made properties [234, 235, 236].

On top of the epitaxial strain imposed by the substrate, there are other sources of strain, such as the presence of crystalline defects that give rise to inhomogeneous strain fields [237]. Also, the strain can be temperature-dependent because of the different thermal expansion coefficientsof the substrate and the thin film.

Independently of the origin of the strain, the coupling between strain and polarization in perovskite oxides is known to be very strong. Lines and Glass considered explicitly the coupling of the soft optic mode to strain in their basic model hamiltonian to account for acoustic anomalies near the ferroelectric transitions (chapter 10 of Ref. [1]). This coupling is similarly present in the first-principles effective Hamiltonian described in Sec. 2.3.2. Also, as discussed in Sec. 3.5, the pioneer first-principles calculations on ferroelectric perovskite oxides by Cohen and Krakauer [238] revealed that tetragonal strain has a considerable effect in lowering the energy of $BaTiO_3$ [238, 93] (see Fig. 2.4) and in $PbTiO_3$ it stabilizes the tetragonal structure over the rhombohedral phase [93].

Many experimental evidences corroborate the strong influence of strain on the structural and ferroelectric properties of perovskite oxides. For example, polarization at room temperature can be induced by strain on $SrTiO_3$ [239], whose bulk, pure, unstressed form remains paraelectric down to 0 K (although it is considered an incipient ferroelectric). The shift of the transition temperature increases with strain, being the largest ever reported up to date. The direction of the spontaneous polarization changes from lying in the plane of the film when a biaxial tensile strain is applied ($DyScO_3$ substrate, strain of about +0.8 %), to out-of-plane polarization under a compressive strain [$(LaAlO_3)_{0.29} \times (SrAl_{0.5}Ta_{0.5}O_3)_{0.71}$ substrate (LSAT), strain of about -0.9 %].

Another striking experimental example is given by strained $BaTiO_3$ thin films. At growth temperature (typically of the order of 700 °C), $BaTiO_3$ is stable in a simple cubic lattice with an experimental lattice parameter of 4.00 Å [240]. However, when it is epitaxially grown on top of cubic (001) $SrTiO_3$ substrate [241, 242, 243], whose lattice constant is 3.905 Å then $BaTiO_3$ is not allowed to adopt the configuration that minimizes its energy at the bulk. Under the compressive misfit strain imposed by the substrate, Yoneda and coworkers [241] have observed that the tetragonal phase is stabilized with a tetragonality larger than in bulk, and that the ferroelectric phase is maintained well beyond the bulk phase transition temperature (130 °C); even at 600 °C, an spontaneous polarization could still be observed in the hysteresis loops. On top of the large shift in the paraelectric-to-ferroelectric phase transitions temperature, a large enhancement of the remanent polarization (at least 250 % higher than in bulk $BaTiO_3$ single crystals) has been reported for $BaTiO_3$ grown on $GdScO_3$ and $DyScO_3$ [236].

Similar enhancements of the transition temperature have been observed for $PbTiO_3$ thin films grown on insulating (001) $SrTiO_3$ substrates [244].

Under the light of these experiments, it seems that physical quantities can be tuned to a desired value by controlling the lattice mismatch between a ferroelectric film and its substrate, providing an alternative method to the traditional substitution and alloying of the A and B cations. This can be practically achieved by choosing the appropriate substrate so that the development of new substrates that enable the growth of uniformly strained films has become an important research topic [245]. However, this strain-engineering has also its limitations since epitaxial strain will only be preserved up to the critical thickness for the appearance of misfit dislocations.

When the misfit strain and the roughness of the substrate [243] are not too large, the thin film will keep growing coherently with the substrate as long as the elastic energy required to strain the film and match the substrate is smaller than the energy required for the appearance of misfit dislocations. Beyond this limit, dislocation defects will appear to reduce the strain by allowing the structure to relax to the bulk geometry (zero-strain state), therefore imposing an upper limit to beneficial strain effects such as polarization enhancement for instance [246, 247]. Dislocation densities of $\sim 10^{11}$ cm$^{-2}$ are usual in epitaxial ferroelectrics grown on lattice-mismatched substrates, and the resulting inhomogeneous strain smears out the ferroelectric phase transitions [245]. An estimate for the critical thickness for misfit dislocations can be obtained using the Matthews-Blakeslee formula [248], based on the inhomogeneous elastic theory. Since the critical thickness at which dislocations begins to form is inversely proportional to lattice-mismatch, a lower mismatch is desirable to allow the growth of strained, low dislocation density, ferroelectric thin films [239]. However, due to kinetic barriers, coherent epitaxial films might be observed in a metastable form even well above the theoretically estimated critical thickness for misfit dislocations, provided an small enough growth temperature [236, 249] (Fig. 4.9).

Effects of strain on the structural, ferroelectric, dielectric and piezoelectric properties have been studied theoretically, both from a phenomenological Devonshire-Ginzburg-Landau theory and from first-principles. In order to isolate these effects, typically infinite bulk materials with homogeneous strain tensor in short-circuit electrical boundary conditions (zero internal electric field) are simulated. Thus, no inhomogeneous source of strain, nor surface or interface related effect are taken into account (Fig. 4.10) . Usually perovskite oxides are supposed to grow in a monodomain configuration on (001) cubic substrates, although Koukhar and coworkers [250], and Zembilgotov *et al.* [251] have already considered the possibility of dense domain structures and anisotropic in-plane strains, respectively. Assuming a cubic substrate, the 2D-clamping lowers the symmetry of the paraelectric phase from cubic to tetragonal, even in the paraelectric phase. Six possible phases are allowed by symmetry that, following the notation by Pertsev *et al.* [252], are referred to as: (*i*) the paraelectric *p* phase ($\mathcal{P}_1 = \mathcal{P}_2 = \mathcal{P}_3 = 0$), (*ii*) the tetragonal *c* phase ($\mathcal{P}_3 \neq 0$, $\mathcal{P}_1 = \mathcal{P}_2 = 0$), (*iii*) the *a* phase, with the polarization along an in-plane cartesian direction ($\mathcal{P}_1 \neq 0$, $\mathcal{P}_2 = \mathcal{P}_3 = 0$), (*iv*) the *ac* phase, where the polarization lies on one of the faces ($\mathcal{P}_1 \neq 0$, $\mathcal{P}_3 \neq 0$, $\mathcal{P}_2 = 0$), (*v*) the orthorrombic *aa* phase ($\mathcal{P}_1 = \mathcal{P}_2 \neq 0$, $\mathcal{P}_3 = 0$), and (*vi*) the monoclinic *r* phase ($\mathcal{P}_1 = \mathcal{P}_2 \neq 0$, $\mathcal{P}_3 \neq 0$).

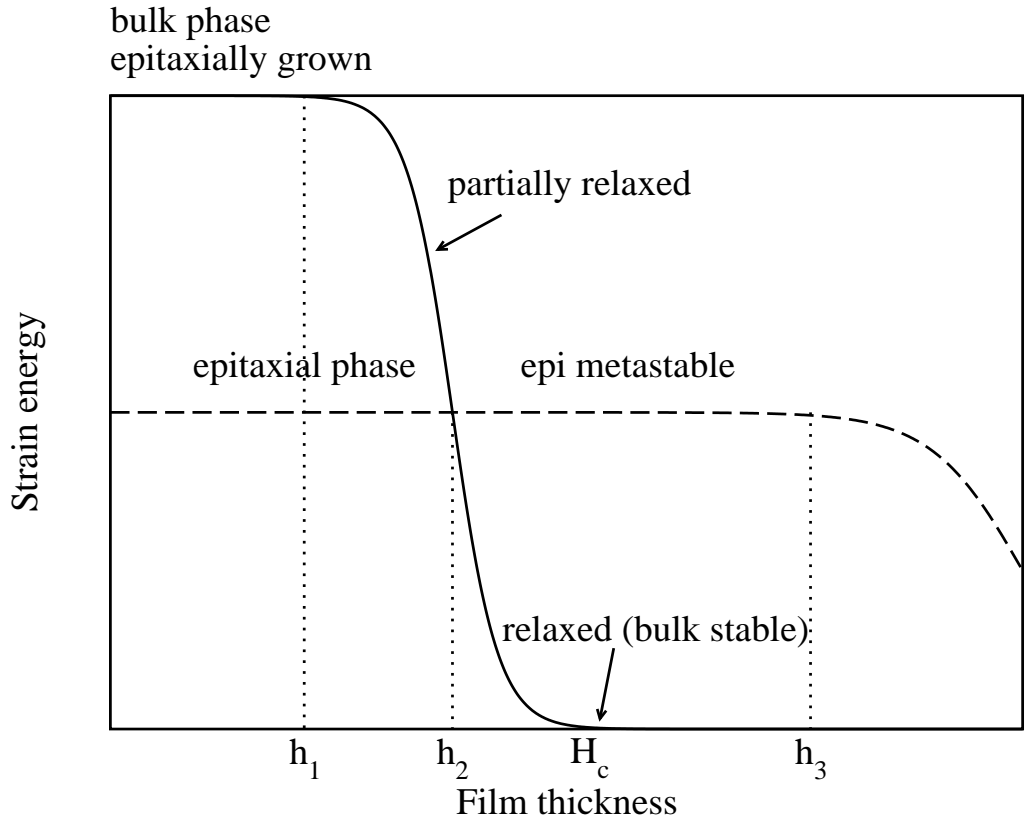

Figure 4.9: Schematic representation of the strain energy during epitaxial growth. Due to the smaller lattice mismatch with the substrate, the epitaxial phase is more stable than an hypothetical bulk phase epitaxially grown on the same substrate upto a critical thickness for misfit dislocation $h_1$. Beyond this thickness, the strain energy for the bulk structure relaxes, and at a thickness $h_2$, crosses the strain energy of the epitaxial structure, which ceases to be the most stable phase. Above a critical thickness $H_c$, the bulk phase is fully relaxed. However, the epitaxial phase might still survive in a metastable phase due to the existence of kinetic energy barriers up to a thickness $h_3$, whose value might be as large as 1000 Å. Similar to Fig. 1 in Ref. [249]

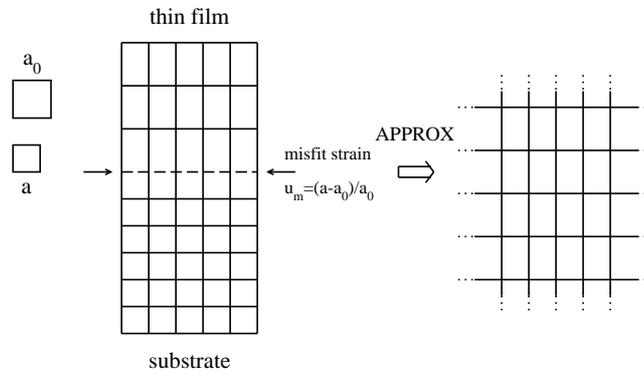

Figure 4.10: Schematic representation of the epitaxial growth of a thin film on a substrate. Left: a perovskite oxide of lattice constant $a_0$ is epitaxially grown on top of a thick substrate of lattice constant $a$. Middle: the substrate imposes its lattice constant at the interface (dashed line). The misfit strain $u_m$, which is strongly coupled with the polarization, induces a deformation of the unit cell of the thin film in the out-of-plane direction (elongation or shrinking). Right: in the simulations, the effect of the homogeneous strain is isolated by simulating a bulk (infinite, the periodical repition is symbolized by the dotted lines) material with homogeneous strain in short circuit. No surface or interface effect are considered. Courtesy of O. Diéguez.

### 4.3.2  Devonshire-Ginzburg-Landau phenomenological calculations.

In a seminal paper, Pertsev and coworkers introduced the effect of strain on epitaxial ferroelectric thin films into a Devonshire-Ginzburg-Landau (DGL) type phenomenological theory [252]. They identified properly the mechanical boundary conditions for a thin film epitaxially grown on a substrate: on the one hand, the lattice matching between the film and a thick substrate implies that the in-plane strains of the film at the film/substrate interface are totally controlled by the substrate (itself assumed to be sufficiently thick); on the other hand, since there are no tractions on the free surface, the stresses acting out-of-plane must vanish, i. e., under the strain conditions imposed by the substrate the epitaxial layer will minimize the elastic energy by elongation or compression of the lattice vectors along $z$. Therefore, the system is subject to mixed mechanical boundary conditions: fixed in-plane strain and fixed out-of-plane stress.

Pertsev and coworkers then applied the DGL phenomenological approach described in Sec. 2.3.1, considering an appropriate thermodynamic potential $\tilde{G}$ (function of temperature, polarization and misfit strain) [252, 253], compatible with the mixed mechanical boundary conditions. To determine the equilibrium thermodynamical state, the thermodynamic potential $\tilde{G}$ is minimized at a given misfit strain and temperature with respect to the polarization. The produced phase diagrams mapping the equilibrium structure of the ferroelectric thin film as a function of the misfit strains and temperature, so called "Pertsev diagrams", have been very succesful in the interpretation of experimental data on thin films. Pertsev diagrams have been developed for BaTiO$_3$ [252, 254] (see Fig. 4.11), PbTiO$_3$ [252], SrTiO$_3$ [255, 256] and single domain epitaxial Pb(Zr$_x$Ti$_{1-x}$)O$_3$ solid solution [257]. The results have been used to explain the change in the order of the phase transitions from the paraelectric state to a single domain ferroelectric phase (from first-order in bulk to second-order in thin films), or the enhancement of the transition temperature experimentally observed in thin films under compressive strains [241] (for BaTiO$_3$ this theory predicts that a strain of only 1 % should produce an increase of transition temperature comparable to unstrained Pb(Zr,Ti)O$_3$ solid solutions).

However, in some cases, the phase diagrams deduced from such DGL approach must be taken with care. They produce excellent results when they are applied around the regime at which the parameters were fitted, but their predictions become less accurate when trying to extrapolate to distant strain-temperature regimes. In particular, a small change in the set of parameters can lead to very different predictions for the low temperature phase transitions (Fig. 4.11). Due to the scattering in the reported model parameters, "range of transitions" rather than clean phase boundaries are plotted in same cases [236, 239]. The larger the misfit strain, the wider (more uncertain) the range of transitions.

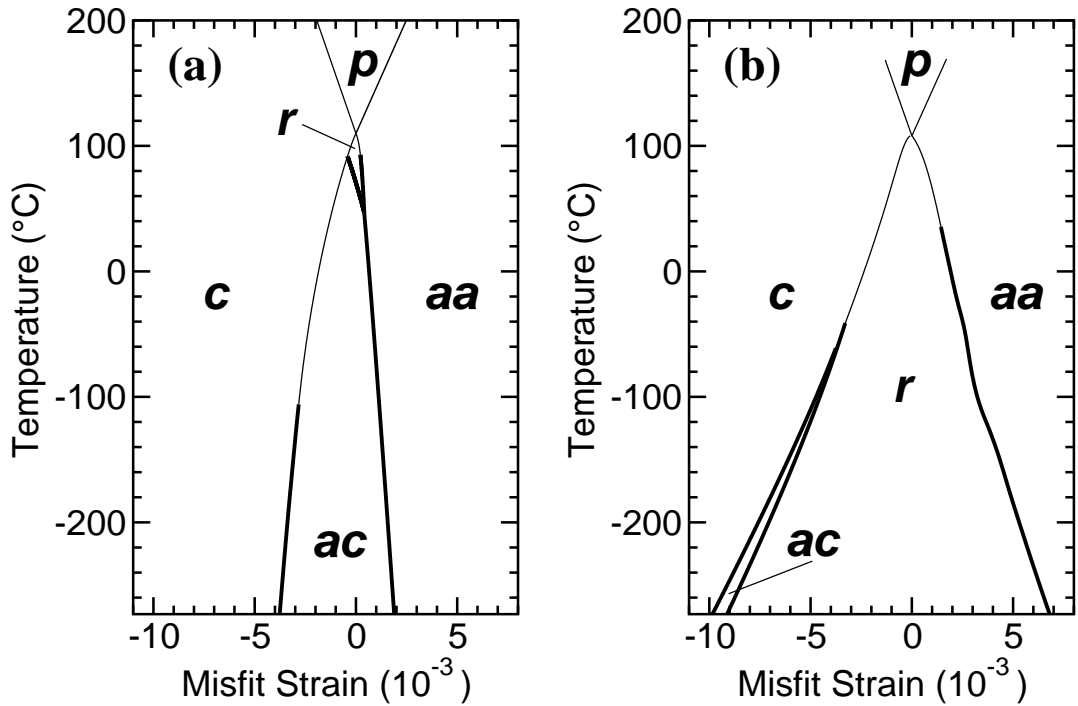

Figure 4.11: Misfit-strain temperature phase diagram ("Pertsev-diagram") for BaTiO$_3$ thin films on a (001)-cubic substrate, obtained using a Devonshire-Ginzburg-Landau phenomenological theory with two different sets of parameters. For panel (a), they were taken from Ref. [252], and from Ref. [254] for panel (b). They predict exactly the same behaviour at high temperature, close to the bulk Curie temperature, but there are drastic differences at lower T. Taken from Ref. [258].

### 4.3.3   First-principles calculations.

Full first-principles calculations provide a parameter-free framework and a wealth of information at the atomic level that allows to supplement efficiently the uncertainties of the DGL approach. Phase stability versus misfit strain from first-principles has been reported for $BaTiO_3$ [258], $PbTiO_3$ [259], $SrTiO_3$ [260], and more recently, for $CaTiO_3$, $KNbO_3$, $NaNbO_3$, $PbZrO_3$, and $BaZrO_3$ [261].

Diéguez and coworkers [258] have performed DFT simulations in order to obtain the Pertsev diagrams for epitaxial $BaTiO_3$ at zero-temperature. In these simulations, the role of epitaxial strain on the structural properties is isolated by systematically seeking the ground state structure of a five-atom unit cell of the bulk perovskite, contrained to several different in-plane lattice constants. The Ti and the O atoms are displaced from the centrosymmetric positions according to one of the six possibles low symmetry phases described above. This symmetry-broken unit cell was the starting point for a relaxation of the atomic positions and of the out-of-plane components of the lattice vectors, while preserving the initial symmetry. The most stable phase at each in-plane strain is found by identifying the minimum energy phase between the six possible symmetries [2]

Within this approach, at T = 0, $BaTiO_3$ undergoes a sequence of second-order phase transitions, $c \rightarrow r \rightarrow aa$, whereas no $ac$ phase is detected at any range of strain, at odds with the phenomenological calculation. An extension to finite temperature was done by means of an effective hamiltonian as the one described in Sec. §.2.3.b. The obtained phase diagrams, shown in Fig. 4.12, share the same topology as the Devonshire-Ginzburg-Landau ones at high T. At lower temperatures, it resembles Fig. 4.11(b), although no $ac$ phase is detected. The main advantage of first-principles methods is that the right sequence of phase transitions at 0 K is established, whereas phenomenological theories can give spurious results in these regimes, far away of the strain-temperature range where the parameters were fitted. The main drawback of these kind of simulations is the understimation of the transition temperatures, a well-known defficiency of the model hamiltonian approach [139].

The condition of having a bulk periodic cell only subject to homogeneous strain was relieved by Lai *et. al.* [262]. They included explicitly in the simulations the presence of a surface/interface, and the finite thickness of the samples (5 and 7 atomic planes in their simulations). Under ideal short circuit boundary conditions, the four phase point (where the $p$, $aa$, $c$, and $r$ phases meet in the phase diagram) is not located at zero-misfit strain, and the diagram is not symmetric with respect to this point. The key to understand the differences with respect to the results shown in Fig. 4.12 is the fact that the surface/interface induces an enhancement of the normal component of the polarization at the surface [31]. Therefore, a tensile strain has to be applied in order to compensate for this increase. Even a tetragonal $c$-phase can exist for low-enough tensile strain. Increasing the thickness of the film results in a reduction of the asymmetry of the phase diagrams.

Schimizu and Kawakubo [263] and Neaton *et al.* [264, 35] checked the large polarization-strain coupling in $c$-phase epitaxial $BaTiO_3$ from first-principles. The out-of-plane polarization was found to be enhanced by nearly 70 % of the bulk value when $BaTiO_3$ is supposed to be grown on top of $SrTiO_3$. A concomitant increase of the out-of-plane lattice constant (tendency to keep the unit cell volume constant) was observed.

A similar approach to that of Diéguez *et al.* was adopted by Bungaro and Rabe [259] to study epitaxially strained $[001]$-$(PbTiO_3)_1/(PbZrO_3)_1$ superlattices and pure $PbTiO_3$. The superlattices showed a strain-driven second-order (continuous) phase transitions from the $c$ phase (stables at large in-plane compressive misfit strains), to a monoclinic $r$ and finally to the $aa$ phase (stable at high tensile strain) [Fig. 4.13(a)]. In the $r$ phase the polarization is almost constant in magnitude, and rotates from $z$ to $x + y$ continuously (notation of the coordinate axis as in Sec. 4.2). For pure $PbTiO_3$ a discontinuous transition from an out-of-plane to in-plane polarization if only tetragonal strain is allowed [Fig. 4.13(b)]

---

[2] At a given fixed strain, the system adopts a structure that minimizes its energy. Rigorously, the same mixed mechanical boundary conditions as in the phenomenological approach still apply in the first-principles calculations (fixed strain in $x$ and $y$ and fixed stress in $z$). In this case, it would be appropriate to minimize a hybrid energy functional

$$\tilde{F} = E + \sigma_{zz}\eta_{zz}. \qquad (4.10)$$

However, most of the first-principles results only consider the case of $\sigma_{zz} = 0$, i.e. "free $zz$ stress", so that minimizing $\tilde{F}$ is the same as minimizing the energy.

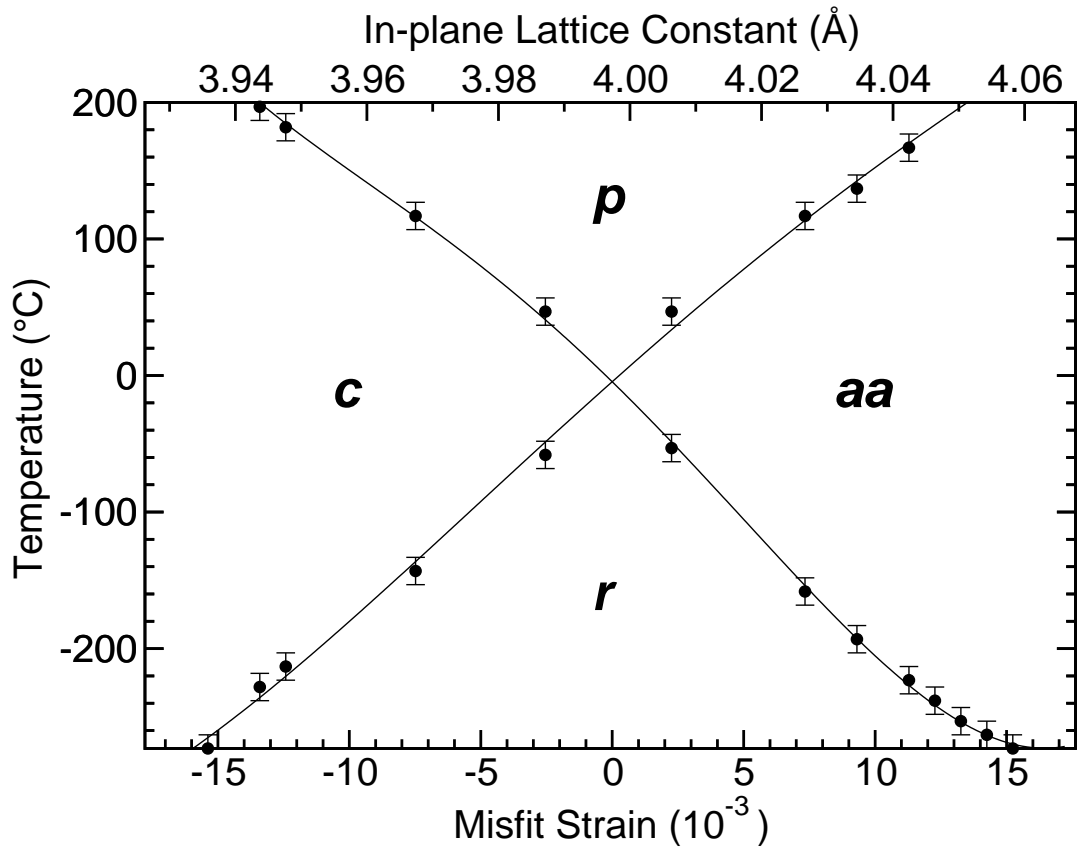

Figure 4.12: Misfit-strain temperature phase diagram of BaTiO$_3$ thin films grown on a (001)-cubic substrate, obtained using a first-principles based model hamiltonian approach [70, 71]. Taken from Ref. [258]

was obtained. Once the coupling with shear strain is properly included, a phase transition $c \rightarrow r \rightarrow aa$ is predicted, in agreement with the results of the thermodynamic theory [252], although with a much smaller region of stability for the $r$ phase.

For $SrTiO_3$ [260] the same kind of continuous phase transition from a ferroelectric tetragonal phase with polarization along [001] to nonpolar tetragonal phase and finally to an orthorombic structure, with polarization along [110] is reproduced [Fig. 4.14(a)], consistent with the phenomenological approach [255, 256]. The only differences are the range of stability of the paraelectric phase, wider in the first-principles calculations, and the absence of a ferroelectric phase with [100] polarization under tensile strain. Both differences were attributed to the lack of oxygen-octahedron rotation in the DFT results. An almost linear dependency of the tetragonality with strain is observed [Fig. 4.14(b)], in good agreement with experimental results [265]. The substantial increase of c/a in the paraelectric phase is remarkable. It means that the epitaxial strain produces an elongation or shrinking of the unit cell, and can be considered as a first source of tetragonality as predicted by the macroscopic theory of elasticity. The lattice contribution to the static dielectric response diverges near the critical strains for the second-order phase transitions reported earlier [Fig. 4.14(c)], as a result of the softening of the corresponding in-plane and out-of-plane transverse zone-center optical phonons [266] at that strains, and show nearly perfect inverse power law behaviour. It is noticeable how the dielectric constant in the paraelectric region is always above 100.

Effects of hydrostatic strains on the soft mode frequency of $SrTiO_3$ have also been reported by Schimizu [267]. The lowest frequency mode softens with the increase of the lattice constant, while the other modes also decrease, but with a smaller ratio. The effect of hydrostatic strain on the pattern of displacements is very weak. Transposing the first-principles results to a phenomenological model, an almost linear dependence of the inverse of the permittivity with strain was predicted in the range of strains investigated (from -0.2 % to +0.6%), in good agreement with experimental measurements [268].

In a way similar to thin films, epitaxial strains also play a major role in ferroelectric multilayers [35]. This discussion is however postponed to Sec. 5.

## 4.4   Electrical boundary conditions.

### 4.4.1   Snapshot on the experimental context.

A third main factor monitoring the ferroelectric behavior of ferroelectric thin films with polarization perpendicular to the surface is the electrostatic energy that strongly depend on the electrical boundary conditions. In order to introduce this subject, let us start with a brief survey of the experimental context within which ferroelectric thin films are usually studied and of the main experimental results that have been reported.

The non-volatile and switchable polarization of ferroelectric thin films make them attractive for information data storage. For this purpose, the figure of merit is the magnitude and stability of the switchable polarization $\Delta\mathcal{P}$. Measurements of the thickness variation of the magnitude of $\Delta\mathcal{P}$ in the thin and ultrathin regime are therefore of particular interest. Unfortunately, a direct experimental measurement of the switchable portion of the spontaneous polarization is made from the correlation with the current that flows through a resistor in a Sawyer-Tower circuit (chapter 4 of Ref. [1]) and, in the ultrathin limit this current is convoluted with and dominated by leakage. For a long time, well defined hysteresis loops were not available for ultrathin thicknesses, and indirect methods such as the measurements of the pyroelectric coefficient [269], the piezoelectric force microscopy [270, 271], scanning probe microscopy [272], the x-ray diffraction techniques [244, 273, 274, 275, 226], Ramann scattering [198, 199, 200], or calorimetry and specific heat measurements [276] have been applied to gain some insight on the behaviour of the films. Only recently, a new research avenue to study hysteresis loops has been opened using an atomic force microscopy (AFM) in pulse switching mode, in the so-called PUND sequence (positive-up-negative-down) [277]. The pulse method is less likely to be convoluted by leakage and non-linear dielectric effects. Transients currents of films down to 40 Å have been measured [278].

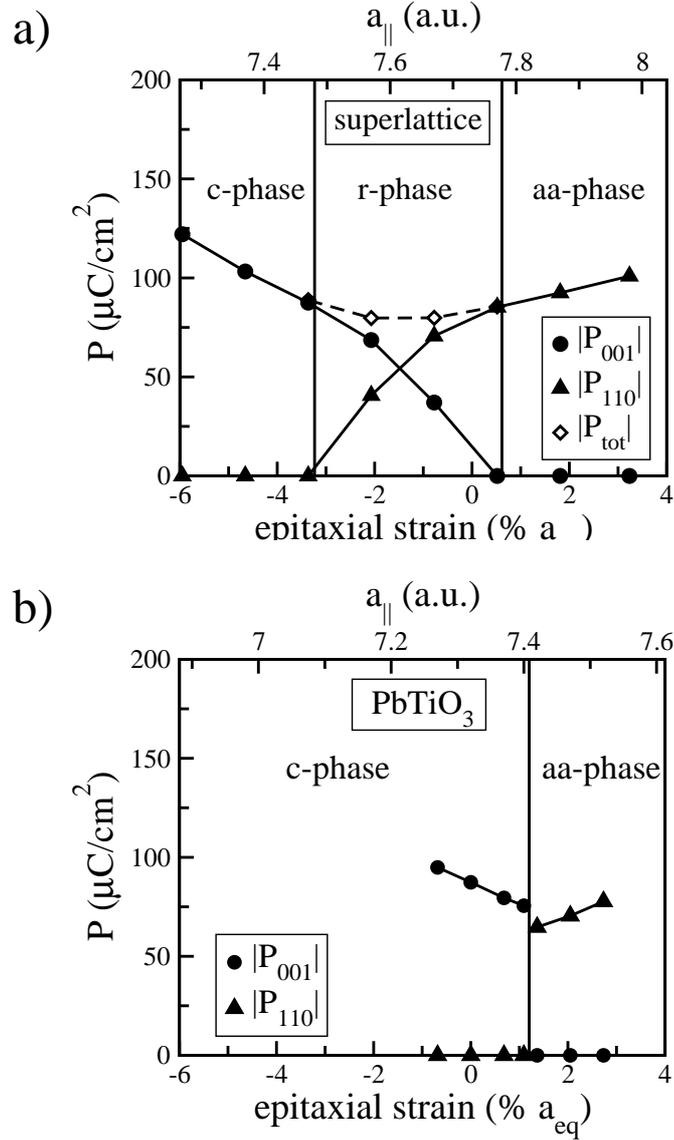

Figure 4.13: Polarization and range of stability of the different epitaxial phases versus misfit strain for a [001]-(PbTiO$_3$)$_1$/(PbZrO$_3$)$_1$ superlattice (a) and strained pure PbTiO$_3$ (b). For the superlattice both the components of the polarization and the total magnitude are represented. Solid lines represent phase transitions. The scale in both axis is the same to facilitate comparison. From Ref. [259]

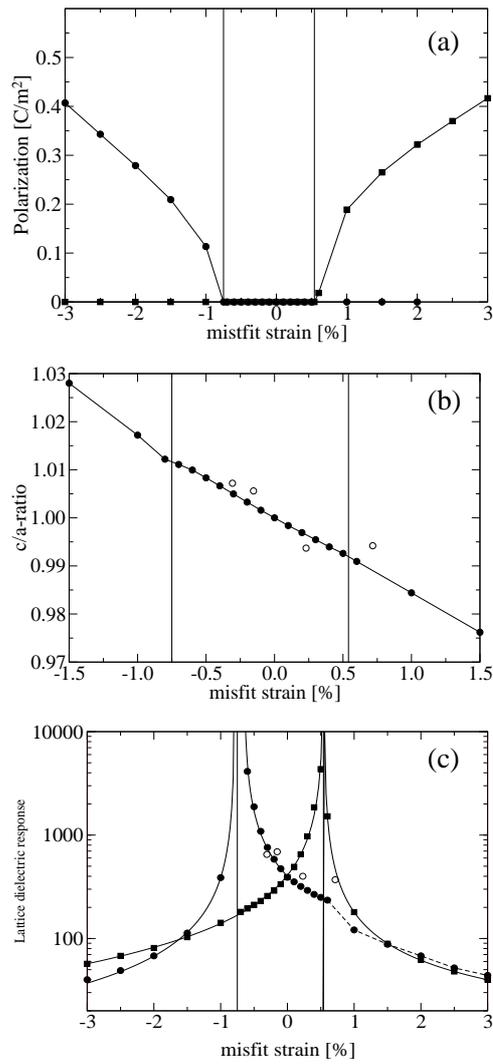

Figure 4.14: Polarization and range of stability of the different epitaxial phases (a), tetragonality (b), and lattice contribution to the dielectric response of epitaxially strained SrTiO₃ films as a function of the misfit strain. In panel (a), solid circles and squares represent polarization along [001] and [110], respectively. In panel (c), solid circles and squares denote $\epsilon_{33}$ and $\epsilon_{11}$ respectively. Open circles in panels (b) and (c) correspond to experimental data from [265]. From Ref. [260]

Early experimental works on this topic were more involved with the detection of a signal that could be related to the existence of a polar distortion than in quantitative measurements of the order parameter. Piezoelectric measurements were between the first attempts tried. Muruyama *et al.* [270] grew a 10 nm-thick *c*-axis oriented epitaxial $Pb(Zr_{0.25}Ti_{0.75})O_3$ on a $SrRuO_3/BaTiO_3/ZrO_2/Si$ heterostructure substrate. Looking at the relationship between the mechanical displacement and the voltage of a pulse applied between an AFM tip and the metallic substrate, the authors detected a clear hysteresis loop, a strong evidence that the 10 nm-thick film retained ferroelectricity. Tybell and coworkers [271] continued the miniaturization of the films by growing atomically-flat (small surface roughness over large areas) single-crystalline $Pb(Zr_{0.2}Ti_{0.8})O_3$ films on metallic Nb-doped $SrTiO_3$ single-crystal substrates. Similarly to Maruyama's technique, the piezoelectric response of the samples was measured by the mechanical deformation along the (001) direction of the high-quality film resulting from the application of a voltage between the metallic tip of the AFM and the conducting Nb-$SrTiO_3$ substrate. Also, the residual surface charge of the ferroelectric was quantified by probing the electrostatic interaction between the metallic AFM tip and the charge distribution of the ferroelectric surface (electric field microscopy, EFM) A stable piezoelectric signal was detected for the whole duration of the experiment for thin films down to 40 Å (ten unit cells; see Fig. 4.15).

The interest on the eventual existence of a ferroelectric ground state in thin films was not restricted to perovskite oxides. Bune *et al.* [269] grew a ferroelectric random copolymer [vinylidine fluoride with trifluorethylene, P(VDF-TrFE)] as thin as two monolayer thick (10 Å) on aluminium substrates. Dielectric anomalies, thermal hysteresis and pyroelectric hysteresis loops (see Fig. 4.16) were measured, and the usual mark of the bulk first-order phase transition at 77 °C was observed. A second first-order phase transition in the surface layers, not present in bulk, was detected for the thinnest films at a smaller temperature (20 °C). Due to the absence of strong finite size effects in such systems, they were considered as two-dimensional ferroelectrics.

Although those works showed undisputable evidence about the existence of a ferroelectric ground state with a polarization perpendicular to the surface, nothing could be said about the evolution of the magnitude of the ferroelectric order parameter with thickness. Indeed, it was very difficult to quantify the signal measured with the AFM because the exact electric field experienced in the film was unknown. The thickness dependence of the polarization was afforded later, by a correlation of the polarization with the structural parameters. Lattice vectors and atomic positions can be measured to a high level of accuracy by x-ray diffraction techniques [244, 274, 275, 226]. The out-of-plane lattice parameter $c$ can be related with the spontaneous polarization through the polarization-strain coupling, that is known to be particularly large in perovskite oxides, especially in $PbTiO_3$ [93] and Pb-rich based solid solutions $Pb(Zr,Ti)O_3$.

As it was mentioned in Sec. 4.1, many are the effects that might influence this thickness dependence. The interplay and delicate competition between the compressive strain typically imposed by the substrate, which tends to enhance the ferroelectricity [264, 236], the depolarization field that tends to suppress the latter [279, 34], and the interface and surface effects [280, 281, 282, 283] result in a wealth of different phases. The complexity of the problem and the subtle influence of the different experimental boundary conditions on the final phase diagram can be exemplified by summarizing the recent experimental results on epitaxial $PbTiO_3$ thin films. When $PbTiO_3$ thin films (below 500 Å) grown on top of an insulating $SrTiO_3$ substrate are cooled below the transition temperature $T_c$, satellites appeared around the Bragg peaks in the x-ray scattering profile, indicating the presence of 180° stripe domains [244, 273, 274]. As it will be explained in the following subsection, the domain pattern appears to minimize the electrostatic energy associated with the coupling of the depolarizing field and the polarization [3]. A rich phase diagram, still not yet fully understood, is observed as a function of the film thickness and temperature, including phases with diffuse domain walls with a low stripe period (referred to as $F_\alpha$), high stripe period ($F_\beta$), or even monodomain configurations [275] ($F_\gamma$). When the insulating substrate is replaced by a metallic electrode, such as Nb-doped $SrTiO_3$, the domain pattern is replaced by a monodomain configuration because the depolarizing field is now screening by the compensation charges provided by the electrode, and the film tetragonality decreases very substantially for films thinner than 200 Å [226]. The *c*-axis lattice reduction is accomodated near the film/substrate interface [284].

---

[3]Since the substrate is an insulating material without free charges, the depolarizing field can not be eliminated by compensation of the surface polarization charge.

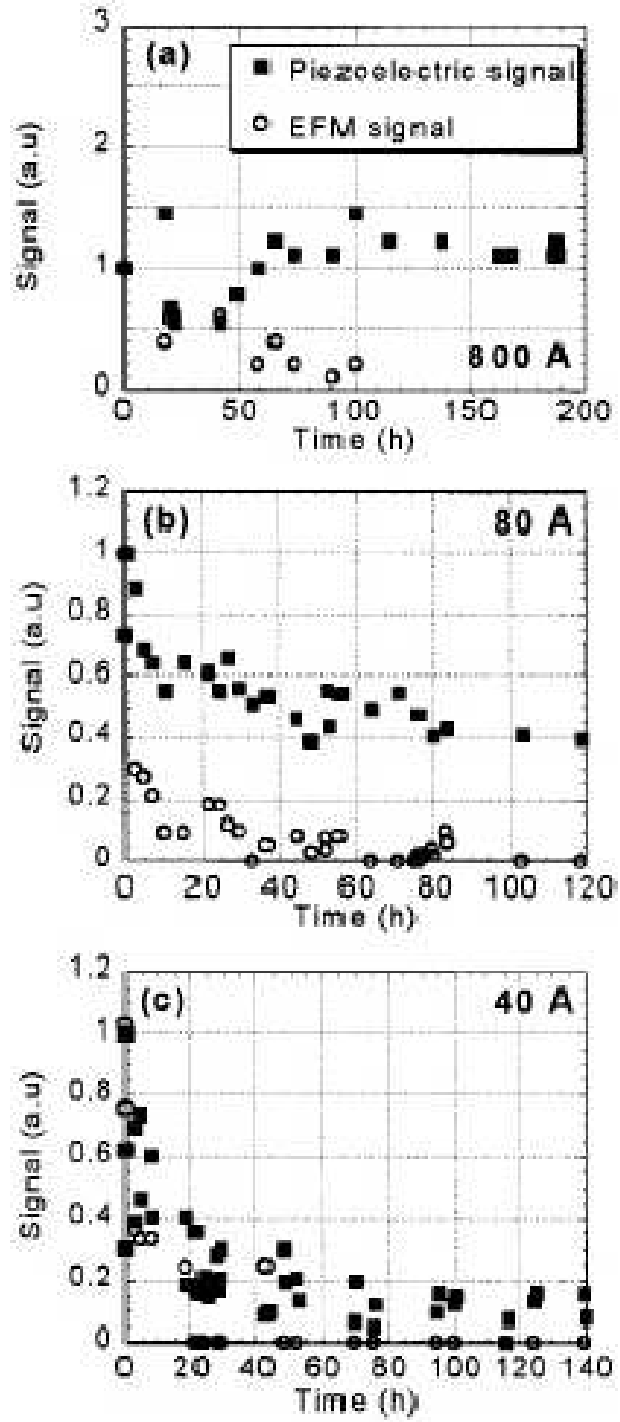

Figure 4.15: Time dependence of the piezoelectric (filled squares) and electric field microscopy (open circles) signals (in arbitrary units, a. u.) of $Pb(Zr_{0.2}Ti_{0.8})O_3$ thin films grown on metallic Nb-doped $SrTiO_3$ for three different thicknesses: 800 Å (a), 80 Å (b) and 40 Å (c). Time in hours. From Ref. [271]

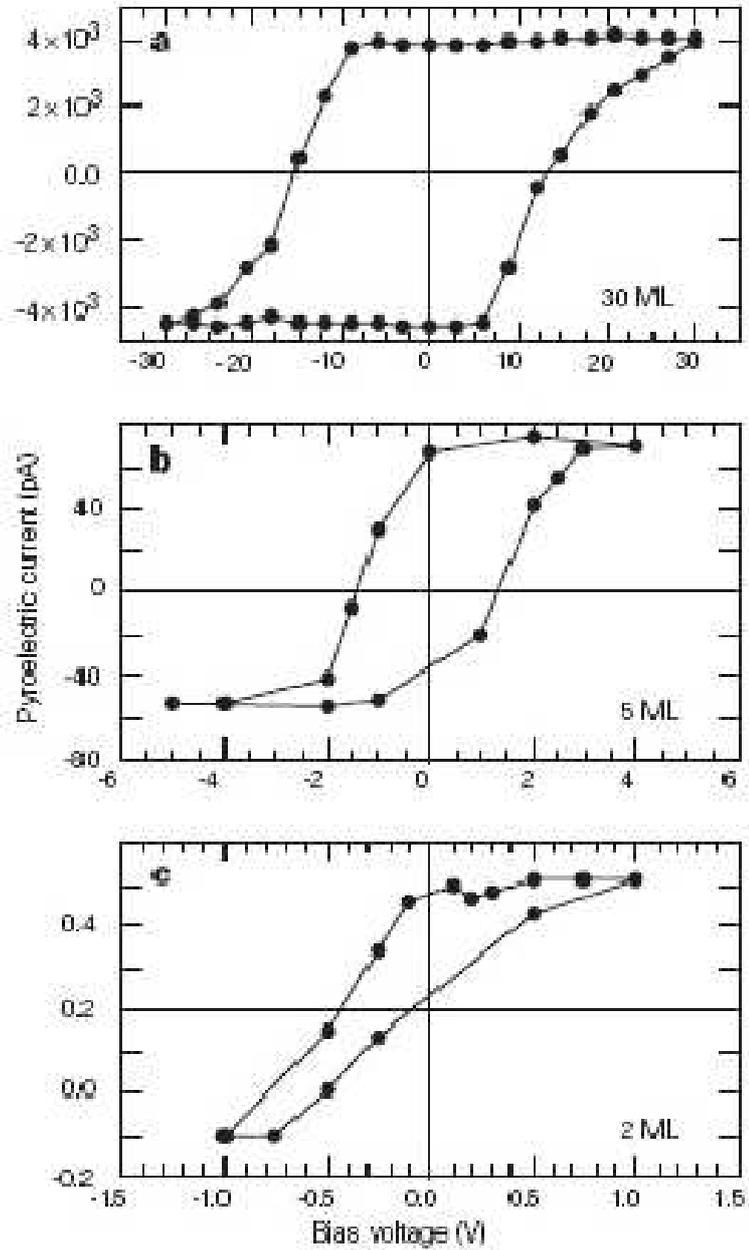

Figure 4.16: Pyroelectric hysteresis loops of a random copolymer grown on aluminium for three different thicknesses: 30 monolayers (ML) (a), 5 ML (b) and 2 ML (c). They show good saturation for the 30 ML, and 5 ML films, and incomplete saturation for the 2 ML film. The saturation value decreases with the thickness. From Ref. [269]

The subtlety of the problem can also be highlighted by the fact that a small change in the deposition technique (replacing off-axis magnetron sputtering by pulsed laser deposition), in the ferroelectric material (replacing $PbTiO_3$ by $PbZr_{0.2}Ti_{0.8}O_3$) and in the electrode (replacing Nb-doped $SrTiO_3$ by $SrRuO_3$, both of them conducting oxides) yields to very different results [226, 278]. Moreover, for the $SrRuO_3/Pb(Zr_{0.2}Ti_{0.8})O_3)/SrRuO_3$ ferroelectric capacitors, an apparent inconsistency also appears between the structural measurements and the measured switchable polarization [278]. On the one hand, there is an enhancement of the tetragonality with respect to the bulk for films thinner than 200 Å(at odds with the pure $PbTiO_3$ heterostructures), and a lack of scaling for the tetragonality in the sub 100 Å regime, suggesting the existence of a ferroelectric polarization in the system [285]. On the other hand, a clear decrease in magnitude of the switchable polarization with decreasing thickness and a substantial reduction of the out-of-plane piezoelectric response for the thinnest films are observed, pointing to a paraelectric ground state with no net polarization.

Complex problems like that are not easy to solve since many factors must be taken into account. Simplified models [244, 275] may result in a disagreement with the data. As it is now explain, first-principles simulations can be of valuable help.

### 4.4.2   The depolarizing field.

From the theoretical point of view, and besides the strain effects already discussed in Sec. 4.3, the efforts have been mainly invested on the intrinsic surface effects [280, 281, 282, 184, 279], and on the origin and effects of the depolarizing field. Three possible mechanisms for the appearance of the depolarizing field have been proposed in the literature: (i) the existence of a passive layer with degraded ferroelectric properties at the ferroelectric/electrode interface (the so-called "dead-layer") [286, 287]; (ii) the spatial variation and inhomogeneity of the polarization distribution, that generates polarization charges $\rho_{pol} = -\nabla\mathcal{P}$, either in the bulk, or near the surfaces [280, 281, 282, 184, 279]; and (iii) the incomplete screening of the depolarizing field by real metallic electrodes [288, 34, 289]. While the works considering (i) and (ii) were based only on the phenomenological Devonshire-Ginzburg-Landau theory, first-principles calculations are throughing some light into (iii) [34, 290, 226, 262, 283]. The effect of the chemistry of the interface on the screening properties starts to be considered also from first-principles [283].

The intrinsic surface effect on the polarization in ferroelectric crystals, of length comparable to the extrapolation length, was first studied by Kretschmer and Binder [280, 281] in the framework of the Devonshire-Ginzburg-Landau phenomenological theory [4] An extra term, quadratic with the polarization values on the film surfaces [282], was added to the Landau expansion of the thermodynamic potential. The coupling with the surface is monitored with the introduction of a new parameter in the boundary conditions, the so-called extrapolation length $\delta$. When $\delta > 0$, the polarization decreases near the film surface relative to the inner region. When $\delta < 0$, the polarization ls larger at the surface than inside the film. The fact that the polarization profile changes with $z$, a property that was assumed to be inherent and completely intrinsic to any ferroelectric material, automatically implies that the gradient (the derivative) of the polarization is not zero. This acts as a source of polarization charge that produces a depolarizing field. This concept was employed by many authors to produce polarization profiles under stress-free boundary conditions [291, 292, 293, 201]. It was extended later to deal with epitaxial thin films [184, 282, 279]. This theory, although simple in its formulation, usually predicts a sharp decay in the ferroelectric properties only appreciable for very small thicknesses, and can not explain many of the evolutions of ferroelectric-related properties experimentally observed [226, 294]. This suggests that other effects beyond the intrinsic surface affect might influence the thickness dependence of the ferroelectric properties.

As we will propose below, the incomplete screening by real metallic electrodes and the existence of a dead-layer at the ferroelectric/electrode interface are, from a fundamental point of view, equivalent models that explain the origin of a residual electric field inside the thin film and, therefore, both of them lead to the same conclusions. On the remaining part of this Section we will focus on the incomplete screening of the depolarizing field as the driving force for the evolution of the ferroelectric properties in thin films.

---

[4]It is noteworthy to point out that the original paper by Kretschmer and Binder [280] contains some typographical errors in the Equations. They have been corrected in more recent works, for instance, in Glinchuk *et al.* [279].

**Basic electrostatics.**

Before starting the review on the recent theoretical works on size effects, few issues about the electrostatic of the problem must be clarified. Determining whether a thin film with an out-of-plane polarization is stable or not is a subtle problem, much more delicate than the situation of a polarization parallel to the surface described in Sec. 4.2.4. The first step to gain some insight into this problem was taken by simulating free-standing slabs [31, 32]. A uniform polarization $\mathcal{P}$ in the ferroelectric slab originates a surface polarization charge density

$$\sigma_{pol} = \mathcal{P} \cdot \hat{n}, \tag{4.11}$$

where $\hat{n}$ is an unitary vector normal to the surface pointing outwards (chapter 10 of Ref. [295]). The surface charge density is equal to the magnitude of the normal component of the polarization inside the material [49], and it has opposite sign at each interface (positive in one and negative in the other). In order to solve the Poisson equation and determine the electric fields generated by the surface charge density, we need to know the *electrical boundary conditions* of the problem. Mathematically, the specification of the electric field at the points of a closed surface [5] (Neumann boundary conditions) defines a unique potential problem, with a well-behaved solution inside the region bounded by the surface (section 1.9 of Ref. [296]).

In our idealized problem of an isolated slab with an uniform and homogeneous polarization perpendicular to the surface, the solution of the electrostatic problem of finding the macroscopic internal field in the ferroelectric thin film, $\mathcal{E}$, can be solved by simple electrostatic arguments. Let us suppose that an external electric field perpendicular to the surface $\mathcal{E}_{ext}$ is applied in the vacuum region. From now, we will only consider electric fields and polarizations normal to the surface and refer to them as scalar quantities. Since, of course, there is no polarization in vacuum the electric displacement [6] there equals the applied electric field, $\mathcal{D} = \mathcal{E}_{ext}$, and is also parallel to $z$. In the absence of free charges, the normal component of the displacement vector is conserved across the ferroelectric/vacuum interface, so

$$\mathcal{E} + 4\pi\mathcal{P} = \mathcal{E}_{ext}. \tag{4.14}$$

Two extreme situations can be considered for a more detailed discussion. A first case of interest is that of vanishing external electric field, $\mathcal{E}_{ext} = 0$ [Fig. 4.17(a)]. In this case, the displacement vector vanishes and the internal field inside the slab amounts to $\mathcal{E} = -4\pi\mathcal{P}$. In other words, in the absence of an external electric field, the surface polarization charge gives rise to an internal field that points in the opposite direction than the polarization, so it tends to restore the paraelectric configuration (due to this, this field is called the *depolarizing field*). The energy coupling of this internal field with the polarization is huge in magnitude and positive in sign, large enough to destabilize any state with an out-of-plane polarization. Under these circumstances, any atomic relaxation would end with the atoms back in the centrosymmetric non-polar configuration.

A second case of particular interest corresponds to a vanishing internal electric field $\mathcal{E} = 0$ [Fig. 4.17(b)]. Since no field opposes now to the polarization, the slab might undergo a ferroelectric phase transition and

---

[5]Here, surface refers to the boundary of a given volume in space, and not the physical surface of a material.

[6]The electric displacement $\mathcal{D}$ is defined (in cgs units) as

$$\mathcal{D} = \mathcal{E} + 4\pi\mathcal{P}, \tag{4.12}$$

and in SI units as

$$\mathcal{D} = \epsilon_0\mathcal{E} + \mathcal{P}, \tag{4.13}$$

where $\epsilon_0$ is the vacuum permittivity. Throughout this chapter we will use the cgs system of units.

a spontaneous polarization $\mathcal{P}_s$ might appear in the slab. For such a configuration, the electric displacement $\mathcal{D}$ does no more vanish. These are the conditions of a bulk DFT calculation on a five-atom unit cell where, due to the periodic boundary conditions imposed for the electrostatic potential, the macroscopic internal field vanishes on a unit cell. If we want to apply these conditions to a ferroelectric slab, an external field is required in the vacuum region. The continuity of the normal component of the electric displacement gives us the value for $\mathcal{E}_{ext}$,

$$\mathcal{E}_{ext} = 4\pi \mathcal{P}_s. \qquad (4.15)$$

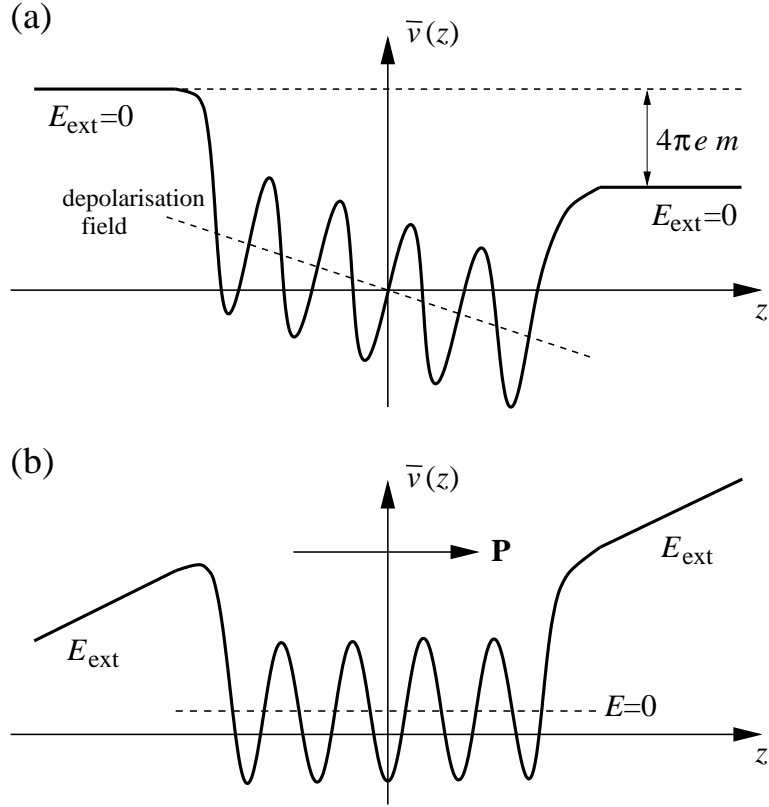

Figure 4.17: Schematic representation of the planar averaged electrostatic potential $\overline{v}(z)$ (full line) of a slab with a polarization perpendicular to the surface under (a) $\mathcal{D} = 0$ boundary conditions (equivalent to a vanishing external electric field), and (b) a vanishing internal electric field $\mathcal{E} = 0$. Planar averages are taken on the $(x, y)$ planes parallel to the surface. Dashed lines represent an average over unit cells of the planar averages. An estimate for the macroscopic internal field inside the slab can be obtained from their slope. $m$ stands for the dipole moment parallel to the surface normal and $e$ for the electron charge. From Ref. [32].

Previous examples show clearly how a polarization perpendicular to the surface of a free-standing slab can be sustained or not depending on the electrical boundary conditions, and emphasizes the crucial role played by them in determining the thin-film ground state.

**Ideal systems : perfect screening**

Ghosez and Rabe developed a first-principles based microscopic model hamiltonian (see Sec. 2.3). for $PbTiO_3$ free-standing slab under stress-free and perfect short-circuit (perfectly conducting sheets as metallic electrodes) boundary conditions [31]. Within this model, where the internal field is zero $\mathcal{E} = 0$, the ground state of films of thickness down to three unit cells were ferroelectric, with a polarization perpendicular to the surface in good agreement with the experimental results for Pb(Zr,Ti)O$_3$ thin films [270, 271]. Even more, the spatial variation of the polarization profile showed an enhancement of the polarization at the surface layers (see Fig. 4.18). However no information about the atomic relaxation at the surface was included in the model hamiltonian, and this effect might modify the picture of the polarization in the surface layer.

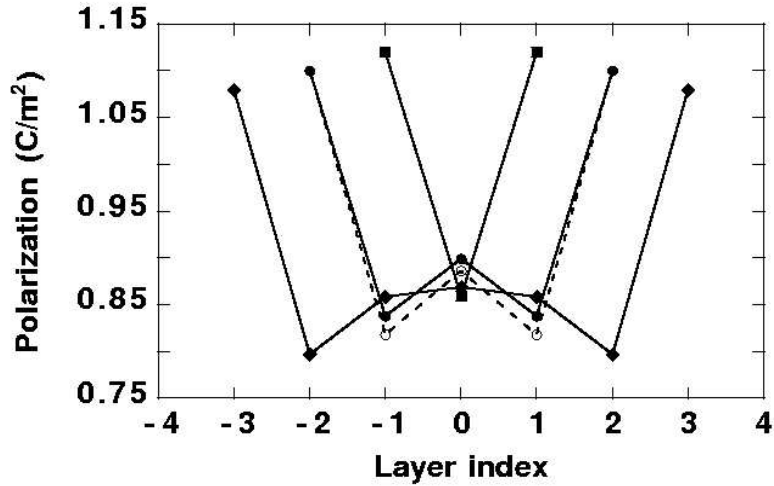

Figure 4.18: Profile of the layer by layer polarization of a stress-free PbTiO$_3$ free-standing slab under perfect short circuit boundary conditions, obtained using a first-principles based model hamiltonian. Results for slabs of three different thicknesses (3, 5 and 7 unit cells) are represented. In every case, there is an enhancement of the polarization at the surface unit cell (bulk value 0.83 C/m$^2$). From Ref. [31].

To simulate the effect of atomic relaxations full first-principles simulations must be carried out. Most of these *ab-initio* calculations are performed within a plane-wave implementation of the DFT, and assume a supercell geometry that is periodic in all three directions. That means that the boundary conditions are neither of $\mathcal{D} = 0$, nor $\mathcal{E} = 0$ (the limit cases schematized in Fig. 4.17), but an intermediate case as shown in Fig. 4.19(a). A fictitious uniform electric field, artifact of the calculation, appears in the vacuum to ensure the continuity of the potential at the supercell boundary. The strength of this field depends on the material, on the geometry of the slabs, and on the amount of vacuum that separates the periodic replicas (the larger the vacuum region, the weaker the field). This unphysical field, whose related potential is not just a constant but is linear, might introduce spurious effects on the results because the selfconsistency of the charge density is achieved in such unphysical situation [49].

Neugebauer and Scheffler [215] have introduced a dipole correction to cancel the effects of the artificial field, lately modified by Bengtsson [216] to derive the correct expression for the electrostatic energy. A dipole layer in the vacuum region produces an electrostatic potential that is linear within the supercell, and discontinuous at the border [ramp-shaped potential of Fig. 4.19(b)]. Since the jump in the potential is placed within the vacuum region, where the wavefunctions are essentially zero, it should not affect the computed properties [7]. The strength of the correction-field can be monitored by the magnitude of the compensating dipole.

Playing with the amplitude of the dipole correction, Meyer and Vanderbilt [32] studied the field-induced change in the structural relaxations at the surface of BaTiO$_3$ and PbTiO$_3$ free standing slabs, and the eventual existence of a ferroelectric ground state with a polarization perpendicular to the surface.

---

[7]Some problems might arise, however, if the slope of the ramp-shaped potential is too large. In such cases, the potential in vacuum might drop below the Fermi level of the material in a region close to the supercell border, and it could be populated with electrons transferred from the surface. This effect imposes an upper limit on the magnitude of the dipole correction that can be introduced in the simulation [32].

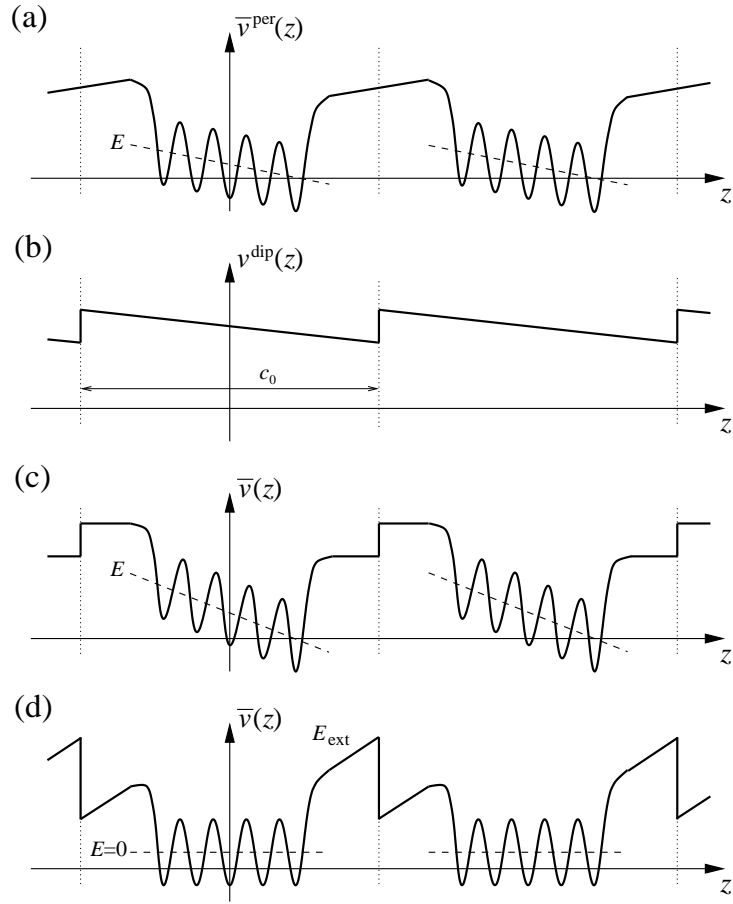

Figure 4.19: Schematic representation of the planar averaged electrostatic potential of a slab with a non-vanishing polarization perpendicular to the surface under short circuit boundary conditions (a). A fictitious field to guarantee the continuity of the potential at the boundary of the supercell (dotted lines) appears in the supercell. This unphysical field might be compensated by a dipole field (b). By tuning the magnitude of the dipole correction properly, adding or substracting certain amounts to the periodically repeated potential, we can simulate a slab under $\mathcal{D} = 0$ (c), or $\mathcal{E} = 0$ (d) electrical boundary conditions. From Ref. [32].

To discriminate between paraelectric and ferroelectric behaviour of the slab, Meyer and Vanderbilt [32] analyzed the influence of the electric displacement $\mathcal{D}$ on the internal electric field $\mathcal{E}$. A simple thermodynamic model tells us that for a paraelectric system, the elastic Gibbs energy $G$ (the most convenient thermodynamic potential while working with the typical bulk constraints) is roughly quadratic with the electric displacement, while it displays the typical double-well shape for a ferroelectric material. [Fig. 4.20(a)]. The derivative of the free energy with respect to the displacement field is, by definition, the internal electric field (chapter 3 of Ref. [1])

$$\mathcal{E}(\mathcal{D}) = \frac{\partial G}{\partial \mathcal{D}}. \tag{4.16}$$

As it can be shown in Fig. 4.20(b), the function $\mathcal{E}(\mathcal{D})$ is very different depending on the situation, changing from linear in the paraelectric case to more complicated anharmonic shape in a ferroelectric slab. For a given external field normal to the surface (or electric displacement, since $\mathcal{D} = \mathcal{E}_{ext}$ in vacuum, and it is conserved across the ferroelectric/vacuum interface) controlled by the magnitude of the dipole corrections, the internal field can be calculated by averaging the computed electrostatic potential (output of the first-principles calculations) inside the material. The shape of the first-principles $\mathcal{E}(\mathcal{D})$ function reveals whether the slab is ferroelectric or not. For BaO-terminated $BaTiO_3$ slab, and $TiO_2$ terminated $PbTiO_3$ slab, Meyer and Vanderbilt [32] found strong evidence for a ferroelectric ground-state with the polarization perpendicular to the surface, in agreement with effective Hamiltonian calculations of Ghosez and Rabe [31]. For the other two studied surfaces ($TiO_2$-terminated $BaTiO_3$ and PbO-terminated $PbTiO_3$ slabs), it was not possible to determine the internal electric field with a very high accuracy, so the analysis was not clear enough to make an undisputable statement, although both of them were very close to a ferroelectric instability, also in good agreement with experimental results on thin films perovskite oxides [270, 271].

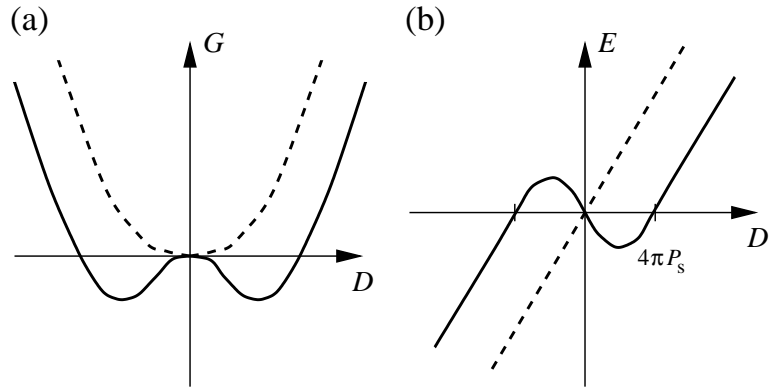

Figure 4.20: (a) Elastic Gibbs energy, $G$ as a function of the displacement field $\mathcal{D}$ for a paraelectric (dotted line) and a ferroelectric (full line) material. The internal electric field is defined as the derivative of $G$ with respect $\mathcal{D}$. Corresponding relations for the internal electric field as a function of the displacement for the two different materials are shown in panel (b). From Ref. [32].

The previous theoretical approaches were a step forward in the understanding of the problem, but both of them relied on approximations that can be relaxed in a more realistic simulation. In particular (i) the *mechanical boundary conditions*, that have shown to strongly affect the ferroelectric properties of thin films (as described in Sec. 4.3) were not considered. On the one hand, the model hamiltonian calculation was performed under stress-free boundary conditions, so the in-plane lattice constant (i. e. the in-plane strain) was allowed to relax. On the other hand, Meyer and Vanderbilt set up the lattice constant in the $x$ and $y$ direction to the theoretical equilibrium lattice parameter, computed for the bulk tetragonal phase. And (ii) the *electrical boundary conditions* were empirically introduced modifying by hand the external electric field and *perfect screening* ($\mathcal{E} = 0$ within the film) was assumed.

### Real systems : imperfect screening

The next obvious improvement was to include explicitly in the calculation the existence of a metallic electrode on top of which the thin ferroelectric film is epitaxially grown, and calculate its influence on the atomic relaxations and the polarization at the interface. Such a research line for approaching the simulations to the actual experiments has the degree of freedom of the choice of a particular metal as electrode. Since the atomic structure and the chemical bonding at the interface are expected to play a major role in the screening of the polarization charge, and both of them will depend strongly on the type of electrode, the results may change (at least quantitatively) for different metals, being intrinsically less universal. Conventional ferroelectric capacitors use noble metals (Pt, Ir, Au, Ru) or conductive oxides ($LaNiO_3$, $(La,Sr)CoO_3$, $SrRuO_3$) as electrodes. While the former form Schottky blocking contact, thus limiting the leakage current [190, 191, 193], conducting oxides electrodes form nonblocking Poole-Frenkel contacts [190, 191, 192, 193, 194] and are the solution to the problem of fatigue (loss of polarization due to bipolar cycling) and imprint (preference of a ferroelectric capacitor for one polarization state) [297].

Some first-principles DFT-based calculations on transition-metal/perovskite interfaces were due to Oleinik *et al.* [298] on $Co/SrTiO_3/Co$ heterostructures, and Rao and coworkers [299] on $5d$ transition metals on $BaTiO_3$. However, those works were only concerned by the atomic and electronic structures at the interface, and did not enter into the discussion of ferroelectric instabilities. Both of them agree on the fact that the most stable structure corresponds to the metal atom lying on top of the interfacial oxygen atoms. The calculated electronic structure for the $Co/SrTiO_3/Co$ interface shows an exchange coupling between the interface Co and Ti atoms, mediated by O, which results in an induced magnetic moment on the interfacial Ti atoms, antiparallel to the magnetic moment of Co. Regarding the transition metal/$BaTiO_3$ interfaces, the most remarkable results are the appearance of metal induced gap states (MIGS) [300] on $BaTiO_3$ due to the penetration of the $5d$ metal states into the perovskite, and the strong dependence of the Schottky barrier with the metal deposited (absence of Fermi energy pinning). Only recently, Na Sai and coworkers [283] have studied the influence of a noble metal electrode (Pt) in the stabilization of the ferroelectric phase (the results will be discussed below).

Regarding conductive oxides, strontium ruthenate ($SrRuO_3$) is one of the most extended electrodes for ferroelectric thin films capacitors [301, 302] because of the superior fatigue and leakage characteristics exhibited with respect to those made of noble metals [303]. In a typical ferroelectric capacitor heterostructures it is usually grown on top of $SrTiO_3$ [see Fig. 4.21(a)]. Under this epitaxial growth condition, $SrRuO_3$ adopts a pseudocubic perovskite structure [304, 305, 306]. That makes the atomic arrangement along the [001] direction similar to that in the ferroelectric perovskite, allowing the growth of high crystalline quality thin films, with coherent and sharp interfaces between the ferroelectric and the electrode. $SrRuO_3$ thin films show a good metallic behaviour (density of carriers of the order of $10^{22}$ electrons/$cm^3$ [307]), with low isotropic resisitivity, excellent chemical and thermal stability, and small surface roughness [304]. Although its ground state is ferromagnetic, the Curie temperature (around 140 K for thin films [304]) is so low in comparison with the usual device conditions that it can be considered as a typical paramagnet in the calculations.

Junquera and Ghosez [34] performed full first-principles DFT calculations on a realistic ferroelectric capacitor made of $BaTiO_3$ thin films in between two $SrRuO_3$ electrodes in short circuit [Fig. 4.21(b)]. The basic unit cell, periodically repeated in space, had a generic formula $[(SrO − RuO_2)_n − SrO/TiO_2 − (BaO − TiO_2)_m]$, where $n$ (respectively $m$) stands for the number of unit cells of $SrRuO_3$ (respectively $BaTiO_3$). In all the simulations $n$ was fixed to five, a thickness large enough to avoid interaction between periodic replicas of the interface. The thickness of the ferroelectric thin film was monitored by changing $m$, from two to ten. For odd $n$ and even $m$ the structure presents two mirror symmetry planes located on the central $RuO_2$ and $TiO_2$ layers. Due to the volatility of the $RuO_2$ layer, a $SrO/TiO_2$ interface is expected experimentally and simulated theoretically. The whole heterostructure was supposed to be grown on top of a thick $SrTiO_3$ substrate. The $SrTiO_3$ substrate was not explicitly treated in the calculation. However, its mechanical effect on the whole stack was implicitly included by constraining the in-plane lattice constant of the whole capacitor to the bulk cell parameter of $SrTiO_3$. As it was seen in Sec. 4.3, the main effect of this compressive strain at the bulk level is the stabilization of a ferroelectric tetragonal phase with a polarization along the [001] direction (the previously so-called $c$ phase). The change in the structure of strained $BaTiO_3$ is accompanied by a significant enhancement of the polarization [236, 263, 264, 35], whose magnitude changed from 24 $\mu C/cm^2$, assuming a cubic cell, to 31 $\mu C/cm^2$ in the strained tetragonal cell [8]. The compressive 2D-clamping also lowers the symmetry of the epitaxial $SrRuO_3$ from cubic to tetragonal. As it was done for the case of the free-standing slabs (Sec. 4.2), a starting *reference* ionic configuration was defined by piling up truncated bulk strained materials, using as building blocks the tetragonal unit cells of both $BaTiO_3$ and $SrRuO_3$. Atomic coordinates were then relaxed, imposing a mirror symmetry plane on the central $TiO_2$ plane. The symmetry constraint prevents the appearance of ferroelectricity, and the resulting relaxed structures correspond to the optimized *paraelectric* states (lowest possible energy with centrosymmetric symmetry). Starting from these relaxed symmetric paraelectric positions, $BaTiO_3$ atoms were then displaced along $z$ following the pattern of displacement of the bulk soft zone-center ferroelectric mode, $\xi$ (see Section 2.3), calculated for the same tetragonal geometry as the one used to build the reference supercell. Only a uniformly polarized monodomain configuration was considered ($\xi_i = \xi$, identical everywhere), while the $SrRuO_3$ atoms were kept fixed at their positions in the paraelectric heterostructures. Figure 4.22 shows the evolution of the free energy of the supercell with respect to the paraelectric phase, as a function of the ferroelectric distortion $\xi$, and the number of unit cells of $BaTiO_3$, $m$. A change in the behaviour of the energy is observed for a thickness around 6 unit cells (24 Å). For thicker films, the energy decreases when the atoms move along the bulk soft-mode pattern. This indicates that there exists at least one ferroelectric state energetically more stable than the paraelectric phase, in good agreement with experimental results for the same thicknesses [271, 308]. In contrast, for thinner films, the energy is minimized for the paraelectric configuration. This fact alone is not enough to conclude anything about the absence of ferroelectricity, since only a small region of the whole configuration space is analyzed in Fig. 4.22. To further prove this point, a full first-principle relaxation of all the atomic positions (including those of $SrRuO_3$) for the $m = 2$ capacitor structure starting from a ferroelectric configuration was performed, with the result that the atoms moved back to the paraelectric positions.

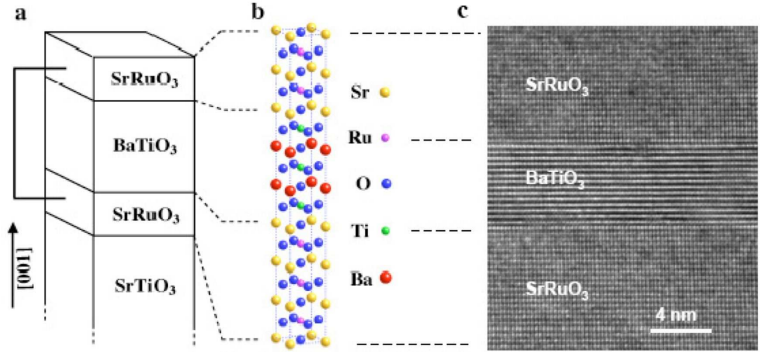

Figure 4.21: Schematic view of a typical ferroelectric capacitor (a), and atomic representation of the periodically repeated unit cell (for two unit cells of $BaTiO_3$, $m = 2$) used in the simulations (b). An experimental transmission electron microsocopy (TEM) image of the structure is shown in (c). The epitaxy is such that $SrTiO_3(001)\|SrRuO_3(001)\|BaTiO_3(001)$, and $SrTiO_3\langle 100\rangle \|SrRuO_3\langle 100\rangle \|BaTiO_3\langle 100\rangle$. Panels (a) and (b) from Ref. [34]. Panel (c) courtesy of J. Rodríguez-Contreras.

This last work brought back to the forefront of fundamental research the old idea of the modification of the ferroelectric properties in thin films by a residual depolarizing field, due to the *imperfect screening* of the surface polarization charges by real metallic electrodes. This model was originally developed by Batra and coworkers in the early seventies, on the basis of a phenomenological Devonshire-Ginzburg-Landau theory [309, 310, 311, 288, 312, 313]. The origin of the residual depolarizing field can be easily understood revisting the external field introduced empirically by Ghosez and Rabe [31] and Meyer and Vanderbilt [32] to neutralize the bare depolarizing field in free-standing slabs, and analyzing the physical source of such a field in a real capacitor. We have already explained how, in the absence of an external electric field, a polarization perpendicular to the surface of an isolated free-standing slab induces surface polarization charges $\sigma_{pol}$ [Eq. (4.11)], and a huge depolarizing field [9] $\mathcal{E}_d = -4\pi\mathcal{P}$ [Fig. 4.23(a)]. Therefore, the presence of compensation charges $\sigma_{com}$, provided by electrodes, is required in order to stabilize the ferroelectric ground state. In an ideal case, assuming *perfect metallic electrodes* [Fig. 4.23(b)], the compensation of the ferroelectric polarization charge would be perfect. That means no net charge at the interface, and the center of gravity of $\sigma_{pol}$ and $\sigma_{com}$ that coincide (both of them would lie on a sheet right at the interface between the metal and the ferroelectric electrode). In this hypothetical case, the depolarizing field would be fully compensated, and the macroscopic field inside the ferroelectric thin film would vanish. However, in the case of *real electrodes*, the compensation of the polarization charges will be always incomplete. The compensation charge provided by the electrodes to screen the polarization charge spreads over a finite distance within the electrode. Due to this spatial dispersion, an interface dipole density arises at the ferroelectric/electrode interfaces. The interface dipole density, which has the same sign at each ferroelectric/electrode interface, is responsible for an offset of the electostatic potential in the same direction at each interface [Fig. 4.23(c)]. To establish short-circuit boundary condition, an amount of charge must flow from one interface to the other, creating a residual depolarizing field inside the thin film. Let us express this depolarizing field as $\mathcal{E}_d = -4\pi\alpha\mathcal{P}$, where $\alpha$ is referred to as the *screening factor* and takes a value between 0 (perfect screening) and 1 (no screening at all). Its value can be used to monitor the electrical boundary condition from open circuit to short-circuit and any situation in between [290]. Batra and coworkers proposed slightly different expressions for the effective screening length and the residual depolarizing field depending on whether the electrode has a semiconductor [309, 288] or metallic [311, 288] character (see Table 4.3). However, these models rely on the knowledge of several material-dependent parameters (dielectric constants of both the ferroelectric and the electrode, average density and effective mass of the carriers, etc.) hampering their applicability, or even giving questionable results depending on the chosen values for the parameters. Also, the dependency of the screening length on the type of the electrode/ferroelectric interface is not taken into account.

---

[8]The c-strain was fixed to its $\mathcal{P} = 0$ value and not allowed to relax with $\mathcal{P}$ in this calculation, which explains that the value is slightly smaller than that independently reported by Neaton and Rabe.

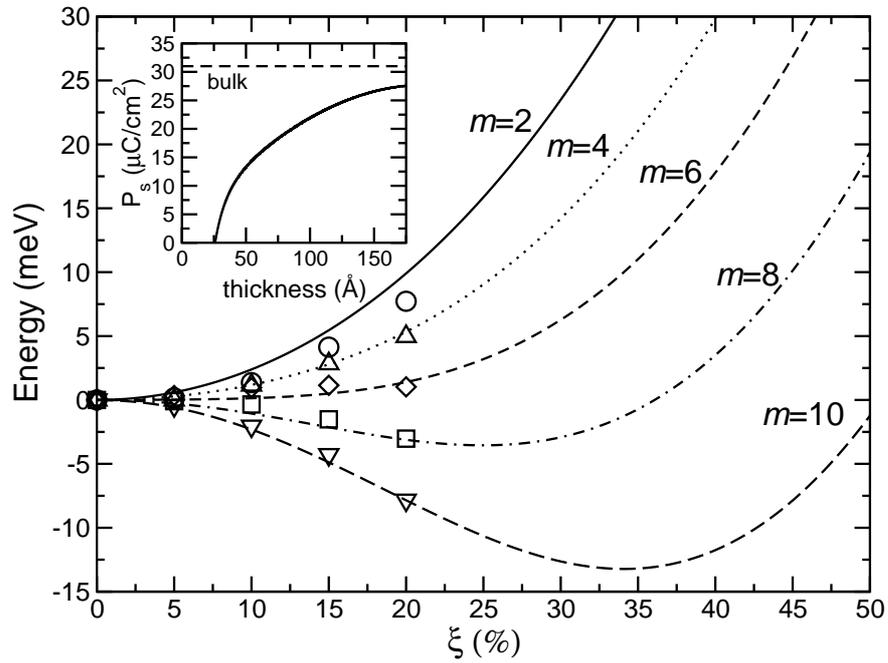

Figure 4.22: First-principles (symbols) and electrostatic model results (lines) for the evolution of the energy, with respect the paraelectric phase, as function of the soft-mode distortion $\xi$, at different thickness of the BaTiO$_3$ layer: $m = 2$ (circles, full lines), $m = 4$ (triangles up, dotted lines), $m = 6$ (diamons, short-dashed lines), $m = 8$ (squares, dot-dashed lines), $m = 10$ (triangles down, long-dashed lines). The magnitude of $\xi$ is reported as a percentage of the bulk soft-mode displacements of BaTiO$_3$ atoms ($\xi = 1$ corresponds to the distortion of the bulk tetragonal ferroelectric phase). Inset: evolution of the polarization as a function of the thickness of the BaTiO$_3$ layer. The horizontal line represents the bulk spontaneous polarization for strained BaTiO$_3$. From Ref. [34].

The novelty of the Junquera-Ghosez approach is that the depolarizing field was computed for the first time from first-principles (without any empirical parameter or input from other previous calculation) from the slope of the nanosmoothed average [314, 315, 316] of the electrostatic potential in the ferroelectric. In the calculation, the transfer of charge from one side of the interface to the other, that determines the surface dipole density, and ultimately the depolarizing field, is treated self-consistently. So the electrons are allowed to reaccomodate to the specific new environment at the ferroelectric/electrode interface. However it is not always feasible to perform such demanding calculations, and a simple model is desirable to make quick and reasonable estimations of the depolarizing field.

In order to determine the screening factor $\alpha$, and consequently the magnitude of the depolarizing field $\mathcal{E}_d$, the first step is to map our complicated interface into a simplified system, made of uniform sheets of charge in vacuum, as shown in Fig. 4.24. The second step will be to adjust the model on the first-principles data. We will assume that the surface charge $\sigma_{pol}$ induced by the polarization $\mathcal{P}$ lies in a sheet right at the interface between the ferroelectric and the electrode, whereas the compensation charges provided by the electrode $\sigma_{com}$ are displaced from the ferroelectric-electrode interface, located on sheets at a distance $\lambda^{eff}$ from the interface, the so-called *effective screening length* [10].

---

[9]Hereafter, we refer to the deoplarizing field as $\mathcal{E}_d$ to emphasize that is it opposed to the polarization.

[10]The effective screening length might be related but is *not* equal to the Thomas-Fermi screening length of the electrode.

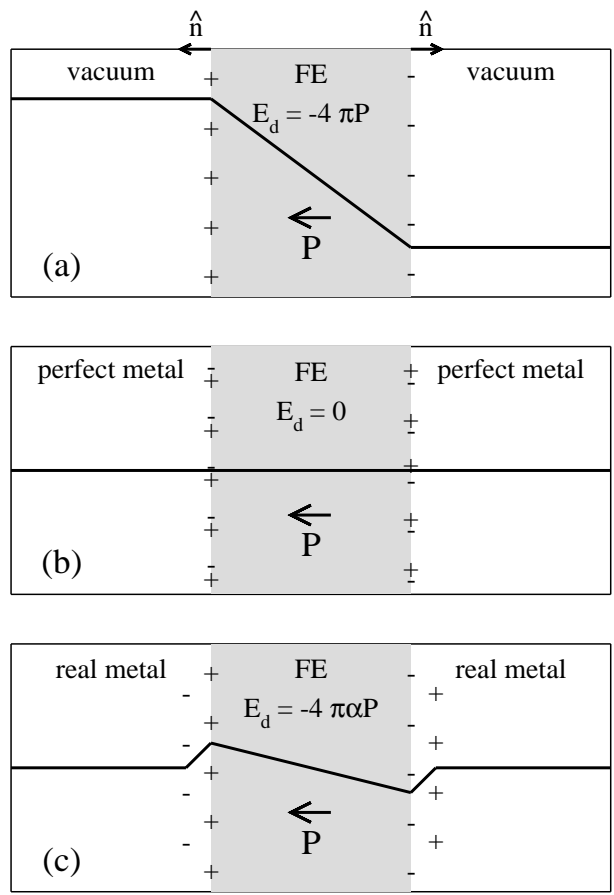

Figure 4.23: Depolarizing fields for an isolated free-standing slab in the absence of an external electric field (a), and for a ferroelectric capacitor with a perfect screening (ideal metal) (b) and a real metallic electrode (c) under short circuit boundary conditions. The ferroelectric thin film (grey region) is denoted by FE. $E_d$ and P stand for depolarizing field and polarization, respectively.

| Author | Electrode | Effective screening length $\lambda$ | Boundary conditions | Depolarizing field |
|---|---|---|---|---|
| I. P. Batra and B. D. Silverman [309] | semiconducting | $\lambda = \sqrt{\frac{\epsilon_e kT}{4\pi e^2 n_e}}$ | short-circuit | $\mathcal{E}_d = -4\pi\mathcal{P}\left[1 - \theta\left(\lambda, d\right)\right]$ $\theta\left(\lambda, d\right) = \left[1 + \frac{2\lambda}{\epsilon_e d} csch\left\{\frac{L-d}{2\lambda}\right\}\left(cosh\left\{\frac{L-d}{2\lambda}\right\} - 1\right)\right]^{-1}$ |
| R. R. Mehta *et al.* [311] | metallic | $\lambda = \sqrt{\frac{\epsilon_e}{3e^2}\left(\frac{3}{8\pi}\right)^{2/3}\frac{h^2}{m_{eff}}\frac{1}{n_0^{1/3}}}$ | short-circuit | $\mathcal{E}_d = -\left(\frac{\mathcal{P}}{\epsilon_F}\right)(1-\theta)$ $\theta = \frac{\epsilon_e/\lambda}{2\epsilon_F/d + \epsilon_e/\lambda}$ |
| M. Dawber *et al.* [289] | metallic | $\lambda = \sqrt{\epsilon_e}\lambda_{TF}$ | bias $V$ | $\mathcal{E}_d = \frac{V + 8\pi\mathcal{P}_s\frac{\lambda}{\epsilon_e}}{d + \epsilon_F\left(2\frac{\lambda}{\epsilon_e}\right)}$ |
| J. Junquera and Ph. Ghosez [34] | metallic | fitted from first-principles | short-circuit | $\mathcal{E}_d = -\frac{8\pi\mathcal{P}\lambda}{d}$ |
| A. M. Bratkovsky and A. Levanyuk [287] | "dead-layer" | thickness of the dead-layer $\lambda$ | short-circuit | $\mathcal{E}_d = -\frac{8\pi\mathcal{P}\frac{\lambda}{\epsilon_e}}{d + \epsilon_F\left(2\frac{\lambda}{\epsilon_e}\right)}$ |

Table 4.3: Different formulations for the for the depolarizing field and the effective screening length reported in the Literature. $\mathcal{P}_s$ stands for the spontaneous polarization. $d$ represents the thickness of the thin film and $L$ is used for the length of the thin film plus the length of the metallic electrodes. $\epsilon_e$ and $\epsilon_F$ stand for the dielectric constant of the electrode and the thin ferroelectric film respectively. $\lambda_{TF}$ is the Thomas-Fermi screening length. $k$ represents the Boltzmann constant, and $T$ the temperature. $e$ is the electronic charge. $n_e$ and $m_{eff}$ are, respectively, the density and the effective mass of the carriers in the electrode. $n_0$ is the average electron density under the conditions of zero field. For the dead-layer model we have assumed a monodomain configuration in the ferroelectric thin film under short-circuit boundary conditions. Units in cgs, except for the paper of Mehta *et al.* [311] in SI.

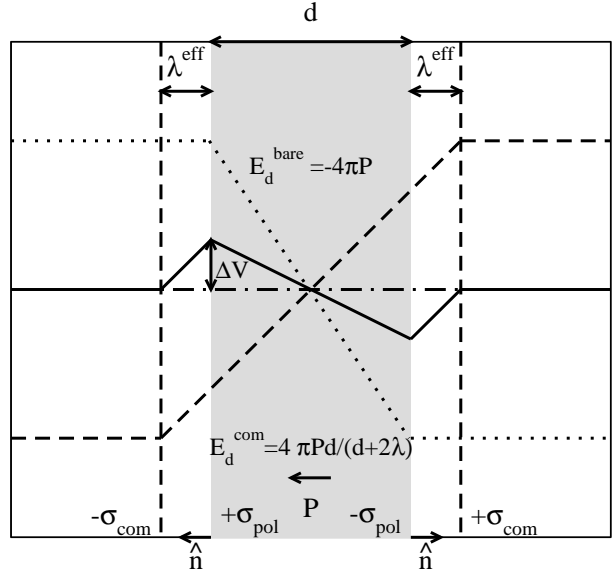

Figure 4.24: Simplified system to model a symmetric metal/ferroelectric/metal ferroelectric capacitor. Dotted line represents the profile of the electrostatic potential generated by the polarization charges ($\sigma_{pol}$), located at the electrode/ferroelectric interface. Dashed line represents the profile of the compensation potential generated by charges in the electrode ($\sigma_{com}$), supposed to lie on sheets at a distance $\lambda^{eff}$ from the interface (vertical dashed lines). Solid line is the residual uncompensated electrostatic potential. d stands for the thickness of the ferroelectric thin-film (grey region), and $\Delta V$ for the potential drop across the interface.

Under a vanishing external electric field, the bare macroscopic field generated by the polarization charges inside the ferroelectric thin film would be the same as in the case of an isolated free-standing slab

$$\mathcal{E}_d^{bare} = -4\pi\mathcal{P}, \tag{4.17}$$

and the corresponding potential jump across the film would be equal to $4\pi\mathcal{P}d$, where $d$ is the film thickness.

The macroscopic field must vanish inside the metallic electrode and, under short-circuit boundary conditions, the potential generated by the metallic sheets has to compensate the potential drop in a distance $d + 2\lambda^{eff}$. Therefore, the compensating field $\mathcal{E}^{com}$ will be given by

$$\mathcal{E}^{com} = \frac{4\pi d\mathcal{P}}{d + 2\lambda^{eff}}. \tag{4.18}$$

The residual macroscopic depolarizing field inside the ferroelectric thin film, $\mathcal{E}_d$, can be finally obtained from the sum of both fields,

$$\begin{aligned}
\mathcal{E}_d &= \mathcal{E}_d^{bare} + \mathcal{E}^{com} \\
&\approx -\frac{8\pi\lambda^{eff}\mathcal{P}}{d},
\end{aligned} \tag{4.19}$$

where we have supposed that $\lambda^{eff} \ll d$. It is worthy to note that this expression for the depolarizing field is the same as the one that would be obtained supposing an exponential-decaying charge density, $\rho(z)$, in the electrodes [289] of the form

$$\rho(z) = \frac{\mathcal{P}}{\lambda^{eff}} exp\left(-\frac{z}{\lambda^{eff}}\right).$$ (4.20)

As it was mentioned in the Introduction of this Section, the existence of a "dead" or "passive" dielectric layer at the interface with the electrodes has also been proposed to explain the apperance of a macroscopic electric field inside the thin film. Within this model, assuming that the film is uniformly polarized and under short circuit boundary conditions, the homogeneous contribution of the electrostatic energy per unit volume $\tilde{F}_h$ takes the form (from Eq. (11) of Ref. [287]) [11]

$$\frac{\tilde{F}_h}{A\,d} = \frac{8\pi \mathcal{P}_s \lambda}{\epsilon_e d + \epsilon_F 2\lambda} \frac{\mathcal{P}_s}{2} = \frac{8\pi \mathcal{P}_s \frac{\lambda}{\epsilon_e}}{d + \epsilon_F 2\frac{\lambda}{\epsilon_e}} \frac{\mathcal{P}_s}{2}.$$ (4.21)

From Eq. (4.21) we infer a value of the electric field inside the thin film that is exactly the same as the one obtained by Dawber *et al.* [289] from the incomplete screening model. This fact suggest a formal equivalency of the two models. What is called the thickness of the dead-layer in one of them is replaced by the effective screening length in the other. The larger the screening length or, equivalently, the larger of the thickness of the dead-layer, the worse the screening of the depolarizing field and the larger the field inside the thin film. Therefore, both models yield to the same results (for instance both of them predict the breaking of the system in 180° stripe domains for large enoguh fields).

The screening factor inferred from Eq. (4.19) takes the form $\alpha = \frac{2\lambda^{eff}}{d}$. For $d \to \infty$ (bulk material) or $\lambda^{eff} \to 0$ (perfect compensation of the depolarization charge by metallic electrodes), the depolarizing field vanishes and we would expect to recover the bulk behaviour in the thin film [288, 283].

All the quantum effects and the chemistry at the interface are hidden in the only parameter of the model, the *effective screening length* $\lambda^{eff}$. It will depend on the ferroelectric material, the electrode, and the atomic structure and the particular orientation of the interface. In short, on any variable that might alter the chemical bonding or the transfer of charge at the interface. From a practical point of view, the effective screening length is fitted from a single first-principles calculation for a given capacitor structure. The thickness of the ferroelectric thin film, $d$, is known from the number of layers of the ferroelectric perovskite included in the simulations. The depolarizing field, $\mathcal{E}_d$, is obtained from the slope of the nanosmoothed averaged electrostatic potential, output of the first-principles calculations [314, 315]. Considering $\xi$ and $\mathcal{E}$ as independent degrees of freedom, the polarization is given by

$$\begin{aligned} \mathcal{P} &= \sum_\kappa \frac{\partial \mathcal{P}}{\partial \xi}\Big|_{\mathcal{E}_d=0} \xi + \frac{\partial \mathcal{P}}{\partial \mathcal{E}_d}\Big|_\xi \mathcal{E}_d \\ &= (1/\Omega_0)\bar{Z}^* \xi + \chi^\infty \ \mathcal{E}_d \\ &= \mathcal{P}^0 + \chi^\infty \ \mathcal{E}_d, \end{aligned}$$ (4.22)

where $\bar{Z}^*$ is the Born effective charge associated to the atomic displacement pattern $\xi$ (Sec. 2.3.2), $\chi^\infty$ is the optical electric susceptibility of the ferroelectric, and $\Omega_0$ refers to the unit cell volume of the ferroelectric. The first term $\mathcal{P}^0$ is the zero-field polarization that arises from the displacement of the ions from their position in the unpolarized structure, whereas the second term is the field induced electronic polarization, typically one order of magnitude smaller than the former with metallic electrodes. In what follows, we will therefore make the approximate $\mathcal{P} \approx \mathcal{P}^0$.

Replacing all these magnitudes in Eq. (4.19), we can get an estimate of the effective screening length $\lambda^{eff}$ by imposing to the model to reproduce the slope of the macroscopic electrostatic potential computed at the DFT level. For the $SrRuO_3/BaTiO_3/SrRuO_3$ interface, it turned out to be $\lambda^{eff} = 0.23$ Å. This value was checked to be independent of the amplitude of $\xi$ and of the film thickness $d$, attesting that the simple model with a single parameter is enough to encapsulate the essential physics. Therefore, in practice, once it has been calculated for a single interface, it can be used independently of the thickness and the induce polarization[12]

---

[11]In order to make the notation compatible with the one used throughout this chapter, we have replaced in the original equation the thickness of the thin film $l$ by $d$, the thickness of the dead layer $d/2$ by $\lambda$, the dielectric constant of the thin film $\epsilon_e$ by $\epsilon_F$, and the dielectric constant of the "dead-layer" $\epsilon_g$ by $\epsilon_e$.

## 4.4.3   Dependence of the depolarizing field on different factors.

Our simple model highlights how the amplitude of the depolarizing field depends on: (i) the effective screening length, $\lambda^{eff}$; (ii) the magnitude of the induced polarization, $\mathcal{P}$ ; and (iii) the film thickness of the ferroelectric thin film, $d$.

The dependence of $\lambda^{eff}$ on the interface chemistry has been confirmed by Sai and coworkers [283] who have performed first-principles calculations on Pt/PbTiO$_3$/Pt and SrRuO$_3$/PbTiO$_3$/SrRuO$_3$ ferroelectric capacitors. In contrast with the work of Junquera and Ghosez [34], they have carried out a full atomic relaxation of the coordinates of the whole heterostructure. A slope in the averaged electrostatic potential was found in both Pb-based capacitors, although the magnitude of the the field was much smaller for the Pt-electrodes, suggesting a value of $\lambda^{eff}$ approaching to zero, and a negligible internal field within the ferroelectric. Therefore, PbTiO$_3$ films with Pt electrodes are ferroelectric for all thickness down to $m = 1$, even with an enhancement of the polarization with respect to the bulk value for the thinnest case, in good agreement with the model hamiltonian results under the same electrical boundary conditions [31]. The slope of the averaged electrostatic potential with SrRuO$_3$, points to a partial compensation of the PbTiO$_3$ surface polarization charge. The screening is weaker than in the previous case, and the polarization of the PbTiO$_3$ thin film roughly halves the bulk value. Remarkably, the relaxation of the SrRuO$_3$ atoms change significantly the value of $\lambda^{eff}$; keeping the metallic atoms fixed at the ideal positions results in a complete suppression of the ferroelectricity in PbTiO$_3$ thin films for either termination. This fact indicates that the chemical bonding at the interface plays an important role for the significant reduction of the depolarizing field.

First-principles results for the SrRuO$_3$/BaTiO$_3$/SrRuO$_3$ ferroelectric capacitor confirm the prediction of the model with respect the dependency on the ionic polarization and the thickness. Figure 4.25(a) represents the dependence of the residual depolarizing field with the induced polarization at constant thickness ($m = 6$) of the ferroelectric thin film. From the profile of the macroscopic electrostatic potential across the ferroelectric capacitor, we can see how $\Delta V$ depends linearly with the induced polarization, monitored here by the amplitude of the atomic displacement along the bulk soft-mode, $\xi$.

In Fig. 4.25(b) we plot the thickness dependence of the depolarizing field at constant value of the ferroelectric distortion ($\xi = 0.15$). From the profile of the electrostatic potential, we can draw how the larger the thickness of the ferroelectric layer, the longest the distance to compensate the potential drop across the interface, itself a quantity almost independent of $d$. That results in smaller slopes of the smoothed electrostatic potential for increasing $m$, that is, in smaller depolarizing fields.

The fact that $\Delta V$ is not exactly constant for a given ferroelectric distortion with the change of the thickness can be explained by the interaction between the polarization and the depolarizing field. In a first-principles calculation both magnitudes are not independent: the induced polarization induced by the atomic displacement of the atoms following the bulk soft-mode and fixing their coordinates along this path, generates an electric field that would be able to displace the electronic clouds, modifying in this way the original induced polarization [Eq. (4.22)]. This effect, treated self-consistently in first-principles, will be more remarkable in the smallest supercells, whith larger depolarizing fields. The field-induced polarization, neglected when assuming $\mathcal{P} \approx \mathcal{P}^0$, will point in the direction of the field, reducing the original induced polarization, and therefore the potential drop across the interface. As previously stated, this constitutes however a small correction that does not significantly affect the value of $\lambda^{eff}$.

---

[12]The small value of the effective screening length, one order of magnitude smaller than the length of the Ru $d$-orbitals present in the SrRuO$_3$ electrode, can be explained by the fact that the magnitude of the dipole surface density at the interface is reduced by the interpenetration of the electrode and the ferroelectric wave functions. In other words, the polarization charge might be screened in part within the ferroelectric material due to the exponential decay of the electrode wavefunctions within the ferroelectric material and vice versa [316].

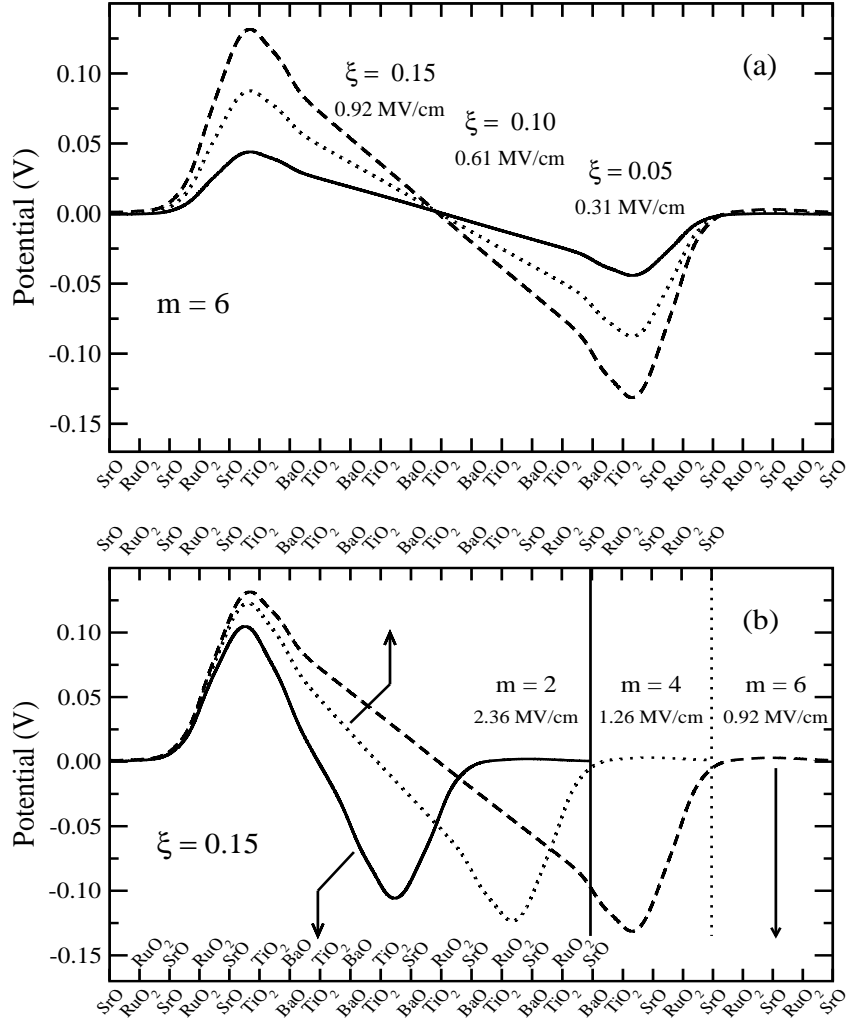

Figure 4.25: Dependence of the depolarizing field with the induced polarization (a) and the thickness of the thin film (b). In panel (a) the profile of the nanosmoothed electrostatic potential, obtained from a first-principle calculation in a $SrRuO_3/BaTiO_3/SrRuO_3$ ferroelectric capacitor in short circuit, is represented for different values of the ferroelectric soft-mode distortion: $\xi = 5$ % (full line), $\xi = 10$ % (dotted line), and $\xi = 15$ % (dashed line). The thickness of the ferroelectric layer was kept fixed at $m = 6$. In panel (b) the amplitude of the ferroelectric distortion was kept fixed at $\xi = 15$ %, while the thickness of the $BaTiO_3$ layer was changed from $m = 2$ unit cells (full lines), $m = 4$ unit cells (dotted lines), $m = 6$ unit cells (dashed lines). The magnitude of the depolarizing field, in MV/cm is indicated for comparison.

### 4.4.4   Electrostatic energy in thin films.

In the presence of a residual depolarizing electric field the energy of the thin film $E$ can be approximated by

$$E\left(\mathcal{P}^0\right) = U\left(\mathcal{P}^0\right) + E_{elec}\left(\mathcal{P}^0\right), \tag{4.23}$$

where $U$ is the internal energy under zero field. Assuming that the interface effects and the modification of the chemical bonding at the ferroelectric/metal interface is hidden in the screening length parameter and, thus, only enters through the magnitude of the depolarizing field $\mathcal{E}_d$, it can be approximated to first order by the bulk internal energy $U_{FE}$,

$$U\left(\mathcal{P}^0\right) = m\ U_{FE}\left(\mathcal{P}^0\right), \tag{4.24}$$

where $m$ stands for the number of unit cells of the perovskite present in the thin films. $U_{FE}$ can be computed from bulk DFT calculations under zero electric field, and its dependence with $\mathcal{P}^0$ has typically the shape of a double well like the one showed in Fig. 1.1.

The electrostatic energy, second term in Eq. (4.23), can be approximated by [112, 125]

$$E_{elec}\left(\mathcal{P}^0\right) = m\Omega_0\left(-\mathcal{E}_d \cdot \mathcal{P}^0 - \frac{1}{2}\chi^\infty \mathcal{E}_d^2\right). \tag{4.25}$$

The depolarizing field can be computed from first-principles by taking the average of the electrostatic potential [314, 315], or can be estimated from the expression given in the previous section $\mathcal{E}_d = -4\pi\alpha\mathcal{P}$, where $\alpha = 2\lambda^{eff}/d$. In this last case, Eq. 4.25 can be mapped into another expression only dependent on the zero-field polarization. Assuming that $\lambda^{eff} << d$, then $\alpha$ tends to vanishes, and to leading order in $\alpha$

$$E\left(\mathcal{P}^0\right) = U\left(\mathcal{P}^0\right) + \frac{8\pi\lambda^{eff}\left(\mathcal{P}^0\right)^2}{d}. \tag{4.26}$$

The electrostatic energy (second term) is positive, meaning that the effect of the depolarizing field is to suppress the ferroelectric instability through a renormalization of the quadratic term of the double well. Eq. (4.26) has been implemented by Kornev *et al.* [290], and Lichtensteiger *et al.* [226] in order to extend the first-principles model hamiltonian to thin films.

The ground state of the thin film is found by a minimization of the energy with respect the zero-field polarization, $\min_{\mathcal{P}^0} E\left(\mathcal{P}^0\right)$.

This model has been tested trying to reproduce the first-principles calculations for the SrRuO$_3$/BaTiO$_3$/SrRuO$_3$ ferroelectric capacitor [34] showed in Fig. 4.22. Since the atomic displacements were constrained in the calculations to follow the bulk soft-mode pattern of displacement, only a line minimization along the corresponding direction was simulated with the model. As can be seen, the simple model reproduces quite accurately the complex first-principles results.

### 4.4.5   Effects of the depolarizing field.

We have seen in Eq. (4.25) that, to leading order in the depolarizing field, the electrostatic energy takes the form $-\mathcal{E}_d \cdot \mathcal{P}^0$. Basically, there are to ways to minimize this contribution to the energy: a reduction of the polarization or a reduction of the depolarizing field.

The first one is the decrease of the ionic polarization, while keeping the system in a monodomain configuration. From the data reported in Fig. 4.22 we can draw how, even well above the critical thickness for ferroelectricity, the depolarizing field contributes to a substantial lose of the ground state polarization: the minimum of the $E$ versus $\xi$ curve is not at $\xi = 1$, as it is in bulk [see Fig. 1.1(b)], but displaced to smaller distortions. The spontaneous polarization of the thin film only saturates slowly to the bulk value for films of increasing size (see inset of Fig. 4.22). A direct comparison of this model with the experiment can be done due to the very recent polarization versus electric field loops on high-quality

SrRuO$_3$/BaTiO$_3$/SrRuO$_3$ heterostructures by Kim *et al.* [294, 317, 318, 319]. For thick enough films, where the influence of the depolarizing field is negligible, the remanent polarization is enhanced by strain with respect to the bulk unstrained value. For a 300 Å-thick film it amounted to 36 $\mu$C/cm$^2$, larger than the bulk value of around 26 $\mu$C/cm$^2$ (A similar effect was also reported by Yanase *et al.* [308]). But the polarization is systematically reduced with thickness due to the increasing importance of $\mathcal{E}_d$. The experimental value dropped to 10 $\mu$C/cm$^2$ for a 50 Å-thick films. The evolution is in good agreement with the previous depolarizing field model, while the intrinsic model predicts a much sharper decay, noticeable only for films thinner than 100 Å.

Since the polarization is strongly coupled to strain in perovskite oxides [93], a reduction of the polarization must be accompanied by a concomitant reduction of the tetragonality of the system. This point was investigated experimentally on monodomain PbTiO$_3$ thin films epitaxially grown on Nb-doped SrTiO$_3$. A global reduction of the $c/a$ ratio was observed from x-ray diffraction [226], and interface local strained regions with a reduced $c$-axis parameter were detected with annular dark field transmission electron microscopy [284]. These experimental results were reproduced theoretically [226] (see Fig. 4.26). The first-principles model hamiltonian developed by Waghmare and Rabe [72] for bulk PbTiO$_3$ was extended to thin films. Following Eq. (4.26), an extra term to the hamiltonian, that accounted for the electrostatic energy, was added. The effective screening length was taken as a parameter of the calculation. The best fit with the experimental data was obtained for a $\lambda^{eff} = 0.12$ Å, smaller than the one adjusted from first-principles results on a BaTiO$_3$/SrRuO$_3$ interface (0.23 Å). This suggests a better screening for the PbTiO$_3$-based heterostructure, in good agreement with the trends observed by first-principles (a better screening in PbTiO$_3$/SrRuO$_3$ than in a BaTiO$_3$/SrRuO$_3$ interface has been recently reported [283]). The DGL intrinsic model predicts a much sharper decay, with no substantial decrease in the polarization and tetragonality predicted above 50 Å. The results favour the incomplete screening of the depolarizing field as the driving force for a global reduction of the polarization in perovskite ultrathin films.

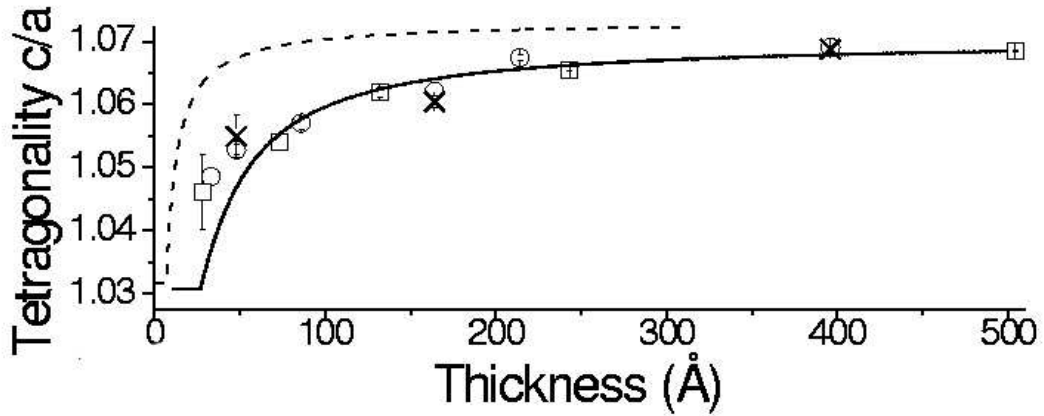

Figure 4.26: Evolution of the $c/a$ ratio with the film tickness for monodomain PbTiO$_3$ thin films grown epitaxially on top of Nb-doped SrTiO$_3$. Circles and squares correspond, respectively, to tetragonality of samples grown at 280 Å /h and 110 Å /h growth rates. Crosses correspond to samples coated with a Au top electrode. Dashed line is the phenomenological theory prediction supposing a ratio between the extrapolation length and the correlation length $\delta/\xi = 1.41$ [282]. Solid line corresponds to the first-principles based model hamiltonian results, supposing an extrapolation length of $\lambda^{eff} = 0.12$ Å. Model hamiltonian results have been rescaled to get rid of the bulk overstimation of $c/a$ and focus only on the evolution of tetragonality with thickness. From Ref. [226].

The second way to minimize the electrostatic energy is the reduction of the depolarizing field. We have identified the incomplete screening of the polarization charge density as the driving force for the appearance of the residual depolarizing field. When there are no compensation charges provided by electrodes (i. e., when the ferroelectric material is grown on top of an insulating substrate [244, 274]), or when the compensation charges do not provide an efficient enough screening, the system might break up into 180 ° domains to reduce the magnitude of the surface dipole density. Atomic scale details of the multidomains and their formation mechanism have been studied from first-principles only recently [152, 320, 290, 321, 322, 278].

Atomistic shell models [152] and first-principles model hamiltonian [321] have been applied to study free-standing slabs of $BaTiO_3$ and $Pb(Zr_{0.5}Ti_{0.5})O_3$, respectively. The response of the system was strongly dependent on clamping effects. Above a critical compressive strain, the in-plane polarization disappears and the system breaks into periodic 180° stripe domains. Their period have the same order of magnitude as the film thickness, and increase with increasing thickness. No domain structure could be found for films thinner than four unit cell. As the film thickness decreases, the critical temperature $T_c$ also decreases quickly.

The evolution of the depolarizing field with thickness also explains the apparent contradiction between the structural measurements and the measured switchable polarization observed in $SrRuO_3/Pb(Zr_{0.2}Ti_{0.8})O_3/SrRuO_3$ ferroelectric capacitors [285, 278]. In that case, a sharp polydomain-monodomain phase transition driven by $\mathcal{E}_d$ is theoretically predicted at a thickness around 150 Å (see Fig. 4.27). The phase transitions results from the competition between the compressive strain imposed by the substrate, which tends to stabilize the tetragonal phase and enhance out-of-plane polarization, and the depolarizing field that tends to suppress the latter [13]. As the film gets thinner, the residual field in the ferroelectric grows and, consequently, both the polarization and the tetragonality are progressively reduced. Below a threshold thickness, the systems breaks up into 180° stripe domains (net polarization zero) to minimize the energy associated with $\mathcal{E}_d$. This domain structure has been observed and characterized in ultrathin $PbTiO_3$ thin films grown on insulating substrates [244, 274], but its presence when metallic electrodes are used is surprising. In the phase transition, that has been carefully described by Kornev and coworkers [290], the system undergoes a sequence of partial transformations from a homogeneous tetragonal phase (with all the local polarizations pointing in the same direction) to an inhomogeneous tetragonal phase (when some laterally confined nanodomains with the local polarization pointing in the opposite polarization have nucleated), and finally to a macroscopically non-polar phase (where the nanodomains have grown till the amount of negative polarization equals the positive one). Though the net switchable polarization is zero in this last configuration, each domain exhibits the bulk strained polarization and tetragonality, larger than in the strained sample, explaining the apparent inconsistency of the experimental data. Moreover, since the out-of-plane piezoelectric response is also directly proportional to the polarization inside each domain, its average also tends to cancel for films thinner than the threshold thickness, explaining the drop on $d_{33}$ experimentally observed.

The reason why some systems, such as $PbTiO_3$ on top of Nb-doped $SrTiO_3$, stands in a monodomain configuration while other similar heterostructures, such as $SrRuO_3/Pb(Zr_{0.2}Ti_{0.8})O_3/SrRuO_3$ breaks up into domains remains an open question that requires further clarification.

The combined effect of the depolarizing field and the in-plane strain was considered by Lai *et al.* [262] in the production of Pertsev-like diagrams for $BaTiO_3$ thin films with a first-principles based model hamiltonian. The depolarizing field punishes the appearance of an out-of-plane polarization so all the phase transitions involving a normal component of the polarization to the surface have a lower transition temperature with respect to the ideal short circuit case. Similar phase diagrams have been reported for $Pb(Zr_{0.5}Ti_{0.5})O_3$ thin films under open circuit boundary conditions [322]. Two phases, not present in the bulk phase diagram, were found: a new monoclinic phase, and a vortex stripe domain phase, with the polarization slightly tilted to the $x$-axis (i. e., not forming exactly 180 ° lamellas).

Phase diagrams for $Pb(Zr_{0.5}Ti_{0.5})O_3$ thin films (20 Å thick) under stress-free and open-circuit boundary conditions have been reported by Almahmoud *et al.* [323] using a model hamiltonian approach. The coupling of the surface polarization and the inhomogeneous strain with vacuum, together with some surface-induced modifications to the short-range forces at the surface layers (that turned out to be very important in the determination of the local and macroscopic properties), were explicitly considered in the model hamiltonian. Since the system is not constrained in the plane, a polarization appears always along the [100] or in the [010] direction. The transition temperature is lowered 120 K with respect to the bulk. When the Ti composition decreases, the polarization continuously rotates from the [010] direction to the [110] direction, leading to large piezoelectric and dielectric properties.

---

[13]The same abrupt transition from a monodomain to a polydomain state has been found (with the increase of the passive layer thickness within the dead-layer model [286, 287].

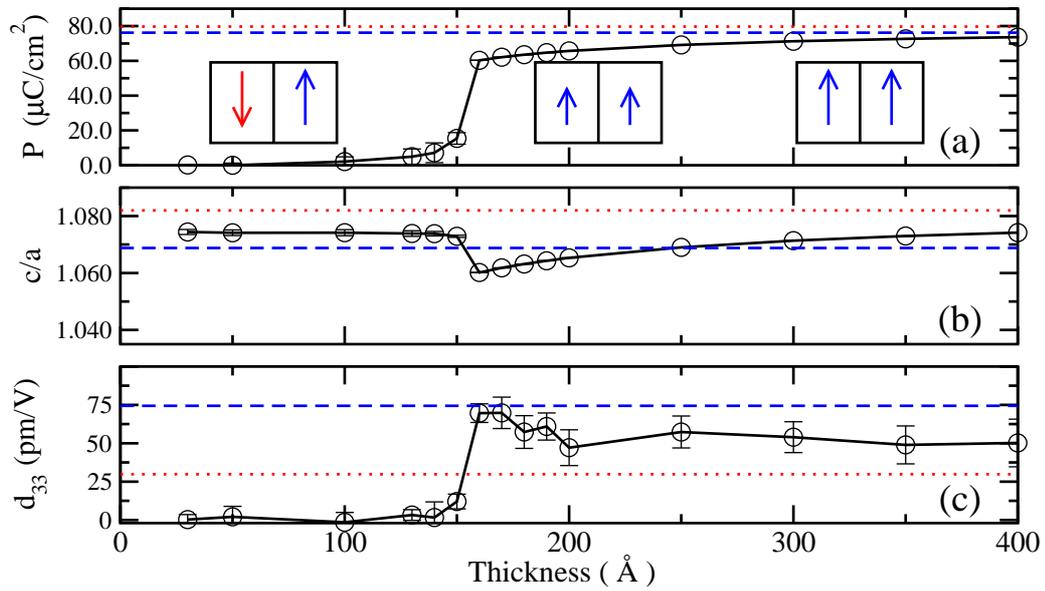

Figure 4.27: Theoretical prediction of the thickness dependence of the normal average polarization $\mathcal{P}$, the tetragonality $c/a$, and the out-of-plane piezoelectric constant $d_{33}$ at room temperature for PbTiO$_3$ thin films grown on a SrRuO$_3$/SrTiO$_3$ substrate. Values of the quantities at the bulk level are represented by dashed lines for the unstrained configuration, and dotted lines for a geometry under the strain imposed by the substrate. The evolution of the domain structure, from a monodomain configuration at large thickness, where the depolarizing field $\mathcal{E}_d$ is small, to a 180° stripe domains to minimize the energy associated with $\mathcal{E}_d$ is represented in the inset. Standard deviations, obtained from averaging over different number of Monte Carlo steps in many simulations, starting with different seeds for the random number generator, are indicated by error bars. From Ref. [278].

One very important issue introduced by the depolarizing fields is the difference between the field measured in the external circuit and that in the film itself. In other words, the field imposed in the experimental setup, that is monitored and might be read in a display of the actual machine, might *not* be the same as the one really applied to the film itself. Both of them are identical only in the case of a vanishing depolarizing field. Dawber *et al.* [289] introduced this correction while observing the thickness dependence of the coercive field $E_c$ in ferroelectric thin film, with the result that the semiempirical Key-Dunn scaling law ($E_c(d) \propto d^{-2/3}$) is fullfilled over five decades in length scale.

## 4.5  Perspectives

For many years it was unclear whether or not ferroelectricity can survive in ultrathin films. From this Section, it appears that mechanical, electrical and chemical boundary conditions are playing a central role in monitoring the ferroelectric properties and must be considered explicitly when discussing this difficult problem. The question of the suppression of the ferroelectricity in ultrathin films cannot therefore be answered "in general" but, instead, must be addressed independently for each individual system, considering together the ferroelectric and its environment.

The characterization and fundamental understanding of finite size effects in ferroelectric ultrathin films has significantly improved during the last years thanks to the appearance of new experimental techniques for the deposition and characterization of high-quality ultrathin films, and to the development of new theoretical methods for simulations at the atomic scale with real predictive capabilities.

Perovskite ultrathin films have shown ferroelectricity at thicknesses down to a few unit-cells, although usually with a polarization reduced with respect to the bulk value. Only very recently, polarization comparable to bulk have been reported for heterostructures with slightly different boundary conditions.

In spite of significant advances in the understanding of various kinds of ferroelectric ultrathin films, many issues still need to be clarified in the future, including the role of defects and vacancies, the consequences of a finite conductivity of the sample, or the origin of the screening in absence of top electrode.

# Ferroelectric superlattices.

## 5.1 Overview.

During the recent years, significant advances have been reported concerning the understanding of the behavior of ferroelectric multilayers [180, 234]. They mainly rely on valuable exchanges and discussions between theorists and experimentalists.

Nowadays, various types of ferroelectric oxide superlattices with nanometric periodicity and atomically sharp interfaces can be grown using different techniques including MBE [324, 325], off-axis magnetron sputtering [326] or pulsed laser deposition [235]. Most recent works deal with *bicolor* [001] – superlattices $(A - B - A - B - A - B)$ in which the ferroelectric compound alternates with a regular insulator [96], an incipient ferroelectric [327, 324, 328] or another ferroelectric material [329] [1]. Some other recent studies also concern *tricolor* [001] – superlattices $(A - B - C - A - B - C)$ [331, 332, 333, 235] breaking the inversion symmetry as they are expected to generate interesting new properties [30].

At the same time, first-principles simulations allow to investigate the properties of these systems in a very accurate way. Available complementary tools includes bare DFT calculations [30, 35], effective Hamiltonian methods [33, 334, 144], shell-model calculations [149] or simple electrostatic models [35, 326]. These techniques revealed particularly useful to isolate the key parameters monitoring the properties of ferroelectric multilayers, to understand experimental results and even to make trustable predictions.

In this Section we first label and discuss the key factors monitoring the physics of ferroelectric multilayers as they were recently identified. We then illustrate how these distinct parameters can interact and compete in selected systems, in order to generate new phenomena and eventually new interesting functional properties.

## 5.2 Three main factors monitoring the ferroelectric properties.

Discussing extensively the behavior of ferroelectric superlattices is a complex issue since, in the same way as previously discussed for thin films, their properties can be influenced by numerous factors. Amongst the latter, it appears however that three are playing a dominant role: the mechanical constraint imposed by the substrate, the macroscopic electrostatics and the atomic relaxation at the interface. If these factors alone cannot account for everything, they already allow to understand the major trends. This is particularly true for almost "ideal" systems such as multilayers with atomically sharp interfaces and nanometric modulation lengths, as recently grown with a control of the structure at the atomic scale.

---

[1] The case of metal/insulator $ABO_3$ multilayers such as $SrVO_3/SrTiO_3$ was also reported [330] but will not be discussed here since it does not directly concern ferroelectricity.

### 5.2.1   Epitaxial strain.

As previously discussed in Sec. 3.5, ferroelectricity is very sensitive to volume and mechanical constraints. In the case of multilayers, the structure is free to relax along the modulation direction (conventionally [001] in this Section) so that, in the same way as for thin films, the main mechanical constraint is the in-plane epitaxial strain imposed by the substrate.

A first key parameter for understanding the behavior of ferroelectric superlattices is therefore *the epitaxial strain*. For instance, independently of other effects, it is predicted that it could produce a polarization enhancement larger than 50% for $BaTiO_3$ layers on $SrTiO_3$ [35]. Consequences of different epitaxial strains on the properties of various ferroelectric compounds have already been discussed in Sec. 4.3. This discussion still applies here. Let us simply recall that, as a general rule, a compressive in-plane strain will favor a *c*-phase with polarization along [001] while a tensile strain will usually favor a *aa*-phase with polarization in-plane. The phase diagram of epitaxially strained $[001] - (PbTiO_3)_1/(PbZrO_3)_1$ superlattice has been computed within DFT by Bungaro and Rabe as illustrated in Fig. 5.1. Various properties of this superlattice versus strain have also been reported and will be discussed in Section 5.5.2.

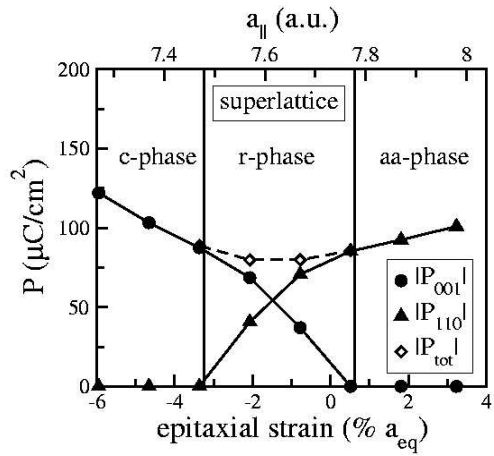

Figure 5.1: Magnitude and components of the polarization as a function of in-plane strain for a $[001] - (PbTiO_3)_1/(PbZrO_3)_1$. (Fig. 2 from Ref. [259] by Bungaro and Rabe)

In the discussion of superlattices, it is therefore mandatory to know if the different layers are still coherent with the substrate, or if the strain constraint has been partly or totally relaxed through the formation of misfit dislocations. The precise value for strain relaxation cannot be easily predicted and strongly depends on the method of synthesis: indeed, under appropriate conditions, it was shown that coherent films can be grown well over the theoretically estimated elastic critical thickness [239, 236]. In most cases, the identification of the strain state therefore relies on experimental data.

In all the studies that will be discussed below, the superlattice was grown on a $SrTiO_3$ substrate, eventually doped with Nb (to be conductive) or covered by a metal oxide electrode as $SrRuO_3$ (coherent with the substrate). In order to play with the epitaxial strain, other substrates (such as MgO) can be considered [329] but were only marginally used.

### 5.2.2   Electrostatics.

A second key parameter that monitors the behavior of ferroelectric multilayers when the polarization is along the modulation direction is the *electrostatic energy* associated to the presence of non-vanishing electric fields in the different layers. For ferroelectric ultra-thin films, the depolarizing field was shown to play a major role in monitoring the ferroelectric properties. A simple electrostatic model was then introduced by Junquera and Ghosez [34] that allows to predict the evolution of the polarization with thickness. This approach can be generalized for thin films [326, 335].

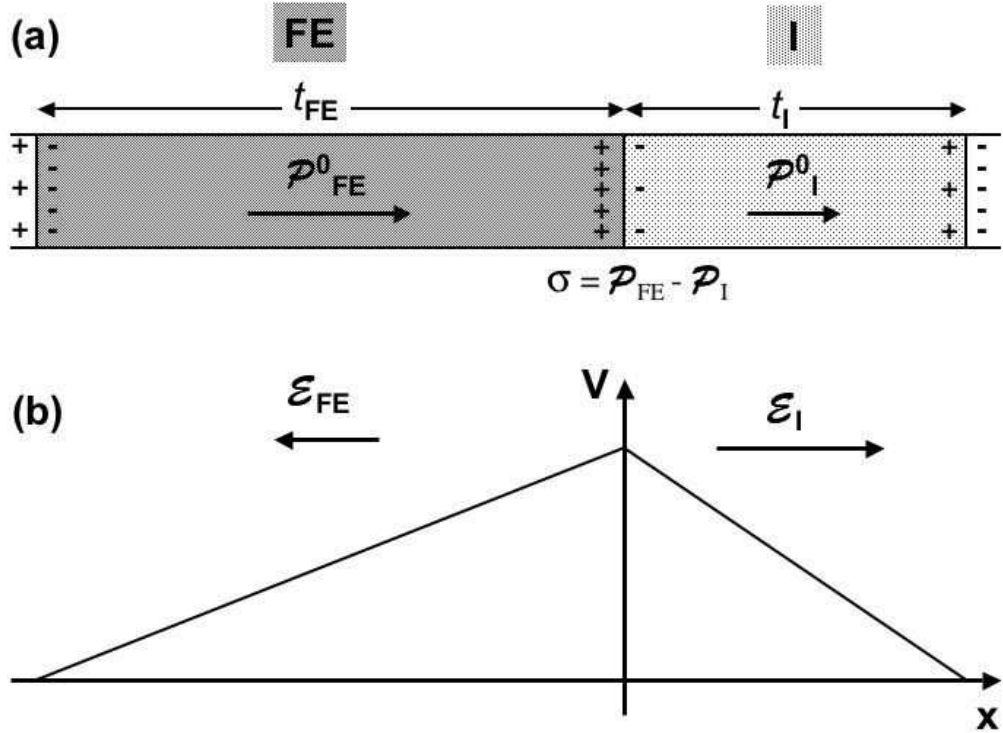

Figure 5.2: (a) Schematic view of the supercell of a bicolor FE/I superlattice in short-circuit. The layer thicknesses and polarizations are equal to $t_{FE}/t_I$ and $\mathcal{P}^0_{FE}/\mathcal{P}^0_I$. A surface charge $\sigma = \mathcal{P}^0_{FE} - \mathcal{P}^0_I$ appears at the interface when the polarizations are not equal. (b) Evolution of the macroscopic electrostatic potential along the structure when $\mathcal{P}^0_{FE} \neq \mathcal{P}^0_I$ and related macroscopic electric fields $\mathcal{E}_{FE}$ and $\mathcal{E}_I$.

Let us consider a [001]–$FE/I$ superlattice in which a ferroelectric ($FE$) material alternates with another insulating ($I$) compound that can be either a regular insulator, an incipient ferroelectric or another ferroelectric as illustrated in Fig. 5.2. The repeated bilayer is made of $n_{FE} + n_I$ basic unit cells and has a total thickness $t_{FE} + t_I = n_{FE}c_{FE} + n_Ic_I$ where $c_i$ refer to the individual lattice constant along the [001] direction. The influence of the substrate is only included through an appropriate choice of the in-plane lattice constant of the whole stack. As in most experiments, we will consider that the system is between metallic electrode in short circuit [2]. This is pratically imposed in the simulation by imposing periodic boundary conditions.

We will further assume that the individual layers are uniformly polarized [3] along the [001] direction. Considering the atomic positions $R_\kappa$ and eventual electric field $\mathcal{E}_i$ as individual parameters, the polarization of layer $i$ is given by [112] [equivalent to Eq. (4.22) but written in terms of individual atomic displacements].

---

[2]We assume here *ideal* metallic electrodes which is reasonable since the total thickness of the superlattice is large compared to the effective screening length.

[3]This is supported by DFT calculations for short modulation [35, 326] lengths, but is less justified when the insulator gets thicker [153].

$$\begin{aligned}
\mathcal{P}_i &= \sum_{\kappa \in i} \frac{\partial \mathcal{P}_i}{\partial R_\kappa}\Big|_{\mathcal{E}=0} \Delta R_\kappa + \frac{\partial \mathcal{P}_i}{\partial \mathcal{E}}\Big|_{R_{\kappa 0}} \mathcal{E}_i \tag{5.1} \\
&= \Omega_i \sum_{\kappa \in i} Z_\kappa^* \Delta R_\kappa + \chi_i^\infty \ \mathcal{E}_i \\
&= \mathcal{P}_i^0 + \chi_i^\infty \ \mathcal{E}_i.
\end{aligned}$$

The first term $\mathcal{P}^0$ is the polarization arising, under the condition of zero electric field, from the displacement of the ions from their position in the unpolarized structure of reference [4]. The second term is the field induced electronic polarization at linear order. In what follows, it will be neglected since it does not play a central role in the discussion. The present approach can however be easily generalized to higher orders in the field.

Generalizing the approach of Junquera and Ghosez [34] for thin films (see Sec. 4.4.4), the stable polarization state of the previous superlattice is the one that minimizes the total energy

$$E(\mathcal{P}_{FE}^0, \mathcal{P}_I^0) = U(\mathcal{P}_{FE}^0, \mathcal{P}_I^0) + E_{elec}(\mathcal{P}_{FE}^0, \mathcal{P}_I^0). \tag{5.2}$$

The first term $U$ is the internal energy in zero field. Neglecting interface corrections discussed in the next Section, it can be estimated, in first approximation, from bulk internal energies of the individual compounds under the same strain constraint

$$U(\mathcal{P}_{FE}^0, \mathcal{P}_I^0) \approx n_{FE} U_{FE}(\mathcal{P}_{FE}^0) + n_I U_I(\mathcal{P}_I^0). \tag{5.3}$$

As in Ref. [326], $U_i(\mathcal{P}_i^0)$ can for instance be deduced directly from DFT calculations under finite fields using the method of Sai *et al.* [112]. The second term $E_{elec}$ arises from the eventual presence of non-vanishing electric fields as it is now discussed.

When a finite polarization $\mathcal{P}_{FE}^0$ appears in the ferroelectric layer, it gives rise to a depolarizing field $\mathcal{E}_{FE}$ (see Fig. 5.2). Because of the short circuit boundary conditions (no potential drop along the whole supercell), a field $\mathcal{E}_I$ will appear in the insulating layer and induce a polarization $\mathcal{P}_I^0$. Both fields and polarizations will mutually adjust in order to satisfy two conditions. First, the continuity of the normal component of the electric displacement field at the interfaces imposes that

$$\mathcal{E}_{FE} + 4\pi \mathcal{P}_{FE}^0 = \mathcal{E}_I + 4\pi \mathcal{P}_I^0. \tag{5.4}$$

Second, for a system under short circuit boundary conditions, the total potential drop along the structure must vanish so that

$$t_{FE}\mathcal{E}_{FE} = -t_I \mathcal{E}_I. \tag{5.5}$$

Combining these equations, we can write

$$\mathcal{E}_{FE} = -4\pi t_I \frac{(\mathcal{P}_{FE}^0 - \mathcal{P}_I^0)}{(t_{FE} + t_I)}, \tag{5.6}$$

$$\mathcal{E}_I = +4\pi t_{FE} \frac{(\mathcal{P}_{FE}^0 - \mathcal{P}_I^0)}{(t_{FE} + t_I)}. \tag{5.7}$$

Opposite finite electric fields will therefore appear in both layers as soon as their polarizations slightly differs. They are related to the appearance of a surface charge $\sigma = \mathcal{P}_{FE}^0 - \mathcal{P}_I^0$ at the interface. They will give rise in each layer $i$ to an electrostatic energy that, to leading order in the field, is equal to $E_{elec} = -t_i \mathcal{E}_i \cdot \mathcal{P}_i^0$. Combining this with Eq. (5.6) and summing over both layers we obtain

$$E_{elec} = +4\pi \frac{t_{FE}t_I}{(t_{FE} + t_I)}(\mathcal{P}_{FE}^0 - \mathcal{P}_I^0)^2. \tag{5.8}$$

---

[4]In order to avoid confusion we should remark here that $\mathcal{P}^0$ is neither the spontaneous polarization nor the remanent polarization of the material.

Introducing now Eq (5.3) and Eq. (5.8 into Eq. (5.2) we obtain as a final expression for the energy

$$E(\mathcal{P}_{FE}^0, \mathcal{P}_I^0) = n_{FE} \ U_{FE}(\mathcal{P}_{FE}^0) + n_I \ U_I(\mathcal{P}_I^0) + 4\pi \frac{t_{FE} t_I}{(t_{FE} + t_I)} (\mathcal{P}_{FE} - \mathcal{P}_I)^2. \tag{5.9}$$

This clearly highlights that different terms compete in order to define the stable polarization state for given $n_{FE}$ and $n_I$. $U_{FE}$ has a typical double-well shape so that a finite $\mathcal{P}_{FE}^0$ will decrease the internal energy. However, polarizing the ferroelectric alone produces a huge electrostatic energy penalty. This one can be avoided by polarizing the insulating material but this also has an energy cost if it is a regular insulator ($U_I$ with single-well shape). As it will be discussed below on practical examples, when the insulator is highly polarizable (incipient ferroelectric or another ferroelectric), the superlattice will sustain a finite polarization in its ground-state, with $\mathcal{P}$ nearly uniform in order to minimize the electrostatic energy. When the insulator becomes less polarizable, both polarizations can differ and will be totally suppressed when the energy cost to polarize the insulator becomes too large [5].

In summary, we have seen in this epigraph how not only the lattice mismatch, reviewed in Sec. 5.2.1, between the different materials that constitute the heterostructure is important, but also the dielectric mismatch, defined as the different susceptibilities between the two materials.

### 5.2.3 Interface atomic relaxation.

A third key parameter expected to play a role in multilayers is the *interface* between alternating layers. Interface atomic relaxations will affect the chemical bonding and therefore the short-range interatomic force constants at the vicinity of the interface. Also the Born effective charges at the interface must differ from their bulk values, locally modifying the long-range dipole-dipole interaction.

Interface effects are properly accounted in DFT calculations and are also included, at least in an approximate way, in effective Hamiltonian and shell-model approaches. Up to now, however, no systematic study of their specific influence on the global behavior of multilayers has been reported. In most cases studied so far, the oxide compounds that are put together are very similar and the interface is expected to be very "gentle". The atomic relaxations (both rumpling and interplane distance changes) at the interface of ferroelectric oxide with other oxides has been observed to have similar features with relaxation at free surfaces (see Sec. 4.2.1) but is globally smaller in amplitude and located in a very sharp region close to the interface. This will be illustrated later for $ABO_3$/$AO$ interfaces (Fig. 5.3).

The relatively minor role played by the interface in many cases is, to some extent, confirmed by the fact that the electrostatic model presented above (that totally neglect interface relaxation through it simple assmption on $U$) applies for many systems down to very short modulation lengths. This might however not be a general rule and this simple electrostatic model is not always enough to reproduce quantitatively DFT data [335].

## 5.3 Ferroelectric/regular insulator superlattices.

Only few studies have been reported concerning ferroelectric/regular insulator superlattices. This might be due to the fact that, from previously discussed electrostatic arguments, they are not a priori well-suited to enhance the ferroelectric properties. We focus here on few cases where the insulating material is a AO alkaline earth oxide of rocksalt structure, with a bandgap typically larger than that of the related $ABO_3$ compound ($E_g$(BaO)= 4.8 eV, $E_g$(SrO)= 5.7 eV) and a still moderate static dielectric constant ($\epsilon^0(BaO) = 31 - 34$, $\epsilon^0(SrO) = 14.4$).

---

[5]This simple model is restricted to *monodomain* configurations (i.e. homogeneous in-plane polarization). When the electrostatic energy becomes too large, formation of 180° alternating domains might be favored as previously discussed for thin films (Sec. 4.4).

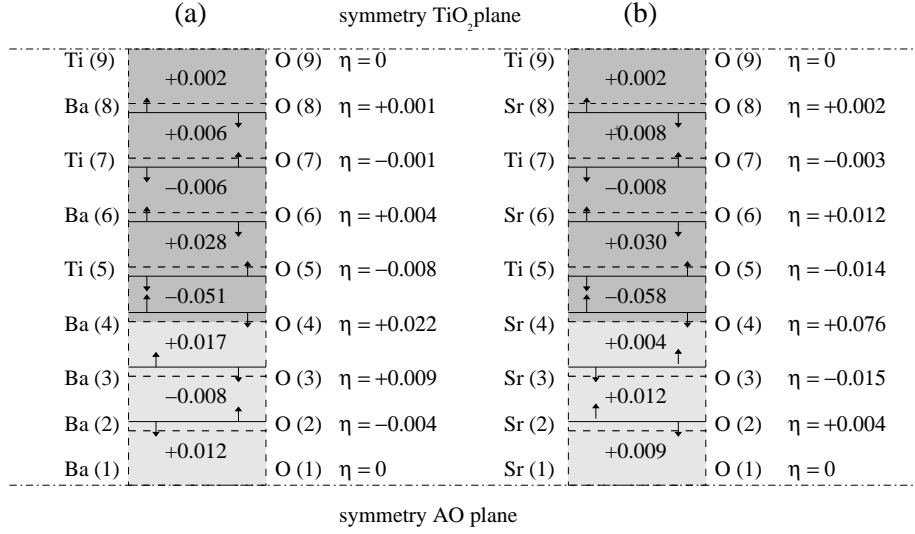

Figure 5.3: Schematic view of the atomic relaxation in the bottom-half unit cell of $(BaTiO_3)_5/(BaO)_6$ (panel a) and $(SrTiO_3)_5/(SrO)_6$ (panel b) supercells. Dashed lines correspond to the *reference* positions of the atomic planes, for strained truncated bulk coherent with the substrate. The full lines are the mean plane position in the relaxed structure. Changes in the interplanar distance are written in Å. The atoms (A or Ti, depending on the layer, at the left and O at the right) move in the direction indicated by the arrow. The rumpling parameter, $\eta$, is expressed in Å (it is defined as half the metal-oxygen distance in each plane, $\eta_i = [\delta_z(M_i) - \delta_z(O_i)]/2$). (Fig. 1 from Ref. [96] by Junquera *et al.*)

### 5.3.1    BaTiO$_3$/BaO and SrTiO$_3$/SrO

A DFT study of $(BaTiO_3)_m/(BaO)_n$ and $(SrTiO_3)_m/(SrO)_n$ superlattices has been reported by Junquera *et al.* [96]. The growth of multilayers in which a cubic perovskite ABO$_3$ compound alternate with the related AO alkaline earth oxide of rocksalt structure is allowed by the fact the A–A distance is relatively similar in both cases. The epitaxy is such that ABO$_3$ (001) ∥ AO (001) and ABO$_3$ ⟨110⟩ ∥ AO ⟨100⟩; i.e. the ABO$_3$ atomic planes are rotated 45° around the (001) AO direction. In their study, they considered a $(AO)_n/(AO-TiO_2)_m$ supercell with $n = 6$ and $m = 5$. Also, the in-plane lattice constant was fixed to that of Si. This was motivated by the fact that the AO/ABO$_3$ interface is particularly relevant for application in electronics since an approach to grow ABO$_3$ perovskite on Si is to include an AO buffer layer (see Sec. 6.2).

The main conclusions of that study are as follows. First, macroscopic elasticity still apply to predict the $c$-strain relaxation resulting from the epitaxial strain imposed by the Si substrate. Second, the atomic relaxation at the interface is much smaller that at free surfaces and located in a sharp region close to the interface (Fig. 5.3). Third, at the level of the electronic structure, there is no interface induced gap state and the interfacial AO layer behaves essentially as in BaTiO$_3$. Finally, none of the structures studied with $n \approx m$ were found to be ferroelectric.

This last feature is not surprising. On the one hand the in-plane compressive epitaxial strain should enhance ferroelectricity. On the other hand, however, the electrical boundary conditions are not favorable for ferroelectricity and will dominantly monitor the ferroelectric behavior. As discussed in Sec. 5.2.2, to avoid a huge electrostatic energy cost for the superlattice, the polarization must be roughly similar in both the AO and ABO$_3$ layers. Even if the internal energy of the ABO$_3$ compound decreases when polarizing it ($\Delta U(ABO_3) < 0$), it costs energy to polarize the alkaline earth oxide ($\Delta U(AO) > 0$). As shown by Bousquet and Ghosez [335], ferroelectricity can be recovered for such systems but requires $m > n$ and, even in that case, the polarization in BaO remains significantly smaller than in $BaTiO_3$ [6]. (a ferroelectric ground state has been predicted also for $BaTiO_3$ grown on top of Ge for a thick enough layer of $BaTiO_3$ [336] using a phenomenological DGL theory).

### 5.3.2 Ruddlesden-Popper $A_{n+1}B_nO_{3n+1}$ structures.

The $A_{n+1}B_nO_{3n+1}$ Ruddelsden-Popper (RP) series [337, 338] of compounds is closely related to the perovskites (Fig. 5.4). It corresponds to a superlattice made from the stacking of AO-terminated $ABO_3$ perovskite [001] slabs of thickness equal to $n$ primitive unit cells and in which adjacent slabs are shifted relative to one another along [110] by $a/2$ ($a$ is the perovskite lattice constant). To some extent, it can be viewed as the limit of $ABO_3$ layers alternating with an ultrathin AO layer (i.e. 1 unit cell). The $n = \infty$ end-member of the RP series correspond to a pure $ABO_3$ compound.

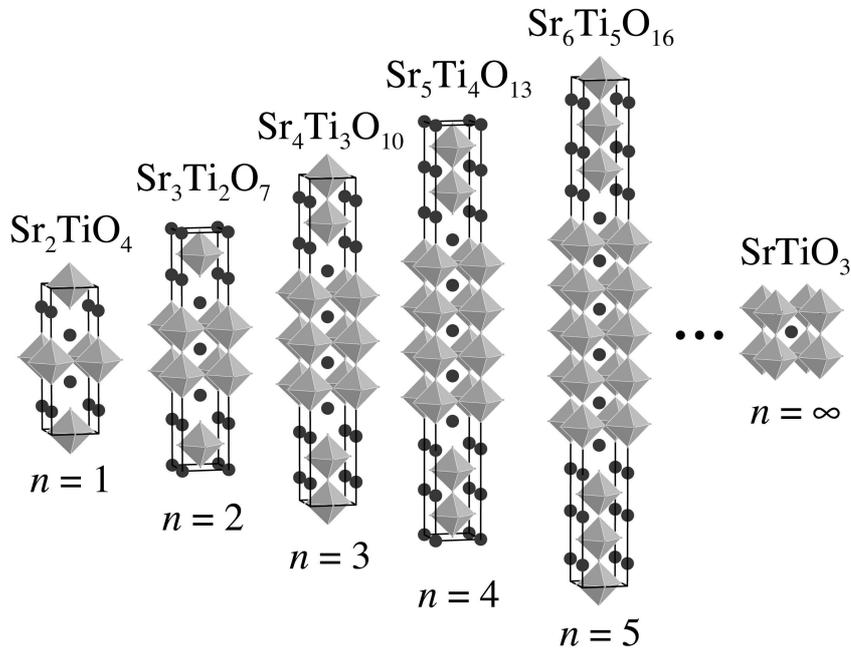

Figure 5.4: Schematic view of the crystal structure of the unit cell for various member of the $Sr_{n+1}Ti_nO_{3n+1}$ Ruddlesden-Popper homologous series. Circles represent Sr atoms, while Ti atoms are at the center of the octahedra with oxygen atoms at each apex. (Fig. 1 of Ref. [339] by Haeni *et al.*)

The existence of different RP titanate phases has been experimentally reported. It concerns essentially some members of the $Sr_{n+1}Ti_nO_{3n+1}$ and $Ca_{n+1}Ti_nO_{3n+1}$ series. None of these appear to display ferroelectricity which is not surprising since neither $SrTiO_3$ nor $CaTiO_3$ are ferroelectric. However some RP compounds exhibit properties potentially more interesting than the related $ABO_3$ compounds for some specific applications. As an example, thin films of the first-members of the $Sr_{n+1}Ti_nO_{3n+1}$ series have been grown by molecular beam epitaxy on a $SrTiO_3$ substrate [339]. It was shown that $Sr_2TiO_4$ has a reasonably large dielectric constant ($\epsilon_{33} = 44 \pm 4$) and a dielectric loss lower than the detection limit. Its large dielectric constant makes it a good candidate as alternative gate dielectric in MOSFETs, with the advantages over $SrTiO_3$ that it has a better lattice match with Si and a higher bandgap (important to reduce leakage currents).

---

[6]In this latter study, it appeared also that simple electrostatic concepts as presented here, although valuable to discuss global trends, are not enough to reproduce quantitatively the behavior of the superlattice.

A DFT study of the structural and dielectric properties of $Sr_2TiO_4$ has been reported by Fennie and Rabe [340]. They computed its static dielectric permittivity tensor and compared their results with experiments on both ceramic powders and epitaxial thin films. They performed a mode by mode analysis and highligthed the high anisotropy of the dielectric response, providing key informations for the analysis of experimental data.

First-principles techniques are also suitable for making prediction on hypothetical compounds. In a recent paper, Fennie and Rabe also investigated within DFT the properties of as-yet hypothetical $Pb_2TiO_4$ [341]. They showed that, contrary to other Sr- and Ca-based RP titanates studied so far, it does display a ferroelectric instability. They predicted that, if it could be grown (for instance using modern epitaxial growth techniques), it would undergo a ferroelectric phase transition to an orthorhombic phase with a spontaneous polarization comparable to that of $PbTiO_3$.

## 5.4   Ferroelectric/incipient ferroelectric superlattices.

Ferroelectric materials can also be combined with incipient ferroelectrics such as $SrTiO_3$ or $KTaO_3$. Ferroelectric/incipient ferroelectric superlattices correspond to a special case of ferroelectric/insulator structures. The situation is unusual since the insulating layer is at the border of the ferroelectric instability and is highly polarizable ($\epsilon^0_{SrTiO_3} = 300$). The internal energy cost to polarize the incipient ferroelectric is therefore much lower than for the case of alkaline earth oxides previously discussed. The system might therefore remain ferroelectric even in extreme limit where the ferroelectric thickness becomes very thin.

### 5.4.1   $BaTiO_3/SrTiO_3$

At present, $BaTiO_3/SrTiO_3$ is one of the system that was the most widely investigated at the experimental level [342, 343, 344, 325, 345, 346, 347, 327] .

Neaton and Rabe performed a DFT study of [001]–$BaTiO_3/SrTiO_3$ superlattices epitaxially grown on a $SrTiO_3$ substrate [35]. They focused on $(BaTiO_3)_{5-n}/(SrTiO_3)_n$ structures with $n = 0$ to 5. The effect of the substrate was only implicitly treated by fixing the in-plane lattice constant to that of bulk $SrTiO_3$ and the superlattice was assumed to remain fully strained independently of the relative thicknesses of the layers. They applied periodic boundary conditions, equivalent to having the superlattice between metallic electrodes in short circuit. They considered polarization along the [100] direction since canting it toward [111] was found to be energetically unfavorable.

Their findings can be summarized as follows. Because of the epitaxial compressive strain, the spontaneous polarization of pure $BaTiO_3$ ($n = 0$) is enhanced by 57 % ($\mathcal{P}_{5/0} = 39.20$ $\mu C/cm^2$) compared to its value in the unconstrained bulk tetragonal phase ($\mathcal{P}^s_{BTO} = 24.97$ $\mu C/cm^2$). When $n$, and so the percentage of $SrTiO_3$, increases the average polarization $\bar{\mathcal{P}}$ of the superlattice progressively decreases. It remains however larger than $\mathcal{P}^s_{BTO}$ up to $n = 3$ and reaches zero only for pure $SrTiO_3$ ($n = 5$). As expected from electrostatic considerations, in each polar superlattice, $SrTiO_3$ is found to be strongly polarized while the polarization of $BaTiO_3$ is partly reduced in such a way that the local polarization [7] is roughly constant along the whole superlattice (see Fig. 5.5).

They developed a simple electrostatic model that properly predicts the evolution of $\bar{\mathcal{P}}$ with the respective thicknesses $t_{BTO} = 5 - n$ and $t_{STO} = n$. It is based on the same electrostatic arguments than in Sec. 5.2.2: (i) short circuit boundary conditions requires the electric fields within each part to be related by $\mathcal{E}_{BTO}t_{BTO} = -\mathcal{E}_{STO}t_{STO}$; and (ii) the displacement field must be preserved at the interface so that $\mathcal{E}_{BTO} + 4\pi\mathcal{P}_{BTO} = \mathcal{E}_{STO} + 4\pi\mathcal{P}_{STO}$. Considering that $\mathcal{P}_{BTO} = \mathcal{P}_{5/0} + \chi^0_{BTO}\mathcal{E}_{BTO}$ and $\mathcal{P}_{STO} = \chi^0_{STO}\mathcal{E}_{STO}$ where $\chi^0_i$ are the static dielectric susceptibilities [8], the average polarization can be

---

[7]The *local* polarization $\mathcal{P}_\lambda$ of cell $\lambda$ is deduced from the displacements and Born effective charges of the atoms $\kappa$ within that cell, $\mathcal{P}_\lambda = (1/\Omega_\lambda)\sum_{\kappa \in \lambda} Z^*_{\lambda,\kappa}\Delta R_{\lambda,\kappa}$, and the polarization enhancement is defined in reference to unconstrained bulk tetragonal $BaTiO_3$ as $\mathcal{P}_\lambda/\mathcal{P}^s_{BTO}$.

[8]These expressions are equivalent to $\mathcal{P}_i = \mathcal{P}^0_i + \chi^\infty_i\mathcal{E}_i$ [Eq. (5.1)] . Writing the static dielectric constant as a sum of ionic and electronic constributions ($\chi^0_i = \chi^R_i + \chi^\infty_i$), we obtain $\mathcal{P}^0_{BTO} = \mathcal{P}_{5/0} + \chi^R_{BTO}\mathcal{E}_{BTO}$, and $\mathcal{P}^0_{STO} = \chi^R_{STO}\mathcal{E}_{STO}$. Instead of being treated explicitly as described in Sec. 5.2.2, ionic degrees of freedom are included here globally into a single fitting

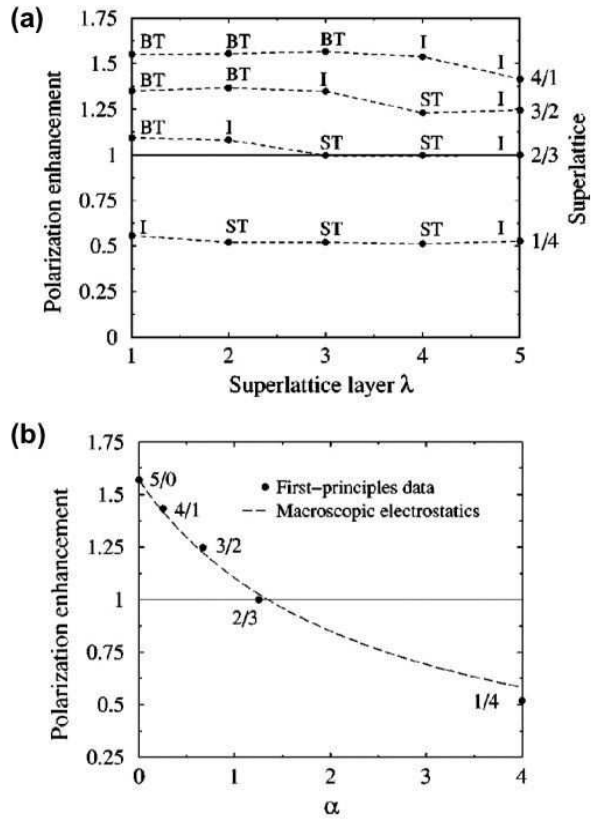

Figure 5.5: (a) Local polarization enhancement ($\mathcal{P}_\lambda/\mathcal{P}_{BTO}^s$) by layer $\lambda$ in various $(BaTiO_3)_n/(SrTiO_3)_m$ superlattices identified as $n/m$. Each layer is labeled BTO ($BaTiO_3$), STO ($SrTiO_3$) or I (interface). (b) Average polarization enhancement ($\overline{\mathcal{P}}/\mathcal{P}_{BTO}^s$) as a function of $\alpha = t_{STO}/t_{BTO}$ as obtained from first-principles DFT calculations and the electrostatic model (with $\epsilon_{BTO}^0/\epsilon_{STO}^0 = 0.4229$). $\mathcal{P}_{BTO}^s = 24.97$ $\mu C/cm^2$. (Fig. 1-2 from Ref. [35] by Neaton and Rabe)

conveniently written as:

$$\bar{\mathcal{P}} = \frac{\mathcal{P}_{5/0}}{1 + (t_{STO}/t_{BTO})(\epsilon_{BTO}^0/\epsilon_{STO}^0)} \tag{5.10}$$

The static dielectric constant ratio appears as a fitting parameter that can be adjusted to reproduce the DFT data. As illustrated in Fig. 5.5, the agreement between the electrostatic model and the DFT results is very good for the best fit, obtained for $(\epsilon_{BTO}^0/\epsilon_{STO}^0) = 0.4229$. This ratio is compatible with the values sometimes assumed for the dielectric constants ($\epsilon_{BaTiO_3}^0 = 160$, $\epsilon_{SrTiO_3}^0 = 300$) [235]. It emphasizes that epitaxial strains and electrostatics dominates the behavior of such kind of superlattices in such a way that a good estimate of $\bar{\mathcal{P}}$ with thickness can already be easily achieved from the only knowledge of the strained bulk spontaneous polarization and individual static dielectric constants.

Recently, Jiang *et al.* [347] and Ríos *et al.* [327] investigated the properties of $(BaTiO_3)_n/(SrTiO_3)_n$ superlattices. Combining x-ray diffraction and second harmonic generation measurements, they found that, below $n = 30$, (i) $BaTiO_3$ is tetragonal with a polarization along the [001] superlattice direction while; (ii) $SrTiO_3$ has a symmetry lower than tetragonal, consistent with a non-zero polarization along [110] direction. These results were puzzling, as they are in contradiction with the findings of Neaton and Rabe [35].

In a recent paper, Johnston *et al.* [97] proposed a possible explanation to the apparent inconsistency between theory and experiment. In the experimental studies, the in-plane lattice constants were reported to be larger than that of bulk $SrTiO_3$. They explored the idea that this increase of in-plane lattice constant might be the responsible responsible for the change of structure. To that end, they explored the properties of $(BaTiO_3)_n/(SrTiO_3)_m$ superlattices when expanding the in-plane lattice constant to 1.01 times the theoretical lattice constant of $SrTiO_3$. In such a case, the $SrTiO_3$ layer is therefore under in-plane tensile strain while the $BaTiO_3$ layer is still under compressive strain. They found that the $SrTiO_3$ layer develop both a polarization along the [001] direction similar to that of $BaTiO_3$ to minimize the electrostatic energy, as well as a non-zero polarization component along [110] lowering the symmetry as experimentally observed.

### 5.4.2 PbTiO$_3$/SrTiO$_3$

$PbTiO_3$/$SrTiO_3$ superlattices with atomically sharp interfaces have be grown experimentally using MBE [324] and off-axis RF magnetron sputtering [326]. In the latter study, Dawber *et al.* [326] considered $(PbTiO_3)_n/(SrTiO_3)_3$ superlattices (with $n = 1$ to 54) on a conductive Nb-SrTiO$_3$ substrate. They combined x-ray diffraction and piezoelectric atomic force microscopy to study the evolution of the ferroelectric polarization with $n$. As previously reported for thin films [226], and due to strain-polarization coupling, the thickness dependence of the $c$-axis parameter, as deduced from x-ray measurements, provides an indirect way to follow the evolution of the polarization with the thickness. The dependence of the out-of-plane lattice constant with the polarization is quadratic $c(\mathcal{P}) = c(0) + \alpha\mathcal{P}^2$, where $c(0)$ is the lattice parameter of the unpolarized phase, and might account for the increase of tetragonality induced by the 2D-clamping [9]. In $(PbTiO_3)_n/(SrTiO_3)_3$ superlattices, $c$-axis measurements are compatible with a progressive reduction of the average polarization as $n$ decreases (Fig. 5.6). A complete suppression of $\mathcal{P}$ is reported for $n = 3$ while ferroelectricity is surprisingly recovered for $n = 1, 2$, both these facts being further confirmed by piezoelectric atomic force microscopy.

In Ref. [326], Dawber *et al.* first applied the electrostatic model of Sec. 5.2.2 to $(PbTiO_3)_n/(SrTiO_3)_3$ superlattices to study the changes of $\mathcal{P}$ and $c$ with $n$. As expected, the polarization is predicted to be very similar in $PbTiO_3$ and $SrTiO_3$ layers and to globally decrease with $n$ in a way compatible with the experimental data. The spontaneous polarization is totally suppressed below $n = 4$ and no recovery is predicted for smaller $n$. DFT calculations are also reported for $n$ ranging from 1 to 7. Apart from the fact that the computed polarization is slightly bigger and that all superlattices are found to be ferroelectric,

---

parameter $\chi^0$.

[9]Even a non-polar phase might exhibit a non-vanishing tetragonality due to the mechanical response to the strain imposed by the substrate.

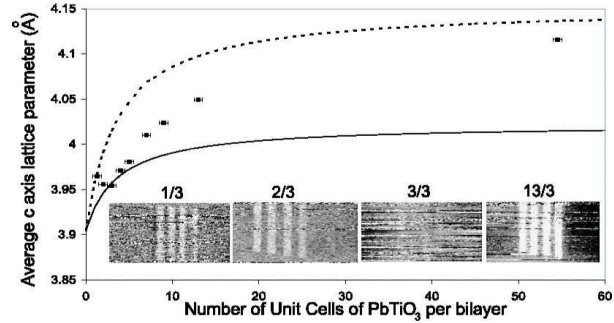

Figure 5.6: Average $c$-axis lattice parameter versus $n$ in $(PbTiO_3)_n/(SrTiO_3)_3$ superlattices as deduced from x-ray measurements. Suppression and recovery are emphasized from comparison with two limiting cases obtained by fixing the $SrTiO_3$ lattice parameter to its paraelectric cubic bulk value and the $c$ lattice parameter of $PbTiO_3$ either to its expected value for a *paraelectric* tetragonal phase coherent with the substrate or to a *fully polarized* bulk tetragonal phase (dashed line). Complementary PFM images are shown as insets and point out the disappearance of the piezoelectric response (white vertical bands) for the 3/3 superlattice only. (Fig. 3 from Ref. [326] by Dawber *et al.*)

the general trends are similar than for the electrostatic model: DFT calculations predict a homogeneous polarization and its *monotonic* decrease as the $PbTiO_3$ volume fraction is reduced. This confirms the dominant role of electrostatic effects for this system. Nevertheless, the recovery of a polarization for $n = 1, 2$ experimentally observed, remains an enigma. It might be related to effects not accounted for in the theoretical simulations such that interlayer mixing or the appearance of a new phase for $PbTiO_3$ similar to that formed under negative pressure [178].

### 5.4.3 $KNbO_3/KTaO_3$

$KTaO_3$ is another incipient ferroelectric material. Specht *et al.* [328] characterized $(KNbO_3)_n/(KTaO_3)_n$ superlattices grown by pulsed laser deposition on $KTaO_3$ substrates. The superlattice modulation length was ranging from 0.8 ($n = 1$) to 33.8 nm ($n = 42$). The $KNbO_3$ layers were found to remain strained in-plane to match the substrate lattice parameter independently of the thickness. They measured the effect of the layer thickness on the ferroelectric phase transition temperature $T_c$. They reported $T_c = 825K$ for $n = 42$, a value much higher than for pure $KNbO_3$ (708 K) that can be related to strain effects. $T_c$ progressively decreases down to 475 K for $n = 6$ and then saturates at this value for smaller $n$. Using Ginzburg-Landau theory, this saturation of $T_c$ was recently explained by a crossover to a regime of strongly coupled ferroelectric layers, coming from the interactions between the ferroelectric domains formed in the ferroelectric layers [348].

The behavior of $[001]$–$(KNbO_3)_n/(KTaO_3)_n$ multilayers on a $KTaO_3$ substrate was also theoretically investigated by Sepliarsky *et al.* [149, 150] from atomic level simulations with shell-model potentials. These first studies are restricted to zero temperature. It is shown by Sepliarsky *et al.* that due to in-plane compressive strain, $KNbO_3$ expands along the perpendicular direction, therefore breaking the rhombohedral symmetry expected at low temperature. The strain is however not strong enough to force the material to become tetragonally polarized and the $KNbO_3$ layers keep both perpendicular ($[001]$) and in-plane ($[110]$) polarizations (see Fig. 5.7). The in-plane polarization has a bulk-like behavior with abrupt jumps at the interfaces: it vanishes in $KTaO_3$ and keeps a constant value (independently of $n$), similar to that of a bulk phase under the same strain conditions, in $KNbO_3$. In contrast, the polarization along the $[001]$ modulation direction is continuous through the interface, with the interior of $KTaO_3$ remaining substantially polarized. Both the absence of coupling for $[110]$ polarizations and the strong coupling for the [001] polarizations agree with what was discussed for BaTiO$_3$/SrTiO$_3$ superlattices.

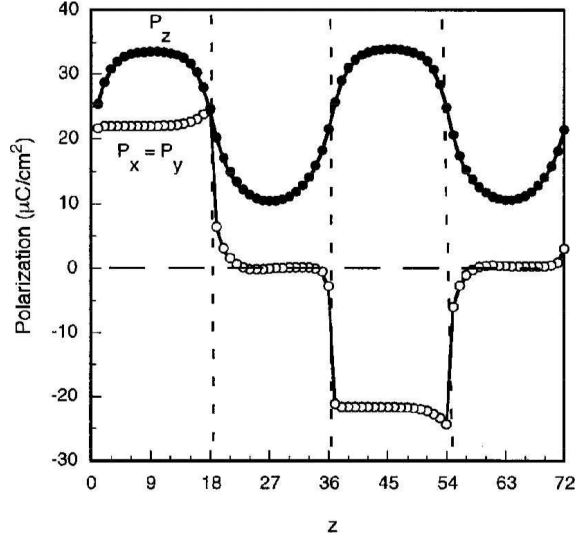

Figure 5.7: Spatial dependence of the polarization in the plane (open circles) and in the [001] modulation direction (solid circles) as computed from atomic level simulations with shell-model potentials for a (KNbO$_3$)$_{18}$/(KTaO$_3$)$_{18}$ superlattice epitaxially grown on a KTaO$_3$ substrate. (Fig. 1 from Ref. [154] by Sepliarsky *et al.*)

The shell-model approach allows to access larger values of $n$ than bare DFT calculations. Doing so, Sepliarsky *et al.* [149, 150] pointed out that the uniformity of the [001] polarization is only expected for short modulation lengths: as $n$ increases, the polarization progressively decreases at the center of the KTaO$_3$ layer, to vanish for $n \approx 80$. The polarization does however not drop abruptly to zero at the interface; independently of $n$, it keeps a value very similar to that of KTa$_{0.5}$Nb$_{0.5}$O$_3$ solid solution under the same strain conditions.

They also reported hysteresis loops. This allowed them to identify distinct behaviors depending of $n$: for $n < 6$ the different KNbO$_3$ layers are strongly coupled while for large modulation lengths $n > 12$ the different KNbO$_3$ layers behave essentially independently. This is qualitatively consistent with the experimental finding of a progressive decrease of $T_c$ with $n$ down to a constant value for $n = 6$ and below [328].

In a subsequent study, Sepliarsky *et al.* [154] directly computed the phase transition temperature of (KNbO$_3$)$_n$/(KTaO$_3$)$_n$ multilayers from molecular dynamics simulations with the same shell-model as before. They report that the Curie temperature for the transition from a polarized to an unpolarized state in the [001] modulation direction decreases linearly with $n$ down to $n = 6$ and then saturates to a constant value for smaller $n$, in qualitative agreement with experimental data [328].

## 5.5 Ferroelectric/ferroelectric superlattices.

In ferroelectric/ferroelectric superlattices, the question is no more to know if the system remains ferroelectric but instead to see if the atomic arrangement can be tuned in order to optimize the functional properties. Multilayers obtained from the stacking of pure ABO$_3$ ferroelectrics have not been investigated in great details. In contrast, multilayers arising from the ordering of ferroelectric alloys have been the subject of various studies, mainly based on the effective Hamiltonian approach that provides access to the functional properties in terms of composition and temperature. Two kind of alloys were explicitly considered: $PbZr_{1-x}^{4+}Ti_x^{4+}O_3$ (PZT) a typical *isovalent* alloy and $PbSc_{1-x}^{3+}Nb_x^{5+}O_3$ (PSN) a typical *heterovalent* alloy. As it is now discussed, the appearance of unusually high functional properties in these ordered structures is directly related to the modification of the phase diagram compared to random alloys and arises in situations where the polarization can rotate easily. Again, in most cases, electrostatic effects are seen to play a major role.

## 5.5.1 PSN superlattices.

Modification of the atomic ordering in *heterovalent* alloys such as PSN [349] can also lead to unexpected structural properties and enhanced functional responses. This was highlighted from effective Hamiltonian simulations by George *et al.* [33] for PSN superlattices [10] made from the $Pb(Sc_{0.5+\nu}Nb_{0.5-\nu})O_3/Pb(Sc_{0.5}Nb_{0.5})O_3/Pb(S$ sequences along the [001] direction as illustrated in Fig. 5.8. Changes in the structural, dielectric and piezoelectric responses were analyzed as a function of the composition and temperature.

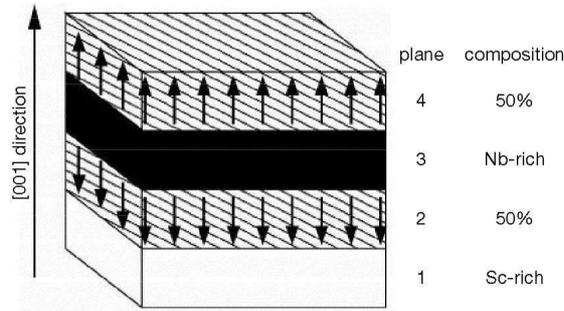

Figure 5.8:
Schematic view of the $Pb(Sc_{0.5+\nu}Nb_{0.5-\nu})O_3/Pb(Sc_{0.5}Nb_{0.5})O_3/Pb(Sc_{0.5-\nu}Nb_{0.5+\nu})O_3/Pb(Sc_{0.5}Nb_{0.5})O_3$ superlattice studied by George *et al.* The internal electric fields acting on the B planes are shown by arrows. (Fig. 1 from Ref. [33] by George *et al.*)

First, simulations at low temperatures (20 K) reveal that the static susceptibility ($\chi_{33}$) and piezo-electric ($d_{34}$) responses are strongly dependent of $\nu$ and exhibit a huge enhancement around $\nu = 0.375$ (Fig. 5.9). The structure and the direction of the polarization also strongly evolves with $\nu$: initially rhombohedral for $\nu = 0$ ($\mathcal{P}$ along pseudocubic [111] directions), the structure is monoclinic up to $\nu = 0.44$ ($\mathcal{P}$ between [111] and [110]) and then becomes orthorhombic ($\mathcal{P}$ along [110]).

Second, simulations in temperature for $\nu = 0.375$ reveals two phase transitions: the structure, tetragonal paraelectric at high temperature, becomes orthorhombic at 400 K ($\mathcal{P}$ along [110]) and then monoclinic at 40 K ($\mathcal{P}$ between [111] and [110]) [11]. As illustrated in Fig. 5.9, the existence of the orthorhombic to monoclinic phase transition results in a huge enhancement of the dielectric and piezoelectric responses in a relatively broad range of temperature around 40 K.

Some insight into these results can be obtained from the effective Hamiltonian. As explained in Sec 2.3.2, the effective Hamiltonian of mixed perovskites consists in an average part, obtained within the virtual crystal approximation, and local corrections for local modes and inhomogeneous strain variables that depend on the atomic arrangement. In particular, the energy correction describing the effect of the atomic configuration on the local mode variable $\xi$ [linear term in $\xi_i$ in Eq. (2.40)] can be reformulated as

$$\Delta E_{loc} = -\sum_i Z^* \xi_i \cdot \left[ \sum_j -\sigma_j S_{ji} \mathbf{e}_{ji} \right] = -\sum_i Z^* \xi_i \cdot \mathcal{E}_i^{int}, \qquad (5.11)$$

where $i$ runs over all cells, $j$ runs over the three nearest neighbors of $i$, $Z^*$ is the Born effective charge associated to $\xi$ and $\mathbf{e}_{ji}$ is the unit vector joining B site $j$ to B site $i$. The variables $\sigma_j$ characterize the ionic configuration: $\sigma_j = 1 (-1)$ corresponds to having Nb (Sc) in cell $j$. $S_{ji}$ are alloy related parameters directly extracted from DFT calculations on small supercells. Interestingly, $\Delta E_{loc}$ can be recasted as an interaction between the dipole moment $Z^* \xi_i$ at site $i$ and an internal electric field $\mathcal{E}_i^{int} = [\sum_j -\sigma_j S_{ji} \mathbf{e}_{ji}]$ induced by the $B$ atoms of site $j$. All the $S_{ji}$ are found to be positive so that the field acting on site $i$ and induced by Sc$^{3+}$ (Nb$^{5+}$) at site $j$ points from $i$ to $j$ (from $j$ to $i$). This is coherent with a simple electrostatic picture since Sc (Nb) are negatively (positively) charged with respect to average B-ion valence of $+4$. For the superlattice, the atomic arrangement gives rise to the internal electric fields as schematize by the arrows in Fig. 5.8. These fields of opposite direction tend to suppress the polarization $\mathcal{P}_z$ along the [001] modulation direction and scale linearly with $\nu$. For the largest values of $\nu$, the internal fields are so large that $\mathcal{P}_z = 0$ at any temperature and the system adopts an orthorhombic ground state. For intermediate value of $\nu$, the fields are smaller and the system can acquire a $\mathcal{P}_z \neq 0$ at a finite temperature resulting in the orthorhombic $\rightarrow$ monoclinic phase transition. The large functional responses in the vicinity of this transition originates in the high modification of $\mathcal{P}_z$ when parameters are slightly modified. In other words, they are directly related to the ease of rotating the polarization [111].

[10]These superlattices are made of four independent layers but do not break the inversion symmetry. They are addressed in this section because of their analogy with previously discussed structures.

[11]This has to be contrasted with disordered PSN that has a unique phase transition from from paraelectric cubic to ferroelectric rhombohedral.

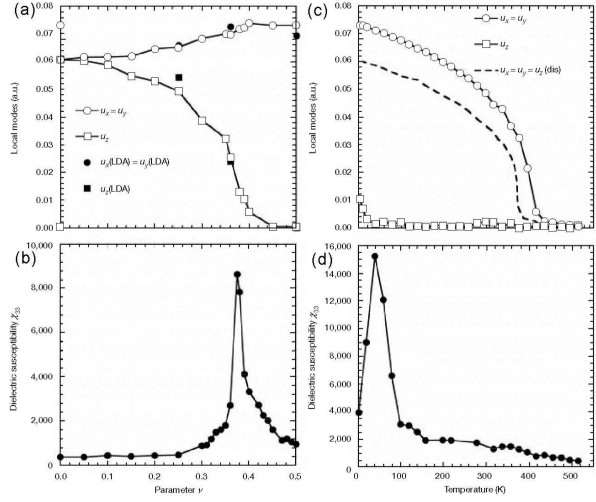

Figure 5.9: (a) Average cartesian coordinates of the local mode (here called $u$ instead of $\xi$) and (b) dielectric susceptibility ($\chi_{33}$) as a function of the $\nu$ parameter for the PSN superlattice of Fig. 5.8 at 20K. (c) Average cartesian coordinates of the local mode and (d) dielectric susceptibility ($\chi_{33}$) as a function of the temperature for $\nu = 0.375$. Curves similar to $\chi_{33}$ have been reported for the piezoelectric coefficient $d_{34}$. (Fig. 2-3 from Ref. [33] by George *et al.*)

The previous study illustrates that fine tuning of the functional properties can be expected from modifications of the atomic ordering in heterovalent ferroelectric alloys, and opens the door to the design of new artificial structures with optimized properties. Usually, this is achieved by proposing different new arrangements and predicting their properties. The fact that electrostatics plays a major role in PSN can guide our choices towards potentially appropriate ordered structures. However, such an approach is still limited by our ability to imagine new structures. It was illustrated by Íñiguez *et al.* that it might be possible to avoid this problem by following the inverse approach: combining the effective Hamiltonian approach with inverse Monte Carlo methods, they demonstrated the possibility of identifying the atomic arrangement leading to prefixed properties in PSN ordered structures [144].

## 5.5.2  PZT superlattices.

Bellaiche *et al.* developed an effective Hamiltonian for PZT alloys, suitable to describe its properties around 50/50 composition. This model allowed to discuss the origin of the high piezoelectric response in the random alloy around the morphotropic phase boundary (MPB) [77]. It also allowed to estimate the effects of short-range order on the properties of PZT [142].

Using this effective Hamiltonian, Huang *et al.* [334] reported huge enhancement of the electromechanical response in $[100] - (\mathrm{PbZr}_{1-x+\nu}\mathrm{Ti}_{x-\nu}\mathrm{O}_3)_1/(\mathrm{PbZr}_{1-x-\nu}\mathrm{Ti}_{x+\nu}\mathrm{O}_3)_1$ superlattices. The computed dependence of the local mode (here called $u$ instead of $\xi$) and piezoelectric coefficient $d_{15}$ with $\nu$ is reported in Fig. 5.10 for $x = 0.485$ and $T = 50K$. Panel (a) points out a sequence of phase transition from so-called $M_A$ monoclinic to triclinic and finally $M_B$ monoclinic, associated to the decrease of the [001] polarization along the modulation direction from a finite value to zero. In a way similar to what was previously discussed for PSN, in panel (b) a peak is observed in the piezoelectric response around the triclinic-$M_B$ phase transition. Again, this feature can be related to the ease of rotating the polarization around that composition. However, in the present case of isovalent substitution, the explanation can no more be found in the difference of formal valence charge of the B cations. It is however suggested that electrostatic effects still play an important role. In spite of their similarities, $\mathrm{PbZrO}_3$ and $\mathrm{PbTiO}_3$ do not polarize the same way, yielding Ti-rich planes with a higher polarization than Ti-poor planes. The difference of polarization between consecutive planes increases with increasing $\nu$. As it costs a huge electrostatic energy to have different polarizations in consecutive planes in the [100] modulation direction, $\mathcal{P}_x$ progressively decreases with $\nu$ and the polarization rotates in plane, explaining the previous results.

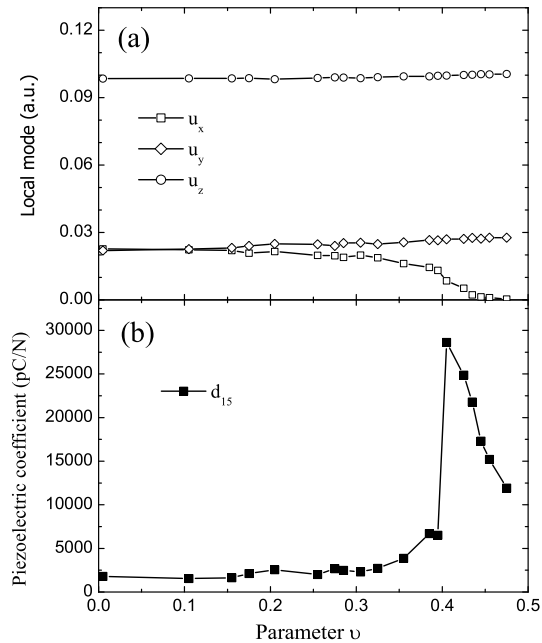

Figure 5.10:   (a) Average cartesian coordinates of the local mode (here called $u$ instead of $\xi$) and (b) piezoelectric coefficient $d_{15}$ as functions of the parameter $\nu$ in $[100] - (\mathrm{PbZr}_{1-x+\nu}\mathrm{Ti}_{x-\nu}\mathrm{O}_3)_1/(\mathrm{PbZr}_{1-x-\nu}\mathrm{Ti}_{x+\nu}\mathrm{O}_3)_1$ superlattices with $x = 0.485$ at 50K. (Fig. 2 of Ref. [334] by Huang *et al.*)

With the same effective Hamiltonian, Kornev and Bellaiche [350] studied $[001] - (PbZr_{1-x_1}Ti_{x_1}O_3)_n/(PbZr_{1-x_2}Ti_{x_2}O_3)_n$ superlattices with $x_1$ and $x_2$ chosen to be located across the MPB while $(x_1+x_2)/2$ lies inside this boundary. Their calculations reveal phase transition sequences that are different from the corresponding sequence in random PZT with the same average composition: the ground state of the superlattices is triclinic and the phase transition sequences are drastically affected by the modulation length $2n$. On top of that, these systems also have deep local minima of monoclinic symmetry that are very close in energy from the global triclinic minima. It is emphasized that this multivalley configuration can generate a nonergodic behavior. These unusual features arise from the atomic ordering. It is proposed that they come from inhomogeneous strain fields resulting from the difference in ionic radius between $Ti^{4+}$(0.605 Å) and $Zr^{4+}$(0.72 Å).

Full DFT calculations on PZT superlattices have finally been reported by Bungaro and Rabe. In Ref. [351], they discuss and compare the lattice dynamics of $[001]$, $[110]$ and $[111]$ $(PbTiO_3)_1/(PbZrO_3)_1$ structures. They point out that unstable polar modes in the tetragonal $[001]$ and $[110]$ lattices are confined in either the Ti or Zr centered layer and display two-mode behavior, while in the $[111]$ case one-mode behavior is observed. In Ref. [259], they investigate the influence of epitaxial in-plane strain on the properties of a $[001] - (PbTiO_3)_1/(PbZrO_3)_1$ superlattice. Between the usual $c$- and $aa$-phases typical for large compressive and tensile strain, they reveal the existence of a monoclinic $r$-phase for moderate in-plane strains (see Fig. 5.1). In the $r$-phase, the polarization progressively rotates from the $[001]$ direction ($c$-phase) to the $[110]$ direction ($aa$-phase) and the dielectric and piezoelectric responses are high. Coherently with what was discussed all along this section, this was assigned to the ease with which the polarization can be changed through rotation within this phase.

## 5.6  Tricolor superlattices.

"Tricolor" superlattices, structured as alternating stack of three components $(ABO_3/A'B'O_3/A''B''O_3)$, have also recently attracted some of attention both theoretically [30, 352] and experimentally [331, 332, 333, 235].

Contrary to most conventional "bicolor" superlattices, tricolor heterostructures breaks the inversion symmetry. As highlighted in a pioneer work of Sai, Meyer and Vanderbilt [30], this might generate interesting properties if we can achieve fine tuning of the symmetry breaking. On the one hand, Sai and coworkers looked at tricolor structures in which successive layers differs by an *isovalent* substitution at the A-site (e.g. $BaTiO_3/SrTiO_3/CaTiO_3$) or at the B-site (e.g. $BaTiO_3/BaZrO_3/BaHfO_3$). They showed that, in such systems, the symmetry breaking is sufficiently small to preserve an asymmetric double-well but sufficiently large for the system to be self-poled. On the other hand, they also considered a tricolor structure obtained from *heterovalent* substitution at the B-site (e.g. $BaSc^{3+}O_3/BaTi^{4+}O_3/BaNb^{5+}O_3$). In this case, the asymmetry is much stronger and yields a single energy minimum. The electrostatic interaction resulting from the different valence charges in the different layers is shown to be a dominant effect. Calling $\delta$ the difference of formal valence charge in successive layers in structures of the type $AB^{(n-\delta)+}O_3/AB''^{n+}O_3/AB''^{(n+\delta)+}O_3$, the symmetry breaking was shown to scale as $\delta^3$. Therefore, the epitaxial growth of structures as $Ba(Sc_{1-y}^{3+}Nb_y^{5+})O_3$ in an alternating sequence of layers with $y=0.5(1-\delta)$, $y=0.5$ and $y=0.5(1+\delta)$ might allow a fine tuning of the symmetry breaking by controlling the concentration variable $\delta$ and potentially optimization of the desired properties of the material.

On the experimental side, strong polarization enhancement (almost 50% with respect to similarly grown pure $BaTiO_3$) have been reported by Lee *et al.* [235] in $(SrTiO_3)_x/(BaTiO_3)_y/(CaTiO_3)_z$ (also labeled as $S_xB_yC_z$) tricolor lattices, in spite of the paraelectric behavior of $SrTiO_3$ and $CaTiO_3$ (see Fig. 5.11). The superlattice was grown using pulsed laser deposition on a $SrRuO_3$ (10nm)/$SrTiO_3$ substrate. The polarization enhancement was assigned to different sources. First, when the $BaTiO_3$ layer thickness does not exceed the combined thickness of the $SrTiO_3$ and $CaTiO_3$, the superlattice keeps an in-plane lattice identical to the $SrTiO_3$ substrate [see Fig. 5.11(a)]. This should result in a strain enhancement of the ferroelectric polarization in $BaTiO_3$ similar to that discussed for $SrTiO_3/BaTiO_3$ superlattices and estimated to more than 50% [35]. Competing with that, there is however the dilution of the ferroelectric material in the paraelectric matrix. Considering the superlattice as an alternance of ferroelectric and paraelectric layers of thicknesses $t_{FE} = t_{BTO}$ and $t_{PE} = t_{STO} + t_{CTO}$ and using the electrostatic model proposed by Neaton and Rabe [35], the average polarization of the multilayer, $\bar{\mathcal{P}}$ can be related to that of the ferroelectric constituent $\mathcal{P}_{FE} = \mathcal{P}_{BTO}$ [12] through [35]

$$\bar{\mathcal{P}} = \mathcal{P}_{FE}[1 + (t_{PE}/t_{FE})(\epsilon_{FE}^0/\epsilon_{FE}^0)]^{-1}, \qquad (5.12)$$

where the static dielectric constants to be considered are $\epsilon_{FE}^0 = \epsilon_{BTO}^0$ and $\epsilon_{PE}^0 = \epsilon_{STO}^0\epsilon_{CTO}^0(t_{STO} + t_{CTO})/(\epsilon_{CTO}^0 t_{STO} + \epsilon_{STO}^0 t_{CTO})$ (deduced from "two capacitors" in series) with $\epsilon_{BaTiO_3}^0 = 160$, $\epsilon_{SrTiO_3}^0 = 300$ and $\epsilon_{CaTiO_3}^0 = 186$. Such a model provides however results partly inconsistent with experimental measurements. It predicts for instance that $\bar{\mathcal{P}}$ scales with the ratio $r = (t_{PE}/t_{FE})$ while it is experimentally shown that it also evolves with $t_{FE}$ when keeping $r$ constant [see Fig. 5.11(b)]. It has therefore been proposed that interfacial effects are playing a role and that $TiO_6$ units in the center of $BaTiO_3$ layers are more polarizable that those at the interface [see Fig. 5.11(c)]. The strongest polarization enhancement is therefore achieved by the proper balancing of two competing requirements: the $BaTiO_3$ layer must be (i) *thick* enough to contain a sufficient amount of non interfacial $TiO_6$ units but (ii) *thin* enough to remain fully strained. In Ref. [235], this is achieved for $S_2B_4C_2$ with an almost 50% increase in polarization over that of $BaTiO_3$ samples [see Fig. 5.11(b)].

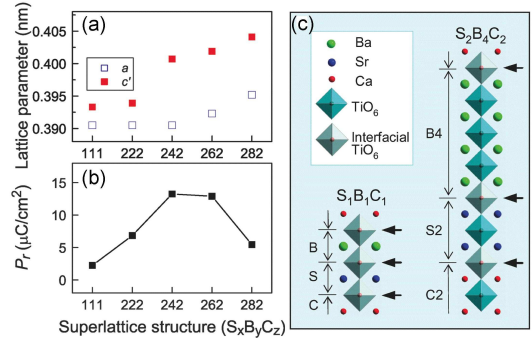

Figure 5.11: (a) In-plane (open squares) and out-of-plane (filled squares) lattice parameters of various tricolor superlattices. $c'$ corresponds to the supercell $c$ divided by the number of perovskite units in the supercell. (b) Remanent polarization $\mathcal{P}_r$ from $\mathcal{E} = \pm 750$ kV/cm loops (except for $S_2B_8C_2$ measured with $\mathcal{E} = \pm 750$ kV/cm). (c) Diagrams of supercells showing the local environment of $TiO_6$ octahedra. Interfacial octahedra are shown indicated by solid black arrows. (From Figure 4 of Ref. [235] by Lee *et al.*)

Recent first-principles calculations by Nakhmanson *et al.* [352] confirm the polarization enhancement in $(SrTiO_3)_x/(BaTiO_3)_y/(CaTiO_3)_z$ multilayers, in qualitative agreement with experimental data (polarizations are overestimated but polarization enhancement factors show similar trends). It is shown that a substantial polarization arise from $CaTiO_3$ and $SrTiO_3$ layers and that the local polarization profile is flat along the whole stack. The largest polarization enhancement is achieved for the highest concentration of $BaTiO_3$ and $CaTiO_3$, assembled in the thickest possible layers that are still consistent with a fully strained state (from experiment limited to $t_{BaTiO_3} \le t_{SrTiO_3} + t_{CaTiO_3}$ [235]). The calculations also highlight the lack of inversion symmetry and the existence of two slightly distinct values of the polarization for up and down states. Experimentally, symmetry inversion breaking was not discussed in Ref. [235] that presents other kind of asymmetry in the experimental geometry. Inversion symmetry breaking in the same tricolor lattice was evidenced by Warusawithana [332] from dielectric measurements.

---

[12] $\mathcal{P}_{BTO}$ refers to the polarization of pure $BaTiO_3$ film under the same strain constraints than in the stack.

## 5.7   Perspectives.

The combination of different oxides in superlattices structures, with a control at the atomic scale, is opening the door to new posibilities in the design of multifunctional materials. Enhancement of properties, such as the polarization or the piezoelectric response, appearance of new phases with new properties not present in bulk, or the tuning of specific properties by just changing the composition of the superlattices (the inverse problem), are now accessible possibilities.

The lattice mismatch (different lattice constants), the dielectric mismatch (different dielectric constants and susceptibilities), and compositional gradients between the materials that constitute the interface are the keys ingredients to be monitored.

The study of superlattices is one of the fields were theory and experiment have gone hand by hand. From these valuables exchanges and discussions, quick significant advances have been obtained and are still expected in the future.

# Perovskites on Si.

## 6.1 Overview.

The search for alternative gate dielectric materials to replace silica ($SiO_2$) in microelectronic devices is one of the grand challenges that the materials science community and the Si-based semiconductor industry are facing at the current time [353]. The figure of merit in a metal-oxide-semiconductor field-effect-transitor (MOSFETs) is the saturation current from the source to the drain. This saturation current is directly proportional to the capacitance $C$ of the gate capacitor. Assuming that the gate capacitor adopts a planar geometry, such as the one shown in Fig. 6.1

$$C = \frac{\epsilon_0 \kappa \ A}{d},$$
(6.1)

where $\epsilon_0$ is the permittivity of vacuum, $A$ is the lateral area of the capacitor, and $\kappa$ and $d$ are, respectively, the static dielectric constant [1] and the thickness of the dielectric material between the Si channel and the gate electrode.

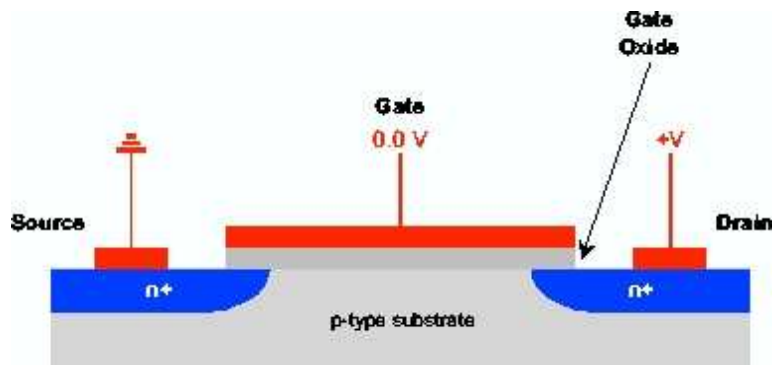

Figure 6.1: Schematic representation of a metal-oxide-semicondutor field-effect-transitor (MOSFET).

The rapid scaling down of the physical gate lengths of the MOSFETs implies a reduction of the area $A$ of the capacitors. In order to maintain a high gate oxide capacitance, a concomitant reduction of the gate dielectric thickness $d$ is required (roughly speaking, the thickness of the gate dielectric has to be directly proportional to the channel length in a MOSFET [354]). However, we have reached a stage where the thickness of the $SiO_2$ layer can no more be decreased because, together with problems in the thickness control, the leakage current would become inacceptably high, producing an increase of the power consumption and a deterioration of the switching characteristics of the transistor. Indeed, the leakage current from the channel to the gate is due to the direct tunneling of carriers, and increases exponentially with the decrease of both the thickness of the gate dielectric and the height of the electrostatic barrier for the electrons through the gate stack. The current roadmap projection (assessed by the International Technology Roadmap for Semiconductors, ITRS [355]) imposes the very short-range choice of an alternative gate dielectric with a good capacitance for a thick-enough layer.

---

[1]The static dielecric constant is called $\kappa$ all along this chapter, in agreement with the literature in the field.

The properties that the new dielectric should meet are well established and have been reviewed recently by Wilk and Wallace [356]. They can be divided into *fundamental material properties*, and *device processing and performance properties*. Amongst the material properties, we can enumerate (i) a higher dielectric constant than amorphous silica ($\kappa_{SiO_2} = 3.9$) that would allow for a greater physical thickness while preserving the capacitance and the electrical properties; (ii) large band gaps and band offsets with Si to prevent tunneling currents (a large value of the Conduction Band Offset, CBO, between Si and the gate dielectric is required to minimize carrier injection into the conduction band, and typically materials with CBO smaller than 1.0 eV are rejected for further applications); (iii) a good thermodynamic stability in contact with the Si substrate, even at the high growth temperatures (around 900 °C); (iv) a good quality of the interface with the Si channel, which means a small number of electrical defects and a low midgap interface state density; and (v) film morphology avoiding the formation of polycrystalline films and grain boundaries. Amongst the device properties, we can cite (vi) a good compatibility with metallic gate electrodes; (vii) a compatiblity with the deposition mechanism during the fabrication process; (viii) reliability.

Many materials satisfy some subsets of the previous criteria, but the identification of a dielectric that addresses *simultaneously* all of the requirements, an might compete with the Si-SiO$_2$ interface is far from trivial. Investigations on oxides like Al$_2$O$_3$, ZrO$_2$ [357, 358, 359, 360], HfO$_2$ [361, 362, 363], Ta$_2$O$_5$, Y$_2$O$_3$, Gd$_2$O$_3$, and TiO$_2$ [364], have thrown encouraging results in the last few years [365, 366]. ABO$_3$ perovskite oxides have also attracted a lot of attention.

The ABO$_3$ compounds have a dielectric constant above 300, one order of magnitude higher than the other candidates. However, they present two major problems that have prevented their application for a long time. First, they are thermodynamically unstable in direct contact with Si (they react to form titanium silicide and alkaline-earth silicate [365, 210]). Second, Robertson and Chen [367], aligning the Charge Neutrality Levels (CNL) [368] of both semiconductors, have estimated a CBO for a Si/SrTiO$_3$ interface of -0.14 eV (SrTiO$_3$ below, that is no barrier at all for the electrons) in very good agreement with experimental results [369]. This prevents, in principle, the use of the titanate as the gate dielectric in electronic devices.

Some important works have been done in the last few years in order to sort out both shortcomings. The efforts have ended up with the construction of some SrTiO$_3$ based MOSFETs. Using a 110 Å-thick SrTiO$_3$ layer as the gate dielectric, Eisenbeiser *et al.* [370] have fabricated a transistor that behaves comparably to a 8 Å-thick SiO$_2$/Si MOSFET. The improvement in transistor performance was very satisfactory, and the leakage currents was two order of magnitudes smaller than in a similar SiO$_2$-based device.

In this section we will review those works concerning the structure at the interface and how the band offset between the Si channel and the perovskite can be tuned and controlled efficiently.

## 6.2 Epitaxial growth of ABO$_3$ perovskites oxides on Si.

In a milestone paper, McKee, Walker and Chisholm [371] grew a commensurate crystalline oxide on silicon (COS) structure using molecular beam epitaxy (MBE). They showed that the growth of ABO$_3$ perovskites oxides on Si can not be simply initiated with an oxide layer, but it must include the stable Si/ASi$_2$/AO interface. Indeed they observed that, after the deposition of 1/4 monolayer at high temperature (850 °C) on top of a (001) Si surface, the alkaline earth atoms have dispaced and replaced Si atoms on the top surface, forming a *c*(4 × 2) silicide. The silicide is fully commensurate and thermodynamically stable in contact with Si, and remains at the interface as part of the overgrowing perovskite structure. Then, a layer of the alkaline-earth oxide can be deposited in thermodynamic equilibrium with the underlying silicon/silicide interface. Finally, either more layers of the alkaline earth oxide, AO, or of the transition metal oxide layer, $BO_2$, can be deposited to form a unit cell of the NaCl-type alkaline-earth oxide structure or one unit cell of the perovskite, respectively. The final structure corresponds to the sequence $Si/ASi_2/AO/ABO_3$, and the epitaxy is such that $ABO_3$ (001) ∥ AO (001) ∥ Si (001), and $ABO_3$ ⟨110⟩ ∥ AO ⟨100⟩ ∥ Si ⟨100⟩, i. e. the $ABO_3$ atomic planes are rotated 45° around the (001) AO direction [372] (see Fig. 6.2). The epitaxial crystalline growth at the oxide/semiconductor interface avoids the formation of defects and ensures the continuity of the dielectric displacement [373]. MBE techniques allow the control of the growing sequence at the submonolayer level preventing grain-boundaries and providing a good quality interface and extremly smooth surface morphology. These COS heterostructures have shown the largest mobility ever reported for an alternative gate field-effect-transistor [373].

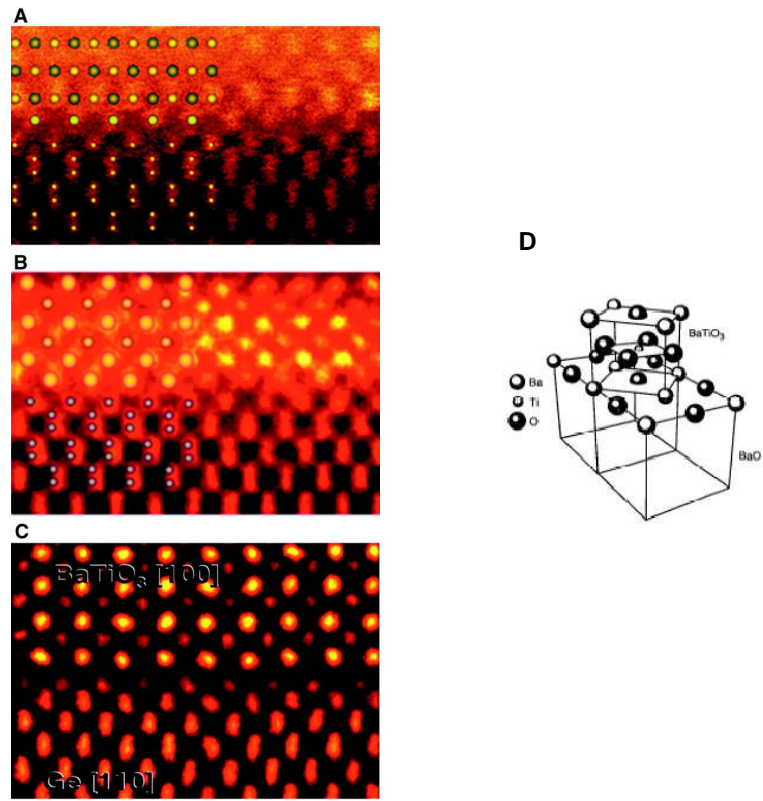

Figure 6.2: Z-contrast images of crystalline oxide on semiconductor (COS) heterostructures. The layer sequencing changes is controlled during growth: (A) three unit cells of the alloy $Ba_{0.725}Sr_{0.275}O$ on Si, (B) two unit cells of $SrTiO_3$ grown on one layer of $Ba_{0.725}Sr_{0.275}O$ on Si, and (C) three unit cells of $BaTiO_3$ on top of Ge. (D) Schematic view of the epitaxial growth of $BaTiO_3$ on top of BaO. The unit cell of the perovskite is rotated 45 ° with respect the (001) direction of BaO. Panels (A)-(C) from Ref. [373]. Panel (D) from Ref. [372].

Först and coworkers [354, 374] have proposed a different interface structure for $SrTiO_3$ on Si. In their proposal, the interfacial layer must (i) provide a covalent bonding environment towards the Si substrate, bridging the two fundamentally different chemical bondings (covalent in Si, mostly ionic in $SrTiO_3$); and (ii) behave as an ionic template compatible with $SrTiO_3$. The interfacial model includes the deposition of 1/2 monolayer Sr atoms on Si(001) surface. Each Sr atoms donates two electrons to the dangling bonds of the surface, saturating them. When the dangling bonds become fulfilled with electrons, the dimer buckling disappears. This $(2 \times 1)$ structure is the only Sr-covered surface without surface states in the band gap of Si, as required for device applications, and is fairly resistant to oxidation. Using molecular dynamics, the deposition of SrTiO$_3$ on top of the Sr-passivated Si surface was performed. The first deposited SrO layer reconstructs significantly for a temperature of 600 K. The subsequents SrO or SrTiO$_3$ layers crystallize into the perfect bulk crystal, and an abrupt interface is obtained between the Si and the high-$\kappa$ material Fig. 6.3(a). The stability of the structure with respect to oxidation was tested by introducing extra oxygen atoms in the calculation, simulating a high oxygen pressure during growth. The extra O atoms bond with the Si surface atoms at their vacant coordination site Fig. 6.3(b). This phase can be formed without growing an interfacial SiO$_2$ layer by oxidizing the substrate.

Zhang *et al.* [375] proposed another two possible interfaces (Fig. 6.4) between Si and SrTiO$_3$. The first is obtained by covering the (001) Si reconstructured surface with half a monolayer of Sr atoms, and then oxidizing the template and connecting it with a TiO$_2$ terminated SrTiO$_3$ slab. The second can be viewed as a connection of the SrO terminated SrTiO$_3$ slab with the unreconstructed Si (001) surface. However, in these structures the conduction band offset between the two materials is very small (around 0.57 eV for the first and -0.13 eV for the second) in good agreement with the simple Tersoff [368] model. These results suggest that SrTiO$_3$ is not likely to replace SiO$_2$ as a gate dielectric, at least as a simple stack.

## 6.3  Band offsets engineering.

The presence of an interfacial ASi$_2$/AO heterostructure between the Si and the perovskite is important not only for the structural stabilization. The heteroepitaxial approach allows the manipulation of the interface band structure as well as of the interface charge by changing the layer sequence of the structure.

As mentioned in the introduction, previous theoretical models estimated the barrier between Si and SrTiO$_3$ to be -0.14 eV. The absence of barrier for electrons would prevent the use of the perovskite oxide as the gate dielectric of our capacitors. However, the presence of the interface heterostructure was missing in their approach.

The band offsets and electrostatic barriers at the interface depend on the amount of charge transferred from one side of the interface to the other after the interfacial hybridization. They are therefore strongly dependent of the structure of the interface. In this context, techniques trying to extrapolate band-offsets from the properties of independent bulk compounds only are not fully satisfactory since they do not properly account for the chemical-bonding-induced charge transfer at the interface. The latter can only be obtained accurately from first-principles calculations on a supercell including both materials.

McKee *et al.* [376] coined the term "Coulomb buffer" to describe an interface-specific region in which the wave function of the junction interacts. It is present even at the monolayer thickness as a distint phase, both thermodynamically and electrodynamically. Its mission in the heterostructure is twofold: (i) it constraints the physical structure at the junction and (ii) it sets the electrical boundary conditions that establish the junction electrostatics and might therefore influence the domain structure of the ferroelectric oxide. An example of this concept was obtained by examining the physical structure and the charge localized in interface states on silicon atoms in the ASi$_2$ interface phase. The most remarkable result is that, by changing the alkaline earth atom in the interfacial ASi$_2$ layer, the band offset can be tuned by more than 0.6 eV as we move down in the periodic table while keeping the same alkaline-earth atom in the AO layer.

Junquera *et al.* performed first-principles simulations on the atomic and electronic properties of the SrO/SrTiO$_3$ and BaO/BaTiO$_3$ interfaces [96]. The atomic and electronic properties of these heterostructures have been shown in Sec. 5.3.1. They showed that, thanks to the wide band gap of the alkaline earth oxides, the inclusion of the AO layer in the heterostructure guarantees the existence of sufficiently high electrostatic barriers (both VBO and CBO $\sim 1$ eV, (see Fig. 6.5) but does not affect the relative position of the Si and ABO$_3$ electronic levels.

In the same spirit, Först and coworkers [354, 374] showed how the oxidized structure displayed in Fig. 6.3 creates a large charge dipole, which increases the conduction band offset by 1.1 eV, in the line

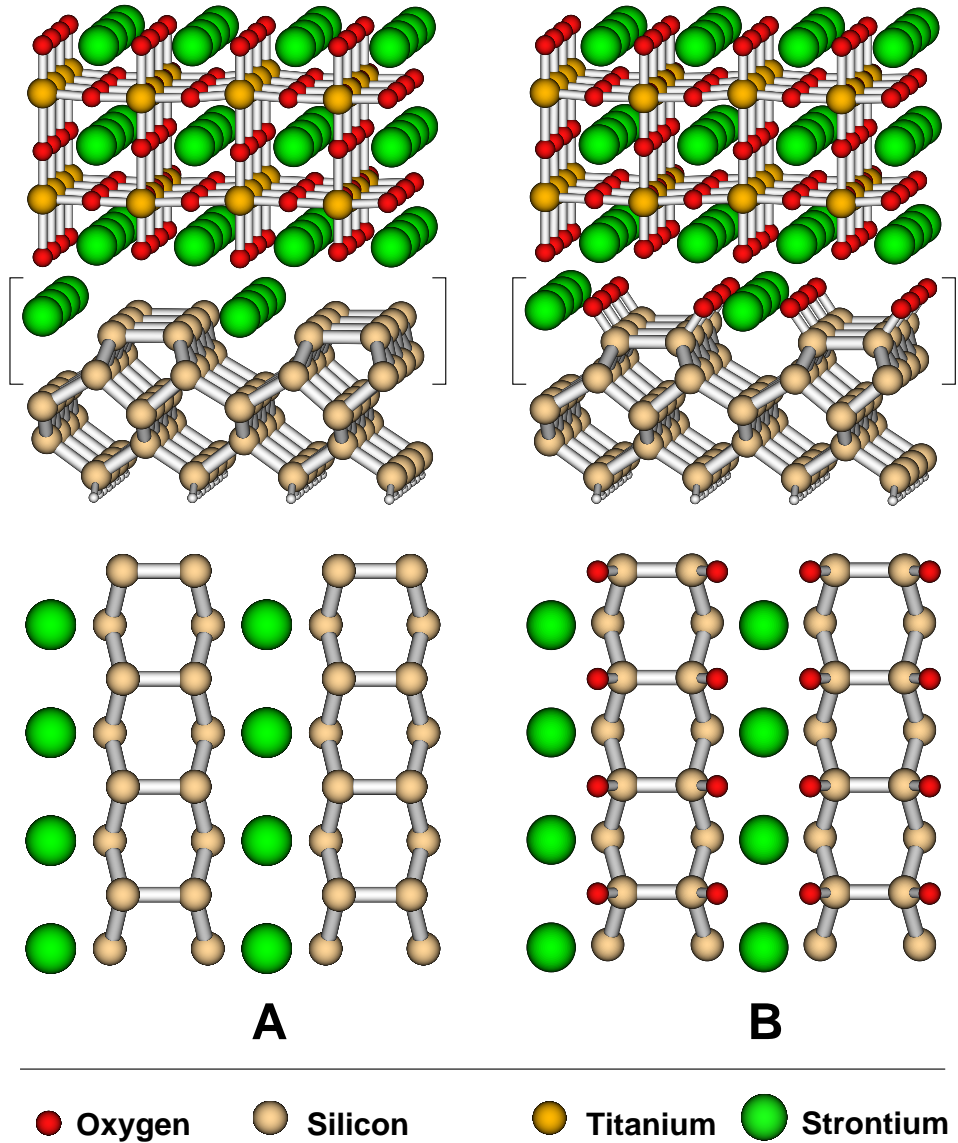

**A**       **B**

● Oxygen     ● Silicon     ● Titanium     ● Strontium

Figure 6.3: Unoxidized (A) and oxidized (B) atomic structures of the SrTiO$_3$/Si interface according to Först *et al.* Top row: lateral view. Bottom row: top view. From Ref. [374].

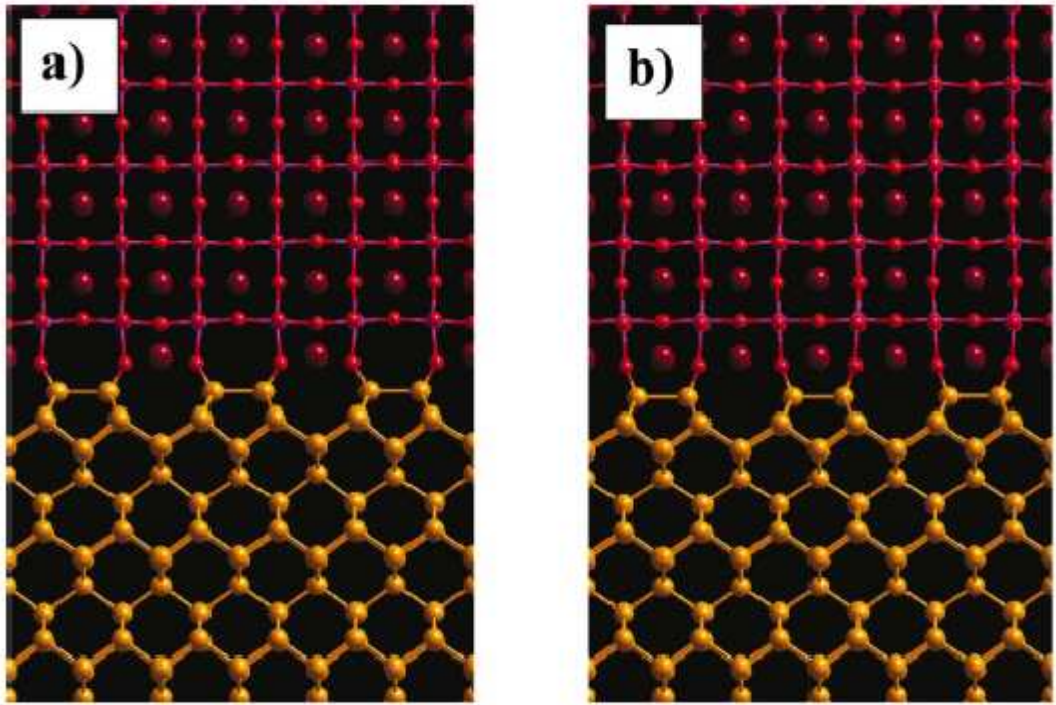

Figure 6.4: $(2 \times 1)$ structure of the Si-SrTiO$_3$ interface with one half of monolayer (a) and one monolayer (b) of Sr at the interface, according to Zhang *et al.* From Ref. [375].

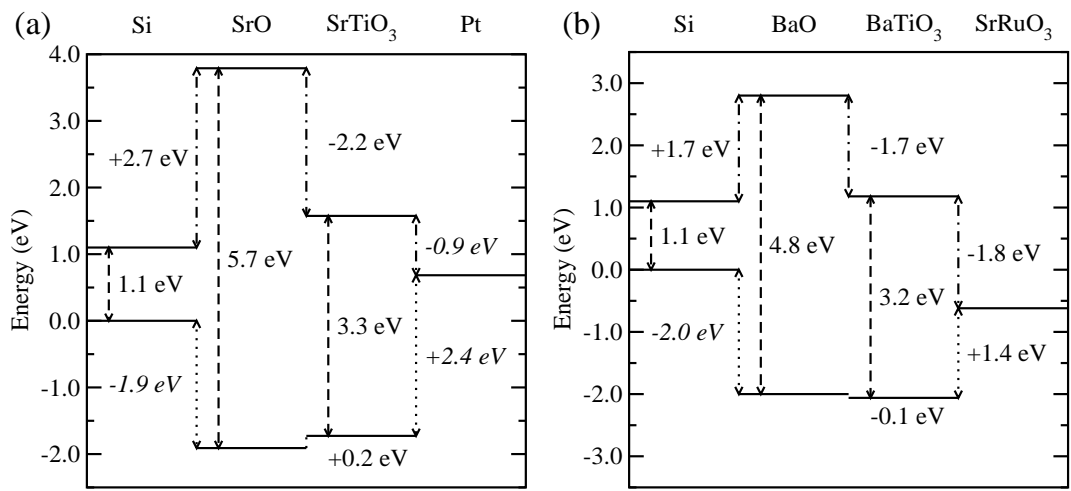

Figure 6.5: Estimation of the valence (dotted lines) and conduction (dot-dashed lines) band offsets for the whole heterostructures Si/SrO/SrTiO$_3$/Pt (panel a), and Si/BaO/BaTiO$_3$/SrRuO$_3$ (panel b). Dashed lines represent the experimental band gaps. Theoretical value for the VBO between Si and AO (in italic) has been taken from Ref. [377] for the Sr interface, and from Ref. [378] for Ba-interface. Theoretical Schottky-barriers between SrTiO$_3$ and Pt (also in italic) have been taken from Ref. [367]. From Ref. [96].

with technological requirements.

## 6.4 Perspective.

$ABO_3$ perovskites are not in the short list of materials selected to most likely replace $SiO_2$ in the next generation of MOSFET devices. However, some specific advantages of the COS structure (such as high mobility, band-offset engineering) could make it attractive for specific applications in the longer term. Also, the growth of $ABO_3$ perovskites directly on semiconductors is a valuable step in order to include these materials and to take advantage of their exceptional functional properties in fully integrated multifunctional devices in the future. In this line, recent calculations have shown how $BaTiO_3$ grown on top of Ge, with a non-zero interface density of states, can substain a normal polarization above a critical thickness of 7 nm [336], and opens the door to combine ferroelectric properties into the semiconductor technology.

# Ferroelectric nanoparticles and nanowires.

## 7.1 Overview.

The previous Sections dealt with different kinds of 2-dimensional ferroelectric systems including thin films, planar heterostructures and multilayers. Going further, we can now ask whether ferroelectricity still exists in systems with even more reduced symmetries, such as quantum dots and nanowires. Despite the great amount of works devoted to ultrathin films, only few studies address the problems of ferroelectricity in these other kinds of confined systems.

## 7.2 Ferroelectric nanoparticles.

Size effects on the ferroelectric phase transition in $PbTiO_3$ ultrafine particles (nearly free from the effect of the substrate) were initially reported by Ishikawa and coworkers [198, 199] and Zhong *et al.* [200]. Using high-resolution transmission electron microscopy (HRTEM) techniques, Jiang *et al.* [379] have found that the nanoparticles might be grown with extremely clean and sharp surfaces. Particles larger than 20 nm formed 90° domains, whose width decreased with decreasing the particle size. A phase transition to a monodomain state was observed for particle sizes smaller than 20 nm. For a given particle size, the shift of the lowest phonon line, of symmetry E(TO1), was measured as a function of temperature using Raman scattering techniques. This frequency shift abruptly drops with the increase of the temperature and the corresponding wave number tends to zero at a critical temperature $T_c$. Figure 7.1 represents the thickness dependence of $T_c$. The results strongly suggested the presence of a critical size, estimated between 9.1 nm and 10.7 nm, below which the ferroelectric state becomes unstable. The same conclusion could be drawn by following the evolution of the specific heat [200] or the tetragonality [200, 199].

Why does ferroelectricity disappear in these nanoparticles? Is it eventually possible to preserve ferroelectricity in these systems? In spite of the disappearance of ferroelectricity, are these nanostructures still suitable materials to be used in memory devices? Some of these questions have been recently addressed using different atomistic appoaches (effective Hamiltonian, shell-model).

Fu and Bellaiche simulated $BaTiO_3$ [380] and $Pb(Zr,Ti)O_3$ [36] quantum dots using a first-principles based model hamiltonian (see Sec. 2.3). In order to simulate isolated nanoparticles, the use of periodic boundary conditions was avoided, so no interactions between periodically repeated replicas of the nanodots were considered. The potential field generated by *every* dipole in the nanoparticle was computed in real space, so the depolarizing field generated by the surface charge density was naturally included in the calculation. Surface-induced atomic relaxations and cell-shape changes (Sec. 4.2.1) were taken into account (at least in an approximate way) by adding a new term in the model hamiltonian that deals with the coupling between the normal modes and inhomogeneous strain at the surface with vacuum. At zero applied electric field, the local polarization in small dots rotate from cell to cell, forming a "vortex-like" pattern (see Fig 7.2). The dot cannot maintain any net surface polarization charge, responsible for the depolarizing field as it was explained in Sec. 4.4.2, and that is the reason why the normal component of the local modes at the dot surfaces tends to vanish. The parallel-to-surface components point in opposite directions at opposite surfaces. The flip is accomodated across the entire dot, balancing the increase of the short-range energy and the lowering of electrostatic energy, and showing clearly the new equilibrium between the short-range and the long-range interactions in these confined systems. The average *magnitude* of the local dipole in each unit cell is comparable to the bulk value, even in a tiny $4 \times 4 \times 4$ rectangular dot (made of four five-atom unit cells along the pseudomorphic [100], [010], and [001] directions), although due to the particular shape of the vortex, the macroscopic polarization vanishes. This structure seems to be stable with respect the local surface environment. Increasing the size of the dot, rather uniform ferroelectric domains start to nucleate. Despite the local ferroelectric ordering, the macroscopic polarization of the entire dot equals to zero. The domain structure resemble the 90° domains experimentally observed in PbTiO₃ nanoparticles [379]. A macroscopically paraelectric nanostructure can be switched to a macroscopically ferroelectric phase by applying a field that aligns all the local dipoles in the direction of the field. The alignment proceeds sequentially, starting from the center of the vortex and ending at the dot surface. The poling field drastically decreases when increasing the size of the dot in the direction of the applied field.

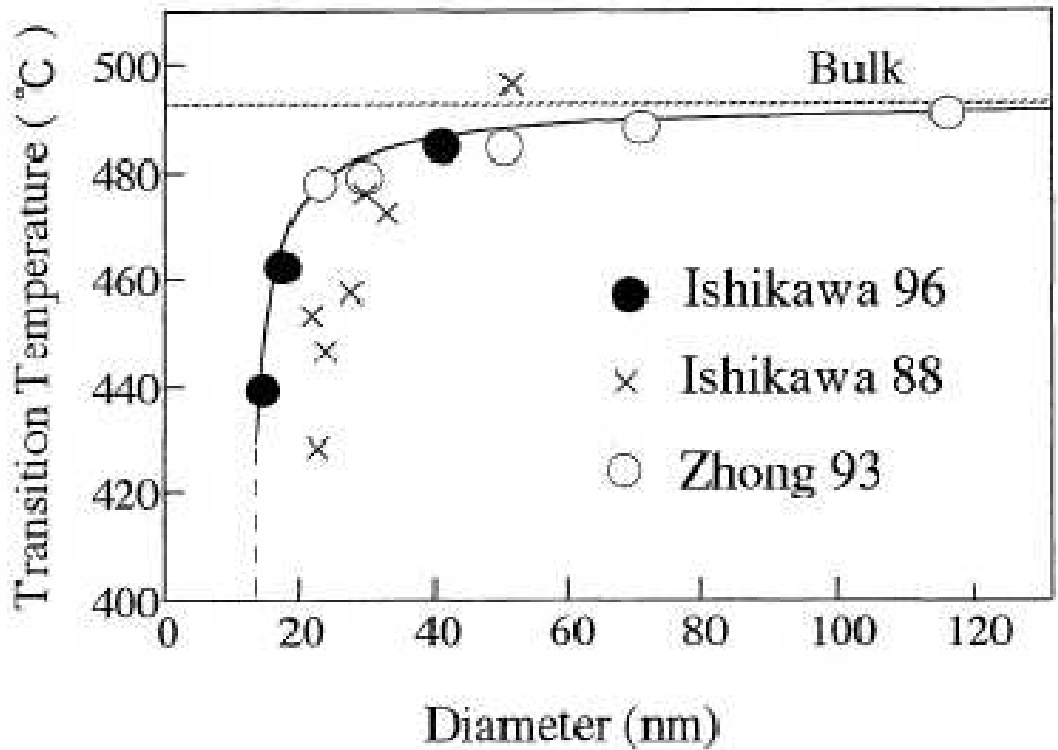

Figure 7.1: Evolution of the paraelectric-to-ferroelectric transition temperature of PbTiO$_3$ ultrafine particles as a function of the particle size. Filled circles correspond to data reported in Ref. [199], crosses to Ref. [198] and open circles to Ref. [200]. The solid line is a fit to the empirical relation T$_c(d) =$ T$_c(\infty) - \frac{C}{d-d_{crit}}$, where T$_c(\infty)$ is the critical temperature of the bulk crystal (dashed line), $d$ stands for the particle size, $d_{crit}$ is the critical diameter below which ferroelectricity disappears, and $C$ is a constant to fit the formula. From Ref. [199].

Although the macroscopic polarization of these quantum dots vanishes at zero applied field, they exhibit a new order parameter, the *toroid moment*, introduced by Naumov *et al.* [36] and defined as

$$\vec{G} = \frac{1}{2N} \sum_i \vec{R}_i \times \vec{p}_i,$$

(7.1)

where $\vec{p}_i$ is the local dipole of cell $i$ located at $\vec{R}_i$, and $N$ is the number of unit cells in the simulation. Since the local dipoles in the vortex can rotate either clockwise or anticlockwise, the toroid moment might adopt two different values pointing in opposite directions, so one bit of information might be stored by assigning one value of the Boolean algebra ("1" or "0") to each of these states.

Naumov and coworkers carried out first-principles based model hamiltonian calculations on Pb(Zr,Ti)O₃ cylindrical nanodisks (oblate form; height smaller than the diameter) and nanorods (prolate form; height larger than the diameter). At low temperature (64 K) and for a given height of 14 five-atoms unit cells, they observed a phase transition from a spin-glass phase (with a vanishing macroscopic dipole and toroid moment although with substantial off-center displacements at each unit cell) to an intermediate non-polar phase with local toroid moments pointing in the $(x, y)$ plane at a critical diameter of 6 unit cells (Fig. 7.3) (the cylindrial $z$ axis is chosen to be along the pseudocubic [001] direction, whith the $x$ and $y$ axis in the [100] and [010] directions, respectively). The toroid moment changes from in-plane to out-of-plane (directed along $z$) at a second critical diameter of 8 unit cells (similar to the vortex-like structure shown on Fig. 7.2). Then, the low temperature toroid moment increases in magnitude as the diameter becomes larger.

Such nanoparticles might potentially present several advantages when used for information data storage. First, it was proposed by Fu and Bellaiche that, in order to switch the toroid moment from one state to the other, a time dependent magnetic field could be applied, instead of an electric field as in thin-film devices. The generation of magnetic fields does not require metallic electrodes, avoiding in this way the problems of the ferroelectric/electrode contacts highlighted in Sec. 4.4. The amplitude of the magnetic field required for switching was theoretically estimated. It is not demonstrated however that such a magnetic switching could be realized in real practical devices. Second, since the nanodisks do not have any net polarization, they do not produce any long-range electric fields, so the "cross-talk" problem is minimized. The toroid state of a single nanoparticle does not modify the state of the neighbors. That means that, third, we can pack the nanodots more compactly, increasing the density for information storage up to five orders of magnitude (60 Tbits/inch²) with respect those ferroelectric or magnetic media currently available. But, on top of this, they are also a challenge from the fundamental point of view, since they represent a possibility to study phase transitions in zero-dimensional systems, where phonons and entropy fluctuations were supposed to conspire to make long-range parallel ordering of dipoles unlikely [381].

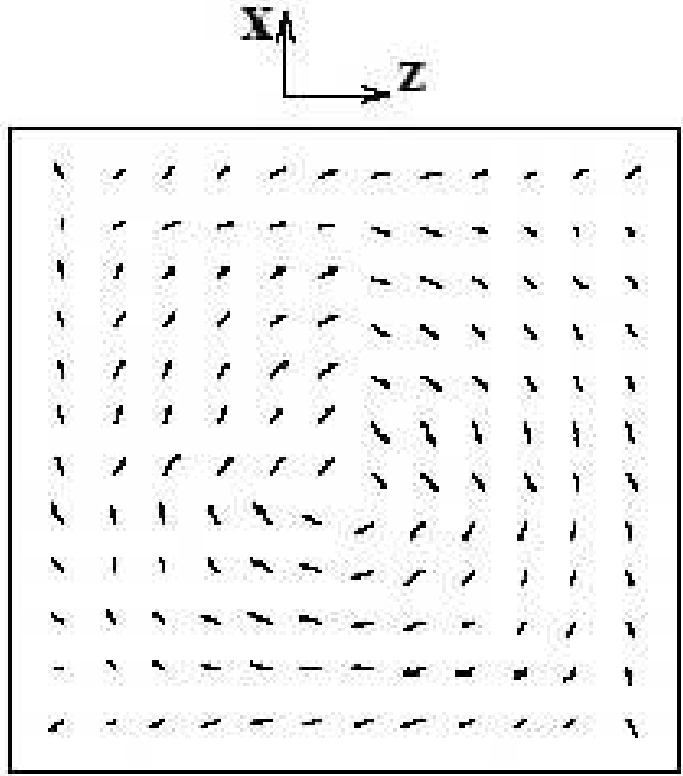

Figure 7.2: "Vortex-like" structure of the local mode displacements of the cells under zero-electric field. The simulated dot was made of 12 five-atoms unit cells along the three direction of space $x$, $y$, and $z$ [$(12 \times 12 \times 12)$ box], and the figure corresponds to a cut along the $y = 6^{th}$ plane. The arrows give the direction of the projected displacement along the $xz$ plane, and the arrow length indicates the projected magnitude. Although the off-center displacements at each unit cell are substantial, the macroscopic polarization vanishes since the local dipoles point in opposite direction at opposite faces. However, the toroid moment of the vortex structure has a well-defined finite value. From Ref. [380].

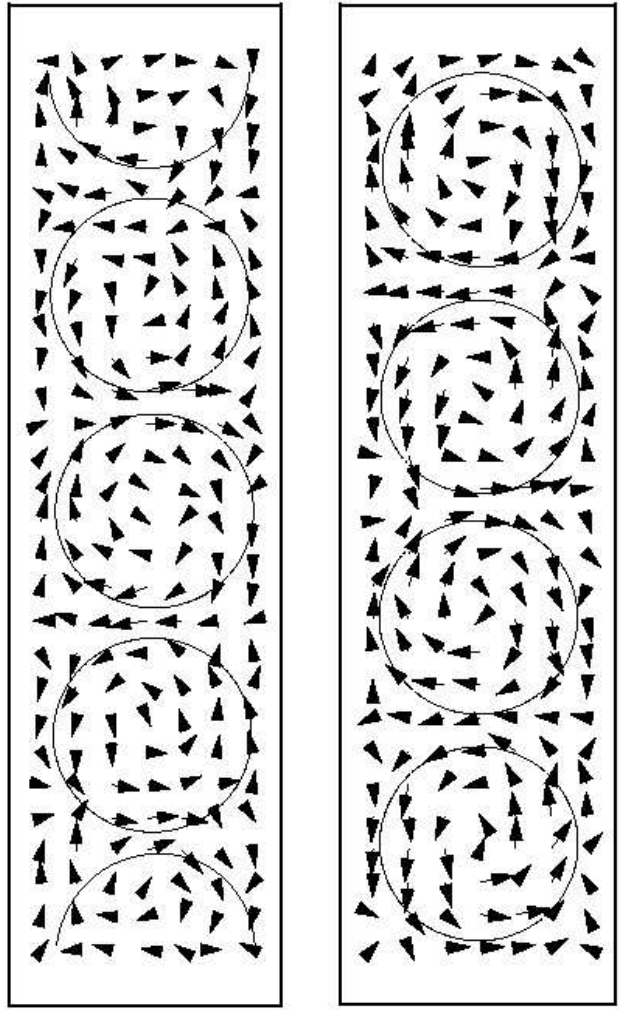

Figure 7.3: Local dipoles in the central $x$ (left) and the $y$ (right) cross section of a nanorod of 7 five-atoms unit cell diameter and 28 five-atoms unit cell high. Substantial off-center occur, although with a null net polarization. Local vortices, pointing in the $xy$ plane, are schematically shown as circles. The global toroid moment tends to cancel. From Ref. [36].

Few other theoretical works also considered ferroelectric nanoparticles. The existence of a critical thickness for small $BaTiO_3$ particles from a model cluster calculation has so been reported by Nakao and coworkers [382]. A $Ba_8Ti_7O_6$ cluster was embedded at the center of a three-dimensional point-charge array, whose size simulated the size of the particle. Analyzing the change in the Mulliken population analysis and the bond-overlap populations as a function of the point charges included in the simulation along the $z$ direction, the authors quantified a critical thickness for ferroelectricity around 12 nm.

The behavior of nanoparticles has also been studied using an atomistic shell-model parametrized from first-principles starting from free standing slabs and reducing progressively tyhe lateral size [155]. For $TiO_2$ symmetrically-terminated free-standing slabs under stress-free boundary conditions, the shell-model predicts that the film breaks into a 180° stripe domain structure below a critical thickness of 36 Å (9 unit cells) to neutralize the depolarizing field in the absence of metallic electrodes [152]. For a 36-Å-thick film, Stachiotti then reduced the lateral dimension, while keeping the same kind of termination for the four lateral faces. For the Ba nanocells (lateral sides terminated in BaO) the total polarization at chains located at the edges or at the faces vanishes, but it recovers the bulk-like behaviour in chains at the center of the dot. The cells breaks up into domains as in bulk, with equal sizes for the up and down regions and the domain wall along the diagonal of the cell. These results suggest that the decrease in the lateral size does not significantly affect the original ferroelectric properties of thin films. For Ti nanocells (lateral faces terminated in $TiO_2$), the chains along the edges and along the faces show a large polarization along $z$, in good agreement with first-principles calculations that showed an enhancement of the in-plane polarization parallel to the surface in $TiO_2$-terminated slabs [207]. This surface effect contributes to increase the polarization inside the nanocell. The systems breaks up into 180 ° domains, but with a different pattern in comparison with the Ba cells. Now, the four lateral sides display a negative polarization, while the nanocell core is polarized positively. Since the $x$, $y$, and $z$ direction are equivalent in a Ti cell the polarization in the core is directed along [111].

It is worth noticing that full DFT calculations on these systems are still missing and might be valuable to confirm on a more rigorous basis the exciting results reported above.

## 7.3   Ferroelectric nanowires.

The growth of ferroelectric nanowires has also been reported recently. $BaTiO_3$ ferroelectric nanowires of diameters ranging from 5 to 60 nm and lengths larger than 10 $\mu m$ have been synthesized by Yun and coworkers [383]. A polarization perpendicular to the nanowire axis was induced and manipulated by an external electric field. Nonvolatile polarization domains, as small as 100 $nm^2$ in size (equivalent to a data storage density of about 1 $Tbit/cm^2$ in a real memory device) were written and reversed repeteadly and independently on these nanowires.

A DFT first-principles study of finite-size effects in $BaTiO_3$ nanowires has also been reported [384], looking explicitly at the dependence of the ferroelectric instability in the direction of the wire axis. Model systems of increasing size were built starting from $BaTiO_3$ clusters that are assemble together into infinite chains along $z$ direction, themselves gathered together into nanowires of diameter $n$ ($n$ corresponds to the number of Ti-O chains combined within the wire). Each of these stoechiometric wire has two BaO surfaces and two $TiO_2$ surfaces (Fig. 7.4).

Due to the low coordination of the atoms at the surface, the relaxed bond lengths for the thinnest wires are much shorter than in bulk. A critical diameter for ferroelectricity along the wire direction was predicted for $n_c \approx 9$. Below $n_c$, the polar distortion along the Ti-O chains is suppressed for the relaxed geometry but can be recovered by applying a tensile strain (Fig. 7.5)

The transition from the paraelectric-to-ferroelectric nanowire is second order with the amplitude of the polar distortions increasing gradually as the equilibrium lattice constant tends to the bulk value (inset of Fig. 7.5). The strong inhomogeneity of the distortions for the different Ti-O chains is in good agreement with the first-principles calculations of $BaTiO_3$ free surfaces, where an enhancement of the polarization for the in-plane polarization is observed for the $TiO_2$ terminated surface, while it is decreased in the BaO-terminated surface [207]. This fact points to a strong influence of the surface on the polarization of the different chains.

The ferroelectric behaviour of the nanowires appears to be monitored by the strong sensivity of ferroelectricity on the unit cell volume [93], itself determined by the low coordination effects at free surfaces.

## 7.4   Perspectives.

At this stage, only few calculations have been reported concerning ferroelectric nanostructures other than two-dimensional systems, mainly because the lack of periodicity make their study computationally very intensive. Their first-principles study constitute however a very exciting challenge for the future since many questions remains concerning their behavior. The interest for ferroelectric nanobjects is still increased by recent experiments that, beyond nanoparticles and nanowires, report the existence of exotic ferroelectric nanostructures such as nanoislands [247] or nanotubes [385, 386, 387].

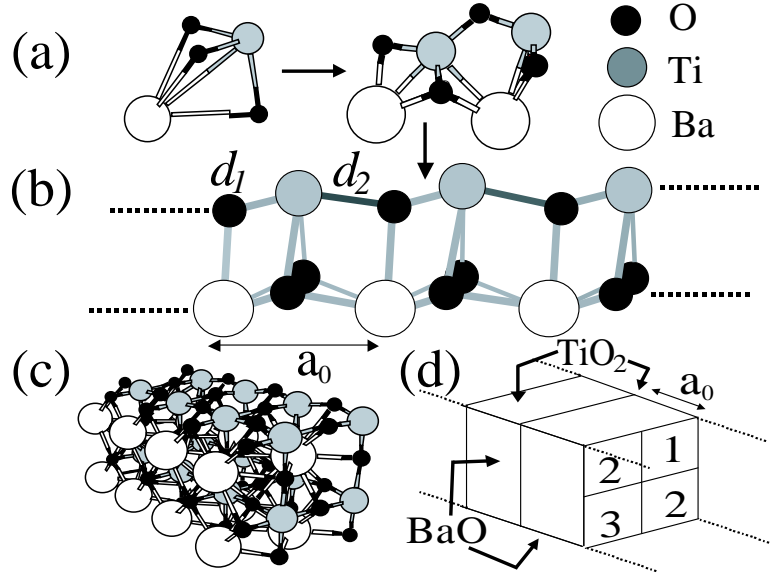

Figure 7.4: BaTiO$_3$ stechiometric wires: BaTiO$_3$ clusters (a) assemble into infinite chains (b), themselves gathered together into nanowires (c) of diameter $n$. Panels (c) and (d) correspond to $n = 4$. From Ref. [384].

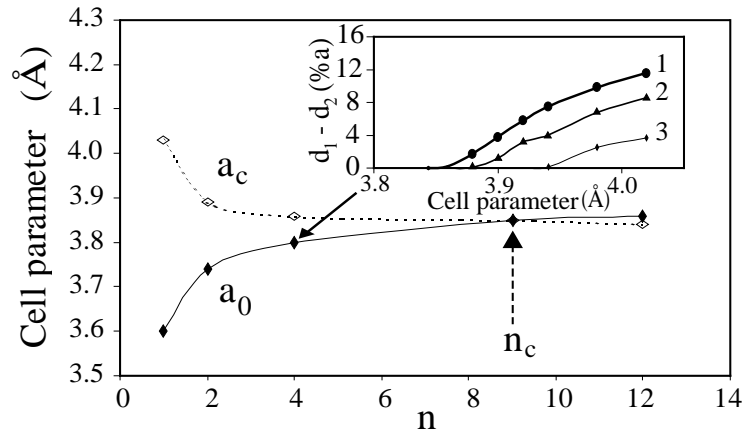

Figure 7.5: Evolution of the equilibrium lattice parameter along $z$ $a_0$, and the critical lattice constant for ferroelectricity $a_c$, as a function of the number of Ti-O chains assembled in the wire $n$. Below a critical diameter $n_c$, $a_0$ is smaller than $a_c$ so the unconstrained wire is paraelectric. Nevertheless, ferroelectricity can be recovered by applying the appropiate tensile strain. The critical diameter for the appearance of ferroelectricity is estimated around $n_c = 9$, where a crossover of $a_0$ and $a_c$ is observed. Inset: local distortions as a function of the cell parameter for a $n = 4$ wire. $d_1$, $d_2$ and the numbers referring to the different chains in the inset are defined in Fig. 7.4. From Ref. [384].

In particular, the recent growth of ferroelectric nanotubes has generated a lot of excitements since these tubes might be suitable for various applications such as pyroelectric detectors, piezoelectric ink-jet printers and memory capacitors. Morrison *et al.* have grown ferroelectric nanotubes of $SrBi_2Ta_2O_9$ (SBT) [385, 386] into a porous Si substrate, with pore diameters ranging between 800 nm and 2 $\mu$m, and a depth of 100 $\mu$m. Using Si with 2 $\mu$m pore diameter as a substrate, a regular array of SBT nanotubes with uniform 200 nm thick tube walls, and a high aspect ratio, defined as the ratio between the depth and the diameter of the tube, were grown. Each individual tube consists in a SBT polycrystalline ceramics with randomly oriented grains, while the walls appear to be single grain thick. Although it was not possible to make electrical contact on the tubes to confirm directly the existence of a polarized state, the ferroelectric character was postulated indirectly. Luo and coworkers [387] have grown $Pb(Zr_{0.48}Ti_{0.52})O_3$ (PZT) and $BaTiO_3$ nanotubes on different silicon and alumina porous templates by first wetting the pore walls using polymeric precursors, and then performing appropriate annealing and selective etching of the porous substrate template. Tubes (see Fig. 7.6) are straight, smooth and with a high aspect ratio. Piezoelectric hysteresis loops for a $Pb(Zr_{0.48}Ti_{0.52})O_3$ tube, with an outer diameter of 700 nm and a wall thickness of 90 nm, showed a sharp switching and a remanent piezoelectric constant of 90 pm/V, comparable with usual values found in PZT thin films. The door is now open to the first-principles modeling of these systems.

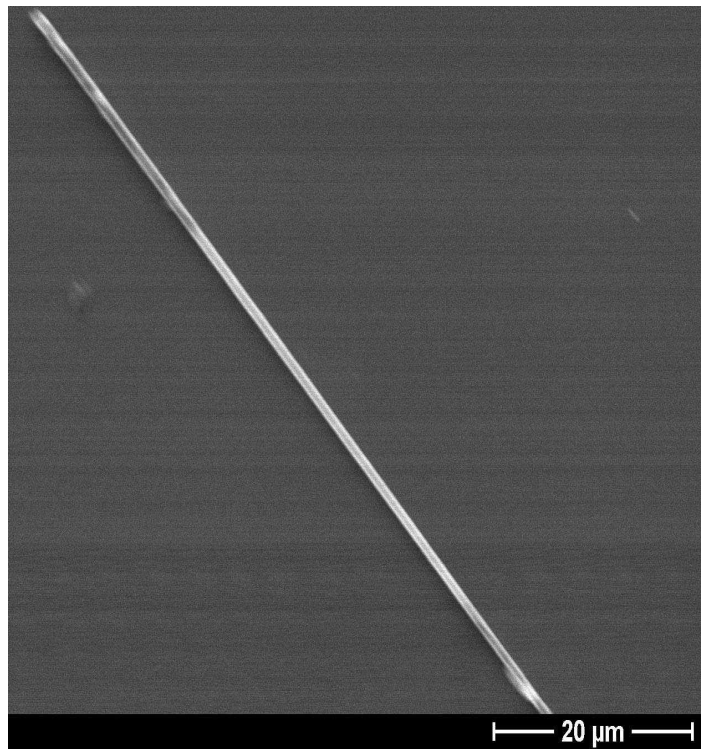

Figure 7.6: Scanning electron microscopy (SEM) image of a single $BaTiO_3$ nanotube grown on a Si substrate. From Ref. [387].

# Conclusions.

In this Chapter, we tried to cover the most important advances that have been reported recently concerning the first-principles modeling of ferroelectric oxides nanostructures. These advances, combined with parallel spectacular experimental achievements in the synthesis, grow and local characterization of almost ideal oxide nanostructures, allowed to improve significantly our fundamental understanding of finite size effects in ferroelectrics.

If significant breakthroughs have been reported, many questions still remains. On the one hand, many systems and geometries are unexplored, and might be investigated in the light of the recent advances. On the other hand, we are at a point where some ferroelectric nanostructures start to be included in practical technological devices. This raises numerous new questions that were not directly discussed in this Chapter. They concern, for instance, the mobility of domains and the dynamics of switching, the role of defects such of oxygen vacancies or misfit dislocations, the consequences of finite electrical conductivity. If these questions have started to be investigated experimentally, most of them were not yet addressed at the first-principles level since they are beyond the scope of present computational capabilities. To some extent, we can hope that the recurrent increase of the computational power will allow to address more complex questions and systems in the future. However, a great challenge is now in the development of new multiscale techniques, that in the spirit of the effective Hamiltonian approach, encapsulate the essential physics and allow to tackle such new complex problems.

For sure, the first-principles modeling of ferroelectric oxide nanostructures is and will remain in the future a very exciting topic for both fundamental and applied researches.

# List of Tables

# List of Figures